\documentclass[11pt,a4paper]{article}
\usepackage{jheppub}
\usepackage[T1]{fontenc}
\usepackage[export]{adjustbox}
\setkeys{Gin}{max width=\linewidth}
\usepackage[export]{adjustbox}
\setkeys{Gin}{max width=\linewidth}
\usepackage{tabularx}
\usepackage{enumitem}

\usepackage{dcolumn}
\usepackage{bm}
\usepackage{booktabs}
\usepackage{makecell}
\usepackage{makebox}
\usepackage{multirow}
\usepackage[figuresright]{rotating}
\usepackage{amstext}
\usepackage{xfrac}
\usepackage{xcolor}

\definecolor{darkblue}{RGB}{0,0,150}
\definecolor{darkred}{RGB}{150,0,0}

\usepackage{amsfonts}
\usepackage{longtable, booktabs}
\usepackage{url}
\usepackage{subfigure}
\usepackage{dsfont}
\usepackage{amsbsy}
\usepackage{amsthm}
\usepackage{esint}
\usepackage{cleveref}
\usepackage{mathrsfs}
\usepackage{color}
\usepackage{extarrows}
\usepackage{datetime}
\usepackage{comment}
\usepackage[super]{nth}
\usepackage{tikz-cd}
\usepackage{float}
\usepackage{braket}
\usepackage{scalerel}
\usepackage{tikz}
\usetikzlibrary{svg.path}

\definecolor{orcidlogocol}{HTML}{A6CE39}
\tikzset{
	orcidlogo/.pic={
		\fill[orcidlogocol] svg{M256, 128c0, 70.7-57.3, 128-128, 128C57.3, 256, 0, 198.7, 0, 128C0, 57.3, 57.3, 0, 128, 0C198.7, 0, 256, 57.3, 256, 128z};
		\fill[white] svg{M86.3, 186.2H70.9V79.1h15.4v48.4V186.2z}
		svg{M108.9, 79.1h41.6c39.6, 0, 57, 28.3, 57, 53.6c0, 27.5-21.5, 53.6-56.8, 53.6h-41.8V79.1z M124.3, 172.4h24.5c34.9, 0, 42.9-26.5, 42.9-39.7c0-21.5-13.7-39.7-43.7-39.7h-23.7V172.4z}
		svg{M88.7, 56.8c0, 5.5-4.5, 10.1-10.1, 10.1c-5.6, 0-10.1-4.6-10.1-10.1c0-5.6, 4.5-10.1, 10.1-10.1C84.2, 46.7, 88.7, 51.3, 88.7, 56.8z};
	}
}

\newcommand\orcidicon[1]{\href{https://orcid.org/#1}{\mbox{\scalerel*{
				\begin{tikzpicture}[yscale=-1, transform shape]
					\pic{orcidlogo};
				\end{tikzpicture}
			}{|}}}}
\tikzset{bold/.style={color=blue,  line width=2pt}}
\tikzset{redop/.style={circle, fill=red}}
\tikzset{blueop/.style={circle, fill=blue}}

\newtheorem{theorem}{Theorem}

\newtheorem{definition}{Definition}[section]

\makeatletter
\AddToHook{package/ltcaption/before}{%
	\expandafter\let\csname longtable\endcsname\undefined
	\expandafter\let\csname endlongtable\endcsname\undefined
}
\makeatother
 \title{Microscopic Constructions of the $BF+AAB$ Topological Field Theory
and Borromean-Rings Braiding in $(3+1)$ Dimensions}
 \author[a]{Yizhou Huang\orcidicon{0009-0008-4080-8174}$^{1}$\note[1]{These authors contributed equally.}}
\author[b,c]{Zhi-Feng Zhang\orcidicon{0009-0005-5267-4108}$^{1}$}
\author[d]{Qing-Rui Wang\orcidicon{0000-0002-4850-8511}$^{2}$\note[2]{Corresponding authors.}}
\author[a]{Peng Ye\orcidicon{0000-0002-6251-677X}$^{2}$}
\affiliation[a]{School of Physics, Sun Yat-sen University, Guangzhou, 510275, China}
\affiliation[b]{Max Planck Institute for the Physics of Complex Systems, N\"othnitzer Stra{\ss}e~38, Dresden~01187, Germany}
\affiliation[c]{Department of Physics, The Hong Kong University of Science and Technology, Clear Water Bay, Kowloon, Hong Kong, China}
\affiliation[d]{Yau Mathematical Sciences Center, Tsinghua University, Haidian, Beijing, 100084, China}
\emailAdd{huangyzh66@mail2.sysu.edu.cn}
\emailAdd{zfzhang@pks.mpg.de}
\emailAdd{wangqr@mail.tsinghua.edu.cn}
\emailAdd{yepeng5@mail.sysu.edu.cn}
\abstract{
Topological order was first studied in fractional quantum Hall states, where anyons reflect electron fractionalization. Recent advances in continuum topological field theories have revealed braiding, fusion, shrinking, and diagrammatic consistency relations in higher dimensions [e.g., \href{https://doi.org/10.1007/JHEP11(2023)210}{Huang, Zhang, Ye, JHEP 11 (2023) 210}; \href{https://doi.org/10.1007/JHEP06(2025)238}{JHEP 06 (2025) 238}].
As the next step of research, these advances call for explicit microscopic lattice constructions that faithfully reproduce the structures and processes described by continuum field theories.
Here we construct, step by step, microscopic counterparts of the $(3+1)$-dimensional $BF$ field theory with an $AAB$ twist and gauge group $(\mathbb{Z}_2)^3$. 
This field theory was introduced in our earlier work on Borromean-Rings braiding [\href{https://doi.org/10.1103/PhysRevLett.121.061601}{Chan, Ye, Ryu, Phys. Rev. Lett. 121, 061601 (2018)}].
We start with the construction of string and thickened membrane operators that create, fuse, shrink, and braid particle and loop excitations, providing microscopic counterparts of Wilson operators in continuum field theories. We derive fusion and shrinking rules and the operators governing particle--loop and Borromean-Rings braiding. We demonstrate selective control of non-Abelian shrinking channels through the internal degrees of freedom of loop operators and establish fusion--shrinking consistency for arbitrary finite groups. An explicit correspondence between all 22 excitations of the $\mathbb{D}_4$ quantum double lattice model and the $BF+AAB$ theory preserves fusion and shrinking rules and particle--loop and Borromean-Rings braiding phases.
These constructions place the continuum results on a microscopic lattice footing with tensor-product local Hilbert spaces and short-range interactions.
}
\compress
\hypersetup{pdftitle={Bridging Continuum Topological Field Theory and Microscopic Constructions for Three-Dimensional Non-Abelian Topological Order},pdfauthor={Yizhou Huang, Zhi-Feng Zhang, Qing-Rui Wang, Peng Ye}}
\begin{document}
\maketitle
\flushbottom

\section{Introduction}
 
\subsection{Background, motivation, and main results}

Since the discovery of the fractional quantum Hall effect (FQH), 
2D\footnote{In this paper, ``$n$D'' refers to $n$-dimensional real space where quantum states or many-body systems are supported, e.g., $3$D ground state, $3$D topological order. ``$(n+1)$D'' specifically refers to $(n+1)$-dimensional spacetime with $n$-dimensional real space.} 
topological orders characterized by long-range entanglement~\cite{zeng2018quantum} have attracted sustained attention. 
Their universal topological data---including fusion rules and braiding statistics---admit a unified description in terms of continuum topological quantum field theories, most notably Chern--Simons field theories. 
For Abelian FQH states, multicomponent $U(1)$ Chern--Simons theories with integer-valued $K$ matrices encode the anyon content, fusion rules, braiding statistics, and quantized Hall responses. 
Non-Abelian FQH states exhibit an even richer field-theoretical structure, formulated in terms of non-Abelian Chern--Simons theories or, equivalently, chiral conformal field theories via bulk--edge correspondence. 
Prominent examples include the $\nu=5/2$ Moore--Read state described by an $SU(2)_2$ (Ising) topological field theory, the $\nu=12/5$ Read--Rezayi state associated with an $SU(2)_3$ theory, and more general parafermionic and higher-rank Chern--Simons theories~\cite{Witten1989, Turaev2016, PhysRevB.42.8133, sarma_08_TQC, wen2004quantum, KITAEV20032, KITAEV20062, PhysRevB.78.115421, PhysRevB.88.115133, simon2023topological}. 
In these cases, the bulk topological field theory captures universal properties such as ground-state degeneracy, non-Abelian braiding, and knot and link invariants. 
Together with microscopic wave functions and many-body Hamiltonians, these field-theoretical descriptions provide a coherent understanding of topological order across length scales.

Despite this progress, establishing a precise and systematic correspondence between microscopic constructions\footnote{The term ``{\textit{microscopic construction}}'' in this paper refers to microscopic model studies at short distances that directly describe the constituents, e.g., electrons or local spins, of many-body systems. Microscopic constructions can proceed through either continuum models or lattice models. For example, FQH systems at short distances are typically continuum systems \textit{that do not involve lattices at all}. By contrast, Kitaev's toric code lattice model microscopically realizes $\mathbb{Z}_2$ topological order. At long wavelengths, quantum many-body systems are treated as novel quantum liquids described by continuum quantum field variables. Fracton order is an exception: it is a non-liquid state described by unconventional types of field theories; see, e.g., Refs.~\cite{PhysRevB.92.235136, zeng2018quantum, PhysRevB.105.195124, PhysRevLett.126.101603, li2026infinitecomponentbffieldtheory, PhysRevB.110.205108}.} 
and continuum field theories remains a central challenge. 
This challenge is especially acute for non-Abelian phases, where the underlying topological structure is exceptionally rich and potentially relevant for topological quantum computation. 
Typically, the long-distance field variables, such as dynamical gauge fields of various differential forms, are sufficiently far removed from microscopic degrees of freedom, such as electrons or local spins, that \textit{identifying their microscopic origins is highly nontrivial}. 
Moreover, the guiding principles in the two descriptions are substantially different. 
In contrast to Landau's Fermi liquid theory for weakly interacting electron gases~\cite{shankarRMP1994}, the continuum field variables of topological order are not derived directly from microscopic constituent operators through perturbative renormalization-group analysis. 
Rather, they are introduced through general principles such as gauge invariance, locality, symmetry, expected response, and topological invariance. 
Because continuum field theories of topological order are often constructed without explicit reference to microscopic Hamiltonians, their physical relevance and microscopic realizability are sometimes questioned, especially from a condensed-matter perspective. 
\textit{Therefore, bridging microscopic constructions and continuum field theories in a systematic and step-by-step manner remains a crucial task for understanding topological phases of matter across length scales.}

While this task is already challenging in two dimensions, for example in identifying effective field theories of FQH plateaus, it remains far less explored in three dimensions. 
In higher spatial dimensions, the landscape of topological order broadens substantially: in $3$D and beyond, point particles are restricted to bosons or fermions~\cite{Leinaas1977, Wilczek1982, PhysRevLett.49.957, PhysRevLett.52.2103}, \textit{but} spatially extended objects such as loops and membranes~\cite{PhysRevX.8.021074, PhysRevX.9.021005} become legitimate topological excitations. 
In this work, we focus on regular unknotted loops; nevertheless, loop excitations can in general carry more complicated knot or link structures~\cite{10.21468/SciPostPhys.14.6.141, PhysRevB.97.085147, doi:10.1142/2324, doi:10.1142/4256, hatcher_algebraic_2001, link_groups_Milnor}, and can exhibit nontrivial statistics themselves~\cite{6k88-w52n, PhysRevB.106.165135, 10.21468/SciPostPhys.14.5.089}.

Recently, $(3+1)$D $BF$ theories~\cite{HANSSON2004497, Horowitz:1989aa, Horowitz:1990aa} have provided a powerful long-distance framework for $3$D topological orders, including gauged symmetry-protected-topological (SPT) responses~\cite{PhysRevB.94.115104, 10.21468/SciPostPhys.14.2.023, PhysRevB.89.045127, PUTROV2017254, PhysRevB.99.235137, PhysRevB.93.205157, PhysRevB.97.085147, PhysRevLett.121.061601, PhysRevResearch.3.023132, bti2, PhysRevB.88.235109}. 
Their actions contain one or more $BF$ topological terms, schematically $B dA$ with $B$ a $2$-form and $A$ a $1$-form gauge field\footnote{The wedge product symbol is omitted for notational convenience. The single-component $BF$ term $\int\frac{k}{2\pi} BdA$ ($k\in\mathbb{Z}$) describes untwisted $\mathbb{Z}_k$ topological order, or equivalently the deconfined phase of $\mathbb{Z}_k$ gauge theory, in which expectation values of Wilson surfaces and loops are $\mathbb{Z}_k$-valued. For this reason, unless otherwise specified, it is customary to say that the gauge group of the $BF$ term is $\mathbb{Z}_k$.\label{footnote_gaugegroup}}, supplemented by various \textit{twisted} topological terms, e.g., $A A dA$, $A A A A$, $A A B$, $dA dA$, and $B B$. 
These twisted terms generate a broad range of three-dimensional topological phenomena, including particle--loop braiding~\cite{PhysRevLett.62.1071, PhysRevLett.62.1221, HANSSON2004497, ALFORD1992251, PRESKILL199050}, multi-loop braiding~\cite{PUTROV2017254, PhysRevLett.113.080403, PhysRevLett.114.031601, PhysRevB.91.035134, PhysRevX.4.041043, PhysRevX.4.031048, PhysRevX.6.021015, PhysRevB.95.245124, kapustin2014, PhysRevB.94.045113}, particle--loop--loop braiding (i.e., \textbf{\textit{Borromean-Rings braiding}})~\cite{PhysRevLett.121.061601}, emergent fermionic statistics~\cite{PhysRevB.99.235137, bti2, Kapustin:2014gua, PhysRevResearch.5.043111}, and nontrivial topological responses~\cite{PhysRevB.94.115104, PhysRevB.89.045127, PhysRevB.88.235109, RevModPhys.83.1057, PhysRevB.95.035149, PhysRevB.96.085125, RevModPhys.88.035001, PhysRevB.99.205120}. 
Within this framework, fractionalization of global symmetry on loop excitations has also been analyzed in Refs.~\cite{PhysRevB.105.205137, PhysRevB.94.245120, PhysRevB.97.125127}, where mixed three-loop braiding formed by external symmetry fluxes and internal loop excitations plays a key role in classifying higher-dimensional symmetry-enriched topological (SET) phases. 
However, not all braiding processes can consistently coexist in a topological order, since certain sets of twisted terms violate gauge invariance. 
By excluding these gauge-inconsistent possibilities, Ref.~\cite{PhysRevResearch.3.023132} enumerates the allowed braiding structures.

In addition to braiding, twisted $BF$ theories reveal a class of topological data absent in two-dimensional topological orders: \textbf{\emph{shrinking rules}}~\cite{Zhang2023fusion}. 
These rules describe how spatially extended excitations can be continuously contracted into lower-dimensional ones. 
Given an $n$-dimensional excitation, for example, one may shrink it into several $k$-dimensional excitations with $k<n$; in a $3$D topological order, a loop excitation can shrink to a point-like particle. 
In our previous work~\cite{Zhang2023fusion}, such shrinking processes were systematically analyzed within $(3+1)$D twisted $BF$ theories, where certain twisted terms were shown to induce non-Abelian fusion and shrinking rules even for Abelian discrete gauge groups (see footnote~\ref{footnote_gaugegroup}).

The above physics of ``braiding$\rightarrow$fusion$\rightarrow$shrinking'' becomes even richer in $4$D topological orders, where topological excitations include particles, loops, and membranes~\cite{10.21468/SciPostPhys.13.3.068, Johnson-Freyd:2022aa, 10.21468/SciPostPhys.15.1.001, PhysRevD.100.085012, PhysRevB.103.064426, 4jww-6b6t}. 
Within $(4+1)$D twisted $BF$ theories~\cite{Zhang:2021ycl, Huang2023}, there are two types of $BF$-type topological terms, $C dA$ and $\tilde B dB$, where $A$ is a $1$-form, $C$ is a $3$-form, and $B$ and $\tilde B$ are distinct $2$-form gauge fields. 
Corresponding twisted terms include $A A A A A$, $A A A dA$, $A dA dA$, $A A C$, $A A A B$, $B B A$, $A dA B$, $A A dB$, and $B C$. 
It was shown in Ref.~\cite{Huang2023} that $B B A$, $A A A B$, $A A C$, $A A A A A$, and $A A A dA$ can lead to non-Abelian shrinking. 
Furthermore, $B B A$ and $A A A B$ produce \emph{hierarchical} shrinking, in which a membrane shrinks first to loops and the resulting loops subsequently shrink to particles.

Importantly, fusion and shrinking are not independent operations. 
In $(3+1)$D twisted $BF$ theories, the derived fusion and shrinking rules obey a consistency condition: fusing first and then shrinking gives the same outcome as shrinking first and then fusing. 
This \textit{\textbf{fusion--shrinking consistency condition}}~\cite{Zhang2023fusion, Huang2023} strongly constrains the fusion and shrinking coefficients, and provides an organizing principle for higher-dimensional topological data. 
In $4$D topological orders, hierarchical shrinking remains compatible with fusion, although shrinking and fusion need not commute at each intermediate step. 
These consistency structures motivate \textbf{\emph{diagrammatic representations}}~\cite{Huang_2025} of fusion and hierarchical shrinking processes, in which algebraic constraints such as \textit{pentagon relations for fusion} and \textit{(hierarchical) hexagon relations for shrinking--fusion} are naturally encoded. 
They point toward underlying higher-categorical structures, extending the well-established diagrammatic description of fusion and braiding in two-dimensional anyon theories; see, e.g., Refs.~\cite{KITAEV20062, simon2023topological}.

While these $(3+1)$D field-theoretical results established over the past years have provided a coherent and predictive continuum framework, their microscopic realization remains largely unexplored. A central question is whether such continuum descriptions admit faithful three-dimensional microscopic counterparts equipped with tensor-product local Hilbert spaces and local interactions. 
\textit{Resolving this question requires more than identifying suitable microscopic degrees of freedom: one must reproduce, at the level of explicit lattice operators, the excitation content, fusion rules, shrinking rules, braiding data, and fusion--shrinking consistency relations predicted by continuum field theory.}

Along this line, several concrete issues lie at the core of this correspondence between microscopic constructions and long-distance topological field theories. 
First, continuum field theory identifies topologically distinct excitations through equivalence relations among gauge-invariant Wilson operators, including \textbf{\textit{particles}}, \textbf{\textit{pure loops}}, and \textbf{\textit{decorated loops}}. 
A microscopic construction must therefore provide creation operators for all such excitations and establish the corresponding equivalence relations directly at the microscopic level. 
Second, the fusion, shrinking, and braiding of loop-like excitations encode intrinsically three-dimensional topological phenomena. 
It is essential to realize these operations microscopically and to determine whether the allowed shrinking channels predicted by field theory are complete, redundant, or controllable. 
Third, the diagrammatic framework derived from continuum field theory imposes fusion pentagon relations and shrinking--fusion hexagon relations as consistency constraints on anomaly-free topological data~\cite{Huang2023, Huang_2025, Zhang2023fusion}. 
Whether these constraints admit a concrete microscopic interpretation in exactly solvable three-dimensional lattice models remains an open question.

Finally, and \textit{most importantly for the present work}, the $(3+1)$D $BF$ field theory with an $AAB$ twist was introduced in Ref.~\cite{PhysRevLett.121.061601} in connection with Borromean-Rings braiding among a particle and two loop excitations. 
However, a microscopic construction of this field theory and the associated non-Abelian topological order has remained elusive. 
This has left open a basic question since its proposal in 2018: can the $BF{+}AAB$ field theory be correctly realized by a local lattice model with a tensor-product local Hilbert space and local interactions?

\begin{figure}
	\centering
	\includegraphics[width=0.90\linewidth, keepaspectratio]{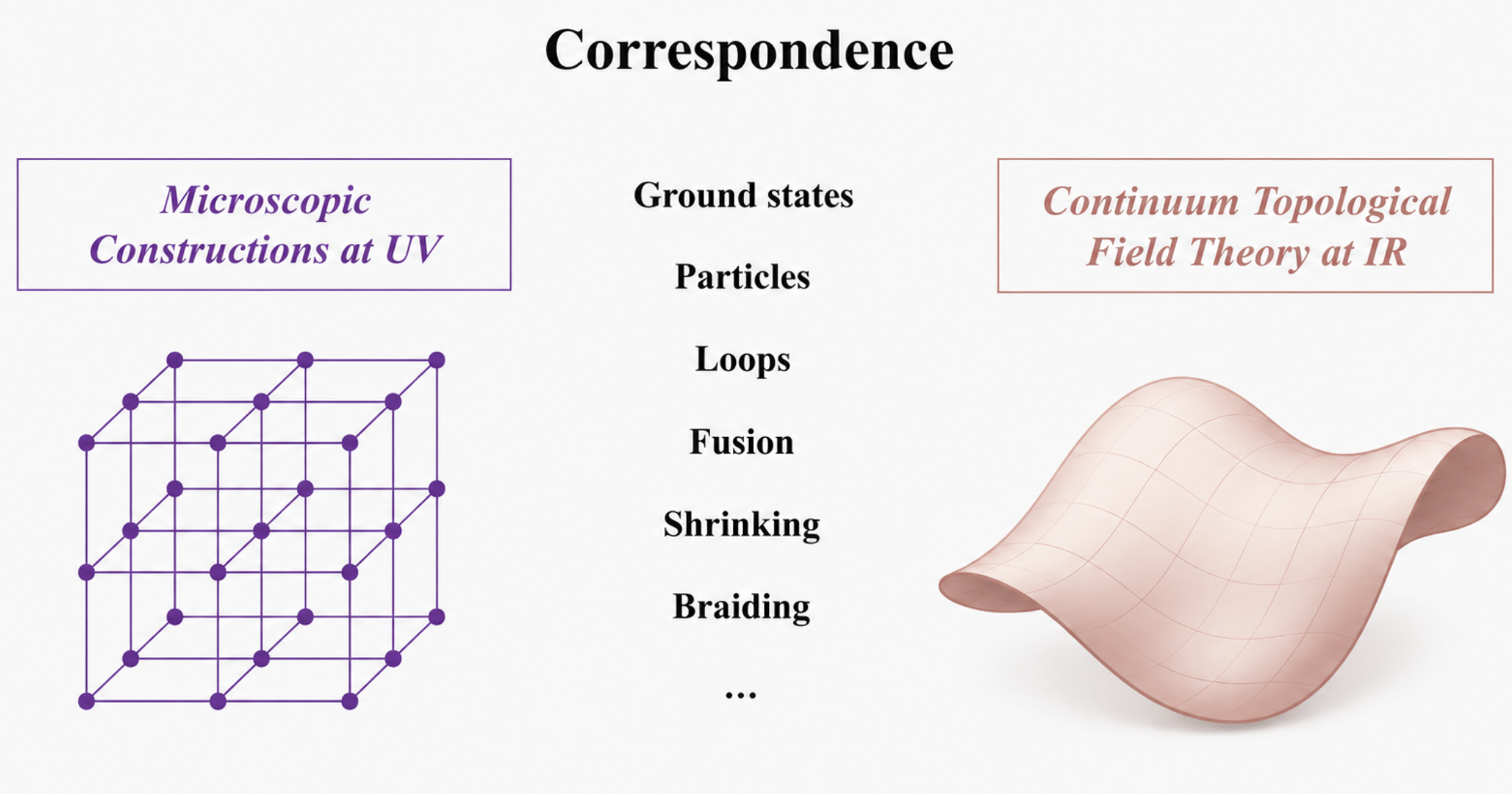}
	\caption{
\textbf{Schematic correspondence between microscopic constructions at ultraviolet scales and continuum topological field theory at infrared scales.} The lattice on the left represents the microscopic degrees of freedom, while the smooth surface on the right is a pictorial representation of the continuum manifold underlying the field-theoretical description. The central list summarizes the matched topological data, including ground states, particles, loops, fusion, shrinking, and braiding.}
	\label{fig_correspond}
\end{figure}

\textit{In this work}, we develop a microscopic construction that realizes all results derived from continuum field theories for three-dimensional non-Abelian topological orders, as illustrated in Fig.~\ref{fig_correspond}. 
We construct ground states, characterize all excitations,\footnote{Throughout this paper, by an ``excitation'' we mean a \emph{simple} (elementary) topological excitation, i.e., a single superselection sector that cannot be decomposed by local operations into several independently movable excitations. In the quantum double model, these are precisely the excitations labeled by one irreducible representation $(C,R)$ of the quantum double algebra $DG$, which are the objects created by the string and membrane operators constructed in Sec.~\ref{Sec.3}. Composites and condensation descendants (such as the Cheshire string of the 3D toric code) are not elementary in this sense and are not considered in this work.} derive fusion and shrinking rules, and compute particle--loop and Borromean-Rings braiding. 
Non-Abelian shrinking channels are shown to be controllable via the internal degrees of freedom of loop operators, and lattice shrinking rules are fully consistent with fusion rules, thus implementing the fusion--shrinking consistency conditions at the operator level. 
Most importantly, by establishing an explicit isomorphism between the excitations of the three-dimensional $\mathbb{D}_4$ quantum double model and those of the $(3+1)$D $BF$ field theory with an $AAB$ twist and gauge group $G=(\mathbb{Z}_2)^3$~\cite{PhysRevLett.121.061601}, we demonstrate that the former provides a concrete microscopic realization of the latter, bridging long-distance topological field theory and exactly solvable lattice models. 

The main body of the paper is organized into five parts. The first four parts develop the microscopic topological data of the lattice model: excitations, fusion, shrinking, and braiding. Together, these results provide the basis for the microscopic construction of the $(3+1)$D $BF$ field theory with an $AAB$ twist presented in the fifth part:  
\begin{enumerate}[label=\roman*.]
\item  
\textbf{\textit{Microscopic construction of particle and loop excitations.} }
In $3$D topological orders, topological excitations include both particles and loops, which are represented in continuum topological field theory by Wilson operators, such as Wilson loops and Wilson surfaces~\cite{Zhang:2021ycl, PhysRevResearch.5.043111, Zhang2023fusion, Huang2023}. 
In the microscopic setting of the $3$D quantum double model with a finite group $G$,  creation operators for particles are constructed as string-like operators. Creation operators for loop excitations are slightly thickened open membrane operators consisting of both a \textit{direct part} and a \textit{dual part}. These operators   carry the conserved topological charge and flux, which are invariant under local operations.

Starting from four elementary edge operators $T_g^{\pm}$ and $L_g^{\pm}$ defined in Eqs.~(\ref{eq_t}) and~(\ref{eq_l}), we construct the cylindrical thickened open membrane operators 
$W_M\left( C, R;c, j;c^{\prime}, j^{\prime} \right)$ in Eq.~(\ref{eq_loop}). 
Here $C$ is a conjugacy class of $G$, and $R$ is an irreducible representation of the centralizer of a representative element in $C$. 
The pair $(C,R)$ gives the topological label of the excitation, while $c,c^{\prime}\in C$ and $j,j^{\prime}=1,\ldots,n_R$ label internal degrees of freedom that can be changed by local operations. 
Accordingly, applying $W_M\left( C, R;c, j;c^{\prime}, j^{\prime} \right)$ to a ground state creates an excitation--antiexcitation pair on the two boundary loops of the cylindrical membrane $M$, with the topological type determined by $(C,R)$. When the conjugacy class $C$ is nontrivial (i.e., $C$ does not contain the identity $e$), the operators 
$W_M\left( C, R;c, j;c^{\prime}, j^{\prime} \right)$ correspond to loop excitations. When the conjugacy class is trivial (i.e., $C$ contains the identity $e$), the centralizer of $e$ is $G$ itself, and thus $R$ becomes an irreducible representation of the group $G$. The operators $W_M\left( C, R;c, j;c^{\prime}, j^{\prime} \right)$ reduce to open string operators $W_L\left( R;j,j^{\prime} \right)$ for particles as shown in Eq.~(\ref{eq_particle}).

This construction provides a microscopic counterpart of the Wilson operators in continuum field theory and forms the foundation for comparing fusion, shrinking, and braiding data in the subsequent sections.

\item   
\textbf{\textit{Microscopic fusion rules and quantum dimensions.}} 
In continuum topological field theory, fusion rules are encoded in correlation functions of Wilson operators~\cite{PhysRevResearch.5.043111, Zhang2023fusion, Huang2023}. 
In the $3$D quantum double model, by contrast, fusion is governed by the representation theory of the quantum double algebra $DG$. 
Although the computational languages are different, the physical operation is the same: two excitations are brought together and the possible resulting topological charges are measured, as illustrated in Fig.~\ref{fig_fusion}.

Using the excitation operators constructed above, we associate each excitation type with a local Hilbert space $V_{(C,R)}$, which serves simultaneously as a representation space of an irreducible representation of $DG$. 
The fusion of two excitations $[C_1,R_1]$ and $[C_2,R_2]$ is then obtained by decomposing the tensor-product representation space
$V_{(C_1,R_1)}\otimes V_{(C_2,R_2)}$ 
into irreducible components, as summarized in Eq.~(\ref{eq_fusion_rules}). 
This provides a direct microscopic realization of the fusion algebra expected from the long-distance theory. 
As concrete illustrations, we work out all fusion rules for $G=\mathbb{Z}_N$ in Eq.~(\ref{eq_ZN_fusion}) and for $G=\mathbb{D}_3$, the dihedral group of order $6$, in Table~\ref{tab_D3_fusion}. 
From the resulting fusion matrices, we further compute the quantum dimensions of the excitations. 
For Abelian groups, all excitations have quantum dimension one, whereas non-Abelian groups support excitations with quantum dimensions greater than one; the full set of quantum dimensions for $G=\mathbb{D}_3$ is listed in Table~\ref{tab_quantum_dimension}.

 \item  
\textbf{\textit{Microscopic shrinking rules and fusion--shrinking consistency.}}
The shrinking processes studied in this paper were first proposed and analyzed in continuum topological field theory, where shrinking a loop excitation means continuously contracting the spatial support of its Wilson operator~\cite{Zhang2023fusion, Huang2023}. 
On the lattice, we realize this process using thickened cylindrical membrane operators that create pairs of loop excitations on their boundary loops. 
By applying the connecting rules of these membrane operators, a thickened cylindrical membrane can be shrunk to an infinitesimally thin membrane, which effectively becomes a string operator whose endpoints carry particle excitations, as illustrated in Fig.~\ref{fig_shrinking}. 
This gives a microscopic counterpart of the continuum shrinking process.

For a loop excitation labeled by $[C,R]$, shrinking maps the associated local Hilbert space $V_{(C,R)}$ to a space $\mathcal{S}\!\left(V_{(C,R)}\right)$ that is spanned by particle sectors. Thus, the resulting shrinking rules are obtained by decomposing this space, as summarized in Eq.~(\ref{eq_DG_shrinking}). 
Equivalently, a particle $[C_e,R^{\prime}]$, where $C_e$ is the conjugacy class of the identity element, appears after shrinking $[C,R]$ when the representation $R^{\prime}$ restricts to $R$ on the centralizer of a representative element in $C$; the corresponding multiplicity is given in Eq.~(\ref{eq_shrinking_multiplicity}). 
We work out the shrinking rules explicitly for $G=\mathbb{Z}_N$ in Eq.~(\ref{eq_Abelian_shrinking}) and for $G=\mathbb{D}_3$ in Table~\ref{tab_shrinking_d3}. 
\textit{For non-Abelian $G$, even loops with trivial charge label $R$ can shrink into nontrivial particle charges, providing a microscopic realization of non-Abelian shrinking.} 
Moreover, by choosing suitable internal degrees of freedom of the membrane operator, we show that individual shrinking channels can be selectively controlled.

Finally, we verify that the microscopic fusion and shrinking rules are not independent: they satisfy the fusion--shrinking consistency condition in Eq.~(\ref{eq_consistent_4D}), previously derived from continuum field theory~\cite{Zhang2023fusion, Huang2023} and organized diagrammatically in Ref.~\cite{Huang_2025}. 
This establishes fusion--shrinking consistency as a concrete lattice-level constraint on the topological data of three-dimensional quantum double models.

 \item  
\textbf{\textit{Microscopic particle--loop and Borromean-Rings braiding.}} 
In continuum topological field theory, braiding statistics is encoded in correlation functions of gauge-invariant Wilson operators supported on nontrivially linked spacetime trajectories~\cite{PhysRevLett.121.061601, PhysRevResearch.3.023132, Zhang:2021ycl}. 
In $3$D topological orders, the presence of loop excitations enables braiding processes beyond the particle--particle braiding familiar from two dimensions, including particle--loop braiding and the more exotic Borromean-Rings braiding. 
We realize these processes microscopically by intertwining particle string operators with thickened cylindrical membrane operators for loops. 

For particle--loop braiding shown in Fig.~\ref{fig_particle_loop_braiding}, dragging a particle $[R^{\prime}]$ around a loop $[C,R]$ produces a braiding-induced operator 
$O_{\left(R^{\prime}\right), \left(C, R\right)}^{\mathrm{PL}}$, derived in Eq.~(\ref{eq_operator_O}).  When either the charge carried by the particle or the flux carried by the loop is Abelian, this operator reduces to a braiding phase $\Gamma^{R^{\prime\ast}}_{ii}\left( c \right)$, as shown in \textbf{Theorem}~\ref{theorem_plphase} and Eq.~(\ref{eq_particle_loop_braiding_phase}). 
When both the charge carried by the particle and the flux carried by the loop are non-Abelian, however, the same braiding process generally acts nontrivially on their internal degrees of freedom.

For Borromean-Rings braiding shown in Fig.~\ref{fig_BR_braiding}, in which a particle winds around two unlinked loops in a collectively linked configuration, we derive the corresponding braiding-induced operator 
$O^{\mathrm{BR}}_{\left(R^{\prime}\right), \left(C_1, R_1\right), \left(C_2, R_2\right)}$ in Eq.~(\ref{eq_BR_operator}). 
We show that this operator is trivial unless the charge carried by the particle and the fluxes carried by the two loops are non-Abelian, as stated in \textbf{Theorem}~\ref{theorem_bring}. 
In special cases satisfying the condition in Eq.~(\ref{eq_BR_phase_condition}), Borromean-Rings braiding reduces to a well-defined nontrivial phase. 
For the $\mathbb{D}_4$ quantum double model, this phase is explicitly evaluated to be $-1$, in agreement with the Borromean-Rings braiding phase of the corresponding $BF{+}AAB$ field theory. 
Thus, the lattice construction not only reproduces continuum braiding phases but also clarifies how braiding acts on the internal degrees of freedom of non-Abelian excitations. Besides the $\mathbb{D}_4$ case, we note that the quaternion group $\mathbb{Q}_8$ also satisfies the condition in Eq.~(\ref{eq_BR_phase_condition}). Therefore, the $\mathbb{Q}_8$ quantum double model serves as another candidate for a microscopic realization of nontrivial Borromean-Rings braiding.

\item  
\textbf{\textit{Exact lattice realization of the $BF$ field theory with an $AAB$ twist.}}   
The $(3+1)$D $BF$ field theory with an $AAB$ twist supports nontrivial Borromean-Rings braiding~\cite{PhysRevLett.121.061601} and provides an important example of three-dimensional non-Abelian topological order, even though its underlying gauge group $(\mathbb{Z}_2)^3$ is Abelian (see footnote~\ref{footnote_gaugegroup}). 
Here ``non-Abelian'' refers to the existence of multiple channels in fusion and shrinking processes. 
Gauge invariance constrains which twisted terms may consistently appear in the $BF$ action, implying that Borromean-Rings braiding can coexist only with a compatible subset of multi-loop braiding processes~\cite{PhysRevResearch.3.023132}. 
Together with the fusion--shrinking consistency conditions derived in Ref.~\cite{Zhang2023fusion}, these continuum results define a highly constrained set of long-distance topological data.

In Sec.~\ref{Sec.6}, we establish a microscopic realization of this continuum theory by constructing an explicit correspondence with the three-dimensional $\mathbb{D}_4$ quantum double model. 
By comparing the excitation spectra and the fusion, shrinking, and braiding data in the two descriptions, we construct an \textit{isomorphism} that   \textit{maps} each excitation in the $BF+AAB$ field theory with gauge group $G=(\mathbb{Z}_2)^3$ \textit{to} a corresponding excitation in the $\mathbb{D}_4$ quantum double model, as shown in Table~\ref{tab_f}. 
This isomorphism preserves fusion, shrinking, particle--loop braiding phases, and Borromean-Rings braiding phases [Eqs.~(\ref{eq_d4aab_fusion}),~(\ref{eq_d4aab_shrinking}),~(\ref{eq_d4aab_PL}), and~(\ref{eq_d4aab_BR})]. 
It therefore demonstrates that the $\mathbb{D}_4$ quantum double model provides a concrete microscopic construction of the $BF$ field theory with an $AAB$ twist and gauge group $G=(\mathbb{Z}_2)^3$. 
\textit{This exact correspondence, summarized in Table~\ref{tab_compare}, shows that the $BF{+}AAB$ field theory, which was introduced in Ref.~\cite{PhysRevLett.121.061601} and subsequently studied in Refs.~\cite{PhysRevResearch.3.023132,Zhang2023fusion,Zhang:2021ycl}, is physically relevant and microscopically realizable in a local lattice model with a tensor-product local Hilbert space and local interactions.}
\end{enumerate}

 \subsection{Outline}
This paper is organized as follows. 

Sec.~\ref{Sec.2} reviews the fusion and shrinking rules derived from field theory and formulates their consistency conditions. 
It also summarizes the diagrammatic representations that encode these processes in Sec.~\ref{sec2.2}. 
Sec.~\ref{Sec.3} constructs excitation operators in the microscopic model. 
We first introduce the $3$D quantum double model, construct its ground states on the three-torus, and compute the \textit{ground-state degeneracy} (GSD) in Sec.~\ref{Sec.3.1}. 
We then define four elementary operators and assemble excitation operators for particles and loops in Secs.~\ref{Sec.3.2} and~\ref{Sec.3.3}. 
Sec.~\ref{sec4} derives the fusion rules. 
In the $3$D quantum double model, fusing two excitations amounts to decomposing the tensor product of irreducible representations of $DG$, the quantum double algebra of the group $G$. 
The general construction is presented in Sec.~\ref{Sec.4.1}, followed by two examples, $G=\mathbb{Z}_N$ and $G=\mathbb{D}_3$, in Secs.~\ref{Sec.4.2} and~\ref{Sec.4.3}. 
Sec.~\ref{sec5} constructs lattice shrinking processes and derives the corresponding shrinking rules. 
A general method for computing shrinking coefficients is given in Sec.~\ref{Sec.5.1}. 
We then work out the examples of $G=\mathbb{Z}_N$ and $G=\mathbb{D}_3$ in Secs.~\ref{Sec.5.2} and~\ref{Sec.5.3}. 
By selecting specific internal degrees of freedom in the operator in Eq.~(\ref{eq_loop}), we demonstrate control over shrinking channels in Sec.~\ref{Sec.5.4}. 
We further verify the fusion--shrinking consistency relations in Sec.~\ref{Sec.5.5}. 
Sec.~\ref{Sec.8} studies particle--loop braiding and Borromean-Rings braiding. 
General braiding processes produce nontrivial operators that can act on the internal degrees of freedom of the excitations. 
We derive the particle--loop and Borromean-Rings braiding-induced operators in Secs.~\ref{Sec.8.1} and~\ref{Sec.8.2}. 
We also determine when these operators reduce to braiding phases. 
Examples with $G=\mathbb{Z}_N$ and $G=\mathbb{D}_3$ are discussed in Sec.~\ref{Sec.8.3}. 
Sec.~\ref{Sec.6} shows that the $3$D $\mathbb{D}_4$ quantum double model provides a microscopic construction of the $BF$ field theory with an $AAB$ twist and gauge group $G=(\mathbb{Z}_2)^3$. 
We first review the Wilson operators, fusion rules, shrinking rules, and braiding statistics of the $BF{+}AAB$ field theory in Sec.~\ref{Sec.6.3}. 
We then obtain the corresponding data in the $\mathbb{D}_4$ quantum double model in Sec.~\ref{Sec.6.2}. 
By comparing the fusion, shrinking, and braiding tables of the two theories, we establish an isomorphism between their excitations in Sec.~\ref{Sec.6.4}. 
This isomorphism bridges the long-distance field-theoretical description and the microscopic lattice construction. 
Finally, Sec.~\ref{sec7} provides a summary and outlook. 
Appendix~\ref{ap3} presents a more advanced discussion of diagrammatics. 
Appendix~\ref{ap2} proves several identities involving the elementary operators introduced in Sec.~\ref{Sec.3}.  
Appendix~\ref{ap4} gives technical details for calculating shrinking rules in the $\mathbb{D}_3$ quantum double model, and Appendix~\ref{ap5} provides technical details on braiding.
Appendix~\ref{ap_notation_summary} summarizes the notation used throughout the paper.

\section{Field-theoretical consistency conditions and diagrammatics of fusion and shrinking rules}\label{Sec.2}
In this section,  we first briefly review the consistency conditions between the fusion and shrinking rules that were obtained from the field-theoretical framework in our previous work. Next,  we review the construction of diagrammatic representations in which fusion and shrinking processes satisfy stringent  consistency conditions. The consistency between fusion and shrinking will be verified in the microscopic lattice model in Sec.~\ref{sec5}. Furthermore,  different unitary symbols can be defined to transform one diagram into another,  as shown in Appendix~\ref{ap3}. These transformations ultimately yield a set of algebraic constraints on higher-dimensional topological orders. Violation of these constraints signals quantum anomalies,  offering a diagnostic tool for identifying anomalies in theoretical models.

A summary of frequently used notation, its physical meaning, and the locations where they first appear is provided in Appendix~\ref{ap_notation_summary}, Table~\ref{tab_notation_summary}.

\subsection{Consistent fusion and shrinking rules}\label{Sec.2.2}
Fusion rules serve as the most fundamental topological data  of a topologically ordered phase. 
Two excitations  
collectively behave as another excitation when they are brought to the same spatial region and their total topological charge and flux are measured,  which is a process referred to as \textit{fusion}. 
In general,  a fusion rule takes the form
\begin{align}
    \mathsf{a}\otimes\mathsf{b}=\bigoplus_{\mathsf{c}}N_{\mathsf{c}}^{\mathsf{ab}}\mathsf{c}\, , \label{eq_fusion}
\end{align}
where $\mathsf{a}$,  $\mathsf{b}$,  and $\mathsf{c}$ denote topological excitations,  and 
$N_{\mathsf{c}}^{\mathsf{ab}}$ is a non-negative integer called the \textit{fusion coefficient}. 
The value of $N_{\mathsf{c}}^{\mathsf{ab}}$ counts the number of topologically distinct \emph{fusion channels} into $\mathsf{c}$. 
If $N_{\mathsf{c}}^{\mathsf{ab}}=0$,  the fusion channel to $\mathsf{c}$ is forbidden. 
A fusion process with exactly one allowed channel is called \textit{Abelian},  whereas one with multiple possible fusion channels is \textit{non-Abelian}.

In 3D topological orders,  the presence of loop-like excitations introduces new types of topological data known as \textit{shrinking rules}~\cite{Huang2023, Huang_2025, Zhang2023fusion},  
which determine how extended excitations can be continuously shrunk into lower-dimensional ones,  typically particles. 
The general form of a shrinking rule reads
\begin{align}
    \mathcal{S}\left(\mathsf{a}\right)=\bigoplus_{\mathsf{b}}S_{\mathsf{b}}^{\mathsf{a}}\mathsf{b}\, , \label{eq_shrinking}
\end{align}
where $\mathcal{S}$ denotes the shrinking operator,  $\mathsf{b}$ labels particle excitations,  
and $\oplus_{\mathsf{b}}$ enumerates all possible outcomes. 
The non-negative integer $S_{\mathsf{b}}^{\mathsf{a}}$ is the \textit{shrinking coefficient},  counting the number of distinct shrinking channels into $\mathsf{b}$. 
If $\mathsf{a}$ is already a particle,  the shrinking process acts trivially,  i.e.,  $\mathcal{S}(\mathsf{a})=\mathsf{a}$. 
As with fusion,  a single shrinking channel corresponds to an Abelian shrinking process,  while multiple channels indicate a non-Abelian one.

In our previous works~\cite{Zhang2023fusion, Huang2023},  
fusion and shrinking rules were systematically analyzed using continuum topological field theory (a brief review of this formalism is provided in Sec.~2 of Ref.~\cite{Huang_2025}). As an illustrative example,  consider a simple untwisted $BF$ theory with the gauge group $G=\prod_{i}\mathbb{Z}_{N_i}$. The topological action is
\begin{align}
    S=\int{\sum_{i}\frac{N_i}{2\pi}B^idA^i}\, , \label{eq_BF_BdA}
\end{align}
where $A^i$ and $B^i$ are $1$- and $2$-form $U\left(1\right)$ gauge fields,  respectively. The gauge transformations are given by $A^i\rightarrow A^i+d\chi^i $ and $ B^i\rightarrow B^i+dV^i$,  where $\chi^i$ and $V^i$ are $0$- and $1$-form gauge parameters satisfying $\int d\chi^{i}\in2\pi\mathbb{Z}$ and $\int dV^{i}\in2\pi\mathbb{Z}$. An excitation $\mathsf{a}$ in our field theory is represented by a gauge-invariant Wilson operator $\mathcal{O}_{\mathsf{a}}$. If $\mathsf{a}$ is a particle carrying $n_i$ units of gauge charge that is minimally coupled to the $A^i$ field,  the Wilson operator $\mathcal{O}_{\mathsf{a}}$ is written as $\mathcal{O}_{\mathsf{a}}=\exp\left(\mathrm{i}\int_{\gamma}n_iA^{i}\right)$,  where $\gamma=S^1$ denotes the closed one-dimensional world-line of the particle $\mathsf{a}$ in (3+1)D spacetime. If $\mathsf{b}$ is a loop carrying $m_j$ units of gauge flux that is minimally coupled to the $B^j$ field,  we have $\mathcal{O}_{\mathsf{b}}=\exp\left(\mathrm{i}\int_{\sigma}m_jB^{j}\right)$,  where $\sigma=S^1\times S^1$ denotes the closed two-dimensional world-sheet of the loop.

In the continuum topological field theory,  fusion of two topological excitations $\mathsf{a}$ and $\mathsf{b}$ is defined as dragging two Wilson operators $\mathcal{O} _{\mathsf{a}}$ and $\mathcal{O} _{\mathsf{b}}$ to the same position. This definition coincides with our microscopic construction of fusion in the lattice model as shown in Sec.~\ref{sec4}. The correlation function $\langle \mathsf{a}\otimes\mathsf{b} \rangle$ is expressed as
\begin{align}
    \langle \mathsf{a}\otimes\mathsf{b} \rangle&=\frac{1}{\mathcal{Z}}\int\mathcal{D} \left[ AB\right] \exp \left( \mathrm{i}S \right) \times \left( \mathcal{O} _{\mathsf{a}}\times \mathcal{O} _{\mathsf{b}} \right) \, ,
\end{align}
where $\mathcal{Z}$ is the partition function,  and $\mathcal{D} \left[ AB\right]$ represents the field configurations. The product of the two operators $\mathcal{O} _{\mathsf{a}}$ and $\mathcal{O} _{\mathsf{b}}$ in the above equation can be decomposed as a sum over other operators:  
\begin{align}
    \langle \mathsf{a}\otimes\mathsf{b} \rangle&=\frac{1}{\mathcal{Z}}\int \mathcal{D} \left[ AB\right] \exp \left( \mathrm{i}S \right) \times \left( \sum_\mathsf{c}{N_{\mathsf{c}}^{\mathsf{a}\mathsf{b}}}\mathcal{O} _{\mathsf{c}} \right)=\langle \bigoplus _{\mathsf{c}}N_{\mathsf{c}}^{\mathsf{a}\mathsf{b}}\mathsf{c} \rangle  \, .
	\label{eq_fusion_TQFT}
\end{align}
This result is the manifestation of the fusion rule Eq.~(\ref{eq_fusion}) in the framework of continuum topological field theory.

In the continuum topological field theory formalism,  shrinking of an excitation $\mathsf{a}$ is defined as contracting only the spatial components of the manifold supporting the Wilson operator $\mathcal{O}_{\mathsf{a}}$,  thereby reducing its spatial dimension while keeping the temporal evolution intact. For example,  this operation contracts the two-dimensional world-sheet $T^2 = S^1 \times S^1$ of a loop to a one-dimensional world-line $S^1$. In Sec.~\ref{sec5},  we will see that our microscopic construction of shrinking in the lattice model coincides with this continuum field theory definition. The correlation function of this process, denoted by $\langle \mathcal{S} \left( \mathsf{a} \right) \rangle$, is written as
\begin{align}
    \langle \mathcal{S} \left( \mathsf{a} \right) \rangle &=\langle \underset{X_1\rightarrow X_2}{\lim}\mathcal{O}_\mathsf{a} \rangle =\underset{X_1\rightarrow X_2}{\lim}\frac{1}{\mathcal{Z}}\int{\mathcal{D} \left[ AB \right] \exp \left( \mathrm{i}S \right) \mathcal{O} _{\mathsf{a}}}\, , \label{eq_9}
\end{align}
where $X_1$ and $X_2\subset X_1$ are the spacetime trajectories of the excitation $\mathsf{a}$ before and after the shrinking process (e.g.,  $X_1 = S^1 \times S^1$ and $X_2 = S^1$ for a loop). Crucially,  the Wilson operator $\mathcal{O}_{\mathsf{a}}$ generally consists of terms involving integration over the entire $X_1$ as well as over its proper submanifolds. Since some of these terms are supported on $X_1$ but cannot be supported on $X_2$,  they do not survive the shrinking process. Only the terms whose support is restricted to a submanifold $X_2$ remain. These remaining terms can be decomposed as a sum of Wilson operators for lower-dimensional excitations such as particles. Consequently,  Eq.~(\ref{eq_9}) can be written as
\begin{align}
    \langle \mathcal{S} \left( \mathsf{a} \right) \rangle&=\sum_{\mathsf{b}}\frac{1}{\mathcal{Z}}\int{\mathcal{D} \left[ AB \right] \exp \left( \mathrm{i}S \right) S_{\mathsf{b}}^{\mathsf{a}}\mathcal{O} _{\mathsf{b}}} =\langle \bigoplus_{\mathsf{b}}{S}_{\mathsf{b}}^{\mathsf{a}}\mathsf{b} \rangle \, , 
	\label{eq_shrinking_TQFT}
\end{align}
where $\oplus_\mathsf{b}$ only enumerates lower-dimensional excitations (particles in 3D). This result serves as the manifestation of the shrinking rule in Eq.~(\ref{eq_shrinking}) in the framework of continuum topological field theory.

A systematic analysis shows that for anomaly-free topological orders,  fusion and shrinking must satisfy a \textit{fusion--shrinking
consistency condition}. 
Specifically,  in 3D topological orders,  the condition reads
\begin{align}
    \mathcal{S}\left( \mathsf{a} \right) \otimes \mathcal{S}\left( \mathsf{b} \right) = \mathcal{S}\left( \mathsf{a}\otimes \mathsf{b} \right)\, , \label{eq_consistent_4D}
\end{align}
meaning that performing fusion before shrinking yields the same result as shrinking first and then fusing. In the language of category theory,  this implies that the shrinking operator is a \emph{tensor functor} that preserves the fusion structure. We will verify that this condition still holds in the lattice model in Sec.~\ref{sec5}.

If we further lift the spatial dimension to 4D,  
topological excitations include not only particles and loops but also membrane-like objects. 
A membrane excitation can undergo a \textit{hierarchical shrinking process},  
in which it first shrinks into loops, which subsequently shrink into particles. 
Correspondingly,  the consistency condition in 3D~(\ref{eq_consistent_4D}) is generalized to
\begin{align}
    \mathcal{S}^2\left( \mathsf{a} \right) \otimes \mathcal{S}^2\left( \mathsf{b} \right)
	= \mathcal{S}^2\left( \mathsf{a}\otimes \mathsf{b} \right)\, , \label{eq_consistent_5D}
\end{align}
where $\mathcal{S}^2$ denotes applying the shrinking operation twice in a  hierarchical shrinking process.

In this paper,  we focus on the consistent fusion and shrinking rules in 3D topological orders. 
Starting from the consistency condition~(\ref{eq_consistent_4D}),  
we can further derive algebraic constraints between the fusion coefficients $N_{\mathsf{c}}^{\mathsf{ab}}$ 
and the shrinking coefficients $S_{\mathsf{b}}^{\mathsf{a}}$. 
Using Eqs.~(\ref{eq_fusion}) and~(\ref{eq_shrinking}),  one finds
\begin{align}
    \mathcal{S}\left( \mathsf{a}\otimes \mathsf{b} \right) 
    &= \mathcal{S}\!\left( \bigoplus_{\mathsf{i}}N_{\mathsf{i}}^{\mathsf{ab}}\mathsf{i} \right)
    = \bigoplus_{\mathsf{i}}N_{\mathsf{i}}^{\mathsf{ab}}\mathcal{S}\!\left( \mathsf{i} \right)= \bigoplus_{\mathsf{i}}N_{\mathsf{i}}^{\mathsf{ab}}\!\left( \bigoplus_{\mathsf{c}}S_{\mathsf{c}}^{\mathsf{i}}\mathsf{c} \right)\nonumber \\
    &= \bigoplus_{\mathsf{c}}\!\left( \sum_{\mathsf{i}}S_{\mathsf{c}}^{\mathsf{i}}N_{\mathsf{i}}^{\mathsf{ab}} \right)\!\mathsf{c}\, ,
\end{align}
and
\begin{align}
    \mathcal{S}\!\left( \mathsf{a} \right)\! \otimes\! \mathcal{S}\!\left( \mathsf{b} \right) 
    &= \!\left( \bigoplus_{\mathsf{i}}S_{\mathsf{i}}^{\mathsf{a}}\mathsf{i} \right) \!\otimes\! \left( \bigoplus_{\mathsf{j}}S_{\mathsf{j}}^{\mathsf{b}}\mathsf{j} \right)= \bigoplus_{\mathsf{i}}\bigoplus_{\mathsf{j}}\!\left[ S_{\mathsf{i}}^{\mathsf{a}}S_{\mathsf{j}}^{\mathsf{b}}\!\left( \mathsf{i}\otimes \mathsf{j} \right) \right]\nonumber \\
    &= \bigoplus_{\mathsf{i}}\bigoplus_{\mathsf{j}}\!\left[ S_{\mathsf{i}}^{\mathsf{a}}S_{\mathsf{j}}^{\mathsf{b}}\!\left( \bigoplus_{\mathsf{c}}N_{\mathsf{c}}^{\mathsf{ij}}\mathsf{c} \right)\! \right] = \bigoplus_{\mathsf{c}}\!\left( \sum_{\mathsf{i}, \mathsf{j}}S_{\mathsf{i}}^{\mathsf{a}}S_{\mathsf{j}}^{\mathsf{b}}N_{\mathsf{c}}^{\mathsf{ij}} \right)\!\mathsf{c}\, .
\end{align}
Comparing the two equations above, we find that the consistency between fusion and shrinking rules enforces a key constraint
\begin{align}
    \sum_{\mathsf{i}}S_{\mathsf{c}}^{\mathsf{i}}N_{\mathsf{i}}^{\mathsf{ab}}
    = \sum_{\mathsf{i}, \mathsf{j}}S_{\mathsf{i}}^{\mathsf{a}}S_{\mathsf{j}}^{\mathsf{b}}N_{\mathsf{c}}^{\mathsf{ij}}\, , \label{eq_84}
\end{align}
where $\mathsf{a}, \mathsf{b}, \mathsf{c}, \mathsf{i}, \mathsf{j}$ label all possible excitations. 
Eq.~(\ref{eq_84}) encodes the algebraic backbone of consistent fusion--shrinking relations and can also be understood through our diagrammatic formulation as shown in Eq.~(\ref{eq_re}) of Appendix~\ref{ap3}.

\subsection{Diagrammatics}\label{sec2.2}
The consistency conditions in Eqs.~(\ref{eq_consistent_4D}) and (\ref{eq_consistent_5D}) in 3D and 4D topological orders inspire diagrammatic representations of consistent fusion and shrinking rules~\cite{Huang_2025}. We can define elementary diagrams for fusion and shrinking processes and treat them as vectors in fusion and shrinking spaces respectively. By combining these elementary diagrams,  we can build more complex diagrams,  and use a series of unitary operations to transform between them. The legitimate forms of these unitary operations are strongly constrained by pentagon equations and (hierarchical) shrinking--fusion hexagon equations,  which reveal algebraic structures of consistent fusion and shrinking rules. In this section,  we briefly review diagrammatic representations of fusion and shrinking rules in 3D. Details about transformations between different complex diagrams and derivations of algebraic constraints are collected in Appendix~\ref{ap3}.

We need to first introduce some notation. We assume that there are only a finite number of topologically distinct excitations in a topological order,  and we collect all of them in a set denoted by $\Phi _{0}^{\mathcal{D}}$,  where the superscript $\mathcal{D}$ denotes the spacetime dimension. In this paper,  we mainly focus on 3D topological orders and thus $\mathcal{D}=3+1$. If we perform a shrinking operation on all excitations in $\Phi _{0}^{3+1}$,  all possible shrinking results form a subset of $\Phi _{0}^{3+1}$,  denoted by $\Phi _{1}^{3+1}$. We can view the shrinking process as a mapping from the set $\Phi _{0}^{3+1}$ to its subset $\Phi _{1}^{3+1}$, 
\begin{align}
    \mathcal{S} :\quad \Phi _{0}^{3+1}\rightarrow \Phi _{1}^{3+1}\, .
\end{align}
The excitations in the set $\Phi _{i}^{3+1}$ have closed fusion rules,  i.e.,  all possible fusion results of fusing two excitations from the set $\Phi _{i}^{3+1}$ still form the set $\Phi _{i}^{3+1}$ itself. This is a consequence of the finite number of topological excitations and has been verified in (3+1)D twisted $BF$ theories ~\cite{Zhang2023fusion, Huang2023}. Thus,  we can also treat a fusion process as a mapping, 
\begin{align}
    \otimes :\, \,  \Phi _{i}^{3+1}\times \Phi _{i}^{3+1}\rightarrow \Phi _{i}^{3+1}\, , \quad i=0, 1\, .
\end{align}
To build a general diagrammatic framework, we allow for fusion and shrinking multiplicities. In this section,  Latin and Greek letters denote excitations and fusion/shrinking channels,  respectively.

\begin{figure}
	\centering
	\includegraphics[scale=0.5, keepaspectratio,max width=\linewidth]{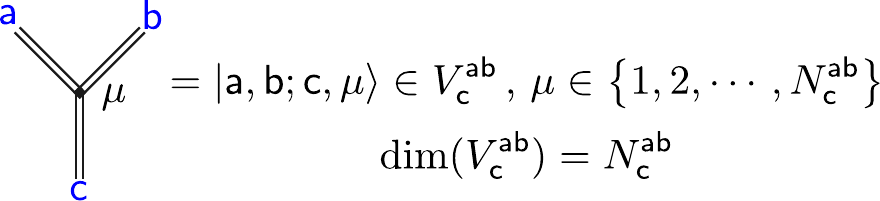}
	\caption{\textbf{Elementary fusion diagrams in 3D.} $\mathsf{a}$,  $\mathsf{b}$ and $\mathsf{c}$ denote excitations and we highlight them in blue. $\mu=\left\{1, 2, \cdots, N_{\mathsf{c}}^{\mathsf{a}\mathsf{b}}\right\}$ labels different fusion channels with the same fusion output $\mathsf{c}$. Double-lines indicate that $\mathsf{a}, \mathsf{b}, \mathsf{c}\in \Phi _{0}^{3+1}$. If we want to draw a fusion diagram that represents fusion in the set $\Phi _{1}^{3+1}$,  we only need to replace all the double lines with single lines. }
	\label{fig_fusion4d}
\end{figure}

Now we can construct the diagrammatics for 3D topological orders. A basic fusion diagram in our 3D diagrammatics is defined as follows. If a fusion process $\mathsf{a}\otimes \mathsf{b}$ has a fusion channel to $\mathsf{c}$, we can represent this fusion channel diagrammatically in Fig.~\ref{fig_fusion4d}. The solid lines are understood as the spacetime trajectories of the excitations,  and here we use double lines to indicate that $\mathsf{a}, \mathsf{b}, \mathsf{c}\in \Phi _{0}^{3+1}$. One can simply replace the double lines with single lines to obtain a fusion diagram that represents a fusion process in the set $\Phi _{1}^{3+1}$. In Fig.~\ref{fig_fusion4d},  we use $\mu=\left\{1, 2, \cdots, N_{\mathsf{c}}^{\mathsf{a}\mathsf{b}}\right\}$ to label different fusion channels to $\mathsf{c}$,  and the fusion diagram can be defined as a vector $\ket{\mathsf{a}, \mathsf{b};\mathsf{c}, \mu} $. A set of orthogonal vectors $\left\{\ket{\mathsf{a}, \mathsf{b};\mathsf{c}, \mu}, \mu=1, 2, \cdots, N_{\mathsf{c}}^{\mathsf{a}\mathsf{b}}\right\} $ spans a fusion space ${V}_{\mathsf{c}}^{\mathsf{ab}}$ with $\text{dim}({V}_{\mathsf{c}}^{\mathsf{ab}})=N_{\mathsf{c}}^{\mathsf{a}\mathsf{b}}$.

\begin{figure}
	\centering
	\includegraphics[scale=0.5, keepaspectratio,max width=\linewidth]{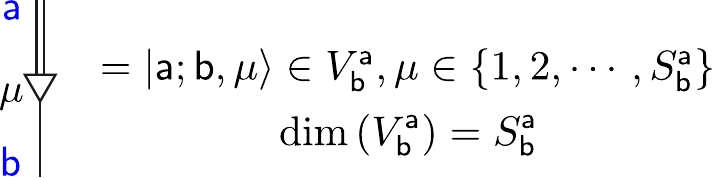}
	\caption{\textbf{Shrinking diagram in 3D.} The double line and single line represent excitations from  $\Phi _{0}^{3+1}$ and $\Phi _{1}^{3+1}$ respectively. Since $\mathsf{a}$ and $\mathsf{b}$ are the input and output of the shrinking process respectively,  $\mathsf{a}$ can be a loop or a particle while $\mathsf{b}$ can only be a particle. We use a triangle to represent a shrinking operation. This diagram can be defined as a vector and an orthogonal set $\left\{ \ket{\mathsf{a};\mathsf{b}, \mu} , \mu =1, 2, \cdots , S_{\mathsf{b}}^{\mathsf{a}} \right\}$ spans a shrinking space  $V_{\mathsf{b}}^{\mathsf{a}}$ with $\mathrm{dim}\left( V_{\mathsf{b}}^{\mathsf{a}} \right) =S_{\mathsf{b}}^{\mathsf{a}}$.
	}
	\label{fig_shrinking_dia}
\end{figure}

A basic shrinking diagram is defined as follows. If $\mathsf{a}\in \Phi _{0}^{3+1}$ and a shrinking process $\mathcal{S}\left(\mathsf{a}\right)$ has a shrinking channel to $\mathsf{b}\in \Phi _{1}^{3+1}$, then we define the corresponding shrinking diagram in Fig.~\ref{fig_shrinking_dia}. We use a triangle to represent a shrinking operation. Note that in a specific diagram,  a double line and a single line may represent the same particle. However,  they carry distinct meanings: the double line means that this particle from $\Phi _{0}^{3+1}$ is treated as the input of the shrinking process,  while the single line means that this particle from $\Phi _{1}^{3+1}$ is the output. We define the shrinking diagram in Fig.~\ref{fig_shrinking_dia} as a vector $\ket{\mathsf{a};\mathsf{b}, \mu}$,  where different $\mu$ label different orthogonal vectors. A set $\left\{ \ket{\mathsf{a};\mathsf{b}, \mu} , \mu =1, 2, \cdots , S_{\mathsf{b}}^{\mathsf{a}} \right\}$ spans a shrinking space denoted by $V_{\mathsf{b}}^{\mathsf{a}}$,  whose dimension is given by $\mathrm{dim}\left( V_{\mathsf{b}}^{\mathsf{a}} \right) =S_{\mathsf{b}}^{\mathsf{a}}$.

\begin{figure}
	\centering
	\includegraphics[scale=0.5, keepaspectratio,max width=\linewidth]{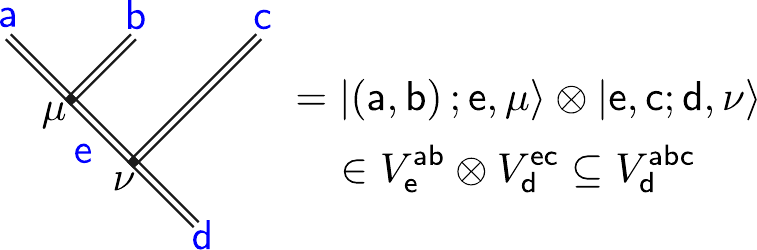}
	\caption{\textbf{Fusing three excitations in 3D.} $\mathsf{a}$,  $\mathsf{b}$,  and $\mathsf{c}$ are ultimately fused into $\mathsf{d}$. The parentheses ``$\left(\mathsf{a}, \mathsf{b}\right)$'' are used only to emphasize that $\mathsf{a}$ and $\mathsf{b}$ fuse together first in the whole three-excitation fusion process and they do not change the vectors. The set of vectors $\left\{\ket{\left( \mathsf{a}, \mathsf{b} \right) ;\mathsf{e}, \mu} \otimes\ket{\mathsf{e}, \mathsf{c};\mathsf{d}, \nu}\right\} $ spans the whole space $V_{\mathsf{d}}^{\mathsf{abc}}$,  where different $\mu$,  $\nu$ and $\mathsf{e}$ label different orthogonal vectors. The space $V_{\mathsf{d}}^{\mathsf{abc}}$ is isomorphic to $\oplus _{\mathsf{e}}(V_{\mathsf{e}}^{\mathsf{ab}}\otimes V_{\mathsf{d}}^{\mathsf{ec}})$. The dimension of $V_{\mathsf{d}}^{\mathsf{a}\mathsf{b}\mathsf{c}}$ is given by $\text{dim}(V_{\mathsf{d}}^{\mathsf{a}\mathsf{b}\mathsf{c}})=\sum_{\mathsf{e}}{N_{\mathsf{e}}^{\mathsf{ab}}}N_{\mathsf{d}}^{\mathsf{ec}}$. Here we only draw the diagram for a fusion process in the set $\Phi _{0}^{3+1}$. One can obtain a diagram for the subset $\Phi _{1}^{3+1}$ by simply replacing all double lines with single lines.
	}
	\label{fig_3fusion4d_1}
\end{figure}

\begin{figure}
	\centering
	\includegraphics[scale=0.5, keepaspectratio,max width=\linewidth]{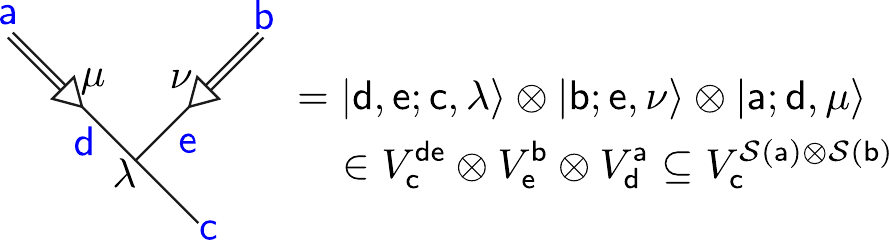}
	\caption{\textbf{Incorporating both fusion and shrinking processes in 3D.} We stack two basic shrinking diagrams and a basic fusion diagram to describe a process $\mathcal{S} \left( \mathsf{a} \right) \otimes \mathcal{S} \left( \mathsf{b} \right) $.
	}
	\label{fig_sa_times_sb}
\end{figure}

We can stack these basic diagrams to construct more complicated diagrams that describe more complicated processes. For example,  by stacking two basic fusion diagrams shown in Fig.~\ref{fig_fusion4d},  we construct a fusion diagram that involves three excitations. If fusing three excitations $\left(\mathsf{a}\otimes \mathsf{b}\right)\otimes \mathsf{c}$ has allowed fusion channels to $\mathsf{d}$, we can depict this process in Fig.~\ref{fig_3fusion4d_1},  which is defined as a vector $\ket{\left( \mathsf{a}, \mathsf{b} \right) ;\mathsf{e}, \mu} \otimes\ket{\mathsf{e}, \mathsf{c};\mathsf{d}, \nu} $,  where $\otimes$ means tensor product. The corresponding space is denoted by $V_{\mathsf{d}}^{\mathsf{a}\mathsf{b}\mathsf{c}}$,  whose dimension is given by
\begin{align}
    \text{dim}(V_{\mathsf{d}}^{\mathsf{a}\mathsf{b}\mathsf{c}})=\sum_{\mathsf{e}}{N_{\mathsf{e}}^{\mathsf{ab}}}N_{\mathsf{d}}^{\mathsf{ec}}\, , \label{eq_dim1}
\end{align}
because $V_{\mathsf{d}}^{\mathsf{a}\mathsf{b}\mathsf{c}}$ is isomorphic to $\oplus _{\mathsf{e}}V_{\mathsf{e}}^{\mathsf{ab}}\otimes V_{\mathsf{d}}^{\mathsf{ec}}$. We can use a similar tensor product construction to incorporate both fusion and shrinking processes in a diagram. For $\mathcal{S} \left( \mathsf{a} \right) \otimes \mathcal{S} \left( \mathsf{b} \right) $, the corresponding diagram is shown in Fig.~\ref{fig_sa_times_sb}. This diagram is understood as a vector $\ket{\mathsf{d}, \mathsf{e};\mathsf{c}, \lambda}\otimes\ket{\mathsf{b};\mathsf{e}, \nu}\otimes\ket{\mathsf{a};\mathsf{d}, \mu}$,  where different $\mathsf{d}$,  $\mathsf{e}$,  $\mu$,  $\nu$,  and $\lambda$ label different orthogonal vectors. The diagram shown in Fig.~\ref{fig_sa_times_sb} describes a process in which $\mathsf{a}$ and $\mathsf{b}$ shrink to $\mathsf{d}$ and $\mathsf{e}$ first in the $\mu$ and $\nu$ channels, respectively, and then $\mathsf{d}$ and $\mathsf{e}$ fuse to $\mathsf{c}$ in the $\lambda$ channel. This set of orthogonal vectors spans a space denoted by $V_{\mathsf{c}}^{\mathcal{S} \left( \mathsf{a} \right) \otimes \mathcal{S} \left( \mathsf{b} \right)}$,  which is isomorphic to $\oplus _{\mathsf{d}, \mathsf{e}}V_{\mathsf{c}}^{\mathsf{de}}\otimes V_{\mathsf{e}}^{\mathsf{b}}\otimes V_{\mathsf{d}}^{\mathsf{a}}$ due to our tensor product construction. Thus,  we have
\begin{align}
    &\mathrm{dim}\left( V_{\mathsf{c}}^{\mathcal{S} \left( \mathsf{a} \right) \otimes \mathcal{S} \left( \mathsf{b} \right)} \right) =\sum_{\mathsf{d}, \mathsf{e}}{N_{\mathsf{c}}^{\mathsf{de}}S_{\mathsf{e}}^{\mathsf{b}}S_{\mathsf{d}}^{\mathsf{a}}}\, .
\end{align}

The consistency condition Eq.~(\ref{eq_consistent_4D}) allows us to introduce a series of unitary operations to transform these complicated diagrams. We can derive important algebraic constraints on these unitary operations known as the pentagon equation and shrinking--fusion hexagon equation. Details of these unitary operations and algebraic constraints are shown in Appendix~\ref{ap3}.

\section{Excitation creation}\label{Sec.3}
To establish a connection between continuum field theory and microscopic lattice constructions,  we need a microscopic model for 3D topological order. The 3D quantum double model serves as an ideal platform. In this section,  we consider the 3D quantum double lattice model and design operators to create topological excitations. We note that topological excitations in 3D differ essentially from their 2D counterparts. While 2D topological orders host point-like anyons that exhibit fractional statistics,  point-like particles in 3D are constrained to be either bosons or fermions. Furthermore,  spatially extended excitations,  i.e.,  loops,  appear in 3D and support nontrivial processes such as fusion, shrinking and braiding. Investigating these particles and loops,  along with their fusion rules, shrinking rules, and braiding statistics,  reveals structures that extend beyond those found in 2D topological orders. These excitations are the cornerstones of the study of fusion, shrinking, and braiding on the lattice in Secs.~\ref{sec4}, \ref{sec5}, and~\ref{Sec.8}.

\subsection{Hamiltonian terms on a cubic  lattice, ground states, and GSD}\label{Sec.3.1}
\begin{figure}
	\centering
	\includegraphics[scale=2.5, keepaspectratio,max width=\linewidth]{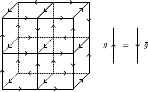}
	\caption{\textbf{Local Hilbert space on a cubic lattice.} Each edge is assigned an arrow and a group element $g\in G$. Reversing the arrow of an edge changes the group element $g$ to its inverse $\bar{g}$.}
	\label{fig_3Dlattice}
\end{figure}

Consider the 3D quantum double model defined on a cubic lattice with periodic boundary conditions. The underlying degrees of freedom are group elements $g \in G$ on edges of the cubic lattice. Each edge is assigned an orientation (an arrow,  see Fig.~\ref{fig_3Dlattice}) that can be chosen arbitrarily. The assignment of edge orientations is not physically observable; it merely fixes a convention for representing group elements. Reversing the arrow on a given edge is equivalent to replacing the group element $g$ on that edge by its inverse $\bar{g}$. All physical observables such as ground-state degeneracy, excitation spectra, fusion rules, shrinking rules, and braiding statistics are invariant under such reassignments, as long as the chosen convention is used consistently. A local state on an edge is denoted by $\ket{g}$ and we set $\braket{g|h}=\delta_{g, h}$. The local Hilbert space is spanned by the basis $\left\{ \ket{g}|g\in G \right\} $ and its dimension is $\left| G \right|$,  the number of elements in $G$. A configuration on the whole lattice is written as $\ket{g_1}\otimes\ket{g_2}\otimes\cdots\otimes\ket{g_N}$,  where $\otimes$ means tensor product and $g_i$ labels the group element on the $i$-th edge,  $i=1, 2, \cdots N$. All possible configurations span the total Hilbert space whose dimension is $\left| G \right|^N$. 

With the local Hilbert space on the edges introduced,  we define four basic operators that act on the Hilbert space of an edge. We will see later that linear combinations of these basic operators give rise to topological excitations. The basic operators (dubbed ``\textbf{$T$-operators}'' and ``\textbf{$L$-operators}'' hereafter) are
\begin{align}
    &T_{g}^{+}=\ket{g}\bra{g}, && T_{g}^{-}=\ket{\bar{g}}\bra{\bar{g}}\, , \label{eq_t} \\
    &L_{g}^{+}=\sum_{k\in G}{\ket{gk} \bra{k}}, && L_{g}^{-}=\sum_{k \in G}{\ket{k\bar{g}} \bra{k}}\, .\label{eq_l}
\end{align}
The operator $T^{+}_{g}$ ($T^{-}_{g}$) acts on $\ket{h}$ as a test for whether the group element on the edge equals $g$ ($\bar{g}$),  being a projection operator selecting $h=g$ ($h=\bar{g}$). The operator $L^{+}_{g}$ acts on $\ket{h}$ as a ``left-multiplication'' operator,  $L^{+}_{g} \ket{h} = \ket{gh}$,  where the group element on the edge is left-multiplied by $g$. In contrast,  the operator $L^{-}_{g}$ acts on $\ket{h}$ as $L^{-}_{g} \ket{h} = \ket{h \bar{g}}$,  where the group element on the edge is right-multiplied by $\bar{g}$. We can verify that these basic operators satisfy the following relations:
\begin{align}
    &T_{g}^{\pm}T_{h}^{\pm}=\delta _{g, h}T_{h}^{\pm}\, , && L_{g}^{\pm}L_{h}^{\pm}=L_{gh}^{\pm}\, , \nonumber \\
    &T_{g}^{+}L_{h}^{+}=L_{h}^{+}T_{\bar{h}g}^{+}\, , && T_{g}^{+}L_{h}^{-}=L_{h}^{-}T_{gh}^{+}\, , \nonumber \\
    &T_{g}^{-}L_{h}^{+}=L_{h}^{+}T_{gh}^{-}\, , && T_{g}^{-}L_{h}^{-}=L_{h}^{-}T_{\bar{h}g}^{-}\, , \nonumber \\
    &\left( T_{g}^{\pm} \right) ^{\dagger}=T_{g}^{\pm}\, , && \left( L_{g}^{\pm} \right) ^{\dagger}=\left( L_{g}^{\pm} \right) ^{-1}=L_{\bar{g}}^{\pm}\, , \nonumber \\
    &\left[L_{g}^{\pm}, L_{h}^{\mp}\right]=0\, ,  &&\sum_g{T_{g}^{\pm}}=L_{e}^{\pm}=1\, .\label{eq_tl}
\end{align}

Before presenting the Hamiltonian of the 3D quantum double model,  we first introduce a plaquette operator $B_{p, s}^{h}$ acting on a plaquette $p$ and a vertex operator $A_v$ acting on a vertex $v$. 

\begin{figure}
	\centering
	\includegraphics[scale=2.3, keepaspectratio,max width=\linewidth]{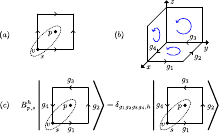}
	\caption{\textbf{The action of a plaquette operator $B^{h}_{p, s}$.} (a) A site $s=\left(v, p\right)$ is a combination of a vertex $v$ and a plaquette $p$. (b) The blue arrows denote the product order of the group elements on the edges of the plaquette. The specific assignment of the product order here is only a matter of convention and has no physical consequences. (c) $B_{p, s}^{h}$ projects the ordered product of all group elements on the edges around $p$,  $g_1g_2g_3g_4$,  to $h\in G$. Notice that the ordered product depends on the choice of the site $s=\left(v, p\right)$,  i.e.,  the choice of vertex $v$. When $h=e$, the condition $g_1 g_2 g_3 g_4 = e$ also implies $g_4 g_1 g_2 g_3 = g_3 g_4 g_1 g_2 = \cdots =e$, so in this situation, the label $s$ is redundant. For convenience, we can introduce a simplified notation $B_p$ via $B^e_{p,s}=B_p$.}
	\label{fig_Bp}
\end{figure}

As shown in Fig.~\ref{fig_Bp},  the action of $B_{p, s}^{h}$ is to measure whether the ordered product of group elements on the edges around the plaquette $p$ equals $h$. If so, the state remains unchanged; otherwise, it is annihilated. The site $s=(v, p)$ defined in Fig.~\ref{fig_Bp}\hyperref[fig_Bp]{(a)} determines the order of multiplication in the product,  which is important because the group elements are generally noncommutative. Specifically,  for a plaquette on the $xy$-plane,  we set its normal direction as the $+z$ direction and set the arrows of the edges to form a counterclockwise loop with respect to the normal direction. Then,  starting from the vertex $v$,  we multiply the group elements along the loop direction. For other plaquettes on $xz$- and $yz$-planes,  we set their normal directions as the $+y$ and $+x$ directions; the orientations of the arrows of the edges are shown in Fig.~\ref{fig_Bp}\hyperref[fig_Bp]{(b)}. This choice of normal direction in a plaquette operator merely fixes a convention for representing ordered products of group elements. Choosing the opposite normal direction for plaquette operators is equivalent to reversing the flux label in the corresponding conjugacy class. The final topological data are independent of these choices, as long as the chosen convention is used consistently.
A plaquette operator $B_{p, s}^{h}$ can always be constructed from the basic operators shown in Eq.~(\ref{eq_t}). For example,  $B_{p, s}^{h}$ in Fig.~\ref{fig_Bp}\hyperref[fig_Bp]{(c)} can be written as 
\begin{align}
    B_{p, s}^{h}=\sum_{g_1, g_2, g_3\in G}{T_{\bar{g}_3\bar{g}_2\bar{g}_1h}^{+}T_{g_3}^{+}T_{g_2}^{+}T_{g_1}^{+}}\, , \label{bp_definition}
\end{align}
where $T_{g_1}^{+}$,  $T_{g_2}^{+}$,  $T_{g_3}^{+}$,  and $T_{\bar{g}_3\bar{g}_2\bar{g}_1h}^{+}$ act on the edges labeled by $g_1$,  $g_2$,  $g_3$,  and $g_4$ respectively. If we take $h$ as the identity element $e$,  the action of $B_{p, s}^{h}$ is independent of the choice of $s$, as shown in Fig.~\ref{fig_Bp}\hyperref[fig_Bp]{(c)}.  \textit{We denote $B_{p, s}^{e}$ as $B_p$ for simplicity}. The $B_p$ term can be understood as a zero-flux condition. But for a general $h\in G$,  the action of $B_{p, s}^{h}$ depends on the choice of the site. If we choose a different vertex $v$ in $s=\left(v, p\right)$,  $B_{p, s}^{h}$ may project the product $g_1g_2g_3g_4$ to another element in the conjugacy class $C_h$ containing $h$.

\begin{figure}
	\centering
	\includegraphics[scale=2.3, keepaspectratio,max width=\linewidth]{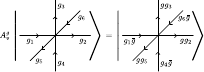}
	\caption{\textbf{The action of the $A_v^{g}$ term on the vertex $v$.} $A_{v}^{g}\ket{g_i}=\ket{gg_i}$ if the arrow points away from $v$,  $A_{v}^{g}\ket{g_i}=\ket{g_i\bar{g}} $ if the arrow points towards $v$.}
	\label{fig_Av}
\end{figure}

A vertex operator $A_v^{g}$ acts on all edges that are attached to the vertex $v$. If the arrow of an edge points away from the vertex $v$,  the group element on the edge is multiplied by $g$ on the left. Otherwise,  the element on the edge is multiplied by $\bar{g}$ on the right,  as shown in Fig.~\ref{fig_Av}. As with the plaquette operator,  a vertex operator $A_v^{g}$ can also be written as a product of the basic operators shown in Eq.~(\ref{eq_l}). As an example,  we write the vertex operator $A_v^{g}$ in Fig.~\ref{fig_Av} as
\begin{align}
    A_{v}^{g}=L_{g}^{-}L_{g}^{+}L_{g}^{+}L_{g}^{-}L_{g}^{+}L_{g}^{-}\, ,
\end{align}
where the three $L_{g}^{+}$ operators act on the edges labeled by $g_2$,  $g_3$,  and $g_5$,  respectively,  while the three $L_{g}^{-}$ operators act on the edges labeled by $g_1$,  $g_4$,  and $g_6$,  respectively.

The Hamiltonian is given by
\begin{align}
    H=-\sum_v{A_v}-\sum_p{B_p}\, , \label{eq_ham}
\end{align}
where $A_v=\frac{1}{\left| G \right|}\sum_{g\in G}{A_{v}^{g}}$. The definition of the $B_p$ term is given after Eq.~(\ref{bp_definition}). The Hamiltonian (\ref{eq_ham}) has been discussed in the literature (see, e.g., Ref.~\cite{PhysRevB.91.075114} ) and here we focus on the excitation operators and on the topological processes implemented using it. Using the $T$- and $L$-operators shown in Eqs.~(\ref{eq_t}) and~(\ref{eq_l}) as well as the properties shown in Eq.~(\ref{eq_tl}),  we can verify that the $A_v$ and $B_p$ operators satisfy
\begin{gather}
	A_{v}^{g}A_{v}^{h}=A_{v}^{gh}\, , \label{eq_Av_mutliplication}
	\\
	A_vA_v=A_v\, , \label{eq5}
	\\
	B_pB_p=B_p\, , 
	\\
	\left[ A_{v}^{g}, A_{v^{\prime}}^{h} \right] =\left[ B_p, B_{p^{\prime}} \right] =\left[ A_v, B_p \right] =0\, .\label{eq7}
\end{gather}
The commutation relations in Eq.~(\ref{eq7}) indicate that all vertex terms $A_v$ commute with all plaquette terms $B_p$, and terms acting on different vertices or different plaquettes also commute with each other. Consequently, the Hamiltonian $H$ is a sum of mutually commuting projectors, which renders the model exactly solvable: all $A_v$ and $B_p$ operators share a common set of eigenstates, and the ground-state subspace consists of states that are simultaneous $+1$ eigenstates of every $A_v$ and $B_p$. The ground state can be obtained as follows. We start from a product state $\ket{\phi}$ that satisfies the zero-flux condition (the $B_p$ term has eigenvalue $+1$) everywhere, i.e., 
$	{B_p}\ket{\phi} =\ket{\phi}, \ \forall p$.
Applying the projector $\prod_v{A_v}$ to $\ket{\phi}$ gives a ground state $\ket{\mathrm{GS}} =\prod_v{A_v}\ket{\phi} \, $.
Using Eq.~(\ref{eq5}) and Eq.~(\ref{eq7}),  we verify that $A_{v}\ket{\mathrm{GS}}=B_p\ket{\mathrm{GS}}=\ket{\mathrm{GS}}$.
\begin{figure}
	\centering
	\includegraphics[scale=1.6, keepaspectratio,max width=\linewidth]{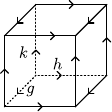}
	\caption{\textbf{The simplest 3D lattice on a $3$-torus manifold.} There are only three independent edges,  three independent plaquettes,  and one independent vertex. The configuration on this lattice is written as $\ket{g, h, k}$.}
	\label{fig_GSD}
\end{figure}

To compute the GSD,  we consider the simplest nontrivial case in which the Hamiltonian is defined on a $3$-torus $T^3$. 
The $3$-torus can be represented by a single cube with periodic boundary conditions as shown in Fig.~\ref{fig_GSD}. A state satisfying the zero-flux condition can be written as $\ket{g, h, k}$,  where commuting group elements $g$,  $h$,  and $k$ denote the three independent group elements on the single cube. The other condition for a ground state is $A_v \ket{GS}=\ket{GS}$. Since there is only one vertex in this cube,  we have
\begin{align}
    \ket{\mathrm{GS}_{g, h, k}}&=A_v\ket{g, h, k} =\frac{1}{\left| G \right|}\sum_{g_0}{A_{v}^{g_0}}\ket{g, h, k}=\frac{1}{\left| G \right|}\sum_{g_0}\ket{g_0g\bar{g}_{0}, g_0h\bar{g}_{0}, g_0k\bar{g}_{0}}\, .
\end{align}
We can also choose states labeled by other commuting elements $\ket{g^{\prime}, h^{\prime}, k^{\prime}}$ to construct other ground states $\ket{\mathrm{GS}_{g^{\prime}, h^{\prime}, k^{\prime}}}$. Group theory ensures that $\ket{\mathrm{GS}_{g, h, k}}$ and $\ket{\mathrm{GS}_{g^{\prime}, h^{\prime}, k^{\prime}}}$ are either the same or orthogonal to each other. By using Burnside's lemma,  we can express the \textit{ground-state degeneracy} (GSD) as
\begin{align}
    \mathrm{GSD}=\frac{1}{\left| G \right|} \sum_{g_0 \in G} \sum_{\left\{g, h, k\right\}}  \delta_{g, g_0 g \bar{g}_0}\delta_{h, g_0 h \bar{g}_0} \delta_{k, g_0 k \bar{g}_0}\, , \label{eq_GSD}
\end{align}
where $\left\{g, h, k\right\}$ is a triplet of commuting group elements. Here,  we give the GSD on a $3$-torus for some simple groups. For an Abelian group $G$,  the GSD is $\left| G \right|^3$. For $G=\mathbb{D}_3$ and $G=\mathbb{D}_4$,  the GSDs are 21 and 92,  respectively. Note that the $BF$ theory with an $AAB$ twist and gauge group $\left(\mathbb{Z}_2\right)^3$ also has $\mathrm{GSD}=92$ on $T^3$~\cite{Zhang2023fusion}. In Sec.~\ref{Sec.6},  we further conclude that the fusion rules, shrinking rules, and braiding statistics of the 3D $G=\mathbb{D}_4$ quantum double model and the $BF$ theory with an $AAB$ twist and gauge group $\left(\mathbb{Z}_2\right)^3$ are isomorphic,  which leads to an exact correspondence between the continuum topological field theory and the microscopic lattice construction.

\subsection{Particle excitations}\label{Sec.3.2}
\begin{figure}
	\centering
	\includegraphics[scale=2.5, keepaspectratio,max width=\linewidth]{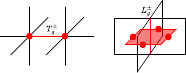}
	\caption{\textbf{Applying $T$- and $L$-operators [Eq.~(\ref{eq_l})] to a ground state.} These basic operators generally do not commute with vertex terms or plaquette terms in the Hamiltonian. Thus,  a $T_{g}^{\pm}$ creates a pair of excited vertices while an $L_{g}^{\pm}$ creates a loop-like excited state.}
	\label{fig_TL}
\end{figure}

In the following,  we proceed to create excitations and investigate their properties. Recall that a vertex term $A_v$ and a plaquette term $B_p$ can be constructed by using several $L_{g}^{\pm}$ terms and $T_{g}^{\pm}$ terms respectively. The ground-state condition is $A_v = B_p =1, \ \forall v, p$. From Eq.~(\ref{eq_tl}) we can see that $T_{g}^{\pm}$ generally does not commute with $L_{g}^{\pm}$,  which means that a $T_{g}^{\pm}$ does not commute with two vertex terms at the endpoints of an edge,  and an $L_{g}^{\pm}$ does not commute with four plaquette terms around an edge. Thus,  as shown in Fig.~\ref{fig_TL},  applying a $T_{g}^{\pm}$ to the ground state creates two excited vertices (i.e.,  violates the condition $A_v =1$) at the endpoints of an edge,  and applying an $L_{g}^{\pm}$ creates four excited plaquettes (i.e.,  violates the condition $B_p =1$) whose centers live on a closed loop in the dual lattice. In this sense,  the topological excitations in the 3D quantum double model can be particles or loops. In Secs.~\ref{Sec.3.2} and~\ref{Sec.3.3},  we illustrate how to construct operators for particles (Eq.~(\ref{eq_particle})) and loops (Eq.~(\ref{eq_loop})) by using $T_{g}^{\pm}$ and $L_{g}^{\pm}$ in our 3D quantum double model. For simplicity,  we say that an operator violates vertex terms or plaquette terms when this operator does not commute with them in the following discussion.

We first consider operators for particles. $T_{L}^{g}$ is defined as an operator that acts on an oriented string $L$\footnote{In the following discussion, we sometimes use two vertices $v_iv_j$ to denote an oriented string from $v_i$ to $v_j$. For example, the subscripts of $T$-operators present in Sec.~\ref{Sec.3.3} each contain two vertices, and such a subscript is to be understood as an oriented string.}. If the string $L$ can be obtained by connecting two shorter strings $L_1$ and $L_2$,  we define the \textit{connecting rule} of $T_{L=L_1\cup L_2}^{g}$ as
\begin{align}
    T_{L=L_1\cup L_2}^{g}=\sum_h{T_{L_2}^{\bar{h}g}T_{L_1}^{h}}.\label{eq_T_connect}
\end{align}
When $L_i$ is a single edge,  then $T_{L_i}^{g}=T_{g}^{+}$ if the arrow of the edge aligns with the orientation of $L_i$,  otherwise $T_{L_i}^{g}=T_{g}^{-}$. Since the operator $T^{h}_{L_1}$ in Eq.~(\ref{eq_T_connect}) acts first, we require that the end of $L_1$ is also the start of $L_2$. In the following discussion,  we use $\partial_0 L$ and $\partial_1 L$ to denote the start and end of an oriented string $L$ respectively,  and we have $\partial_1L_1=\partial_0L_2$. Eq.~(\ref{eq_T_connect}) is designed so that the operator $T_L^g$ faithfully projects onto configurations in which the ordered product of group elements along the entire string $L$ equals $g$. To see that this connecting rule extends the projection property from single edges to arbitrary strings in a consistent way, we consider a state denoted by $\ket{g_1,g_2}$ where ordered products along $L_1$ and $L_2$ are $g_1$ and $g_2$ respectively. Thus, we have
\begin{align}
    \sum_h{T_{L_2}^{\bar{h}g}T_{L_1}^{h}}\ket{g_1,g_2}&=\sum_h{\delta_{\bar{h}g,g_2}\delta_{h,g_1}}\ket{g_1,g_2} =\delta_{\bar{g}_1g,g_2}\ket{g_1,g_2}=\delta_{g,g_1g_2}\ket{g_1,g_2}\,.
\end{align}
The delta function $\delta_{g,g_1g_2}$ enforces the ordered product $g_1g_2$ to be $g$. Otherwise, the state $\ket{g_1,g_2}$ vanishes.

\begin{figure*}
	\centering
	\includegraphics[scale=3.2, keepaspectratio,max width=\linewidth]{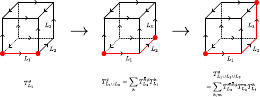}
	\caption{\textbf{Connecting three string operators step by step.} The red lines and dots denote the string and the excited vertices respectively. This connecting rule allows us to freely move particles. }
	\label{fig_move_particle}
\end{figure*}
When we consider connecting more strings as shown in Fig.~\ref{fig_move_particle},  we expect that the order of connection is irrelevant,  i.e., 
\begin{align}
    T_{L_1\cup \left( L_2\cup L_3 \right)}^{g}=T_{\left( L_1\cup L_2 \right)\cup L_3}^{g}\, , \label{eq_connect_order}
\end{align}
where the parentheses in $L_1\cup \left( L_2\cup L_3 \right)$ emphasize that we connect $L_2$ and $L_3$ first. Similarly,  the parentheses in $\left( L_1\cup L_2 \right)\cup L_3$ emphasize that we connect $L_1$ and $L_2$ first. We provide the proof of Eq.~(\ref{eq_connect_order}) in Appendix~\ref{ap2.1}.

Another important property of $T_{L}^{g}$ is that it commutes with $B_{p}$'s and $A_v$'s,  except the $A_v$ terms at the endpoints of $L$. We denote a string by $L=\left( v_0, v_1, \cdots , v_n \right) $ that starts at the vertex $v_0$ and ends at $v_n$,  where an edge is denoted by $v_i v_{i+1}$. By using the algebra of $T$-operators as well as the orthonormality of the basis $\{|g\rangle\}$,  we can verify that $T_{L}^{g}$ commutes with all $B_p$ terms. For the $A_v$ terms,  in Appendix~\ref{ap2.2} we prove that
\begin{align}
    A_{v_i}^{k}T_{L}^{g}=T_{L}^{g}A_{v_i}^{k}\, , \label{eq_31}
\end{align}
where $k\in G$ and $i\ne0, n$. Since $A_{v_i}=\frac{1}{\left| G \right|}\sum_{k\in G}{A_{v_i}^{k}}$,  we conclude that $\left[A_{v_i}, T_{L}^{g}\right]=0$ as long as $v_i$ is a vertex inside the string $L$. However,  $T_{L}^{g}$ does not commute with vertex terms at the endpoints of $L$. In Appendix~\ref{ap2.2},  we derive the commutation relations between $A_{v_{0}}^k $,  $ A_{v_{n}}^k$,  and $T_{L}^{g}$ as follows:
\begin{align}
    A_{v_0= \partial _0L}^{k}T_{L}^{g}&=T_{L}^{kg}A_{v_0= \partial _0L}^{k}\, , \label{eq_32} \\
    A_{v_n= \partial _1L}^{k}T_{L}^{g}&=T_{L}^{g\bar{k}}A_{v_n= \partial _1L}^{k}\, , \label{eq_33}
\end{align}
which means that the operator $T_{L}^{g}$ violates the $A_{v_0}$ and $A_{v_n}$ terms. Therefore,  the operator $T_{L}^{g}$ creates excited vertices at the endpoints of $L$,  as shown in Fig.~\ref{fig_move_particle}.

However, a single $T_{L}^{g}$ operator acting on the ground state does not create an excitation with a definite topological label. The resulting state $T_{L}^{g}\ket{\mathrm{GS}}$ violates $A_{v_0}$ and $A_{v_n}$ (where $v_0=\partial_0L$ and $v_n=\partial_1L$) but carries no additional quantum numbers that distinguish different particle types. From Eqs.~(\ref{eq_32}) and~(\ref{eq_33}) we can see that different group elements $g$ carried by $T_{L}^{g}$ can be continuously connected by $A_{v_0}^{k}$ and $A_{v_n}^{k}$. To obtain excitations with well-defined topological charges that are invariant under local perturbations, we must project onto irreducible representations of the  group $G$ at the endpoints. This projection is achieved by taking appropriate linear combinations of $T_{L}^{g}$ weighted by representation matrix elements, which yields the string operator for a particle labeled by an irrep $R$:
\begin{align}
    W_L\left( R;i, j \right) =\sum_g{\left[ \Gamma _{ij}^{R}\left( g \right) \right] ^{\ast}T_{L}^{g}}=\sum_g{\Gamma _{ij}^{R^{\ast}}\left( g \right) T_{L}^{g}}\, , \label{eq_particle}
\end{align}
where $R$ labels an \textit{irreducible representation} (irrep) of $G$,  $\Gamma$ denotes a matrix of the irrep,  $i, j$ are row and column indices;  $\ast$ and $R^\ast$ respectively denote complex conjugation and the \textit{complex conjugate irrep}. Since a single $T_{L}^{g}$ operator does not commute with the two vertex terms at the endpoints of $L$, the string operator $W_L(R;i,j)$ inherits this property and also fails to commute with them (unless $R$ is the trivial irrep). Thus, the string operator $W_L(R;i,j)$ creates a pair of particles at the vertices $\partial_0L$ and $\partial_1L$. We conclude that in Eq.~(\ref{eq_particle}), the particles on $\partial_0L$ and $\partial_1L$ are characterized by the irreps $R$ and $R^\ast$ respectively. The indices $i$ and $j$ label emergent  internal degrees of freedom on $\partial_0L$ and $\partial_1L$ respectively, and they can be changed by a local action. This will be explained below in Eqs.~(\ref{eq_39}), ~(\ref{eq_40}) and~(\ref{eq_41}).

For a particle living on a vertex $v$,  an arbitrary local action on this particle can be generated by local operators $B_{p, s}^{h}$ and $A_{v}^{g}$ as
\begin{align}
    \sum_{h, g\in G}n_{h, g}B_{p, s}^{h}A_{v}^{g}\, , \label{eq_local_action}
\end{align}
where $v\in s$ and $n_{h, g}$ is a coefficient. Mathematically,  Eq.~(\ref{eq_local_action}) can be viewed as an element of the quantum double algebra $DG$ with basis $\left\{D_{\left(h, g\right)}\equiv B_{p, s}^{h}A_{v}^{g}\right\}$ on a site $s$. Now we illustrate how the local operators change the internal degrees of freedom $i$ and $j$ in Eq.~(\ref{eq_particle}). We label an excited state with two particles as
\begin{align}
    \ket{R;i, j}=W_L\left( R ;i, j \right) \ket{\mathrm{GS}}\, .\label{eq_particle_state}
\end{align}
Applying the local operators $A_{v}^{g}$ and $B_{p, s}^{h}$ to the excited state gives
\begin{align}
    A_{v= \partial _0L}^{g}\ket{R;i, j}&=\sum_k{\Gamma _{ki}^{R}\left( g \right) }\ket{R;k, j}\, , \label{eq_39} \\
    A_{v= \partial _1L}^{g}\ket{R;i, j}&=\sum_k{\Gamma _{kj}^{R^{\ast}}\left( g \right) }\ket{R;i, k}\, , \label{eq_40} \\
    B_{p, s}^{h}\ket{R;i, j}&=\delta_{h, e}\ket{R;i, j}, \ \forall \left(p, s\right)\, , \label{eq_41}
\end{align}
where the derivations are collected in Appendix \ref{ap2.3}. Examining the actions of local operators in Eqs.~(\ref{eq_39}) and~(\ref{eq_40}), we see that $A_{v_0}^g$ transforms the state according to the representation matrix $\Gamma^R(g)$ acting on the row index $i$, while $A_{v_1}^g$ transforms it according to the complex conjugate representation $\Gamma^{R^\ast}(g)$ acting on the column index $j$. When $h \neq e$, the delta function $\delta_{h,e}$ in Eq.~(\ref{eq_41}) evaluates to zero, which means that the state $\ket{R;i,j}$ is an eigenstate of $B_{p,s}^{h}$ with eigenvalue zero. Crucially, the irrep label $R$ is invariant under the action of the local operators. The complex conjugate irrep label $R^\ast$ is therefore also invariant. This shows that the particle at $\partial_0L$ is characterized by $R$, and the particle at $\partial_1L$ is characterized by $R^\ast$. The indices $i$ and $j$ are respectively related to the internal degrees of freedom of the particles on $\partial_0L$ and $\partial_1L$. As we will see in Sec.~\ref{sec4}, these two particles are mutual antiparticles because there always exists a fusion channel to the vacuum (i.e., trivial excitation) when we fuse two particles respectively labeled by $R$ and $R^\ast$, as shown in Eq.~(\ref{eq_fusion_antiparticle}). Physically, this means that the two particles can annihilate into the vacuum.

The particle nature of the excitation can also be diagnosed by the action of the vertex term $A_{v_0=\partial_0L} = \frac{1}{|G|}\sum_{k\in G} A_{v_0}^k$ (or $A_{v_n=\partial_1L}$) on the state $\ket{R;i,j}$. Using Eq.~(\ref{eq_39}), we have
\begin{align}
    A_{v_0}\ket{R;i,j} &= \frac{1}{|G|}\sum_{g\in G} A_{v_0}^g \ket{R;i,j}= \frac{1}{|G|}\sum_{g\in G} \sum_{k} \Gamma_{k i}^{R}(g) \ket{R;k, j} \, .
\end{align}
The orthogonality relation of matrix elements guarantees that $\sum_{g\in G} \Gamma_{k i}^{R}(g)=0$ when the irrep $R$ is nontrivial. Therefore, when $R$ is nontrivial, this sum vanishes, yielding $A_{v_0}\ket{R;i,j}=0$. The same conclusion holds for $A_{v_n}$ at the other endpoint. Consequently, while the ground state satisfies $A_v\ket{\mathrm{GS}}=\ket{\mathrm{GS}}$ (eigenvalue $+1$) for all vertices, the excited state $\ket{R;i,j}$ has $A_v$ eigenvalue $0$ at the string endpoints, indicating the presence of particle excitations. 

Finally,  we give the connecting rule of the string operator $W_{L=L_1\cup L_2}\left( R ;i, j \right)$ for the particle as 
\begin{align}
    W_{L_1\cup L_2}\left( R;i, j \right) &=\sum_k{W_{L_2}\left( R;k, j \right) W_{L_1}\left( R;i, k \right)}\, .\label{eq_move_particle}
\end{align}
Since particles live at the endpoints of strings,  the operator $W_{L_1}\left( R;i, k \right)$ creates a pair of particles at vertices $v_0=\partial_0L_1$ and $v_1=\partial_1L_1$ respectively. This connecting rule allows us to move a particle from $v_1=\partial_1L_1$ to $v_2=\partial_1L_2$. We can similarly move a particle on $v_0=\partial_0L_1$ as well. The proof of Eq.~(\ref{eq_move_particle}) is provided in Appendix~\ref{ap2.4}. Interestingly, this connecting rule allows string operators to be arbitrarily bent on the cubic lattice such that the endpoints of the string can be moved to arbitrary vertices, which is a property missing in the fracton topological order represented by the X-cube model and its generalizations~\cite{PhysRevA.83.042330,PhysRevB.94.235157,PhysRevX.8.031051,PhysRevB.97.165106,PhysRevX.9.021010,PhysRevB.101.245134,PhysRevB.104.235127,PhysRevResearch.4.033111,PhysRevB.107.115169,d8gs-fnwt,PhysRevLett.130.216704}.

\subsection{Loop excitations}\label{Sec.3.3}
In this section,  we proceed to construct membrane operators for loops. Unlike one-dimensional ribbon operators used for anyons in 2D topological orders, an operator that creates a loop excitation in 3D must be a two-dimensional thickened membrane that consists of a dual part and a direct part. The dual part is formed by dual plaquettes and the boundary of the dual part is related to the flux of the loop. Besides nontrivial flux,  a loop can also carry some charge decorations. These charge decorations can freely travel along the boundary of the direct part that is formed by direct plaquettes. 

\begin{figure}
	\centering
	\includegraphics[scale=2.7, keepaspectratio,max width=\linewidth]{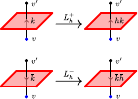}
	\caption{\textbf{The action of $L_{vv^{\prime}}^{h}$ on the edge $vv^{\prime}$ defined in Eq.~(\ref{eq_Lvvph}).} The {\color{red}red} and {\color{blue}blue} regions denote the {\color{red}\textbf{dual part}} and the {\color{blue}\textbf{direct part}} of the thickened membrane respectively. Since $L_{vv^{\prime}}^{h}$ involves only the single edge $vv^{\prime}$,  its dual part is uniquely determined as the dual plaquette intersecting $vv^{\prime}$. For illustration,  we choose the direct part as an infinitesimal direct plaquette $vvvv$ here, i.e., the blue point $v$. However,  the choice of the direct part is not unique. Any direct plaquette $p$ (of finite or infinitesimal area) that contains vertex $v$ can serve as the direct part without altering the form of the thickened membrane operator.}
	\label{fig_L_single_edge}
\end{figure}

We start by considering the simplest thickened membrane as shown in Fig.~\ref{fig_L_single_edge}. The red plaquette is the dual part and cuts the edge $vv^{\prime}$ of the cubic lattice; the blue vertex $v$ serves as the direct part and can be understood as an infinitesimal direct plaquette $vv_1v_2v_3$,  where $v=v_1=v_2=v_3$. Now we consider the action of $L_{vv^{\prime}}^{h}$ on the edge $vv^{\prime}$ defined as (see Fig.~\ref{fig_L_single_edge})
\begin{align}
    \label{eq_Lvvph}
L_{vv^{\prime}}^h =
\begin{cases}
    L_h^+, &\mathrm{if}\ v\rightarrow v^{\prime}\\
    L_h^-, &\mathrm{if}\ v\leftarrow v^{\prime}
\end{cases}.
\end{align}
That is, if the arrow of $vv^{\prime}$ points from the direct part to the dual part,  i.e.,  from $v$ to $v^{\prime}$,  we define $L_{vv^{\prime}}^{h}$ as $L_{h}^{+}$ to change the group element on $vv^{\prime}$. Otherwise,  we define $L_{vv^{\prime}}^{h}$ as $L_{h}^{-}$. Since applying $L_{vv^{\prime}}^{h}$ violates $B_p$ terms around $vv^{\prime}$,  we conclude that there exists a loop-like excited state living on the boundary of the thickened membrane. To create a spatially larger loop excitation,  we need to consider connecting several $L_{vv^{\prime}}^{h}$ terms. 

\begin{figure}
	\centering
	\includegraphics[scale=2.2, keepaspectratio,max width=\linewidth]{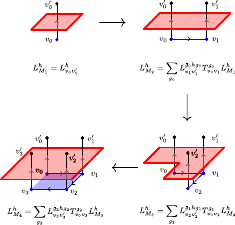}
	\caption{\textbf{Constructing the operator $L_{M_\alpha}^{h}$.} We use {\color{red} red} and {\color{blue} blue} to denote the {\color{red}\textbf{dual part}} and the {\color{blue}\textbf{direct part}} respectively. Note that the choice of direct parts has some arbitrariness because we only demand that $v_i$'s live on the direct parts. Different choices of the direct part lead to the same physical action of the operator $L_{M_\alpha}^{h}$ on the system. For any two thickened membranes $M_i$ and $M_j$ with $M_i \subseteq M_j$, we denote by $M_j \setminus M_i$ the region of $M_j$ that is not in $M_i$. Here,  we choose infinitesimal direct plaquettes $v_0v_0v_0v_0$,  $v_0v_0v_1v_1$,  and $v_1v_1v_2v_2$ as the direct parts of $M_1$,  $M_2\setminus M_1$,  and $M_3\setminus M_2$, respectively. The direct part of $M_4\setminus M_3$ is the plaquette $v_0v_1v_2v_3$,  which has finite area. A loop-like excited state lives on the boundary of the thickened membrane $M_\alpha$,  where $\alpha$ indicates that the dual part cuts $\alpha$ edges. $L_{M_\alpha}^{h}$ is obtained by using a recursive construction. The subscripts $v_0v_1$, $v_0v_2$, $v_0v_3$ of the $T$-operators denote the blue strings on their respective direct parts.}
	\label{fig_connect1_L}
\end{figure}
As shown in Fig.~\ref{fig_connect1_L},  consider creating a loop-like excited state on the boundary of an open thickened membrane $M_\alpha$,  whose dual part cuts $\alpha$ edges. We set the arrows of edges to point from the direct part to the dual part,  and we start by applying $L_{v_0v_0^{\prime}}^{h}$ to $v_0v_0^{\prime}$. We construct an operator $L_{M_\alpha}^{h}$ by using a recursive procedure as follows. The operator $L_{M_1}^{h}$ acts only on one edge $v_0v_0^{\prime}$ and is given by
\begin{align}
    L_{M_1}^{h}&=L_{v_0v_{0}^{\prime}}^{h}\, .
\end{align}
Note that the operator $L_{M_2}^{h}$ cannot be constructed by simply applying another $L_{v_1v_{1}^{\prime}}^{h}$ term to the edge $v_1v_{1}^{\prime}$ because the center of the plaquette $p=v_0^{\prime}v_0v_1v_1^{\prime}$ does not live on the boundary of $M_2$,  which means that the plaquette term $B_{p=v_0^{\prime}v_0v_1v_1^{\prime}}$ must be preserved. To obtain the correct form of the operator $L_{M_2}^{h}$,  we need to connect $L_{M_1}^{h}$ and $L_{v_1v_{1}^{\prime}}^{\bar{g}_0hg_0}$ as
\begin{align}
    L_{M_2}^{h}&=\sum_{g_0}{L_{v_1v_{1}^{\prime}}^{\bar{g}_0hg_0}T_{v_0v_1}^{g_0}L_{M_1}^{h}}=\sum_{g_0}{L_{v_1v_{1}^{\prime}}^{\bar{g}_0hg_0}T_{v_0v_1}^{g_0}L_{v_0v_{0}^{\prime}}^{h}}\, .
\end{align}
The $T_{v_0v_1}^{g_0}$ term and the sum over $g_0$ are introduced to ensure that $L_{M_2}^{h}$ does not violate the plaquette term $B_{p=v_0^{\prime}v_0v_1v_1^{\prime}}$,  as shown in Appendix~\ref{ap2.5}. Note that here we use two vertices in the subscript of a $T$-operator to denote a string (an edge is understood as a string of length $1$) that starts and ends on those two vertices. Following the same idea,  by connecting $L_{M_2}^{h}$ and $L_{v_2v_{2}^{\prime}}^{\bar{g}_1hg_1}$,  we construct the operator $L_{M_3}^{h}$ as
\begin{align}
    L_{M_3}^{h}&=\sum_{g_1}{L_{v_2v_{2}^{\prime}}^{\bar{g}_1hg_1}T_{v_0v_2}^{g_1}L_{M_2}^{h}}\,,
\end{align}
where $T_{v_0v_2}^{g_1}$ acts on an arbitrary string whose endpoints are denoted by $v_0$ and $v_2$.

We can keep constructing operators that involve more edges. Generally,  the \textit{connecting rule} of $L_{M_{\alpha}}^{h}$ is given by
\begin{align}
    L_{M_{\alpha}}^{h}=\sum_g{L_{v_pv_{p}^{\prime}}^{\bar{g}hg}T_{v_0v_p}^{g}L_{M_{\alpha -1}}^{h}}\, , \label{eq_L_connect1}
\end{align}
where $M_\alpha$ is obtained by cutting one more edge $v_pv_{p}^{\prime}$ than $M_{\alpha-1}$ and $T_{v_0v_p}^{g}$ acts on an arbitrary string with endpoints $v_0$ and $v_p$. The vertices $v_0$ and $v_p$ live on the direct part,  and the vertex $v_{p}^{\prime}$ lives on the dual part. If $v_0=v_p$,  then the term $T_{v_0v_p}^{g}$ is defined as $\delta_{g, e}$,  which means that Eq.~(\ref{eq_L_connect1}) now becomes
\begin{align}
    L_{M_{\alpha}}^{h}=\sum_g{L_{v_pv_{p}^{\prime}}^{\bar{g}hg}\delta _{g, e}L_{M_{\alpha -1}}^{h}}=L_{v_pv_{p}^{\prime}}^{h}L_{M_{\alpha -1}}^{h}\, .\label{eq_v0eqvp}
\end{align}
We further provide the \textit{connecting rule} of two operators $L_{M}^{h}$ and $ L_{M^{\prime}}^{\bar{g}hg}$ from Eq.~(\ref{eq_L_connect1}): 
\begin{align}
    L_{M\cup M^{\prime}}^{h}=\sum_g{L_{M^{\prime}}^{\bar{g}hg}T_{v_0v_p}^{g}L_{M}^{h}}\, , \label{eq_L_connect2}
\end{align}
where $M$ and $M^{\prime}$ are two membranes whose direct parts start at $v_0$ and $v_p$ respectively. The two membranes $M$ and $M^{\prime}$ do not overlap with each other but share a common boundary. Their dual parts do not cut the same edges, and their direct parts can always be chosen to share a common boundary without overlapping in area, see Fig.~\ref{fig_connect1_L} as an example. To construct $L_{M}^{h}$ and $L_{M^{\prime}}^{h}$,  we start by applying $L_{h}^{\pm}$ to $v_0v_0^{\prime}$ and $v_pv_p^{\prime}$. 

By using Eq.~(\ref{eq_L_connect2}), we can create any open membrane operator that violates all plaquette terms along the boundary of the dual part of $M$.  Specifically, as shown in Appendix~\ref{ap2.5}, the operator $L_M^{c}$ commutes with the plaquette term $B_{p}$ when the plaquette $p$ lies in the bulk of $M$. However, when $p$ lies on the boundary of $M$, we have the relation $B_{p} L_M^{c} = \delta_{c,e}L_M^{c} B_{p}$ (this relation can be understood as a special case of Eq.~(\ref{eq_B29})), which is analogous to the commutation relations for string operators in Eqs.~(\ref{eq_32}) and~(\ref{eq_33}). Consequently, applying $L_M^{c}$ to the ground state creates a state where the plaquette terms on the boundary have eigenvalue $0$ instead of $+1$, indicating the presence of loop excitations. More precisely, for any boundary plaquette $p$, one finds $B_{p} L_M^{c}\ket{\mathrm{GS}} = 0$, meaning that $L_M^{c}\ket{\mathrm{GS}}$ is an eigenstate of $B_{p}$ with eigenvalue $0$ when $c \ne e$.

\begin{figure}
	\centering
	\includegraphics[scale=1.85, keepaspectratio,max width=\linewidth]{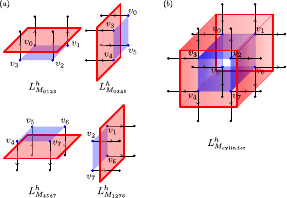}
	\caption{\textbf{Constructing the operator $L_{M_{\mathrm{cylindrical}}}^{h}$.} We use red and blue to denote the dual part and the direct part respectively. We connect four disk-like operators to form a cylindrical operator. For a disk-like membrane,  the subscript implies which vertices are on the direct part,  and the first number in the subscript implies the starting position of the corresponding operator.}
	\label{fig_connect2_L}
\end{figure}

Besides the disk-like operator shown in Fig.~\ref{fig_connect1_L},  we can also construct a cylindrical operator as shown in Fig.~\ref{fig_connect2_L}. We consider four disk-like thickened membranes and label them as $M_{0123}$,  $M_{0345}$,  $M_{4567}$,  and $M_{1276}$,  where the subscripts imply which vertices are on the direct part. For example,  $M_{0123}$ means that the vertices $v_0$,  $v_1$,  $v_2$,  and $v_3$ are on the direct part and we construct the corresponding disk-like operators by applying $L_{h}^{\pm}$ to the edge that contains the vertex $v_0$ first. Subsequently, we connect other $L$-operators that act on the edges containing the vertices $v_1$, $v_2$, and $v_3$. By using Eqs.~(\ref{eq_v0eqvp}) and~(\ref{eq_L_connect2}),  we connect operators $L_{M_{0123}}^{h}$ and $L_{M_{0345}}^{h}$ as
\begin{align}
    L_{M_{0123}\cup M_{0345}}^{h}=\sum_g{L_{M_{0345}}^{\bar{g}hg}\delta _{g, e}L_{M_{0123}}^{h}}=L_{M_{0345}}^{h}L_{M_{0123}}^{h}\, ,
\end{align}
where both $L_{M_{0123}}^{h}$ and $L_{M_{0345}}^{h}$ start at the edge that contains vertex $v_0$,  thus we have $T_{v_0v_p}^{g}=\delta_{g, e}$. Continuing to connect the operator $L_{M_{4567}}^{h}$,  we have:
\begin{align}
    L_{M_{0123}\cup M_{0345}\cup M_{4567}}^{h}&=\sum_g{L_{M_{4567}}^{\bar{g}hg}T_{v_0v_4}^{g}L_{M_{0123}\cup M_{0345}}^{h}} =\sum_g{L_{M_{4567}}^{\bar{g}hg}T_{v_0v_4}^{g}L_{M_{0123}}^{h}L_{M_{0345}}^{h}}\, ,
\end{align}
where the subscript of $M_{4567}$ indicates that the corresponding $L$-operator acts on the edge containing the vertex $v_4$ first. Therefore, the $T$-operator in the above equation acts on the string $v_0v_4$. Finally,  we derive the cylindrical operator by connecting the operator $L_{M_{1276}}^{h}$ as
\begin{align}
    L_{M_{\mathrm{cylindrical}}}^{h}&=\sum_k{L_{M_{1276}}^{\bar{k}hk}T_{v_0v_1}^{k}L_{M_{0123}\cup M_{0345}\cup M_{4567}}^{h}}\nonumber \\
    &=\sum_{g, k}{L_{M_{1276}}^{\bar{k}hk}T_{v_0v_1}^{k}L_{M_{4567}}^{\bar{g}hg}T_{v_0v_4}^{g}L_{M_{0345}}^{h}L_{M_{0123}}^{h}}\, ,
\end{align}
where $M_{\mathrm{cylindrical}}$ is defined as $M_{\mathrm{cylindrical}}= M_{0123}\cup M_{0345}\cup M_{4567}\cup M_{1276}$. The plaquette terms are violated along the two boundaries of the dual part of the cylinder.

As shown in Appendix~\ref{ap2.5},  the connecting rule in Eq.~(\ref{eq_L_connect1}) ensures that plaquette terms that live in the bulk of $M$ are not violated by $L_{M}^{h}$. We also show in Appendix~\ref{ap2.6} that,  among vertex terms,  $A_{v_0}$ does not commute with $L_{M}^{h}$,  if and only if $h$ is not in the center $Z(G)=\left\{ z\in G \mid zg=gz, \forall g\in G \right\}$ of $G$. All other vertex terms commute with $L_{M}^{h}$.

Having derived the connecting rule for the $L$-operators,  we are ready to construct membrane operators for loop excitations which should be labeled by data that are invariant under local perturbations. Topologically, loops are characterized by a pair of topological data $\left(C, R\right)$,  where $C$ is a conjugacy class and $R$ is an irrep of the centralizer of $C$. We write a membrane operator for loops as
\begin{align}
    W_M\left( C, R;c, j;c^{\prime}, j^{\prime} \right) =\sum_{g\in Z_r}{\Gamma _{jj^{\prime}}^{R^{\ast}}\left( g \right) L_{M}^{c}T_{P}^{q_cg\bar{q}_{c^{\prime}}}}\, , \label{eq_loop}
\end{align}
where $M$ is a thickened membrane,  $R^\ast$ denotes the complex conjugate representation,  $c\in C$,  $Z_r$ is the centralizer of a representative $r$ of the conjugacy class $C=\left\{ q_cr\bar{q}_{c} \mid q_c\in G \right\}$. Given a fixed $r\in C$,  for an arbitrary element $c\in C$,  there are generally different choices of $q_c$ satisfying    $c=q_cr\bar{q}_c$. Thus,  we need to preselect $r$ and $q_c$ for a given $C$ and $c$. Different choices of $r$ and $q_c$ do not affect physics. $j$ and $j^{\prime}$ are row and column indices of the matrix $\Gamma ^{R^{\ast}}\left( g \right)$. In the following discussion,  we will see that $c$,  $c^{\prime}$,  $j$ and $j^{\prime}$ are just some local degrees of freedom and they can be changed by local operators. Here, $P$ is an arbitrary open string that lives on the direct part of the thickened membrane $M$. 

We demand that the two endpoints of $P$ reside on the boundary of the direct part of $M$. For example,  consider the membrane $M_4$ shown in Fig.~\ref{fig_connect1_L},  where we use blue to denote the direct part. The boundary of the direct part of $M$ is a closed path $\left(v_0, v_1, v_2, v_3, v_0\right)$,  so choices such as $P=\left(v_0, v_1\right)$,  $P=\left(v_1, v_0, v_3\right)$,  $P=\left(v_0, v_1, v_2, v_3\right)$,  etc.,  are all legitimate. If we consider a cylindrical membrane shown in  Fig.~\ref{fig_connect2_L},  the boundaries of the direct part consist of two closed paths $\left(v_0, v_1, v_6, v_5, v_0\right)$ and $\left(v_2, v_3, v_4, v_7, v_2\right)$. We are allowed to choose $P=\left(v_0, v_3\right)$,  or $P=\left(v_0, v_3, v_2\right)$,  or $P=\left(v_3, v_2, v_7, v_6\right)$,  etc.,  \textit{as long as the two endpoints of $P$ reside on different boundaries}. As in the particle case discussed in Sec.~\ref{Sec.3.2}, a cylindrical membrane operator $W_M(C,R;c,j;c^{\prime},j^{\prime})$ creates a pair of excitations: a loop on $\partial_0M$ and an antiloop on $\partial_1M$. Requiring that the two endpoints of $P$ lie on different boundaries of a cylindrical membrane guarantees that the operator in Eq.~(\ref{eq_loop}) creates a loop--antiloop pair carrying opposite charge decorations.

The conjugacy class $C$ is viewed as the flux of the loop and it is related to the $L_M^c$ term in Eq.~(\ref{eq_loop}). The irrep $R$ is viewed as the charge decoration of the loop and is related to the $T_{P}^{q_cg\bar{q}_{c^{\prime}}}$ term in Eq.~(\ref{eq_loop}). Physically,  a charge decoration is allowed to freely travel along a loop excitation. Thus,  if we consider the thickened membrane $M$ in Eq.~(\ref{eq_loop}) to be a disk-like membrane similar to those in Fig.~\ref{fig_connect1_L},  there will be a pair of charge and anti-charge decorations on a loop excitation. Thus, the total charge decoration carried by this loop sums to the trivial charge decoration. The total topological quantum number of this loop is the same as that of a pure loop (i.e.,  without charge decorations). To encompass decorated loops (i.e., the total charge decorations can be nontrivial),  we need to take $M$ to be a cylindrical membrane as shown in Fig.~\ref{fig_connect2_L},  and let the two endpoints of the string $P$ live on the two boundaries of the direct part of the cylindrical membrane. In this case,  we create a loop--antiloop pair,  and they carry a charge decoration and an anti-charge decoration respectively. Besides,  the operator in Eq.~(\ref{eq_loop}) automatically becomes the string operator in Eq.~(\ref{eq_particle}) for a particle--antiparticle pair when we set $c=e$. In the following discussion,  we only consider $M$ to be a cylindrical membrane.

Finally,  we illustrate how local degrees of freedom in Eq.~(\ref{eq_loop}) are changed by the local operators. For the cylindrical membrane operator $W_M\left( C, R;c, j;c^{\prime}, j^{\prime} \right)$,  we denote the two boundaries of $M$ by $\partial_0M$ and $\partial_1M$. Here we adopt the same notation as for strings: $\partial_0M$ denotes the boundary of $M$ where the construction of the membrane operator begins, and $\partial_1M$ denotes the opposite boundary where the construction finishes. This convention is consistent with the notation $\partial_0L$ and $\partial_1L$ introduced for strings below Eq.~(\ref{eq_T_connect}).

Without loss of generality,  we consider that the $L_M^{c}$ term acts on the edge $v_0v_0^{\prime}\ni p_0$ first (according to our definition, $p_0$ resides on $\partial_0M$),  and the last edge acted on by the $L_M^{c}$ term is denoted by $v_1v_1^{\prime}\ni p_1$ (according to our definition, $p_1$ resides on $\partial_1M$),  where plaquettes $p_0$ and $p_1$ are violated. We set $\partial_0P=v_0$,  $\partial_1P=v_1$,  and require that $v_0$ and $p_0$ form a site $s_0=\left(v_0, p_0\right)$ and that $v_1$ and $p_1$ form a site $s_1=\left(v_1, p_1\right)$. A state for a pair of loops is written as $W_M\left( C, R;c, j;c^{\prime}, j^{\prime} \right) \ket{\mathrm{GS}}$. As shown in Appendix~\ref{ap2.7},  all local actions on the boundary $\partial_0M$ ($\partial_1M$) can be derived by considering the actions of $A_{v_0}^{g}$ ($A_{v_1}^{g}$) and $B_{p_0, s_0}^{h}$ ($B_{p_1, s_1}^{h}$) on the site $s_0$ ($s_1$). On the site $s_0\in \partial_0M$,  the actions of local operators are given by
\begin{align}
    A_{v_0}^{g}W_M\left( C, R;c, j;c^{\prime}, j^{\prime} \right) \ket{\mathrm{GS}} &=\sum_i{\Gamma _{ij}^{R}\left( \bar{q}_{gc\bar{g}}gq_c \right)}W_M\left( C, R;gc\bar{g}, i;c^{\prime}, j^{\prime} \right) \ket{\mathrm{GS}} \, , \label{eq_av} \\
    B_{p_0, s_0}^{h}W_M\left( C, R;c, j;c^{\prime}, j^{\prime} \right) \ket{\mathrm{GS}} &=\delta _{c, h}W_M\left( C, R;c, j;c^{\prime}, j^{\prime} \right) \ket{\mathrm{GS}} \, .\label{eq_bp}
\end{align}
On the site $s_1\in \partial_1M$,  we have
\begin{align}
    A_{v_1}^{g}W_M\left( C, R;c, j;c^{\prime}, j^{\prime} \right) \ket{\mathrm{GS}} &=\sum_i{\Gamma _{ij^{\prime}}^{R^{\ast}}\left( \bar{q}_{gc^{\prime}\bar{g}}gq_{c^{\prime}} \right)}W_M\left( C, R;c, j;gc^{\prime}\bar{g}, i \right) \ket{\mathrm{GS}} \, , \label{eq_av1} \\
    B_{p_1, s_1}^{h}W_M\left( C, R;c, j;c^{\prime}, j^{\prime} \right) \ket{\mathrm{GS}} &=\delta _{\bar{c}^{\prime}, h}W_M\left( C, R;c, j;c^{\prime}, j^{\prime} \right) \ket{\mathrm{GS}} \, .\label{eq_bp1}
\end{align}
Comparing Eqs.~(\ref{eq_av}) and~(\ref{eq_av1}), we see that $A_{v_0}^g$ acts on the state via the representation matrix $\Gamma^R$ evaluated at $\bar{q}_{gc\bar{g}}gq_c$, while $A_{v_1}^g$ acts via its complex conjugate $\Gamma^{R^\ast}$ evaluated at $\bar{q}_{gc^{\prime}\bar{g}}gq_{c^{\prime}}$. Moreover, Eq.~(\ref{eq_bp}) shows that the flux on $\partial_0M$ is labeled by $c\in C$, while Eq.~(\ref{eq_bp1}) shows that the flux on $\partial_1M$ is labeled by $\bar{c}^{\prime}\in \bar{C}$. 
The label $\left(C, R\right)$ is conserved under the actions of local operators,  which justifies the statement: a topological excitation is characterized by a pair of data  $\left(C, R\right)$. We conclude  that the loop on $\partial_0M$ carries the topological label $(C,R)$, and the loop on $\partial_1M$ carries the conjugate label $(\bar{C},R^\ast)$. The local degrees of freedom $c$ ($c^{\prime}$) and $j$ ($j^{\prime}$) can be changed by the local operators on the site $s_0$ ($s_1$),  leading to the conclusion that $c$ ($c^{\prime}$) and $j$ ($j^{\prime}$) label the internal degrees of freedom of the loop living on the boundary $\partial_0M$ ($\partial_1M$). As we will see in Sec.~\ref{sec4}, the two excitations on $\partial_0M$ and $\partial_1M$ are mutual antiexcitations: fusing the loop labeled by $\left(C, R\right)$ and the loop labeled by $\left(\bar{C},R^\ast\right)$ always has a channel to the vacuum (trivial excitation), as shown in Eq.~(\ref{eq_fusion_antiparticle}).

Finally, we comment on the relation between the excitation operators and the plaquette and vertex operators, since both are built from the same elementary operators. The plaquette operator $B^h_{p,s}$ is a product of $T$-operators around a single plaquette, while the vertex operator $A^g_v$ is a product of $L$-operators around a single vertex. Correspondingly, a Wilson line in Eq.~(\ref{eq_particle}) whose closed string wraps a single plaquette $p$ reduces to a combination of the flux projectors: $\sum_h{\Gamma _{ij}^{R^{\ast}}\left( h \right) B_{p,s}^{h}}$, where the string starts and ends on the vertex $v\in s$. The Wilson surface in Eq.~(\ref{eq_loop}) whose dual part cuts all edges that are connected to the single vertex $v$ (i.e., a surface enclosing one dual cube, with the infinitesimal direct part chosen to be $vvvv$ so that the connecting rule in Eq.~(\ref{eq_L_connect1}) does not introduce nontrivial $T$-operators) reduces to the operator $A_{v}^{c}$. This is similar to the familiar statement for the 3D toric code that the closed Wilson line and surface around a single plaquette and a single dual cube reduce to the plaquette and vertex terms, respectively.

\section{Fusion rules and quantum dimensions}\label{sec4}
In Sec.~\ref{Sec.3},  by linearly combining basic $T$- and $L$-operators,  we explicitly construct the operator in Eq.~(\ref{eq_loop}) for topological excitations and obtain the corresponding connecting rules,  which allow us to create,  move,  and deform excitations. To encompass both pure loops and decorated loops,  we choose the thickened membrane $M$ in Eq.~(\ref{eq_loop}) to be cylindrical. In this section,  we proceed to consider a process of fusing topological excitations in the microscopic model. We will first illustrate the general fusion rules in Sec.~\ref{Sec.4.1} and then take  $G=\mathbb{Z}_N$ and  $G=\mathbb{D}_3$ as examples in Sec.~\ref{Sec.4.2} and Sec.~\ref{Sec.4.3} respectively.

\subsection{General microscopic theory}\label{Sec.4.1}
As mentioned in Sec.~\ref{Sec.3.3},  applying the operator in Eq.~(\ref{eq_loop}) to a ground state creates a pair of loops living on the two boundaries of the cylindrical membrane $M$. We write such a state as $W_M\left( C, R;c, j;c^{\prime}, j^{\prime} \right) \ket{\mathrm{GS}}$. Fixing $\left(C, R\right)$ and varying $c, j, c^{\prime}, j^{\prime}$,  we obtain a set of basis states
\begin{align}
    \left\{ W_M\left( C, R;c, j;c^{\prime}, j^{\prime} \right) \ket{\mathrm{GS}} \mid c, c^{\prime}\in C;\, j, j^{\prime}=1, \cdots , n_R \right\}  \, ,
\end{align}
where $C$ denotes a conjugacy class of $G$, $R$ denotes an irrep of the centralizer subgroup of the representative $r\in C$, and $n_R$ is the dimension of the irrep $R$. This set spans the Hilbert space $\mathcal{H} _{\left( C, R \right)}$ of the two loops. From Eqs.~(\ref{eq_av}), ~(\ref{eq_bp}), ~(\ref{eq_av1}) and~(\ref{eq_bp1}),  it follows that for any vector $\ket{w}\in \mathcal{H} _{\left( C, R \right)}$,  we have $\left(\sum_{h, g\in G}n_{h, g}B_{p, s}^{h}A_{v}^{g}\right)\ket{w}\in \mathcal{H} _{\left( C, R \right)}$. Thus,  we  call $\mathcal{H} _{\left( C, R \right)}$ an invariant space under the local actions in Eq.~(\ref{eq_local_action}). Each loop denoted by $\left[C, R\right]$\footnote{Throughout this paper, we use the notation $[C, R]$ with square brackets to denote a topological excitation in the quantum double model, emphasizing that the pair $(C, R)$ labels a specific excitation type. The parentheses alone simply indicate the pair of data, i.e., a conjugacy class $C$ and an irreducible representation $R$ of its centralizer.} corresponds to such an invariant space $\mathcal{H} _{\left( C, R \right)}$,  whose dimension is given by $|C|^2\dim(R)^2$. If we consider all possible choices of $\left(C, R\right)$,  the dimension of the total excitation space $\oplus_{\left( C, R \right)}\mathcal{H} _{\left( C, R \right)}$ becomes
  $
	\sum_{C, R}|C|^2\dim(R)^2 
	= \sum_C |C|^2 \frac{|G|}{|C|} 
	= |G|\sum_C |C|
	=|G|^2.$ 
It follows that each of the two boundaries of $M$ independently contributes a factor of $|G|$ to the dimension of the total excitation space.

Eqs.~(\ref{eq_av}) and~(\ref{eq_bp}) (Eqs.~(\ref{eq_av1}) and~(\ref{eq_bp1})) show that local operators on the boundaries $\partial_0M$ ($\partial_1M$) only change the labels $c$ and $j$ ($c^{\prime}$ and $j^{\prime}$). Thus,  under the local actions,  the Hilbert space $\mathcal{H} _{\left( C, R \right)}$ can be written as
\begin{align}
    \mathcal{H} _{\left( C, R \right)}=V_{\left( C, R \right)}\otimes V_{\left( \bar{C}, R^{\ast} \right)}\, ,
\end{align}
where $\otimes$ denotes the tensor product, $\bar{C}=\left\{ g\bar{c}\bar{g} \mid g\in G \right\} $, and $R^{\ast}$ denotes the complex conjugate representation. $V_{\left( C, R \right)}$ and $V_{\left( \bar{C}, R^{\ast} \right)}$ are local spaces of the loop $\left[C, R\right]$ and the antiloop $\left[\bar{C}, R^{\ast}\right]$ living on $\partial_0M$ and $\partial_1M$ respectively. When we focus on the area near the boundary $\partial_0M$,  we denote the state $W_M\left( C, R;c, j;c^{\prime}, j^{\prime} \right) \ket{\mathrm{GS}}$ in this area as $\ket{ c, j }_{\left(C, R\right)}$ because $c^{\prime}$ and $j^{\prime}$ are invariant under the local actions. $\ket{ c, j }_{\left(C, R\right)}$ only labels the local internal degrees of freedom $c$ and $j$ of the loop $\left[C, R\right]$ living on $\partial_0M$. Thus,  by varying $c$ and $j$ over all allowed values with fixed $c^{\prime}$ and $j^{\prime}$,  we obtain a set of basis states $\left\{\ket{ c, j }_{\left(C, R\right)} \mid c\in C, \, j=1, \cdots, n_R\right\}$,  which spans the local space $V_{\left( C, R \right)}$ of the loop $\left[C, R\right]$. The local space $V_{\left( C, R \right)}$ is invariant under the local actions near the boundary $\partial_0M$. Similarly,  the invariant local space $V_{\left( \bar{C}, R^{\ast} \right)}$ is spanned by $\left\{\ket{ c^{\prime}, j^{\prime} }_{\left(\bar{C}, R^{\ast}\right)} |c^{\prime}\in C, \, j^{\prime}=1, \cdots, n_R\right\}$,  where $\ket{ c^{\prime}, j^{\prime}}_{\left(\bar{C}, R^{\ast}\right)}$ labels the local internal degrees of freedom $c^{\prime}, j^{\prime}$ of the antiloop $\left[\bar{C}, R^{\ast}\right]$ living on $\partial_1M$ and is understood as the state $W_M\left( C, R;c, j;c^{\prime}, j^{\prime} \right) \ket{\mathrm{GS}}$ in the area near the boundary $\partial_1M$. 

The notation above also apply to the particles because Eq.~(\ref{eq_loop}) automatically becomes the string operator in Eq.~(\ref{eq_particle}) for a particle--antiparticle pair when we set $c=e$. In this case,  $\ket{ c, j }_{\left(C, R\right)}$ and $\ket{ c^{\prime}, j^{\prime}}_{\left(\bar{C}, R^{\ast}\right)}$ are simplified as $\ket{ j }_{\left(R\right)}$ and $\ket{ j^{\prime}}_{\left(R^{\ast}\right)}$ and they label the internal degrees of freedom on vertices $\partial_0P$ and $\partial_1P$. 

Consider a site $s=\left(v_0, p_0\right)$ on the boundary $\partial_0M$ and require that the string $P$ starts at $v_0$. Then all possible local actions on the excitation $\left[C, R\right]$ are given by Eq.~(\ref{eq_local_action}),  which form the quantum double algebra $DG$ with basis $\left\{D_{\left(h, g\right)}= B_{p, s}^{h}A_{v}^{g}\right\}$. An irrep of the quantum double $DG$ is also characterized by the pair $\left(C, R\right)$. Mathematically, the local space $V_{\left( C, R \right)}$ of the excitation $\left[C, R\right]$ can be regarded as the \textit{representation space of the irrep}. The action of $DG$ on $V_{\left( C, R \right)}$ is given by
\begin{align}
    \Pi ^{\left( C, R \right)}\left( D_{\left( h, g \right)} \right)\ket{c, j}_{\left( C, R \right)} &=\sum_{i}{\Gamma _{ij}^{R}\left( \bar{q}_{gc\bar{g}}gq_c \right) \delta _{h, gc\bar{g}}\ket{gc\bar{g}, i}_{\left( C, R \right)}}\, , \label{eq_action_DG}
\end{align}
where $	\Pi ^{\left( C, R \right)}$ is the irrep of $DG$. Eq.~(\ref{eq_action_DG}) is essentially the mathematical translation of the physical actions described in Eqs.~(\ref{eq_av}) and~(\ref{eq_bp}) into the language of representation theory. To be more specific, the right-hand side of Eq.~(\ref{eq_action_DG}) encodes the combined effect of $B_{p,s}^h$ and $A_v^g$: the factor $\Gamma_{ij}^{R}(\bar{q}_{gc\bar{g}} g q_c)$ and the summation over $i$ together with the replacement $\ket{c,j}_{\left(C,R\right)} \to \ket{gc\bar{g}, i}_{\left(C,R\right)}$ reflect that $A_v^g$ generally changes both the internal degrees of freedom associated with the flux and charge carried by the excitation; the delta function $\delta_{h, gc\bar{g}}$ comes directly from Eq.~(\ref{eq_bp}), enforcing that the flux after the action of $A_v^g$ must equal $h$.  Thus, Eq.~(\ref{eq_action_DG}) shows that the physical actions of $A_v^g$ followed by $B_{p,s}^h$ on the state precisely realize the representation of the quantum double algebra $DG$ on the local Hilbert space $V_{(C,R)}$.

For $V_{\left( \bar{C}, R^{\ast} \right)}$,  one can similarly regard it as a representation space of an irrep of $DG$ labeled by $\left(\bar{C}, R^{\ast}\right)$. In the following discussion,  we only need to focus on the local space $V_{\left(C, R\right)}$ of the excitation $\left[C, R\right]$ because mathematically,  $V_{\left( \bar{C}, R^{\ast} \right)}$ is just the dual space of $V_{\left(C, R\right)}$. The fusion and shrinking rules derived from $V_{\left( \bar{C}, R^{\ast} \right)}$ are consistent with those derived from $V_{\left(C, R\right)}$.

\begin{figure}
	\centering
	\includegraphics[scale=2.4, keepaspectratio,max width=\linewidth]{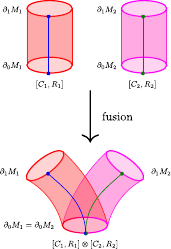}
	\caption{\textbf{Fusing two excitations.} As shown at the top,  we consider the excitations $\left[C_1, R_1\right]$ and $\left[C_2, R_2\right]$ living on the two boundaries $\partial_0M_1$ and $\partial_0M_2$ respectively. Red and purple denote the operators $L_{M_1}^{c_1\in C_1}$ and $L_{M_2}^{c_2\in C_2}$ in Eq.~(\ref{eq_loop}) respectively and blue and green denote the operators $T_{P_1}^{q_{c_1}g\bar{q}_{c_{1}^{\prime}}}$ and $T_{P_2}^{q_{c_2}g\bar{q}_{c_2^{\prime}}}$ in Eq.~(\ref{eq_loop}) respectively. We bring $\left[C_1, R_1\right]$ and $\left[C_2, R_2\right]$ to a common position $\partial_0M_1=\partial_0M_2$ to implement the fusion process. This microscopic realization of fusion corresponds directly to the fusion in the continuum field theory,  where two Wilson operators are brought together at the same location.}
	\label{fig_fusion}
\end{figure}

In our 3D quantum double model,  we move two excitations $\left[C_1, R_1\right]$ and $\left[C_2, R_2\right]$ to the same position $\partial_0M_1=\partial_0M_2$ as shown in Fig.~\ref{fig_fusion} to implement a fusion process. This physical picture of fusion in the lattice model coincides with that in the continuum field theory as shown in Sec.~\ref{Sec.2.2}. When the two excitations are brought together at $\partial_0M_1$,  we denote the state given by
\begin{align}
    W_{M_1}\!\left( C_1, R_1;c_1, j_1;c^{\prime}_1, j^{\prime}_1 \right)\!W_{M_2}\!\left( C_2, R_2;c_2, j_2;c^{\prime}_2, j^{\prime}_2 \right) \ket{\mathrm{GS}} \label{eq_fusion_state}
\end{align}
as the local state
\begin{align}
    \ket{c_1, j_1}_{\left( C_1, R_1 \right)}\otimes\ket{c_2, j_2}_{\left( C_2, R_2 \right)} \label{eq_fusion_local_state}
\end{align}
in the area near the boundary $\partial_0M_1$ because $c^{\prime}_1$,  $j^{\prime}_1$,  $c^{\prime}_2$,  and $j^{\prime}_2$ are invariant under the local actions. The local action of $D_{\left(h, g\right)}=B_{p_0, s_0}^{h}A_{v_0}^{g}$ on the state in Eq.~(\ref{eq_fusion_state}) is given by
\begin{align}
    &B_{p_0, s_0}^{h}A_{v_0}^{g}W_{M_1}\left( C_1, R_1;c_1, j_1;c^{\prime}_1, j^{\prime}_1 \right) \times W_{M_2}\left( C_2, R_2;c_2, j_2;c^{\prime}_2, j^{\prime}_2 \right) \ket{\mathrm{GS}}\nonumber \\
    &=\sum_{i_1}\Gamma _{i_1j_1}^{R_1}\left( \bar{q}_{gc_1\bar{g}}gq_{c_1} \right)W_{M_1}\left( C_1, R_1;gc_1\bar{g}, i_1;c^{\prime}_1, j^{\prime}_1 \right)\nonumber \\
    &\times B_{p_0, s_0}^{g\bar{c}_1\bar{g}h}A_{v_0}^{g} W_{M_2}\left( C_2, R_2;c_2, j_2;c^{\prime}_2, j^{\prime}_2 \right) \ket{\mathrm{GS}} \nonumber \\
    &=\sum_{i_1}\Gamma _{i_1j_1}^{R_1}\left( \bar{q}_{gc_1\bar{g}}gq_{c_1} \right)W_{M_1}\left( C_1, R_1;gc_1\bar{g}, i_1;c^{\prime}_1, j^{\prime}_1 \right)\nonumber \\
    &\times \sum_{i_2}\Gamma _{i_2j_2}^{R_2}\left( \bar{q}_{gc_2\bar{g}}gq_{c_2} \right)W_{M_2}\left( C_2, R_2;gc_2\bar{g}, i_2;c^{\prime}_2, j^{\prime}_2 \right) \times B_{p_0, s_0}^{g\bar{c}_2\bar{c}_1\bar{g}h}A_{v_0}^{g} \ket{\mathrm{GS}}\, .\label{eq_fusion_action}
\end{align}
Note that
$	B_{p_0, s_0}^{g\bar{c}_2\bar{c}_1\bar{g}h}A_{v_0}^{g} \ket{\mathrm{GS}}=\delta_{h, gc_1c_2\bar{g}}\ket{\mathrm{GS}}
	=\sum_{k\in G}\delta_{h\bar{k}, gc_1\bar{g}}\delta_{k, gc_2\bar{g}}\ket{\mathrm{GS}}\, .
 $ 
Thus,  by using Eq.~(\ref{eq_action_DG}),  we conclude that the wave function in Eq.~(\ref{eq_fusion_action}) corresponds to the local state given by
\begin{align}
    &\sum_{k\in G}\Pi ^{\left( C_1, R_1 \right)}\left(D_{\left( h\bar{k}, g \right)}\right)\ket{c_1, j_1}_{\left( C_1, R_1 \right)}\nonumber \\
    &\otimes \Pi ^{\left( C_2, R_2 \right)}\left(D_{\left( k, g \right)}\right)\ket{c_2, j_2}_{\left( C_2, R_2 \right)}\, .\label{eq_local_fusion_action}
\end{align}

From Eq.~(\ref{eq_local_fusion_action}),  we can see that the local Hilbert space on the boundary $\partial_0M_1$ spanned by $\left\{\ket{c_1, j_1}_{\left( C_1, R_1 \right)}\otimes\ket{c_2, j_2}_{\left( C_2, R_2 \right)}\right\}$ can be written as $V_{\left( C_1, R_1 \right)}\otimes V_{\left( C_2, R_2 \right)}$,  which forms a representation space of a tensor product representation $\Pi ^{\left( C_1, R_1 \right)}\otimes \Pi ^{\left( C_2, R_2 \right)}$. Physically, this tensor product space describes all possible joint internal states when two excitations are brought to the same spatial region, and the decomposition below tells us which fused excitations can emerge. Generally,  the tensor product representation decomposes as
\begin{align}
    \Pi ^{\left( C_1, R_1 \right)}\otimes \Pi ^{\left( C_2, R_2 \right)}&=\bigoplus_{\left( C, R \right)}N_{\left( C, R \right)}^{\left( C_1, R_1 \right) \left( C_2, R_2 \right)}\Pi ^{\left( C, R \right)}\, , \nonumber \\
    V_{\left( C_1, R_1 \right)}\otimes V_{\left( C_2, R_2 \right)}&=\bigoplus_{\left( C, R \right)}N_{\left( C, R \right)}^{\left( C_1, R_1 \right) \left( C_2, R_2 \right)}V_{\left( C, R \right)}\, , \label{eq_fusion_rules}
\end{align}
where $\oplus$ is the direct sum that exhausts all possible choices of $\left(C, R\right)$. Since each excitation corresponds to an invariant space $V_{\left( C, R \right)}$ and the tensor product space $V_{\left( C_1, R_1 \right)}\otimes V_{\left( C_2, R_2 \right)}$ decomposes as a direct sum of several invariant spaces,  Eq.~(\ref{eq_fusion_rules}) is interpreted as the fusion rule in Eq.~(\ref{eq_fusion}) of the quantum double,  with the  fusion coefficient $N_{\left( C, R \right)}^{\left( C_1, R_1 \right)\left( C_2, R_2 \right)}$ given by
\begin{align}
    N_{\left( C, R \right)}^{\left( C_1, R_1 \right)\left( C_2, R_2 \right)} &=\frac{1}{\left| G \right|}\sum_{h, g}{tr\left[ \Pi ^{{\left( C_1, R_1 \right)}}\otimes \Pi ^{{\left( C_2, R_2 \right)}}\left( \Delta \left( D_{\left( h, g \right)} \right) \right) \right]}\nonumber \\
    &\times\left\{ tr\left[ \Pi ^{{\left( C, R \right)}}\left( D_{\left( h, g \right)} \right) \right] \right\} ^{\ast}\, , \label{eq_fusion_coe}
\end{align}
where $\ast$ means complex conjugate and we define
$	\Delta \left( D_{\left( h, g \right)} \right) =\sum_{k\in G}{D_{\left( h\bar{k}, g \right)}\otimes D_{\left( k, g \right)}}\, . $ 
If $N_{\left( C, R \right)}^{\left( C_1, R_1 \right)\left( C_2, R_2 \right)}$ does not vanish,  then there are $N_{\left( C, R \right)}^{\left( C_1, R_1 \right)\left( C_2, R_2 \right)}$ channels to fuse $\left[ C_1, R_1\right]$ and $\left[ C_2, R_2\right]$ to $\left[ C, R\right]$. The fusion is Abelian if for the given pair $(C_1,R_1)$ and $(C_2,R_2)$, there is exactly one $(C,R)$ with $N_{\left( C, R \right)}^{\left( C_1, R_1 \right)\left( C_2, R_2 \right)}=1$ and all other fusion coefficients vanish; otherwise, the fusion is non-Abelian, meaning that the result can be a superposition of different fusion channels. A special case of Eq.~(\ref{eq_fusion_coe}) is when $(C_2,R_2) = (\bar{C}_1,R_1^\ast)$. According to the general representation theory of the quantum double $DG$, the tensor product of an irrep $	\Pi ^{\left( C_1, R_1 \right)}$ with its dual $	\Pi ^{\left( \bar{C}_1, R_1^\ast \right)}$ contains the trivial representation exactly once. Consequently, the fusion coefficient for the trivial excitation $[C_e,Id]$ satisfies
\begin{align}
    N_{(C_e,Id)}^{(C_1,R_1),(\bar{C}_1,R_1^\ast)} = 1\, . \label{eq_fusion_antiparticle}
\end{align}
This result confirms that $[C_1,R_1]$ and $[\bar{C}_1,R_1^\ast]$ are mutual antiexcitations because when we consider fusing them, there exists a unique channel to the vacuum (trivial excitation). Besides, the equation above shows that the operator $W_L(R;i,j)$ constructed in Sec.~\ref{Sec.3.2} creates a particle--antiparticle pair, and $W_M(C,R;c,j;c^{\prime},j^{\prime})$ constructed in Sec.~\ref{Sec.3.3} creates a loop--antiloop pair.

\subsection{Examples of Abelian fusion rules}\label{Sec.4.2}
Here,  we provide an example of Abelian fusion rules in the 3D quantum double model. Consider an Abelian group $G=\mathbb{Z}_N$ and denote the generator of $\mathbb{Z}_N$ by $g$. A group element can be written as $g^\alpha$,  and $g^\alpha$ itself forms a conjugacy class. Since the corresponding centralizer is $Z_{g^\alpha}=G$,  and $G$ has $\left| G \right|$ one-dimensional irreps,  we find $\left| G \right|\times \left| G \right|$ excitations labeled by $\left[g^\alpha, R^{\left(\beta\right)}\right]$,  where $\alpha, \beta=0, 1, 2, \cdots, \left| G \right|-1$ and $R^{\left(\beta\right)}$ labels an irrep. We write $\Gamma ^{R^{\left( \beta \right)}}\left( g^{\alpha} \right) =\omega ^{\alpha \beta}$ with $\omega=\exp\left(2\pi \mathrm{i}/N\right)$. Since $G$ is an Abelian group,  the tensor product representation is directly given by 
	 $\Pi^{\left(g^{\alpha_1}, R^{\left(\beta_1\right)}\right)}\otimes \Pi^{\left(g^{\alpha_2}, R^{\left(\beta_2\right)}\right)} 
	= \Pi^{\left(g^{\left(\alpha_1+\alpha_2 \, \, \text{mod}\, \, N\right)}, R^{\left(\beta_1+\beta_2\, \, \text{mod}\, \, N\right)}\right)}\, $
which leads to the fusion rules:
\begin{align}
    \left[g^{\alpha_1}, R^{\left(\beta_1\right)}\right]\otimes\left[g^{\alpha_2}, R^{\left(\beta_2\right)}\right] &=\left[g^{\left(\alpha_1+\alpha_2 \, \, \text{mod}\, \, N\right)}, R^{\left(\beta_1+\beta_2\, \, \text{mod}\, \, N\right)}\right]\, .\label{eq_ZN_fusion}
\end{align}
The corresponding fusion diagram can be obtained by writing $\mathsf{a}$,  $\mathsf{b}$,  and $\mathsf{c}$ as $\left[g^{\alpha_1}, R^{\left(\beta_1\right)}\right]$,  $\left[g^{\alpha_2}, R^{\left(\beta_2\right)}\right]$,  and $\left[g^{\left(\alpha_1+\alpha_2 \, \, \text{mod}\, \, N\right)}, R^{\left(\beta_1+\beta_2\, \, \text{mod}\, \, N\right)}\right]$ respectively,  and writing $\mu=1$ in Fig.~\ref{fig_fusion4d}.

\subsection{Examples of non-Abelian fusion rules}\label{Sec.4.3}
From Eq.~(\ref{eq_action_DG}) we see that when the group $G$ is Abelian,  the irrep of the quantum double $DG$ is one-dimensional,  which indicates that there are only Abelian fusion rules. To obtain non-Abelian fusion rules,  we need to consider a non-Abelian group. Here,  we take the simplest non-Abelian group $G=\mathbb{D}_3$ as an example. The $\mathbb{D}_3$ group has two generators $r$ and $t$,  which satisfy $r^3=t^2=e, \, rt=tr^2$. We label the three conjugacy classes in $\mathbb{D}_3$ as $
	C_e=\left\{ e \right\} \, ,   C_r=\left\{ r, r^2 \right\} \, ,  C_t=\left\{ t, tr, tr^2 \right\} \, ,$ 
where $e$,  $r$,  and $t$ serve as the representatives of the classes $C_e$,  $C_r$,  and $C_t$ respectively. Their corresponding centralizers are $	Z_e=\mathbb{D} _3\, , \quad Z_r=\left\{ e, r, r^2 \right\} \simeq \mathbb{Z} _3\, , \quad Z_t=\left\{ e,t \right\} \simeq \mathbb{Z} _2\, , $
where $\simeq$ denotes isomorphism. To choose a set of $q_c$ in Eq.~(\ref{eq_loop}),  we need to consider $Q_g=G/Z_g$ as follows $Q_e=\left\{ e \right\} \, , \quad Q_r=\left\{ e, t \right\} \, , \quad Q_t=\left\{ e, r, r^2 \right\} \, .$
A natural choice for $\mathbb{D}_3$ is $q_e=q_r=q_t=e$,  $q_{r^2}=t$,  $q_{tr}=r$,  and $q_{tr^2}=r^2$. Thus,  there are eight topological excitations and we label them as
\begin{align}
    &\left[ Id \right]\, , &&\left[ A \right] \, , &&\left[ B \right] \, , &&\left[ C_r, Id \right] \, , \nonumber \\
    &\left[ C_r, \omega \right] \, , &&\left[ C_r, \omega ^2 \right] \, , &&\left[ C_t, Id \right] \, , &&\left[ C_t, - \right] \, ,
\end{align}
where we write the particles $\left[ C_e, Id \right]$,  $\left[ C_e, A \right]$,  and $\left[ C_e, B \right]$ as $\left[ Id \right]$,  $\left[ A \right]$,  and $\left[ B \right]$ respectively for simplicity. $Id$ is the $1d$ trivial representation,  $A$ and $B$ are the nontrivial $1d$ and $2d$ representations of $\mathbb{D}_3$ respectively. $\omega$ and $\omega^2$ are two nontrivial $1d$ representations of $\mathbb{Z}_3$ respectively. $-$ is the nontrivial $1d$ representation of $\mathbb{Z}_2$. We choose representation matrices of the generators of $\mathbb{D}_{3}$ to be
 $\Gamma ^A\left( t^nr^m \right) =\left( -1 \right) ^n\, ,\, \Gamma ^-\left( t \right) =-1\, ,\, \Gamma ^B\left( t \right) =\left( \begin{matrix}
		0&		1\\
		1&		0\\
	\end{matrix} \right) \, ,\, \Gamma ^B\left( r \right) =\left( \begin{matrix}
		\omega&		0\\
		0&		\omega ^2\\
	\end{matrix} \right)   \,,\,\Gamma ^{\omega}\left( r^m \right) =\omega ^m\, ,\,\Gamma ^{\omega ^2}\left( r^m \right) =\omega ^{2m}\,  $, 
where $\omega =\exp \left( 2\pi \mathrm{i}/3 \right)$. By using Eq.~(\ref{eq_fusion_coe}),  we derive all fusion rules as shown in Table~\ref{tab_D3_fusion}. For an excitation $\mathsf{a}$,  we can define a matrix $N_{\mathsf{a}}$ with matrix elements $\left( N_{\mathsf{a}} \right) _{\mathsf{bc}}=N_{\mathsf{c}}^{\mathsf{ab}}$. Finally, the largest eigenvalue $d_{\mathsf{a}}$ of the matrix $N_{\mathsf{a}}$ provides the \textit{quantum dimension} of $\mathsf{a}$. All quantum dimensions of excitations are shown in Table~\ref{tab_quantum_dimension}. 
\begin{table*}
	\caption{\label{tab_D3_fusion}Fusion table for the $\mathbb{D}_3$ quantum double model. There are eight excitations in the $\mathbb{D}_3$ quantum double model. $\left[ Id \right]$ and $\left[ A \right]$ are Abelian particles. $\left[ B \right]$ is a non-Abelian particle. $\left[ C_r, Id \right]$,  $\left[ C_r, \omega \right] $,  $\left[ C_r, \omega ^2 \right]$,  $\left[ C_t, Id \right] $,  and $\left[ C_t, - \right]$ are non-Abelian loops.}
	\centering
	\par\smallskip\hrule\smallskip
		\begin{adjustbox}{max width=\linewidth,max totalheight=.82\textheight}
\begin{tabular}{ccccccccc}
			\noalign{\vspace{4pt}}
			$\otimes$ & $\left[ Id \right]$ & $\left[ A \right] $ & $\left[ B \right]$ & $\left[ C_r, Id \right] $ & $\left[ C_r, \omega \right]$ & $\left[ C_r, \omega ^2 \right]$ & $\left[ C_t, Id \right]$ & $\left[ C_t, - \right]$ \\[4pt]
			\hline\noalign{\vspace{4pt}}
			$\left[ Id \right]$ & $\left[ Id \right]$ & $\left[ A \right] $ & $\left[ B \right]$ & $\left[ C_r, Id \right] $ & $\left[ C_r, \omega \right]$ & $\left[ C_r, \omega ^2 \right]$ & $\left[ C_t, Id \right]$ & $\left[ C_t, - \right]$ \\[8pt]
			$\left[ A \right] $ & $\left[ A \right] $ & $\left[ Id \right]$ & $\left[ B \right]$ & $\left[ C_r, Id \right] $ & $\left[ C_r, \omega \right]$ & $\left[ C_r, \omega ^2 \right]$  & $\left[ C_t, - \right]$ & $\left[ C_t, Id \right]$ \\[12pt]
			$\left[ B \right]$ & $\left[ B \right]$ & $\left[ B \right]$ & \makecell{$\left[ Id \right]$\\$\oplus\left[ A \right]$\\$\oplus\left[ B \right] $} & \makecell{$\left[ C_r, \omega \right]$\\$\oplus\left[ C_r, \omega ^2 \right]$} & \makecell{$\left[ C_r, Id \right]$\\$\oplus\left[ C_r, \omega ^2 \right]$} & \makecell{$\left[ C_r, Id \right]$\\$\oplus\left[ C_r, \omega \right]$} & \makecell{$\left[ C_t, Id \right]$\\$\oplus\left[ C_t, - \right]$} & \makecell{$\left[ C_t, Id \right]$\\$\oplus\left[ C_t, - \right]$} \\[20pt]
			$\left[ C_r, Id \right] $ & $\left[ C_r, Id \right] $ & $\left[ C_r, Id \right] $ & \makecell{$\left[ C_r, \omega \right]$\\$\oplus\left[ C_r, \omega ^2 \right]$} & \makecell{$\left[ C_r, Id \right]$\\$\oplus\left[ A \right]$\\$\oplus\left[ Id \right]$} & \makecell{$\left[ C_r, \omega ^2 \right]$\\$\oplus\left[ B \right]$} & \makecell{$\left[ C_r, \omega \right]$\\$\oplus\left[ B \right]$} & \makecell{$\left[ C_t, Id \right]$\\$\oplus\left[ C_t, - \right]$} & \makecell{$\left[ C_t, Id \right]$\\$\oplus\left[ C_t, - \right]$} \\[20pt]
			$\left[ C_r, \omega \right]$ & $\left[ C_r, \omega \right]$ & $\left[ C_r, \omega \right]$ & \makecell{$\left[ C_r, Id \right]$\\$\oplus\left[ C_r, \omega ^2 \right]$} & \makecell{$\left[ C_r, \omega ^2 \right]$\\$\oplus\left[ B \right]$} &\makecell{$\left[ C_r, \omega \right] $\\$\oplus\left[ A \right]$\\$\oplus\left[ Id \right]$} & \makecell{$\left[ C_r, Id \right]$\\$\oplus\left[ B \right]$} & \makecell{$\left[ C_t, Id \right]$\\$\oplus\left[ C_t, - \right]$} & \makecell{$\left[ C_t, Id \right]$\\$\oplus\left[ C_t, - \right]$} \\[20pt]
			$\left[ C_r, \omega ^2 \right]$ & $\left[ C_r, \omega ^2 \right]$ & $\left[ C_r, \omega ^2 \right]$ & \makecell{$\left[ C_r, Id \right]$\\$\oplus\left[ C_r, \omega \right]$} & \makecell{$\left[ C_r, \omega \right]$\\$\oplus\left[ B \right]$} & \makecell{$\left[ C_r, Id \right]$\\$\oplus\left[ B \right]$} & \makecell{$\left[ C_r, \omega ^2 \right] $\\$\oplus\left[ A \right]$\\$\oplus\left[ Id \right]$} & \makecell{$\left[ C_t, Id \right]$\\$\oplus\left[ C_t, - \right]$} & \makecell{$\left[ C_t, Id \right]$\\$\oplus\left[ C_t, - \right]$} \\[20pt]
			$\left[ C_t, Id \right]$ & $\left[ C_t, Id \right]$ & $\left[ C_t, - \right]$ & \makecell{$\left[ C_t, Id \right]$\\$\oplus\left[ C_t, - \right]$} & \makecell{$\left[ C_t, Id \right]$\\$\oplus\left[ C_t, - \right]$} & \makecell{$\left[ C_t, Id \right]$\\$\oplus\left[ C_t, - \right]$} & \makecell{$\left[ C_t, Id \right]$\\$\oplus\left[ C_t, - \right]$} & \makecell{$\left[ Id \right]$\\$\oplus\left[ B \right]$\\$\oplus\left[ C_r, Id \right]$\\$\oplus\left[ C_r, \omega \right]$\\$\oplus\left[ C_r, \omega ^2 \right]$} & \makecell{$\left[ A \right]$\\$\oplus\left[ B \right]$\\$\oplus\left[ C_r, Id \right]$\\$\oplus\left[ C_r, \omega \right]$\\$\oplus\left[ C_r, \omega ^2 \right]$}\\[28pt]
			$\left[ C_t, - \right]$ & $\left[ C_t, - \right]$ & $\left[ C_t, Id \right]$ & \makecell{$\left[ C_t, Id \right]$\\$\oplus\left[ C_t, - \right]$} & \makecell{$\left[ C_t, Id \right]$\\$\oplus\left[ C_t, - \right]$} & \makecell{$\left[ C_t, Id \right]$\\$\oplus\left[ C_t, - \right]$} & \makecell{$\left[ C_t, Id \right]$\\$\oplus\left[ C_t, - \right]$} & \makecell{$\left[ A \right]$\\$\oplus\left[ B \right]$\\$\oplus\left[ C_r, Id \right]$\\$\oplus\left[ C_r, \omega \right]$\\$\oplus\left[ C_r, \omega ^2 \right]$} & \makecell{$\left[ Id \right]$\\$\oplus\left[ B \right]$\\$\oplus\left[ C_r, Id \right]$\\$\oplus\left[ C_r, \omega \right]$\\$\oplus\left[ C_r, \omega ^2 \right]$} \\[24pt]
		\end{tabular}
\end{adjustbox}
	\par\smallskip\hrule\smallskip
\end{table*}
\begin{table*}
	\caption{\label{tab_quantum_dimension}Quantum dimension table for the $\mathbb{D}_3$ quantum double model. The quantum dimension of $\mathsf{a}$ is defined as the largest eigenvalue of the matrix $N_{\mathsf{a}}$,  whose elements are given by $\left( N_{\mathsf{a}} \right) _{\mathsf{bc}}=N_{\mathsf{c}}^{\mathsf{ab}}$. We can obtain all the fusion coefficients in Table~\ref{tab_D3_fusion}.}
	\centering
	\renewcommand{\arraystretch}{1.5}  
	\par\smallskip\hrule\smallskip
		\begin{adjustbox}{max width=\linewidth,max totalheight=.82\textheight}
\begin{tabular}{ccccccccc}
			Excitation & $\left[ Id \right]$ & $\left[ A \right] $ & $\left[ B \right]$ & $\left[ C_r, Id \right] $ & $\left[ C_r, \omega \right]$ & $\left[ C_r, \omega ^2 \right]$ & $\left[ C_t, Id \right]$ & $\left[ C_t, - \right]$ \\
			\hline
			Quantum dimension & 1 & 1 & 2 & 2 & 2 & 2 & 3 & 3 \\
		\end{tabular}
\end{adjustbox}
	\par\smallskip\hrule\smallskip
\end{table*}

\section{Shrinking rules and fusion--shrinking consistency}\label{sec5}
Shrinking rules describe how extended loop excitations can be continuously contracted into point-like particles. These rules represent a new type of topological data unique to three-dimensional (and higher-dimensional) topological orders. In our previous works~\cite{Zhang2023fusion, Huang2023, Huang_2025}, we first systematically derived these rules from continuum field theory, together with their consistency relations with fusion and their diagrammatic representations. Our field-theoretical description of the shrinking process is briefly reviewed in Sec.~\ref{Sec.2.2}. In this section,  we explicitly construct shrinking processes and calculate shrinking rules in the microscopic model. We begin by presenting the general formulation of shrinking rules in Sec.~\ref{Sec.5.1},  followed by concrete examples for the 3D $G=\mathbb{Z}_N$ and $G=\mathbb{D}_3$ quantum double models in Sec.~\ref{Sec.5.2} and Sec.~\ref{Sec.5.3} respectively. Furthermore,  by appropriately choosing internal degrees of freedom in loop operators,  we can control shrinking channels as demonstrated in Sec.~\ref{Sec.5.4}. Finally,  in Sec.~\ref{Sec.5.5},  we verify that the consistency condition in Eq.~(\ref{eq_consistent_4D}) between fusion and shrinking rules derived from continuum field theory also holds in our microscopic lattice model.

\subsection{General microscopic theory}\label{Sec.5.1}
\begin{figure*}
	\centering
	\includegraphics[scale=1.8, keepaspectratio,max width=\linewidth]{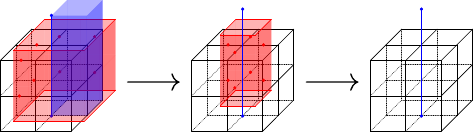}
	\caption{\textbf{Shrinking process on a lattice.} Red and blue denote the dual and direct parts of the cylindrical membrane $M$ respectively. A red dot on an edge marks that the edge is cut by the dual part. All such edges are acted on by the operator $L_{M}^{c}$. The term $T_{P}^{q_cg\bar{q}_{c^{\prime}}}$ acts on an arbitrary string $P$ lying on the direct part, with its endpoints residing on two different boundaries. Here we choose the two vertices marked by blue dots as the endpoints of $P$. The flux of the loop resides on the boundaries of the dual part (red membrane) and we can shrink it to the vacuum. Meanwhile, the direct part (blue membrane) is simultaneously shrunk to a blue string, which cannot be further shrunk. After such a shrinking process,  the endpoints of the blue string generally behave as a superposition of particle states. }
	\label{fig_shrinking}
\end{figure*}

Consider a cylindrical membrane $M$ that creates a loop--antiloop pair as shown in Fig.~\ref{fig_shrinking}. The red membrane supports the $L_{M}^{c}$ operator and the blue string supports the $T_{P}^{q_cg\bar{q}_{c^{\prime}}}$ operators. Using the connecting rule in Eq.~(\ref{eq_L_connect2}) and the fact that the operator $L_c^{\pm}$ has the inverse $L_{\bar{c}}^{\pm}$,  we can shrink a membrane $M\cup M^{\prime}$ to a smaller one $M$  (by smaller we mean $M$ cuts fewer edges) as
\begin{align}
    L_{M}^{c}=\left[\sum_{g}L_{M^{\prime}}^{\bar{g}\bar{c}g}T^{g}_{v_0v_p}\right]L_{M\cup M^{\prime}}^{c}\, , \label{eq_72}
\end{align}
where the direct parts of $M$ and $M^{\prime}$ start at $v_0$ and $v_p$ respectively. When the dual parts of $M^{\prime}$ and $M\cup M^{\prime}$ cut the same edges,  Eq.~(\ref{eq_72}) allows us to shrink a cylindrical membrane to an infinitesimally thin one,  which behaves as a string operator. In Fig.~\ref{fig_shrinking},  this shrinking process shrinks the flux of the loop to the vacuum,  but the blue string is unaffected.  The microscopic realization of shrinking corresponds to the shrinking in the continuum field theory,  as shown in Sec.~\ref{Sec.2.2}. Even if we consider a loop without any charge decoration,  after shrinking, the endpoints of the blue string may still behave as nontrivial particles. To be more specific,  after the shrinking operation (denoted by $\mathcal{S}$),  the operator in Eq.~(\ref{eq_loop}) becomes
\begin{align}
    \mathcal{S} \left( W_M\left( C, R;c, j;c^{\prime}, j^{\prime} \right) \right)     &=\sum_{g\in Z_r}{\Gamma _{jj^{\prime}}^{R^{\ast}}\left( g \right) T_{P}^{q_cg\bar{q}_{c^{\prime}}}}\, .\label{eq_shrink_WM}
\end{align}
If we choose the irrep $R$ to be the trivial representation which corresponds to zero charge decoration,  the operator on the right-hand side is
 $
    \sum_{g\in Z_r}{\Gamma ^{Id}\left( g \right) T_{P}^{q_cg\bar{q}_{c^{\prime}}}}=\sum_{g\in Z_r}{ T_{P}^{q_cg\bar{q}_{c^{\prime}}}}\, , 
$
where $Id^{\ast}=Id$. This operator generally does not commute with the vertex terms $A_{v_0=\partial_0P}$ and $A_{v_1=\partial_1P}$. Thus,  applying this operator to a ground state creates excited vertices $v_0=\partial_0P$ and $v_1=\partial_1P$. To be more specific, the linear combination of $T$-operators in the above equation can be expressed in terms of the particle creation operators $W_P\left( R^{\prime};i, i^{\prime} \right)$, therefore yielding a nontrivial superposition of particle excitations.

In Sec.~\ref{sec4},  we have shown that the set of basis states $\left\{ W_M\left( C, R;c, j;c^{\prime}, j^{\prime} \right) \ket{\mathrm{GS}} \right\} $ spans the Hilbert space $\mathcal{H} _{\left( C, R \right)}$ of a pair of excitations on the two boundaries $\partial_0M$ and $\partial_1M$ of the cylindrical membrane $M$. After the shrinking process,  the set of basis states becomes $\left\{ \mathcal{S} \left( W_M\left( C, R;c, j;c^{\prime}, j^{\prime} \right) \right) \ket{\mathrm{GS}} \right\}$,  where $c, c^{\prime}\in C $ and $ j, j^{\prime}=1, \cdots , n_R$. This set of basis states spans the Hilbert space 
\begin{align}
    \mathcal{S} \left( \mathcal{H} _{\left( C, R \right)} \right) =\mathcal{S} \left( V_{\left( C, R \right)} \right) \otimes \mathcal{S} \left( V_{\left( \bar{C}, R^{\ast} \right)} \right) \, .
\end{align}
At the boundary $\partial_0M$, the local space $\mathcal{S}\left(V_{\left( C, R \right)}\right)$ is spanned by $\left\{\mathcal{S}\left(\ket{ c, j }_{\left( C, R \right)}\right) \right\}$,  where $c\in C$ and $j=1,  \cdots, n_R$. Under the action of the quantum double $DG$ (i.e.,  the local action in Eq.~(\ref{eq_local_action})),  the space $\mathcal{S}\left(V_{\left( C, R \right)}\right)$ decomposes as a direct sum of its invariant subspaces:
\begin{align}
    \mathcal{S}\left(V_{\left( C, R \right)}\right)=\bigoplus_{R^{\prime}}S_{\left( C_e, R^{\prime} \right)}^{\left( C, R \right)}V_{\left( C_e, R^{\prime} \right)}\, , \label{eq_DG_shrinking}
\end{align}
where $S_{\left( C_e, R^{\prime} \right)}^{\left( C, R \right)}$ is the shrinking coefficient, $C_e=\left\{ e \right\} $, and $R^{\prime}$ denotes an irrep of $G$. The equation above is interpreted as the shrinking rule in Eq.~(\ref{eq_shrinking}). Each allowed invariant subspace $V_{\left( C_e, R^{\prime} \right)}$ corresponds to a particle $\left[R^{\prime}\right]$ (we write $\left[C_e, R^{\prime}\right]$ as $\left[R^{\prime}\right]$ for simplicity) and the coefficient  $S_{\left( C_e, R^{\prime} \right)}^{\left( C, R \right)}$ counts the number of independent shrinking channels from the loop $\left[ C, R \right]$ to the particle $\left[R^{\prime}\right]$.

To derive the explicit form of the shrinking coefficient,  we express the membrane operator $\mathcal{S} \left( W_M\left( C, R;c, j;c^{\prime}, j^{\prime} \right) \right)$ as a linear combination of the operators for particles:
\begin{align}
    \label{eq:sh}
\mathcal{S} \left( W_M\left( C, R;c, j;c^{\prime}, j^{\prime} \right) \right)
&=\sum_{R^{\prime}, i, i^{\prime}} x(R^{\prime}, i, i^{\prime}) W_P\left( R^{\prime};i, i^{\prime} \right),
\end{align}
where $x\left( R^{\prime}, i, i^{\prime} \right)$ is an undetermined coefficient,  and $R$ and $R^{\prime}$ denote irreps of the centralizer $Z_r$ and the full group $G$,  respectively.  Substituting the explicit forms of $W_M$ and $W_P$ yields
\begin{align}
    \sum_{h\in Z_r}{\Gamma _{jj^{\prime}}^{R^{\ast}}\left( h \right) T_{P}^{q_ch\bar{q}_{c^{\prime}}}}&=\sum_{R^{\prime}, i, i^{\prime}}{x\left( R^{\prime}, i, i^{\prime} \right)}\sum_{g\in G}{\Gamma _{ii^{\prime}}^{R^{\prime \ast}}\left( g \right) T_{P}^{g}}\, .
\end{align}
Substituting $h=\bar{q}_cgq_{c^{\prime}}$ into the left-hand side of the above equation,  we obtain
\begin{align}
    \sum_{g\in q_cZ_r\bar{q}_{c^{\prime}}}\!\!{\Gamma _{jj^{\prime}}^{R^{\ast}}\left( \bar{q}_cgq_{c^{\prime}} \right)\! T_{P}^{g}}=\!\sum_{R^{\prime}, i, i^{\prime}}{\!x\left( R^{\prime}, i, i^{\prime} \right)}\!\sum_{g\in G}{\Gamma _{ii^{\prime}}^{R^{\prime\ast}}\left( g \right) \!T_{P}^{g}}.
\end{align}
By comparing the coefficients of the $T_P^g$ operator,  we obtain the equation
\begin{align}
    \delta_{g\in q_cZ_r\bar{q}_{c^{\prime}}}{\Gamma _{jj^{\prime}}^{R^{\ast}}\left( \bar{q}_cgq_{c^{\prime}} \right) }=\sum_{R^{\prime}, i, i^{\prime}}{x\left( R^{\prime}, i, i^{\prime} \right)}{\Gamma _{ii^{\prime}}^{R^{\prime\ast}}\left( g \right) }\, , \label{eq_90}
\end{align}
where $\delta_{g\in q_cZ_r\bar{q}_{c^{\prime}}}$ is defined as 1 if $g\in q_cZ_r\bar{q}_{c^{\prime}}$ and 0 otherwise. Using the Schur orthogonality relation,
\begin{align}
    \label{eq:Schur}
    \frac{1}{\left| G \right|}\sum_g{\Gamma _{ii^{\prime}}^{R^{\prime\ast}}\left( g \right) \Gamma _{jj^{\prime}}^{R^{\prime\prime}}\left( g \right)}=\frac{1}{\mathrm{dim}\left( R^{\prime} \right)}\delta _{R^{\prime}, R^{\prime\prime}}\delta _{ij}\delta _{i^{\prime}j^{\prime}}\, ,
\end{align}
we obtain an expression for $x\left( R^{\prime}, i, i^{\prime} \right)$ as
\begin{align}
    \sum_{g\in G}{\sum_{R^{\prime\prime}, j, j^{\prime}}{x\left( R^{\prime\prime}, j, j^{\prime} \right)}}\Gamma _{jj^{\prime}}^{R^{\prime\prime\ast}}\left( g \right) \Gamma _{ii^{\prime}}^{R^{\prime}}\left( g \right) &=\sum_{R^{\prime\prime}, j, j^{\prime}}{x\left( R^{\prime\prime}, j, j^{\prime} \right)}\frac{\left| G \right|\delta _{R^{\prime\prime}, R^{\prime}}\delta _{i, j}\delta _{i^{\prime}, j^{\prime}}}{\mathrm{dim}\left( R^{\prime} \right)}\nonumber \\
    &=\frac{\left| G \right|}{\mathrm{dim}\left( R^{\prime} \right)}x\left( R^{\prime}, i, i^{\prime} \right)\, .\label{eq_92}
\end{align}
Using Eqs.~(\ref{eq_90}) and~(\ref{eq_92}),  we have
\begin{align}
    x\left( R^{\prime}, i, i^{\prime} \right) &=\frac{\mathrm{dim}\left( R^{\prime} \right)}{\left| G \right|}\sum_{g\in G}{\delta _{g\in q_cZ_r\bar{q}_{c^{\prime}}}\Gamma _{jj^{\prime}}^{R^{\ast}}\left( \bar{q}_cgq_{c^{\prime}} \right) \Gamma _{ii^{\prime}}^{R^{\prime}}\left( g \right)}\nonumber \\
    &=\frac{\mathrm{dim}\left( R^{\prime} \right)}{\left| G \right|}\sum_{h\in Z_r}{\Gamma _{jj^{\prime}}^{R^{\ast}}\left( h \right) \Gamma _{ii^{\prime}}^{R^{\prime}}\left( q_ch\bar{q}_{c^{\prime}} \right)}\, ,
\end{align}
where we have used $h=\bar{q}_cgq_{c^{\prime}}$ to obtain the second line. By decomposing $\Gamma _{ii^{\prime}}^{R^{\prime}}\left( q_ch\bar{q}_{c^{\prime}} \right)$ as $\sum_{k, l}{\Gamma _{ik}^{R^{\prime}}\left( q_c \right) \Gamma _{kl}^{R^{\prime}}\left( h \right) \Gamma _{li^{\prime}}^{R^{\prime}}\left( \bar{q}_{c^{\prime}} \right)}$,  we finally obtain the explicit form of $x\left( R^{\prime}, i, i^{\prime} \right)$:
\begin{align}
    \quad x\left( R^{\prime}, i, i^{\prime} \right) &=\frac{\mathrm{dim}\left( R^{\prime} \right)}{\left| G \right|}\sum_{k, l}{\Gamma _{ik}^{R^{\prime}}\left( q_c \right) \Gamma _{li^{\prime}}^{R^{\prime}}\left( \bar{q}_{c^{\prime}} \right)}
    \sum_{h\in Z_r}{\Gamma _{jj^{\prime}}^{R^{\ast}}\left( h \right) \Gamma _{kl}^{R^{\prime}}\left( h \right)}\, . \label{eq:x}
\end{align}
This gives the unique solution for the coefficient in the shrinking rule in Eq.~(\ref{eq:sh}).

Let us now examine which particle excitation $\left[R^{\prime} \right]$ can appear after shrinking a loop excitation $[C,  R]$ in Eq.~(\ref{eq:sh}),  where $C$ is a conjugacy class of $G$ and $R$ is an irrep of the centralizer $Z_r$. 
Shrinking a loop excitation discards the information about the conjugacy class $C$ that labels its flux. Therefore,  we need to understand how a representation $R $ of $Z_r$ decomposes into representations $R^{\prime} $ of $G$. \textit{In other words,  we should view the representation $R$ of the subgroup $Z_r$ as a representation $R^{\prime}$ of the full group $G$. }

A natural candidate is the \emph{induction functor} $\operatorname{Ind}_{Z_r}^G$,  which maps an irrep $R$ of the subgroup $Z_r$ to a direct sum of irreps $R^{\prime}$ of $G$. The induction functor is adjoint to the \emph{restriction functor} $\operatorname{Res}_{Z_r}^G$. This implies that a particle excitation $R^{\prime}$ appears after shrinking the loop excitation $[C,  R]$ if the representation $R^{\prime}$ of $G$ restricts to $R$ on the subgroup $Z_r$.

This is indeed confirmed by the last summation $\sum_{h\in Z_r}{\Gamma _{jj^{\prime}}^{R^{\ast}}\left( h \right) \Gamma _{kl}^{R^{\prime}}\left( h \right)}$ in Eq.~(\ref{eq:x}). We can first decompose $R^{\prime}$ of $G$ into irreps of $Z_r$; applying Eq.~(\ref{eq:Schur}) then yields a delta function. We therefore conclude that $x(R^{\prime}, i, i^{\prime})$ vanishes unless $R$ is contained in $\operatorname{Res}^G_{Z_r}(R^{\prime})$. Consequently,  the loop excitation $[C, R]$ can shrink to the particle excitation $\left[R^{\prime}\right]$ if and only if $R$ is a subrepresentation of $\operatorname{Res}^G_{Z_r}(R^{\prime})$,  or equivalently,  by Frobenius reciprocity,  if $R^{\prime}$ is a subrepresentation of $\operatorname{Ind}^G_{Z_r}(R)$.

This result yields the direct sum decomposition of the space $\mathcal{S}\left(V_{(C, R)}\right)$ into the particle excitations $[ R^{\prime}]$,  where the irrep $R^{\prime}$ of $G$ appears with the multiplicity given by its occurrence in the induced representation $\operatorname{Ind}_{Z_r}^{G}(R)$. Explicitly,  we have
\begin{align}
    \mathcal{S} \left( V_{(C, R)} \right) = \bigoplus_{R^{\prime} \in \operatorname{Irr}(G)} m_R(R^{\prime}) \cdot V_{( R^{\prime})}\,.\label{eq_decom_SV}
\end{align}
The non-negative integral multiplicity $m_R(R^{\prime})$ can be calculated from the inner product of the corresponding characters: 
\begin{align}
    m_R(R^{\prime}) = \langle \chi_{R^{\prime}},  \operatorname{Ind}_{Z_r}^G(\chi_R) \rangle_G = \langle \operatorname{Res}_{Z_r}^G(\chi_{R^{\prime}}),  \chi_R \rangle_{Z_r},\label{eq_shrinking_multiplicity}
\end{align}
where the inner product of two group functions $f_1$ and $f_2$ is defined as
\begin{align}
    \braket{f_1,f_2}_G=\frac{1}{\left| G \right|}\sum_{g\in G}{f_1\left( g \right) ^{\ast}f_2\left( g \right)}\,.
\end{align}
The decomposition in Eq.~(\ref{eq_decom_SV}) corresponds to the shrinking rule
\begin{align}
    \mathcal{S} \left( [C, R] \right) = \bigoplus_{R^{\prime} \in \operatorname{Irr}(G)} m_R(R^{\prime}) \cdot [ R^{\prime}]. \label{eq_induced_rep}
\end{align}

\subsection{Examples of Abelian shrinking rules}\label{Sec.5.2}
To illustrate the shrinking rules more specifically,  we take $G=\mathbb{Z}_N$ and $G=\mathbb{D}_3$ as two examples. We first consider $G=\mathbb{Z}_N$. As mentioned in Sec.~\ref{Sec.4.2},  an excitation can be labeled by $\left[g^\alpha, R^{\left(\beta\right)}\right]$,  where $g$ is the generator of the group $G=\mathbb{Z}_N$,  $\alpha, \beta=0, 1, 2, \cdots, \left| G \right|-1$ and $R^{\left(\beta\right)}$ labels an irrep of $G=\mathbb{Z}_N$. The representation matrix is given by $\Gamma ^{R^{\left( \beta \right)}}\left( g^{\alpha} \right) =\omega ^{\alpha \beta}$,  where $\omega=\exp\left(2\pi \mathrm{i}/N\right)$. Thus,  the operator for excitation $\left[g^{\alpha}, R^{\left(\beta\right)}\right]$ is given by
\begin{align}
    W_M\left( g^{\alpha}, R^{\left(\beta\right)} \right) &=\sum_{\gamma =0}^{N-1}{\omega ^{-\beta \gamma}L_{M}^{g^{\alpha}}T_{P}^{g^{\gamma}}}=\sum_{\gamma =0}^{N-1}{\exp\left(\frac{-\mathrm{i}2\pi\beta\gamma}{N}\right)L_{M}^{g^{\alpha}}T_{P}^{g^{\gamma}}}\, .
\end{align}
Since $c,  j,  c^{\prime}, j^{\prime}$ only have one possible choice,  we omit these labels and simply write $W_M\left( g^{\alpha}, R^{\left(\beta\right)};g^{\alpha}, 1;g^{\alpha}, 1 \right)$ as $W_M\left( g^{\alpha}, R^{\left(\beta\right)} \right)$ in the above equation. We use Eq.~(\ref{eq_shrink_WM}) to shrink the flux part of $\left[g^{\alpha}, R^{\left(\beta\right)}\right]$ and we obtain
\begin{align}
    \mathcal{S}\left(W_M\left( g^{\alpha}, R^{\left(\beta\right)} \right)\right)&=\sum_{\gamma =0}^{N-1}{\exp\left(\frac{-\mathrm{i}2\pi\beta\gamma}{N}\right)T_{P}^{g^{\gamma}}}=W_M\left( e, R^{\left(\beta\right)} \right)\, ,
\end{align}
where $W_M\left( e, R^{\left(\beta\right)} \right)$ is exactly the operator for the particle $\left[e, R^{\left(\beta\right)}\right]$. From the above equation,  we directly conclude that all shrinking rules in the 3D $G=\mathbb{Z}_N$ quantum double model are Abelian and take the form
\begin{align}
    \mathcal{S}\left(\left[g^\alpha, R^{\left(\beta\right)}\right] \right)=\left[e, R^{\left(\beta\right)}\right]\, .\label{eq_Abelian_shrinking}
\end{align}
The corresponding shrinking diagram is obtained by writing $\mathsf{a}$ and $\mathsf{b}$ as $\left[g^\alpha, R^{\left(\beta\right)}\right]$ and $\left[e, R^{\left(\beta\right)}\right]$ respectively,  and writing $\mu=1$ in Fig.~\ref{fig_shrinking_dia}.

\subsection{Examples of non-Abelian shrinking rules}\label{Sec.5.3}
Now we consider the shrinking rules when $G=\mathbb{D}_3$. Unlike the Abelian case,  some excitations in the 3D $G=\mathbb{D}_3$ quantum double model are non-Abelian,  which means that there are some internal degrees of freedom in the corresponding operator. For example,  according to Eq.~(\ref{eq_loop}) and the data about $\mathbb{D}_3$ shown in Sec.~\ref{Sec.4.3},  operators with different degrees of freedom for a pure loop (i.e.,  without charge decoration) $\left[ C_r, Id \right]$ are given by
\begin{align}
    W_M\left( C_r, Id;r;r \right) &=\sum_{g\in Z_r}{\Gamma ^{Id}\left( g \right)L_{M}^{r}T_{P}^{g}}=L_{M}^{r}T_{P}^{e}+L_{M}^{r}T_{P}^{r}+L_{M}^{r}T_{P}^{r^2}\, , \label{eq_rr} \\
    W_M\left( C_r, Id;r^2;r \right) &=\sum_{g\in Z_r}{\Gamma ^{Id}\left( g \right)L_{M}^{r^2}T_{P}^{tg}}=L_{M}^{r^2}T_{P}^{t}+L_{M}^{r^2}T_{P}^{tr}+L_{M}^{r^2}T_{P}^{tr^2}\, , \label{eq_r2r} \\
    W_M\left( C_r, Id;r^2;r^2 \right) &=\sum_{g\in Z_r}{\Gamma ^{Id}\left( g \right)L_{M}^{r^2}T_{P}^{tgt}}=L_{M}^{r^2}T_{P}^{e}+L_{M}^{r^2}T_{P}^{r^2}+L_{M}^{r^2}T_{P}^{r}\, , \label{eq_r2r2} \\
    W_M\left( C_r, Id;r;r^2 \right) &=\sum_{g\in Z_r}{\Gamma ^{Id}\left( g \right)L_{M}^{r}T_{P}^{gt}}=L_{M}^{r}T_{P}^{t}+L_{M}^{r}T_{P}^{tr^2}+L_{M}^{r}T_{P}^{tr}\, , \label{eq_rr2}
\end{align}
where we omit the row and column indices $j=j^{\prime}=1$ of the one-dimensional matrix. The internal degrees of freedom of the operator are labeled by the elements in the conjugacy class $C_r=\left\{ r, r^2 \right\}$. 

Although $\left[ C_r, Id \right]$ does not carry any charge decoration,  after the shrinking process,  the remaining operator may still violate the vertex terms at the two ends of the string $P$. Focusing on the local space on $\partial_0M$, we fix $c^{\prime}=r$ and vary $c\in C_r$ to obtain a set of basis states in $V_{\left(C_r, Id\right)}$. When $c=r$,  after the shrinking process,  the flux part of $\left[ C_r, Id \right]$ vanishes and the operator shown in Eq.~(\ref{eq_rr}) becomes
\begin{align}
    \mathcal{S} \left( W_M\left( C_r, Id;r, r \right) \right) =T_{P}^{e}+T_{P}^{r}+T_{P}^{r^2}\, .
\end{align}
Notice that operators for particles $\left[Id\right]$ and $\left[A\right]$ are given by
\begin{align}
    W_P\left( Id \right) &=T_{P}^{e}+T_{P}^{r}+T_{P}^{r^2}+T_{P}^{t}+T_{P}^{tr}+T_{P}^{tr^2}\, , \\
    W_P\left( A \right) &=T_{P}^{e}+T_{P}^{r}+T_{P}^{r^2}-T_{P}^{t}-T_{P}^{tr}-T_{P}^{tr^2}\, ,
\end{align}
where $W_P\left( Id \right)$ and $W_P\left( A \right)$ are abbreviations for $W_P\left( Id;1,1 \right) $ and $W_P\left(  A;1,1 \right) $. Comparing the above equations,  we have
\begin{align}
    \mathcal{S} \left( W_M\left( C_r, Id;r, r \right) \right) =\frac{1}{2}\left[ W_P\left( Id \right) +W_P\left( A \right) \right] \, .\label{eq_shrink_r}
\end{align}
On the other hand,  when $c=r^2$,  after the shrinking process,  the operator shown in Eq.~(\ref{eq_r2r}) similarly becomes
\begin{align}
    \mathcal{S} \left( W_M\left( C_r, Id;r^2, r \right) \right) =\frac{1}{2}\left[ W_P\left( Id \right) -W_P\left( A \right) \right] \, .\label{eq_shrink_r2}
\end{align}

To interpret these results as shrinking rules,  we note that before the shrinking process,  the space $V_{\left( C_r, Id \right)}$ of $\left[ C_r, Id \right]$ is spanned by $\left\{\ket{r}_{\left( C_r, Id \right)}, \ket{r^2}_{\left( C_r, Id \right)}\right\}$,  where we omit $j=j^{\prime}=1$. According to the discussion in Sec.~\ref{sec4},  $\ket{r}_{\left( C_r, Id \right)}$ and $\ket{r^2}_{\left( C_r, Id \right)}$ correspond to the states $W_M\left( C_r, Id;r, r \right)\ket{\mathrm{GS}}$ and $W_M\left( C_r, Id;r^2, r \right)\ket{\mathrm{GS}}$ near $\partial_0M$. Eqs.~(\ref{eq_shrink_r}) and~(\ref{eq_shrink_r2}) indicate that after the shrinking process,  the set of basis states transforms as $\mathcal{S}\left(\ket{r}_{\left( C_r, Id \right)}\right)=\frac{1}{2}\left[\ket{1}_{\left( Id \right)}+\ket{1}_{\left( A \right)}\right]$ and $\mathcal{S}\left(\ket{r^2}_{\left( C_r, Id \right)}\right)=\frac{1}{2}\left[\ket{1}_{\left( Id \right)}-\ket{1}_{\left( A \right)}\right]$,  where we omit $c=e$ and $C_e$ for simplicity,  $\ket{1}_{\left( Id \right)}$ and $\ket{1}_{\left( A \right)}$ correspond to the states $W_P\left( Id \right)\ket{\mathrm{GS}}$ and $W_P\left( A \right)\ket{\mathrm{GS}}$ near $\partial_0M$. Now the space $V_{\left( C_r, Id \right)}$ becomes $\mathcal{S}\left(V_{\left( C_r, Id \right)}\right)$,  which is spanned by $\left\{\frac{1}{2}\left[\ket{1}_{\left( Id \right)}+\ket{1}_{\left( A \right)}\right]\, , \frac{1}{2}\left[\ket{1}_{\left( Id \right)}-\ket{1}_{\left( A \right)}\right]\right\}$. Since the sets of basis states $\left\{\ket{1}_{\left( Id \right)}\right\}$ and $\left\{\ket{1}_{\left( A \right)}\right\}$ span one-dimensional local spaces $V_{\left( Id \right)}$ and $V_{\left( A \right)}$ that are invariant under the action of the quantum double $DG$,  the space $\mathcal{S}\left(V_{\left( C_r, Id \right)}\right)$ decomposes as $\mathcal{S}\left(V_{\left( C_r, Id \right)}\right)=V_{\left( Id \right)}\oplus V_{\left( A \right)}$,  which gives the shrinking rule 
\begin{align}
    \mathcal{S} \left( \left[ C_r, Id \right] \right) =\left[ Id \right] \oplus \left[ A \right] \, .\label{eq_shrinking1}
\end{align}
This result is consistent with Eq.~(\ref{eq_induced_rep}) since $\mathrm{Ind}_{Z_r}^{\mathbb{D}_3}\left(Id\right)$ can be decomposed as $\mathrm{Ind}_{Z_r}^{\mathbb{D}_3}\left(Id\right)=Id\oplus A$.

One can also fix $c^{\prime}=r^2$ and vary $c\in C_r$ to obtain the basis in $V_{\left(C_r, Id\right)}$,  i.e.,  consider the operators shown in Eqs.~(\ref{eq_r2r2}) and~(\ref{eq_rr2}),  because physically the local actions near $\partial_0M$ cannot distinguish $W_M\left( C, R;c, j;c^{\prime}, j^{\prime} \right)$ and $W_M\left( C, R;c, j;c^{\prime\prime}, j^{\prime\prime} \right)$. By using the same method,  we will derive the same shrinking rule Eq.~(\ref{eq_shrinking1}). 

Similarly,  we can derive all the shrinking rules in the 3D $\mathbb{D}_3$ quantum double model as shown in Appendix~\ref{ap4}. All the shrinking rules are collected in Table~\ref{tab_shrinking_d3}. Although the five loops in the $\mathbb{D}_3$ quantum double model are all non-Abelian excitations,  only $\left[ C_r, Id \right]$,  $\left[ C_t, Id \right]$,  and $\left[ C_t, - \right]$ have non-Abelian shrinking rules,  while $\left[ C_r, \omega \right]$ and $\left[ C_r, \omega ^2 \right]$ have Abelian shrinking rules. 
\begin{table}
	\caption{\label{tab_shrinking_d3}Shrinking table for the 3D $\mathbb{D}_3$ quantum double model. One can also use Eq.~(\ref{eq_induced_rep}) to directly obtain these results.}
	\centering
	\setlength{\extrarowheight}{2pt}
    \renewcommand{\arraystretch}{1.5}  
    \par\smallskip\hrule\smallskip
	\begin{adjustbox}{max width=\linewidth,max totalheight=.82\textheight}
\begin{tabular}{cccccc}
		Loop & $\left[ C_r, Id \right]$ & $\left[ C_r, \omega \right]$ & $\left[ C_r, \omega ^2 \right]$ & $\left[ C_t, Id \right]$ & $\left[ C_t, - \right]$  \\
		\hline
		$\mathcal{S}\left(\text{Loop}\right)$ & $\left[ Id \right] \oplus \left[ A \right]$ & $\left[ B \right]$ & $\left[ B \right]$ & $\left[ Id \right] \oplus \left[ B \right]$ & $\left[ A \right] \oplus \left[ B \right]$  \\
	\end{tabular}
\end{adjustbox}
    \par\smallskip\hrule\smallskip
\end{table}

\subsection{Controlling non-Abelian shrinking channels}\label{Sec.5.4}
In the continuum field theory~\cite{Zhang2023fusion, Huang2023},  non-Abelian shrinking rules obtained from correlation functions generally specify only which shrinking channels are allowed. However,  in our microscopic lattice model,  we can not only derive general shrinking rules as in Table~\ref{tab_shrinking_d3},  but also further investigate how to determine which channels are selected in a specific shrinking process. In this section,  by tuning the internal degrees of freedom in Eq.~(\ref{eq_loop}),  we demonstrate control over these non-Abelian shrinking channels.

Consider a general loop $\left[C, R\right]$ in the quantum double model with a finite group $G$. In Sec.~\ref{Sec.4.1},  we have mentioned that $\left[C, R\right]$ is associated with a local Hilbert space $V_{\left(C, R\right)}$ spanned by $ \left\{\ket{ c, j }_{\left(C, R\right)} \right\}$,  where $\ket{ c, j }_{\left(C, R\right)}$ labels the local degrees of freedom in the state $W_M\left( C, R;c, j;c^{\prime}, j^{\prime} \right) \ket{\mathrm{GS}}$ near the boundary $\partial_0M$. Since all the operators $W_M\left( C, R;c, j;c^{\prime}, j^{\prime} \right) $ and the corresponding basis states $\ket{ c, j }_{\left(C, R\right)}$ carry conserved labels $\left(C, R\right)$,  arbitrary linear combinations of these operators 
\begin{align}
    \sum_{c, j}m\left(c, j\right)W_M\left( C, R;c, j;c^{\prime}, j^{\prime} \right) \label{eq_general_operator_loop}
\end{align}
and the corresponding linear combinations of the basis states
\begin{align}
    \ket{\Psi}_{\left(C, R\right)}=\sum_{c, j}m\left(c, j\right)\ket{ c, j }_{\left(C, R\right)} \label{eq_general_state_loop}
\end{align}
also carry the conserved labels $\left(C, R\right)$,  where $m\left(c, j\right)$ is a coefficient. Thus,  the creation operator for the loop $\left[C, R\right]$ does not necessarily take the form of $W_M\left( C, R;c, j;c^{\prime}, j^{\prime} \right) $ but rather is generally given by Eq.~(\ref{eq_general_operator_loop}). Correspondingly,  the general state for the loop $\left[C, R\right]$ is given by Eq.~(\ref{eq_general_state_loop}). Changing the set of coefficients from $\left\{m\left(c, j\right)\right\}$ to $\left\{m^{\prime}\left(c, j\right)\right\}$ in Eq.~(\ref{eq_general_operator_loop}) is equivalent to rotating the vector $\ket{\Psi}_{\left(C, R\right)}$ in the space $V_{\left(C, R\right)}$ to $\ket{\Psi^{\prime}}_{\left(C, R\right)}$:
\begin{align}
    \ket{\Psi^{\prime}}_{\left(C, R\right)}=\sum_{c, j}m^{\prime}\left(c, j\right)\ket{ c, j }_{\left(C, R\right)}\in V_{\left(C, R\right)}\, .
\end{align}
Suppose the basis state $ \ket{ c, j }_{\left(C, R\right)} $ is shrunk to a linear combination of particle states
\begin{align}
    \mathcal{S}\left(\ket{ c, j }_{\left(C, R\right)}\right)=\sum_{i, R^{\prime}}n\left(i, R^{\prime}\right)\ket{i}_{\left(R^{\prime}\right)}\, ,
\end{align}
where $R^{\prime}$ is an irrep of $G$, $i=1, \cdots, n_{R^{\prime}}$, and $n\left(i, R^{\prime}\right) $ is a coefficient. For a non-Abelian loop,  if we perform the shrinking operation on the states $\ket{\Psi}_{\left(C, R\right)}$ and $\ket{\Psi^{\prime}}_{\left(C, R\right)}$,  we obtain
\begin{align}
    \mathcal{S}\left(\ket{\Psi}_{\left(C, R\right)}\right)&=\sum_{c, j}m\left(c, j\right)\mathcal{S}\left(\ket{ c, j }_{\left(C, R\right)}\right) =\sum_{c, j}\sum_{i, R^{\prime}}m\left(c, j\right)n\left(i, R^{\prime}\right)\ket{i}_{\left(R^{\prime}\right)}\label{eq_channel1}
\end{align}
and 
\begin{align}
    \mathcal{S}\left(\ket{\Psi^{\prime}}_{\left(C, R\right)}\right)&=\sum_{c, j}m^{\prime}\left(c, j\right)\mathcal{S}\left(\ket{ c, j }_{\left(C, R\right)}\right) =\sum_{c, j}\sum_{i, R^{\prime}}m^{\prime}\left(c, j\right)n\left(i, R^{\prime}\right)\ket{i}_{\left(R^{\prime}\right)}\label{eq_channel2}
\end{align}
respectively. The resulting states in Eqs.~(\ref{eq_channel1}) and~(\ref{eq_channel2}) can be different superpositions of particle states. By tuning the coefficients $\left\{m\left(c, j\right)\right\}$ in Eq.~(\ref{eq_general_operator_loop}) properly, we can obtain a state satisfying $\mathcal{S}\left(\ket{\Psi}_{\left(C, R\right)}\right)\in V_{\left(R^{\prime}\right)}$,  which means that the loop selects a definite shrinking channel leading to the particle $\left[R^{\prime}\right]$ in a shrinking process.

To better illustrate the control over non-Abelian shrinking channels,  we take the $\mathbb{D}_3$ quantum double model as an example. According to Table~\ref{tab_shrinking_d3},  the three non-Abelian loops in the $\mathbb{D}_3$ quantum double model,  i.e.,  $\left[ C_r, Id \right]$,  $\left[ C_t, Id \right]$ and $\left[ C_t, - \right]$,  have non-Abelian shrinking rules. We start by considering the loop $\left[ C_r, Id \right]$. By linearly combining the two operators $W_M\left( C_r, Id;r, r \right)$ and $W_M\left( C_r, Id;r^2, r \right)$ as
\begin{align}
    &\left[ W_M\left( C_r, Id;r, r \right) \pm W_M\left( C_r, Id;r^2, r \right) \right] \nonumber \\
    &=\left[ \sum_{g\in Z_r}{\Gamma ^{Id}\left( g \right) L_{M}^{r}T_{P}^{g}}\pm \sum_{g\in Z_r}{\Gamma ^{Id}\left( g \right) L_{M}^{r^2}T_{P}^{tg}} \right] \nonumber \\
    &=L_{M}^{r}T_{P}^{e}+L_{M}^{r}T_{P}^{r}+L_{M}^{r}T_{P}^{r^2}\pm L_{M}^{r^2}T_{P}^{t}\pm L_{M}^{r^2}T_{P}^{tr}\nonumber \\
    &\pm L_{M}^{r^2}T_{P}^{tr^2}\label{eq_select}
\end{align}
and then applying it to a ground state,  we obtain a vector $\ket{r}_{\left( C_r, Id \right)}\pm\ket{r^2}_{\left( C_r, Id \right)}\in V_{\left( C_r, Id \right)}$. Thus,  the operator given by Eq.~(\ref{eq_select}) is a legitimate operator for excitation $\left[ C_r, Id \right]$. The shrinking process will remove the flux part,  i.e., 
\begin{align}
    &\mathcal{S} \left( \left[ W_M\left( C_r, Id;r, r \right) \pm W_M\left( C_r, Id;r^2, r \right) \right] \right) \\
    \nonumber
	=&T_{P}^{e}+T_{P}^{r}+T_{P}^{r^2}\pm T_{P}^{t}\pm T_{P}^{tr}\pm T_{P}^{tr^2}=\begin{cases}
		W_P\left( Id \right) &		+\\
		W_P\left( A \right) &		-\\
	\end{cases}\, .
\end{align}
Thus,  we have
\begin{align}
    \mathcal{S} \left( \ket{r}_{\left( C_r, Id \right)}\pm\ket{r^2}_{\left( C_r, Id \right)}\right) =\begin{cases}
		\ket{1}_{\left( Id \right)}\in V_{\left(Id\right)} &		+\\
		\ket{1}_{\left( A \right)} \in V_{\left(A\right)} &		-\\
	\end{cases}\, .
\end{align}
The result above means that by choosing certain degrees of freedom,  we can choose the shrinking channel to $\left[ Id \right]$ or $\left[ A \right]$. We can use the shrinking diagram shown in Fig.~\ref{fig_shrinking_dia} to describe these different channels. When we choose $+$ in Eq.~(\ref{eq_select}),  we write $\mathsf{a}$ and $\mathsf{b}$ as $\left[ C_r, Id \right]$ and $\left[ Id \right]$ respectively in Fig.~\ref{fig_shrinking_dia}. When we choose $-$ in Eq.~(\ref{eq_select}),  we write $\mathsf{a}$ and $\mathsf{b}$ as $\left[ C_r, Id \right]$ and $\left[ A \right]$ respectively in Fig.~\ref{fig_shrinking_dia}.

Similarly,  we illustrate how to control other non-Abelian shrinking channels.  We can write the corresponding sets of operators for $\left[ C_t, Id \right]$ and $\left[ C_t, - \right]$ as \[
\left\{W_M\left( C_t, Id;t, t \right), W_M\left( C_t, Id;tr, t \right), W_M\left( C_t, Id;tr^2, t \right)\right\}
\] and \[
\left\{\!W_M\!\left( C_t, -;t, t \right) , \!W_M\!\left( C_t, -;tr, t \right), \!W_M\!\left( C_t, -;tr^2, t \right)\!\right\}
\] respectively. The explicit forms of the operators above can be found in Appendix~\ref{ap4}. For $\left[ C_t, Id \right]$,  by linearly combining the operators as 
\begin{align}
    W_M\!\left( C_t, Id;t, t \right)\!+\!W_M\!\left( C_t, Id;tr, t \right)\!+\!W_M\!\left( C_t, Id;tr^2, t \right)\!, \label{eq_116}
\end{align}
and
\begin{align}
    W_M\left( C_t, Id;t, t \right)-W_M\left( C_t, Id;tr, t \right), \label{eq_117}
\end{align}
we can control the shrinking channel to $\left[Id\right]$ and $\left[B\right]$ respectively. One can verify that after the shrinking process,  the remaining $T$-operators in Eqs.~(\ref{eq_116}) and~(\ref{eq_117}) can be written as linear combinations of the operators for $\left[Id\right]$ and $\left[B\right]$ respectively. Likewise,  for $\left[ C_t, - \right]$,  by linearly combining the operators as
\begin{align}
    W_M\left( C_t, -;t, t \right) \!+\!W_M\left( C_t, -;tr, t \right)\!+\!W_M\left( C_t, -;tr^2, t \right),
\end{align}
and
\begin{align}
    W_M\left( C_t, -;t, t \right) -W_M\left( C_t, -;tr, t \right),
\end{align}
we can control the shrinking channel to $\left[A\right]$ and $\left[B\right]$ respectively.

\subsection{Consistency relations between fusion and shrinking}\label{Sec.5.5}
\begin{figure}
	\centering
	\includegraphics[scale=2, keepaspectratio,max width=\linewidth]{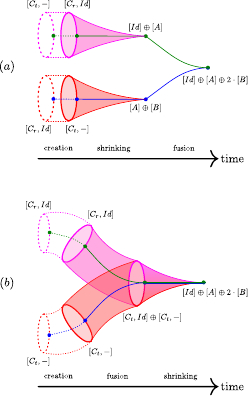}
	\caption{\textbf{Consistency relations between the fusion and shrinking rules.} We create two pairs of loops by applying the cylindrical membrane operators in Eq.~(\ref{eq_loop}). In the $\mathbb{D}_3$ quantum double model,  any excitation is self-dual,  i.e.,  the corresponding antiexcitation is itself. After creating the two pairs of loops,  we can use the connecting rules in Eqs.~(\ref{eq_move_particle}) and~(\ref{eq_L_connect2}) to move or deform the loops. (a) First,  we obtain $\left[ Id \right] \oplus \left[ A \right]$ and $\left[ A \right] \oplus \left[ B \right]$ by shrinking two excitations $\left[C_r, Id\right]$ and $\left[C_t, -\right]$ respectively. Then we obtain $\left[ Id \right] \oplus \left[ A \right] \oplus 2\cdot \left[ B \right] $ by fusing $\left[ Id \right] \oplus \left[ A \right]$ and $\left[ A \right] \oplus \left[ B \right]$. (b) First,  we obtain $\left[ C_r, Id \right] \otimes \left[ C_t, - \right]$ by fusing two excitations $\left[C_r, Id\right]$ and $\left[C_t, -\right]$. Then,  we obtain $\left[ Id \right] \oplus \left[ A \right] \oplus 2\cdot \left[ B \right] $ by shrinking $\left[ C_r, Id \right] \otimes \left[ C_t, - \right]$. These two processes shown in (a) and (b) produce the same results.}
	\label{fig_consistency}
\end{figure}

A systematic check using the fusion table in Table~\ref{tab_D3_fusion} and the shrinking table in Table~\ref{tab_shrinking_d3} demonstrates that all fusion and shrinking processes satisfy the consistency condition in Eq.~(\ref{eq_consistent_4D}) derived from the continuum topological field theory. For example,  as shown in Fig.~\ref{fig_consistency},  if we consider shrinking two excitations $\left[C_r, Id\right]$ and $\left[C_t, -\right]$ followed by fusion in the $\mathbb{D}_3$ quantum double model,  we have
\begin{align}
    \mathcal{S} \left( \left[ C_r, Id \right] \right) \otimes \mathcal{S} \left( \left[ C_t, - \right] \right) &=\left( \left[ Id \right] \oplus \left[ A \right] \right) \otimes \left( \left[ A \right] \oplus \left[ B \right] \right) =\left[ Id \right] \oplus \left[ A \right] \oplus 2\cdot \left[ B \right] \, .
\end{align}
If we first fuse $\left[C_r, Id\right]$ and $\left[C_t, -\right]$,  and then perform the shrinking process,  we have
\begin{align}
    \mathcal{S} \left( \left[ C_r, Id \right] \otimes \left[ C_t, - \right] \right) &=\mathcal{S} \left( \left[ C_t, Id \right] \oplus \left[ C_t, - \right] \right) =\left[ Id \right] \oplus \left[ A \right] \oplus 2\cdot \left[ B \right] \nonumber \\
    &=\mathcal{S} \left( \left[ C_r, Id \right] \right) \otimes \mathcal{S} \left( \left[ C_t, - \right] \right) \, .
\end{align}
We can similarly verify the consistency condition for other excitations. 

Another important property of the shrinking rules obtained from the continuum topological field theory is that the shrinking process preserves quantum dimensions of excitations,  i.e., 
\begin{align}
    d_{\mathsf{a}}=\sum_{\mathsf{b}}{S_{\mathsf{b}}^{\mathsf{a}}d_{\mathsf{b}}}\, , \label{eq_consistent_quantum_dimension}
\end{align}
where $d_{\mathsf{a}}$ and $d_{\mathsf{b}}$ are quantum dimensions of excitations $\mathsf{a}$ and $\mathsf{b}$ respectively. By using the quantum dimensions in Table~\ref{tab_quantum_dimension} and the shrinking rules in Table~\ref{tab_shrinking_d3},  we verify that Eq.~(\ref{eq_consistent_quantum_dimension}) holds in our 3D quantum double model. As an illustration,  consider the excitation $\left[C_r, Id\right]$ in the $\mathbb{D}_3$ quantum double model,  which has $d_{\left[C_r, Id\right]}=2$. Its shrinking rule yields $\left[ Id \right] \oplus \left[ A \right]$,  where $d_{\left[ Id \right]}=d_{\left[ A \right]}=1$,  thus satisfying Eq.~(\ref{eq_consistent_quantum_dimension}).

Finally, we explain that Eq.~(\ref{eq_consistent_4D}) is not special to $\mathbb{D}_3$ but applies to the 3D quantum double model with an arbitrary finite group $G$. The shrinking operation admits a simple interpretation in group representation theory. As discussed in Sec.~\ref{Sec.4.1}, the local action on an excitation is described by the quantum double algebra $DG$, whose basis is $\left\{D_{(h,g)}=B_{p,s}^{h}A_{v}^{g}\right\}$, and each excitation type corresponds to an irreducible representation of $DG$ labeled by $(C,R)$. Particle excitations $\left[R^{\prime}\right]$ are the special case of a trivial flux, and their labels are exactly the irreducible representations of the group $G$, i.e., of the group algebra $C[G]$, which is generated by the operators $A_v^g$. Shrinking a loop means forgetting its flux while retaining its charge, and this is precisely the operation of restricting a representation of $DG$ to the smaller algebra $C[G]$. Restricting an irreducible representation $(C,R)$ of $DG$ to $C[G]$ decomposes it into irreducible representations $R^{\prime}$ of $G$, and the multiplicities of this decomposition are the shrinking coefficients as shown in Eq.~(\ref{eq_shrinking_multiplicity}). An important property of group representations is that restriction is compatible with taking tensor products, i.e., taking the tensor product of two representations and then restricting gives the same result as restricting each representation first and then taking the tensor product. Since fusion is the tensor product of representations and shrinking is restriction, it follows immediately that $\mathcal{S}\left(\mathsf{a}\otimes \mathsf{b}\right)=\mathcal{S}\left(\mathsf{a}\right)\otimes\mathcal{S}\left(\mathsf{b}\right)$ for arbitrary $G$. Moreover, restricting a representation only discards its flux part and does not change the dimension of the representation space, so the total quantum dimension is preserved, i.e., Eq.~(\ref{eq_consistent_quantum_dimension}) also holds for arbitrary $G$. The explicit checks for $G=\mathbb{D}_3$ presented above are concrete illustrations of this general statement.

\section{Particle--loop Braiding and Borromean-Rings braiding}\label{Sec.8}
In Sec.~\ref{sec4} and Sec.~\ref{sec5},  we have established a complete microscopic framework for the fusion and shrinking rules in the 3D quantum double model,  and verified that the fusion and shrinking rules satisfy the consistency condition. In this section,  we extend our analysis to the third fundamental topological operation---braiding. Unlike 2D topological orders, which have only particle--particle statistics, 3D topological orders have richer braiding structures because of the presence of spatially extended loop excitations. In this section,  we investigate not only the familiar particle--loop braiding in Sec.~\ref{Sec.8.1} but also an exotic process called Borromean-Rings braiding in Sec.~\ref{Sec.8.2},  where a particle winds around two unlinked loops in a collectively linked configuration. We microscopically construct operators that implement these braiding processes,  derive associated braiding-induced operators in Eqs.~(\ref{eq_operator_O}) and~(\ref{eq_BR_operator}),  and analyze their behavior in both Abelian and non-Abelian settings. Finally,  in Sec.~\ref{Sec.8.3},  we derive all braiding phases in the $\mathbb{Z}_N$ and $\mathbb{D}_3$ quantum double models as explicit examples.

\subsection{General consideration of particle--loop braiding}\label{Sec.8.1}
In general,  braiding topological excitations implements a unitary transformation on the Hilbert space of the physical system. These braiding operations can serve as quantum gates in topological quantum computation,  where quantum information is stored in non-local degrees of freedom of topological excitations and processed by braiding them. As a result,  an initial wave function typically evolves into a superposition of different states after braiding. However,  there exist special scenarios where a braiding process does not alter the quantum state but only multiplies it by a phase factor,  i.e.,  the wave function returns to itself up to a complex number of unit modulus. Such a phase factor,  often referred to as the braiding phase,  encodes essential information about the topological order. Although we obtain the general form of the unitary transformation caused by the particle--loop braiding (see Eq.~(\ref{eq_operator_O})) and the Borromean-Rings braiding (see Eq.~(\ref{eq_BR_operator})) in this work,  we mainly focus on the case where the braiding operation reduces to a phase factor in the following discussion. A more detailed discussion of the case in which braiding genuinely changes the wave function (i.e.,  implements a unitary transformation rather than merely a phase) will be presented in future work.

\begin{figure*}
	\centering
	\includegraphics[scale=2.7, keepaspectratio,max width=\linewidth]{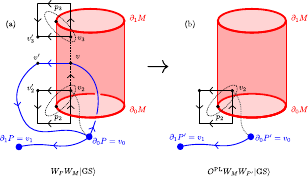}
	\caption{\textbf{Particle--loop braiding process on a lattice.} The red membrane and the blue string represent the operators for loops and particles respectively. For simplicity,  we omit the direct part of the cylindrical membrane operator because the $T$-operator living on the direct part automatically commutes with the string operator for particles. Thus,  the $T$-operator in $W_M$ does not affect the particle--loop braiding process. The cylindrical membrane operator $W_M$ for loops starts and ends on the sites $s_{2}=\left(v_{2}, p_{2}\right)$ and  $s_{3}=\left(v_{3}, p_{3}\right)$ respectively. (a) We apply the cylindrical membrane operator $W_M$ first followed by the string operator $W_{P}$,  where the string $P$ starts from the vertex $\partial_0P=v_{0}$,  winds around the loop living on $\partial_0M$,  and finally ends on the vertex $\partial_1P=v_{1}$. We omit most of the edges that are cut by the membrane $M$ and only draw the starting edge $v_{2}v^{\prime}_{2}$,  the ending edge $v_{3}v^{\prime}_{3}$,  and the edge $vv^{\prime}$ where $M$ and $P$ intersect with each other. The paths $v_{2}v$ and $vv_{3}$ represented by black dashed lines are two arbitrary paths on the direct part of the membrane $M$. The path $v_{0}v_{2}$ represented by a gray dashed line is an arbitrary path such that the closed path $v_{0}v_{2}\cup v_{2}v\cup vv_{0}$ does not link with any loop excitation. (b) By introducing the operator $O^{\mathrm{PL}}$,  we change the order of applying $W_M$ and $W_P$ such that $W_P$ acts to the ground state first as shown in Eq.~(\ref{eq_introduce_O}). Since $W_P\ket{\mathrm{GS}}=W_{P^{\prime}}\ket{\mathrm{GS}}$,  now we can deform the string $P$ to $P^{\prime}$,  where $P^{\prime}$ starts from $v_{0}$ and ends on $v_{1}$ without winding around the loop on $\partial_0M$. We conclude that the particle--loop braiding process leads to an operator $O^{\mathrm{PL}}$ as shown in Eq.~(\ref{eq_operator_O}).}
	\label{fig_particle_loop_braiding}
\end{figure*}
We first consider a particle--loop braiding process as shown in  Fig.~\ref{fig_particle_loop_braiding}. The procedure is as follows.

\begin{enumerate}
	\item We create a pair of loops $\left[C, R\right]$ and $\left[\bar{C}, R^{\ast}\right]$ on the boundaries $\partial_0 M$ and $\partial_1 M$ respectively by applying a cylindrical membrane operator $W_M\left( C, R;c, j;c^{\prime}, j^{\prime} \right)$ to a ground state $\ket{\mathrm{GS}}$.

	\item We create a pair of particles $\left[R^{\prime}\right]$ and $\left[{R^{\prime\ast}}\right]$ on $\partial_0 P$ and $\partial_1 P$ respectively by applying a string operator $W_P\left( R^{\prime};i, i^{\prime} \right)$. The string $P$ in Fig.~\ref{fig_particle_loop_braiding} is composed of a closed string $v_{0}v\cup vv^{\prime}\cup v^{\prime}v_{0}$ and an open string $v_{0}v_{1}$,  which physically means that the particle $\left[R^{\prime}\right]$ is dragged to braid around the loop $\left[C, R\right]$ along the path $v_{0}v\cup vv^{\prime}\cup v^{\prime}v_{0}$. The first two steps are depicted in Fig.~\ref{fig_particle_loop_braiding}\hyperref[fig_particle_loop_braiding]{(a)}.

	\item Since the string $P$ is not allowed to cross any excitation during its deformation,  we cannot simply deform the string $P$ shown in Fig.~\ref{fig_particle_loop_braiding}\hyperref[fig_particle_loop_braiding]{(a)} to the open string $P^{\prime}=v_0v_1$ shown in Fig.~\ref{fig_particle_loop_braiding}\hyperref[fig_particle_loop_braiding]{(b)} without resulting in other consequences. Here $P^{\prime}$ does not intersect with $M$,  which means that $W_M\left( C, R;c, j;c^{\prime}, j^{\prime} \right)$ and $W_{P^{\prime}}\left( R^{\prime};i, i^{\prime} \right)$ commute with each other. To capture the effect of the particle--loop braiding in Fig.~\ref{fig_particle_loop_braiding}\hyperref[fig_particle_loop_braiding]{(a)},  we introduce a braiding-induced operator $O_{\left(R^{\prime}\right), \left(C, R\right)}^{\mathrm{PL}}$ such that 
	\begin{align}
    &W_P\left( R^{\prime};i, i^{\prime} \right)W_M\left( C, R;c, j;c^{\prime}, j^{\prime} \right)\ket{\mathrm{GS}}\nonumber \\
    &=O_{\left(R^{\prime}\right), \left(C, R\right)}^{\mathrm{PL}}W_M\left( C, R;c, j;c^{\prime}, j^{\prime} \right)W_P\left( R^{\prime};i, i^{\prime} \right)\ket{\mathrm{GS}}\, , \label{eq_introduce_O}
\end{align}
	where the string $P$ intersects with the membrane $M$ at the edge $vv^{\prime}$. For simplicity,  we write $O_{\left(R^{\prime}\right), \left(C, R\right)}^{\mathrm{PL}}$,  $W_P\left( R^{\prime};i, i^{\prime} \right)$,  and $W_M\left( C, R;c, j;c^{\prime}, j^{\prime} \right)$ as $O^{\mathrm{PL}}$,  $W_P$,  and $W_M$ respectively when no confusion arises in the following discussion. Here we assume that the arrow of $vv^{\prime}$ aligns with the orientation of the string operator $W_P$. By using the operator $O^{\mathrm{PL}}$,  we change the order of applying $W_M$ and $W_P$ in this step.

	\item Note that the string operator $W_P$ acts on the ground state first on the right-hand side of Eq.~(\ref{eq_introduce_O}). We can freely remove the closed part $v_{0}v\cup vv^{\prime}\cup v^{\prime}v_{0}$ of the entire string $P$ because for any $T$-operator,  we have
	\begin{align}
    T_P^g\ket{\mathrm{GS}}&=\sum_{h}T_{v_0v_1}^{\bar{h}g}T_{v_{0}v\cup vv^{\prime}\cup v^{\prime}v_{0}}^h\ket{\mathrm{GS}}=\sum_{h}T_{v_0v_1}^{\bar{h}g}\delta_{e, h}\ket{\mathrm{GS}}=T_{v_0v_1}^{g}\ket{\mathrm{GS}}\, .
\end{align}
	After removing the closed part $v_{0}v\cup vv^{\prime}\cup v^{\prime}v_{0}$,  the string $P$ is deformed to $P^{\prime}$ in Fig.~\ref{fig_particle_loop_braiding}\hyperref[fig_particle_loop_braiding]{(b)}. Thus,  we have
	\begin{align}
    &W_PW_M\ket{\mathrm{GS}}=O^{\mathrm{PL}}W_MW_{P^{\prime}}\ket{\mathrm{GS}}\, .\label{eq_braid_operator}
\end{align}
	By comparing this with the state $W_MW_{P^{\prime}}\ket{\mathrm{GS}}$ without braiding,  we conclude that the braiding-induced operator $O^{\mathrm{PL}}$ captures the effect of the particle--loop braiding in the 3D quantum double model. 
\end{enumerate}

Recall that in Sec.~\ref{sec4},  we have mentioned that a topological excitation $\left[C, R\right]$ can generally be created by applying an arbitrary linear combination of the operators $W_M\left( C, R;c, j;c^{\prime}, j^{\prime} \right)$ to the ground state $\ket{\mathrm{GS}}$. The set $\left\{W_M\left( C, R;c, j;c^{\prime}, j^{\prime} \right)\ket{\mathrm{GS}}\right\}$ spans the local space $V_{\left(C, R\right)}$ in the region near the boundary $\partial_0 M$. Thus,  braiding a particle $\left[R^{\prime}\right]$ around a loop $\left[C, R\right]$ should be generally understood as a mapping 
\begin{align}
    O^{\mathrm{PL}}\, \, :\, \, V_{\left(R^{\prime}\right)}\otimes V_{\left(C, R\right)}\rightarrow V_{\left(R^{\prime}\right)}\otimes V_{\left(C, R\right)}\, .
\end{align}

In Appendix~\ref{ap5.1},  we show that the braiding-induced operator $O^{\mathrm{PL}}$ is given by
\begin{align}
    O_{\left(R^{\prime}\right), \left(C, R\right)}^{\mathrm{PL}}=\sum_{g, h\in G}T^{g}_{v_{0}v_{2}}B^{h}_{p_{2}, s_{2}}A^{g\bar{h}\bar{g}}_{v_{0}}\, , \label{eq_operator_O}
\end{align}
where $v_{0}$ is the starting vertex of the string $P$,  $s_{2}=\left(v_{2}, p_{2}\right)$ is the starting site of the membrane $M$ as shown in Fig.~\ref{fig_particle_loop_braiding}. If the string $P$ and the membrane $M$ intersect on the edge $vv^{\prime}$, then the path $v_{0}v_{2}$ together with $v_{2}v$ and $vv_{0}$ forms a closed path that does not link with any loop excitation. The right-hand side of Eq.~(\ref{eq_operator_O}) does not explicitly involve the labels $R^{\prime}$, $C$, or $R$. However, the subscripts on the left-hand side are essential because they specify the spatial positions where the operator acts. In particular, the term $A^{g\bar{h}\bar{g}}_{v_{0}}$ acts on the vertex $v_{0}$ associated with the particle, the term $B^{h}_{p_{2}, s_{2}}$ acts on the plaquette $p_{2}$ associated with the loop's flux, and the $T$-operator $T^{g}_{v_{0}v_{2}}$ acts on the string $v_{0}v_{2}$ that connects the particle $[R^{\prime}]$ and the loop $[C,R]$. Thus, the subscripts $(R')$ and $(C,R)$ serve to indicate that $O_{\left(R^{\prime}\right), \left(C, R\right)}^{\mathrm{PL}}$ acts on the subspace $V_{(R')} \otimes V_{(C,R)}$, i.e., on the specific spatial region where the particle and the loop are located. When the excitations participating in the braiding process are clear, we will omit the subscripts and write $O^{\mathrm{PL}}$ for simplicity.

From Eq.~(\ref{eq_operator_O}) we can also see that $O^{\mathrm{PL}}$ does not act on the $T$-operators carried by $W_M$. Since charge decorations of loops are related to these $T$-operators,  \textit{we conclude that the particle--loop braiding process is independent of charge decorations of loops}. As a result, it is sufficient to consider pure loops when computing particle--loop braiding processes.

Fig.~\ref{fig_particle_loop_braiding}\hyperref[fig_particle_loop_braiding]{(a)} depicts braiding the particle $\left[R^{\prime}\right]$ around the loop $\left[C, R\right]$ in the counterclockwise direction,  where the orientation of the string $P$ points from the direct part of the membrane $M$ to the dual part of the membrane $M$ on the edge $vv^{\prime}$. We can similarly obtain that when the particle $\left[R^{\prime}\right]$ braids around the loop $\left[C, R\right]$ in the clockwise direction (i.e.,  the orientation of $P$ points from the dual part of $M$ to the direct part of $M$ on $vv^{\prime}$),  Eq.~(\ref{eq_braid_operator}) becomes
\begin{align}
    &W_PW_M\ket{\mathrm{GS}}=\left(O^{\mathrm{PL}}\right)^{-1}W_MW_{P^{\prime}}\ket{\mathrm{GS}}\, ,
\end{align}
where
\begin{align}
    \left(O^{\mathrm{PL}}\right)^{-1}=\sum_{g, h\in G}T^{g}_{v_{0}v_{2}}B^{h}_{p_{2}, s_{2}}A^{gh\bar{g}}_{v_{0}}\, .
\end{align}

In this work, we are particularly interested in cases where the braiding-induced operator reduces to a phase. To state this result precisely, we introduce two definitions of Abelian charge and Abelian flux.

\begin{definition}[Abelian charge]
	\label{def:abelian_charge}
	The charge carried by an excitation $[C,R]$ is called \textbf{Abelian} if the irrep $R$ is one-dimensional; otherwise, the charge is called \textbf{non-Abelian}.
\end{definition}

\begin{definition}[Abelian flux]
	\label{def:abelian_flux}
	The flux carried by an excitation $[C,R]$ is called \textbf{Abelian} if the corresponding conjugacy class $C$ contains only a single element, i.e., $|C|=1$; otherwise, the flux is called \textbf{non-Abelian}.
\end{definition}

A particle corresponds to the case $C = C_e$ (the conjugacy class of the identity), so its flux is trivial and the overall Abelian nature of the particle is determined solely by its charge: a particle is Abelian if and only if its charge is Abelian. For a loop excitation ($C \neq C_e$), a pure loop carries trivial charge decoration ($R = Id$), hence its overall Abelian nature is determined entirely by its flux: a pure loop is Abelian if and only if its flux is Abelian. A decorated loop, however, is Abelian only if both its flux and its charge decoration are Abelian; if either the flux is non-Abelian or the charge decoration is non-Abelian, the decorated loop is non-Abelian.

We state the following theorem.
\begin{theorem}
	In a particle--loop braiding process, if either the charge carried by the particle or the flux carried by the loop is Abelian, then this braiding only produces a phase $\Gamma^{R^{\prime\ast}}_{ii}\left( c \right)$,  where $R^{\prime}$ is the irrep that corresponds to the particle,  $c$ is an arbitrary element of the conjugacy class $C$ that corresponds to the loop,  $i=1, \cdots, n_{R^{\prime}}$.\label{theorem_plphase}
\end{theorem}
\begin{proof}
	Denote the particle and the loop involved in the particle--loop braiding as $\left[R^{\prime}\right]$ and $\left[C, R\right]$ respectively. Note that in $O^{\mathrm{PL}}$,  only the factor $A^{g\bar{h}\bar{g}}_{v_{0}}$ acts on the operator for $\left[R^{\prime}\right]$. Using Eq.~(\ref{eq_39}),  we have
	\begin{align}
    A^{g\bar{h}\bar{g}}_{v_{0}}W_{P^{\prime}}\left( R^{\prime};i, i^{\prime}\right)\ket{\mathrm{GS}} &=\sum_j\Gamma_{ji}^{R^{\prime}}\left( g\bar{h}\bar{g} \right) W_{P^{\prime}}\left( R^{\prime};j, i^{\prime}\right)\ket{\mathrm{GS}}\nonumber \\
    &=\sum_j\Gamma_{ij}^{R^{\prime\ast}}\left( gh\bar{g} \right) W_{P^{\prime}}\left( R^{\prime};j, i^{\prime}\right)\ket{\mathrm{GS}}\, .\label{eq_AWP}
\end{align}
	If the charge carried by the particle is Abelian (see \textbf{Definition}~\ref{def:abelian_charge}), then the representation matrix is just a number and we have 
	 $
		\Gamma^{R^{\prime\ast}}\left( gh\bar{g} \right)=\Gamma^{R^{\prime\ast}}\left( g \right)\Gamma^{R^{\prime\ast}}\left( h \right)\Gamma^{R^{\prime\ast}}\left( \bar{g} \right)=\Gamma^{R^{\prime\ast}}\left( h \right)\, $, 
	where $\left|\Gamma^{R^{\prime\ast}}\left( h \right)\right|=1$ and we have omitted $i=i^{\prime}=1$. Thus,  Eq.~(\ref{eq_AWP}) becomes
	\begin{align}
    A^{g\bar{h}\bar{g}}_{v_{0}}W_{P^{\prime}}\left( R^{\prime};i, i^{\prime}\right)\ket{\mathrm{GS}}=\Gamma^{R^{\prime\ast}}\left( h \right)W_{P^{\prime}}\left( R^{\prime};i, i^{\prime}\right)\ket{\mathrm{GS}}\, .
\end{align}
	Now,  the operator $O^{\mathrm{PL}}$ acts on the space $V_{\left(R^{\prime}\right)}\otimes V_{\left(C, R\right)}$ as
	\begin{align}
    O^{\mathrm{PL}}&=\sum_{g, h\in G}T^{g}_{v_{0}v_{2}}B^{h}_{p_{2}, s_{2}}\Gamma^{R^{\prime\ast}}\left( h \right) =\sum_{g, h\in G}T^{g}_{v_{0}v_{2}}\delta_{h\in C}\Gamma^{R^{\prime\ast}}\left( h \right)=\sum_{g\in G}T^{g}_{v_{0}v_{2}}\Gamma^{R^{\prime\ast}}\left( c \right)\nonumber \\
    &=\Gamma^{R^{\prime\ast}}\left( c \right)\, , \label{eq_p_Ab}
\end{align}
    where we have used Eq.~(\ref{eq_bp}) to obtain the second equality. $\delta_{h\in C}=1$ when $h\in C$,  otherwise $\delta_{h\in C}=0$.
	
	Now suppose that the flux of the loop is Abelian (see \textbf{Definition}~\ref{def:abelian_flux}), i.e., the corresponding conjugacy class $C$ contains only one element $c$,  and $c$ must commute with any group element $g$. Thus,  Schur's lemma guarantees that
	$
		\Gamma^{R^{\prime\ast}}_{ii^{\prime}}\left( gc\bar{g} \right)=\delta_{i, i^{\prime}}\Gamma^{R^{\prime\ast}}_{ii^{\prime}}\left( c \right)\, $,	where $\Gamma^{R^{\prime\ast}}_{ii}\left( c \right)=\Gamma^{R^{\prime\ast}}_{jj}\left( c \right)$ and $\left|\Gamma^{R^{\prime\ast}}_{ii}\left( c \right)\right|=1$. Thus,  similar to Eq.~(\ref{eq_p_Ab}),  we find that the operator $O^{\mathrm{PL}}$ acts on the space $V_{\left(R^{\prime}\right)}\otimes V_{\left(C, R\right)}$ as
	\begin{align}
    O^{\mathrm{PL}}=\Gamma^{R^{\prime\ast}}_{ii}\left( c \right)\, , \label{eq_l_Ab}
\end{align}
    where we have chosen $i=i^{\prime}$. When $i\ne i^{\prime}$,  the operator $O^{\mathrm{PL}}=0$,  which means that this process is forbidden. Eq.~(\ref{eq_l_Ab}) also shows that the braiding phase $\Gamma^{R^{\prime\ast}}_{ii}\left( c \right)$ is independent of the charge decoration carried by the loop, confirming the earlier observation that it suffices to consider pure loops when computing particle--loop braiding processes (see the discussion below Eq.~(\ref{eq_operator_O})).
\end{proof}
When both the particle and the loop are Abelian,  it is obvious that the particle--loop braiding between them produces a phase. This theorem further shows that if a particle--loop braiding process involves an Abelian excitation and a non-Abelian excitation,  then the braiding still produces a phase 
\begin{align}
    \exp \left(\mathrm{i}\theta_{\left(R^{\prime}\right), \left(C, R\right)}^{\mathrm{PL}}\right)=\Gamma^{R^{\prime\ast}}_{ii}\left( c \right)\, , \label{eq_particle_loop_braiding_phase}
\end{align}
where the particle is denoted by $\left[R^{\prime}\right]$,  the loop is denoted by $\left[C, R\right]$,  $c\in C$. The braiding phase in Eq.~(\ref{eq_particle_loop_braiding_phase}) is independent of the internal degrees of freedom associated with the non-Abelian excitation. If the particle $[R^{\prime}]$ is non-Abelian while the flux carried by the loop is Abelian, the phase $\Gamma^{R^{\prime\ast}}_{ii}(c)$ takes the same value for all indices $i$, i.e., $\Gamma^{R^{\prime\ast}}_{ii}(c) = \Gamma^{R^{\prime\ast}}_{jj}(c)$ for any $i,j = 1,\dots,n_{R^{\prime}}$. If instead the loop $[C,R]$ carries a non-Abelian flux while the particle is Abelian, the phase satisfies $\Gamma^{R^{\prime\ast}}(c) = \Gamma^{R^{\prime\ast}}(gc\bar{g})$ for all $g \in G$, meaning it depends only on the conjugacy class $C$ and not on the specific representative $c$. In either case, the braiding phase is well-defined and does not depend on the choice of internal labels. Only when both the charge carried by the particle and the flux carried by the loop are non-Abelian does the particle--loop braiding genuinely change the wave function $W_MW_{P^{\prime}}\ket{\mathrm{GS}}$ rather than just producing a phase. The form of the wave function $O^{\mathrm{PL}}W_MW_{P^{\prime}}\ket{\mathrm{GS}}$ depends on the internal degrees of freedom in $W_M$ and $W_{P^{\prime}}$.

\subsection{General consideration of Borromean-Rings braiding}\label{Sec.8.2}
\begin{figure*}
	\centering
	\includegraphics[scale=2.7, keepaspectratio,max width=\linewidth]{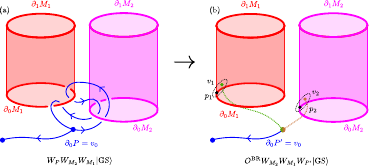}
	\caption{\textbf{Borromean-Rings braiding on a lattice.} $W_P$,  $W_{M_1}$,  and $W_{M_2}$ denote the creation operators for $\left[R^{\prime}\right]$,  $\left[C_1, R_1\right]$,  and $\left[C_2, R_2\right]$ respectively. (a) The trajectory of the particle $\left[R^{\prime}\right]$ and the two loops form the Borromean-Rings structure,  i.e.,  the closed part of the string $P$,  the red boundary $\partial_0M_1$,  and the purple boundary $\partial_0M_2$ are pairwise unlinked yet collectively linked. We denote the starting sites of $M_1$ and $M_2$ as $s_{1}=\left(v_{1}, p_{1}\right)$ and $s_{2}=\left(v_{2}, p_{2}\right)$ respectively. $v_0$ denotes the starting vertex of the string $P$. (b) By introducing the operator $O^{\mathrm{BR}}$ shown in Eq.~(\ref{eq_BR_operator}),  we change the order of applying $W_P$,  $W_{M_2}$,  and $W_{M_1}$ such that $W_P$ acts on the ground state first. In other words,  we have $W_{P}W_{M_2}W_{M_1}\ket{\mathrm{GS}}=O^{\mathrm{BR}}W_{M_2}W_{M_1}W_{P}\ket{\mathrm{GS}}$. Then,  we can freely deform the string $P$ to the open string $P^{\prime}$ by using $W_P\ket{\mathrm{GS}}=W_{P^{\prime}}\ket{\mathrm{GS}}$. The wave function $W_{M_2}W_{M_1}W_{P^{\prime}}\ket{\mathrm{GS}}$ corresponds to the state before the braiding process. We conclude that the operator $O^{\mathrm{BR}}$ reduces to the identity unless the charge carried by the particle and the fluxes carried by the two loops are all non-Abelian. In some special cases such as $G=\mathbb{D}_4$,  the Borromean-Rings braiding of non-Abelian excitations carrying non-Abelian charges and fluxes can produce a nontrivial braiding phase.}
	\label{fig_BR_braiding}
\end{figure*}

We now study microscopically the Borromean-Rings braiding process previously proposed through field-theoretical analysis. As shown in Fig.~\ref{fig_BR_braiding},  we first create two loops $\left[C_1, R_1\right]$ and $\left[C_2, R_2\right]$ by applying operators $W_{M_1}$ ($W_{M_1}$ is an abbreviation for $W_{M_1}\left( C_1, R_1;c_1, j_1;c^{\prime}_1, j^{\prime}_1 \right)$) and $W_{M_2}$ ($W_{M_2}$ is an abbreviation for $W_{M_2}\left( C_2, R_2;c_2, j_2;c^{\prime}_2, j^{\prime}_2 \right)$) to a ground state $\ket{\mathrm{GS}}$. Then we create a pair of particles $\left[R^{\prime}\right]$ and $\left[R^{\prime\ast}\right]$ by subsequently applying a string operator $W_{P}$ ($W_{P}$ is an abbreviation for $W_{P}\left( R^{\prime};i, i^{\prime} \right)$). Starting from the vertex $v_0$,  we drag the particle $\left[R^{\prime}\right]$ to braid around the two loops,  and finally have it return to the vertex $v_0$. Thus,  the corresponding string $P$ is composed of a closed string and an open string. The closed part of $P$ is understood as the trajectory of the particle $\left[R^{\prime}\right]$ and the open part of $P$ connects $\left[R^{\prime}\right]$ and its antiparticle $\left[R^{\prime\ast}\right]$. The closed part of $P$ and the two loops form the Borromean-Rings structure,  in which the three components are pairwise unlinked yet collectively linked. 

To investigate this exotic Borromean-Rings braiding process,  we need to derive an operator $O^{\mathrm{BR}}_{\left(R^{\prime}\right), \left(C_1, R_1\right), \left(C_2, R_2\right)}$ resulting from this braiding process. Similar to Eq.~(\ref{eq_braid_operator}),  we write
\begin{align}
    &W_PW_{M_2}W_{M_1}\ket{\mathrm{GS}}=O^{\mathrm{BR}}W_{M_2}W_{M_1}W_{P^{\prime}}\ket{\mathrm{GS}}\, ,
\end{align}
where $O^{\mathrm{BR}}$ is an abbreviation for $O^{\mathrm{BR}}_{\left(R^{\prime}\right), \left(C_1, R_1\right), \left(C_2, R_2\right)}$. In Appendix~\ref{ap5.2},  we prove that
\begin{align}
    O^{\mathrm{BR}}&=\sum_{\substack{c_1, c_2, \\g_1, g_2\in G}}B_{p_{2}, s_{2}}^{c_2}B_{p_{1}, s_{1}}^{c_1}T_{v_0v_{2}}^{g_2\bar{c}_2}T_{v_0v_{1}}^{\left( g_2:\bar{c}_2 \right) g_1c_1} \times A_{v_0}^{\left( g_2:\bar{c}_2 \right) \left( g_1:\bar{c}_1 \right) \left( g_2:c_2 \right) \left( g_1:c_1 \right)}\, , \label{eq_BR_operator}
\end{align}
where we denote the starting sites of $M_1$ and $M_2$ as $s_{1}=\left(v_{1}, p_{1}\right)$ and $s_{2}=\left(v_{2}, p_{2}\right)$ respectively. \textit{For simplicity,  we define the notation $g:h$ as $gh\bar{g}$.} The right-hand side does not explicitly involve the labels $R^{\prime}$, $R_1$, $R_2$, $C_1$, or $C_2$. However, the subscripts on the left-hand side are essential because they specify the spatial positions where the operator acts. In particular, the $B_{p_1,s_1}^{c_1}$ and $B_{p_2,s_2}^{c_2}$ terms act on the starting sites $s_1$ and $s_2$ of the two membrane operators, respectively, associated with the fluxes of the two loops. The $T$-operators $T_{v_0v_{2}}^{g_2\bar{c}_2}$ and $T_{v_0v_{1}}^{\left( g_2:\bar{c}_2 \right) g_1c_1}$ act on strings $v_0v_2$ and $v_0v_1$ respectively, where $v_0v_2$ connects the particle $[R^{\prime}]$ and the loop $[C_2,R_2]$, and $v_0v_1$ connects the particle $[R^{\prime}]$ and the loop $[C_1,R_1]$. The term $A_{v_0}^{\left( g_2:\bar{c}_2 \right) \left( g_1:\bar{c}_1 \right) \left( g_2:c_2 \right) \left( g_1:c_1 \right)}$ acts on the vertex $v_0$ where the particle resides. Thus, the subscripts $(R^{\prime})$, $(C_1,R_1)$, and $(C_2,R_2)$ indicate that $O^{\mathrm{BR}}_{\left(R^{\prime}\right), \left(C_1, R_1\right), \left(C_2, R_2\right)}$ acts on the space $V_{(R^{\prime})} \otimes V_{(C_1,R_1)} \otimes V_{(C_2,R_2)}$, i.e., on the specific spatial region where the particle and the two loops are located. When the excitations participating in the braiding process are clear, we will omit the subscripts and write $O^{\mathrm{BR}}$ for simplicity.

As in particle--loop braiding, since $O^{\mathrm{BR}}$ does not act on the $T$-operators in $W_{M_1}$ and $W_{M_2}$, \textit{we conclude that the Borromean-Rings braiding process is independent of the charge decorations carried by loops.} 

Not all theories have nontrivial Borromean-Rings braiding. We state the following theorem.
\begin{theorem}
	The Borromean-Rings braiding of $\left[R^{\prime}\right]$,  $\left[C_1, R_1\right]$,  and $\left[C_2, R_2\right]$ is trivial unless the charge carried by the particle and the fluxes carried by the two loops are all non-Abelian.\label{theorem_bring}
\end{theorem}
\begin{proof}
	 As in particle--loop braiding, if the charge is Abelian (\textbf{Definition}~\ref{def:abelian_charge}) or either flux is Abelian (\textbf{Definition}~\ref{def:abelian_flux}), then
	\begin{align}
    \Gamma ^{R^{\prime\ast}}_{ii^{\prime}}\!\left( \left( g_1:\bar{c}_1 \right) \!\left( g_2:\bar{c}_2 \right)\! \left( g_1:c_1 \right) \!\left( g_2:c_2 \right)  \right)=\Gamma ^{R^{\prime\ast}}_{ii^{\prime}}\!\left( e  \right)=\delta_{i, i^{\prime}}\, .
\end{align}
	Then by choosing $i=i^{\prime}$,  the operator $O^{\mathrm{BR}}$ acts on the space $V_{\left(R^{\prime}\right)}\otimes V_{\left(C_1, R_1\right)}\otimes V_{\left(C_2, R_2\right)}$ as
	\begin{align}
    O^{\mathrm{BR}}&=\sum_{\substack{c_1, c_2, \\g_1, g_2\in G}}B_{p_{2}, s_{2}}^{c_2}B_{p_{1}, s_{1}}^{c_1}T_{v_0v_{2}}^{g_2\bar{c}_2}T_{v_0v_{1}}^{\left( g_2:\bar{c}_2 \right) g_1c_1} =\sum_{\substack{c_1, c_2, \\g_2\in G}}B_{p_{2}, s_{2}}^{c_2}B_{p_{1}, s_{1}}^{c_1}T_{v_0v_{2}}^{g_2\bar{c}_2}\nonumber \\
    &=\sum_{\substack{c_1, c_2\in G}}B_{p_{2}, s_{2}}^{c_2}B_{p_{1}, s_{1}}^{c_1}=1\, .\label{eq_trivial_BR}
\end{align}
	 The operator $O^{\mathrm{BR}}$ reduces to the identity. We conclude that nontrivial Borromean-Rings braiding processes are possible only when the charge carried by the particle and the fluxes carried by the two loops are non-Abelian.
\end{proof}
This theorem shows that if a Borromean-Rings braiding process involves any Abelian charge or flux,  this braiding process reduces to the identity operator and it is independent of the choices of degrees of freedom associated with the non-Abelian excitations.

Generally,  the operator $O^{\mathrm{BR}}$ implements a unitary transformation on the space $V_{\left(R^{\prime}\right)}\otimes V_{\left(C_1, R_1\right)}\otimes V_{\left(C_2, R_2\right)}$ when the charge and two fluxes are non-Abelian. However,  if 
\begin{align}
    \left( g_1:\bar{c}_1 \right) \left( g_2:\bar{c}_2 \right) \left( g_1:c_1 \right) \left( g_2:c_2 \right)=z\ne e\, , \, \, \, \forall g_1, g_2\in G\, , \label{eq_BR_phase_condition}
\end{align}
then the Borromean-Rings braiding can produce a nontrivial braiding phase because the group element $z$ satisfying the equation above must commute with all group elements in $G$. Thus,  $\Gamma ^{R^{\prime\ast}}_{ii^{\prime}}\left( z  \right)=\delta_{i, i^{\prime}}\omega\left(z\right)$,  where $\omega\left(z\right)=\Gamma ^{R^{\prime\ast}}_{ii}\left( z  \right)$. The Borromean-Rings braiding phase is now given by
\begin{align}
    \exp \left(\mathrm{i}\theta^{\mathrm{BR}}_{\left(R^{\prime}\right), \left(C_1, R_1\right), \left(C_2, R_2\right)}\right)=\omega\left( z \right)\, .\label{eq_BR_braiding_phase}
\end{align}

For most non-Abelian groups,  the condition in Eq.~(\ref{eq_BR_phase_condition}) is not satisfied. Thus,  when the charge and two fluxes carried by the three excitations are non-Abelian,  the Borromean-Rings braiding implements a unitary transformation on the space $V_{\left(R^{\prime}\right)}\otimes V_{\left(C_1, R_1\right)}\otimes V_{\left(C_2, R_2\right)}$ and changes the initial wave function $W_{M_2}W_{M_1}W_{P^{\prime}}\ket{\mathrm{GS}}$. Only some special cases such as the $\mathbb{D}_4$ quantum double model can satisfy the condition in Eq.~(\ref{eq_BR_phase_condition}). In this case,  although the charge and two fluxes carried by the three excitations are all non-Abelian,  the Borromean-Rings braiding of the three excitations only produces a phase that is independent of the internal degrees of freedom associated with excitations. The details of the $\mathbb{D}_4$ quantum double model are shown in Sec.~\ref{Sec.6.2}.

Interestingly, besides $\mathbb{D}_4$, the quaternion group $\mathbb{Q}_8$ also satisfies the condition in Eq.~(\ref{eq_BR_phase_condition}). 
The eight group elements of the quaternion group are labeled by $\left\{\pm e,\,\pm\mathcal{I},\,\pm\mathcal{J},\,\pm\mathcal{K}\right\}$. 
The multiplication rules are given by $\pm eg=\pm g ,\,\forall g\in \mathbb{Q}_8$, $\mathcal{I}^2 = \mathcal{J}^2 = \mathcal{K}^2 = -e$, $\mathcal{IJ} = \mathcal{K}$, $\mathcal{JK} = \mathcal{I}$, $\mathcal{KI} = \mathcal{J}$, and all other products follow from these relations. 
There are five conjugacy classes in $\mathbb{Q}_8$: $C_e=\left\{e\right\}$, $C_{-e}=\left\{-e\right\}$, $C_{\mathcal{I}}=\left\{\pm\mathcal{I}\right\}$, $C_{\mathcal{J}}=\left\{\pm\mathcal{J}\right\}$, and $C_{\mathcal{K}}=\left\{\pm\mathcal{K}\right\}$. 
Choose, for instance, the two conjugacy classes $C_1=C_{\mathcal{I}}$ and $C_2=C_{\mathcal{J}}$ with $c_1 = \mathcal{I}$, $c_2 = \mathcal{J}$ in Eq.~(\ref{eq_BR_phase_condition}). 
Then for any $g_1, g_2 \in \mathbb{Q}_8$, one computes $(g_1 : c_1) = g_1 \mathcal{I} g_1^{-1} = \pm \mathcal{I}$ and similarly $(g_2 : c_2) = \pm \mathcal{J}$. 
The product in Eq.~(\ref{eq_BR_phase_condition}) satisfies 
\begin{align}
    (g_1 : \bar{c}_1)(g_2 : \bar{c}_2)(g_1 : c_1)(g_2 : c_2) &=(\mp \mathcal{I})(\mp \mathcal{J})(\pm \mathcal{I})(\pm \mathcal{J}) =\left(-e\right)^2 \mathcal{I}\mathcal{J}\mathcal{I}\mathcal{J}=-e\,.
\end{align} 
Thus the product equals the central element $-e$, independent of the choices of $g_1, g_2$, and therefore satisfies Eq.~(\ref{eq_BR_phase_condition}) with $z = -e$.  
Consequently, the $\mathbb{Q}_8$ quantum double model also yields a nontrivial Borromean-Rings braiding phase $\exp(\mathrm{i}\theta^{\mathrm{BR}}) = -1$ for appropriate non-Abelian excitations.

\subsection{Examples of particle--loop and Borromean-Rings braiding}\label{Sec.8.3}
We first consider the Abelian group $G=\mathbb{Z}_N$ as an example. From Sec.~\ref{Sec.4.2} we can see that a loop and a particle are labeled by $\left[g^\alpha, R^{\left(\beta\right)}\right]$ and $\left[R^{\left(\gamma\right)}\right]$ respectively,  where $g$ is the generator of $\mathbb{Z}_N$,  $\alpha, \beta, \gamma=0, 1, 2, \cdots, \left| G \right|-1$ and $R^{\left(\beta\right)}$ labels an irrep of $\mathbb{Z}_N$. The representation matrix is given by $\Gamma ^{R^{\left( \beta \right)}}\left( g^{\alpha} \right) =\omega ^{\alpha \beta}$,  where $\omega=\exp\left(2\pi \mathrm{i}/N\right)$.

According to the discussion in Sec.~\ref{Sec.8.1},  braiding the particle $\left[R^{\left(\gamma\right)}\right]$ around the loop $\left[g^\alpha, R^{\left(\beta\right)}\right]$ produces a phase
\begin{align}
    \exp \left(\mathrm{i}\theta_{\left(R^{\left(\gamma\right)}\right), \left(g^\alpha, R^{\left(\beta\right)}\right)}^{\mathrm{PL}}\right)&=\left[\Gamma ^{R^{\left( \gamma \right)}}\left( g^{\alpha} \right)\right]^{\ast} =\exp\left(-\frac{2\pi \mathrm{i}}{N}\alpha\gamma\right)\, .
\end{align}
Since all excitations are Abelian and carry Abelian charges and fluxes,  according to the discussion in Sec.~\ref{Sec.8.2},  any Borromean-Rings braiding process is trivial.

Then,  we consider the non-Abelian group $G=\mathbb{D}_3$ as an example. Recall that in Sec.~\ref{Sec.4.3},  the two generators are denoted by $r$ and $t$,  which satisfy $r^3=t^2=e, \, rt=tr^2$. The three particles are denoted by $\left[ Id \right]$,  $\left[ A \right]$,  and $\left[ B \right]$ respectively. The five loops are denoted by $\left[ C_r, Id \right]$,  $\left[ C_r, \omega \right]$,  $\left[ C_r, \omega ^2 \right]$,  $\left[ C_t, Id \right]$,  and $\left[ C_t, - \right]$ respectively. From Eqs.~(\ref{eq_operator_O}) and~(\ref{eq_BR_operator}) we can see that the operators $O^{\mathrm{PL}}$ and $O^{\mathrm{BR}}$ do not depend on charge decorations on loops. Thus,  we only need to consider the pure loops $\left[ C_r, Id \right]$ and $\left[ C_t, Id \right]$ in the particle--loop braiding and the Borromean-Rings braiding. 

We first derive the Abelian particle--loop braiding phases. According to the discussion in Sec.~\ref{Sec.8.1},  there are two Abelian particle--loop braiding processes corresponding to braiding $\left[ A \right]$ with $\left[ C_r, Id \right]$ and braiding $\left[ A \right]$ with $\left[ C_t, Id \right]$. Their phases are given by
\begin{align}
    \exp \left(\mathrm{i}\theta_{\left(A\right), \left(C_r, Id\right)}^{\mathrm{PL}}\right)&=\left[\Gamma^{A}\left( r \right)\right]^{\ast}=1\, , \\
    \exp \left(\mathrm{i}\theta_{\left(A\right), \left(C_t, Id\right)}^{\mathrm{PL}}\right)&=\left[\Gamma^{A}\left( t \right)\right]^{\ast}=-1\, .
\end{align}

The result of braiding the non-Abelian particle $\left[ B \right]$ with the non-Abelian loops $\left[ C_r, Id \right]$ or $\left[ C_t, Id \right]$ is not a phase,  but a unitary transformation given by Eq.~(\ref{eq_operator_O}). As for the nontrivial Borromean-Rings braiding,  the only possible choices are given by $\left[R^{\prime}\right]=\left[ B \right]$,  $\left[C_{i}, R_{i}\right]=\left[ C_r, Id \right], \left[ C_t, Id \right]$,  $i=1, 2$. We can verify that when $c_1, c_2\in C_r$,  we have $\left( g_1:\bar{c}_1 \right) \left( g_2:\bar{c}_2 \right) \left( g_1:c_1 \right) \left( g_2:c_2 \right)=e$. When $c_1\in C_t$,  $c_2\in C_r$ or when $c_1\in C_t$,  $c_2\in C_t$,  the group element $\left( g_1:\bar{c}_1 \right) \left( g_2:\bar{c}_2 \right) \left( g_1:c_1 \right) \left( g_2:c_2 \right)$ depends on the choice of $g_1$ and $g_2$. Thus,  they do not satisfy Eq.~(\ref{eq_BR_phase_condition}),  which means that the Borromean-Rings braiding in the $\mathbb{D}_3$ quantum double model cannot produce a nontrivial phase. To be more specific,  the Borromean-Rings braiding between $\left[ B \right]$,  $\left[ C_r, Id \right]$,  and $\left[ C_r, Id \right]$ is trivial. The Borromean-Rings braiding between $\left[ B \right]$,  $\left[ C_t, Id \right]$,  and $\left[ C_t, Id \right]$,  as well as the Borromean-Rings braiding between $\left[ B \right]$,  $\left[ C_t, Id \right]$,  and $\left[ C_r, Id \right]$,  are non-Abelian and they lead to the unitary transformation shown in Eq.~(\ref{eq_BR_operator}).

\section{Microscopic construction of the $BF$ field theory with an $AAB$ twist}\label{Sec.6}
\subsection{Overview}
Within the framework of continuum field theory,  the (3+1)D $BF$ field theory with an $AAB$ twist unveils a hidden layer of quantum statistical behavior in 3D topological phases. Unlike conventional particle--loop braiding,  where a phase arises from the direct winding of a particle around a loop,  this theory supports braiding processes in which a particle moves around two unlinked loops without winding around either~\cite{PhysRevLett.121.061601}. The resulting trajectory,  together with the loops,  forms intricate Brunnian structures such as the Borromean-Rings,  i.e.,  a configuration in which all three components are pairwise unlinked yet collectively linked. Remarkably,  the nontrivial Borromean-Rings braiding indicates that this theory exhibits non-Abelian topological order,  despite the underlying gauge group $G=\prod_i\mathbb{Z}_{N_i}$ being Abelian. It is important to note that Borromean-Rings braiding is not compatible with all multi-loop braiding processes,  because a legitimate topological action cannot be written down to describe certain combinations of these processes~\cite{PhysRevResearch.3.023132}. This incompatibility stems from the requirement that the action for the twisted $BF$ theory must be gauge-invariant. Building on these results,  Ref.~\cite{Zhang2023fusion} systematically computes all fusion and shrinking rules in the $BF$ theory with an $AAB$ twist and confirms the presence of non-Abelian fusion and shrinking rules. Moreover,  these rules are shown to satisfy the consistency condition in Eq.~(\ref{eq_consistent_4D}),  which strongly constrains the associated fusion and shrinking coefficients.
\begin{table*}
	\caption{\label{tab_compare}Checklist for the comparison \textit{between} the $BF$ field theory with an $AAB$ twist and gauge group $G=(\mathbb{Z}_2)^3$ \textit{and} the 3D $\mathbb{D}_4$ quantum double model. We list the methods of deriving the GSD,  excitations,  fusion,  shrinking,  and braiding in both theories. Their topological data agree.}
	\centering
	\renewcommand{\arraystretch}{1.3}  
    \par\smallskip\hrule\smallskip
	\begin{adjustbox}{max width=\linewidth,max totalheight=.82\textheight}
\begin{tabular}{ccc}
		\noalign{\vspace{3pt}}
		& \parbox{6.7cm}{\raggedright $BF+AAB$ theory with gauge group $G=(\mathbb{Z}_2)^3$} & Microscopic construction 
		\\[7pt]
        \hline\noalign{\vspace{5pt}}
		GSD & \parbox{6.7cm}{\raggedright Computed from the topological partition functions. See Ref.~\cite{Zhang2023fusion}.} & \parbox{6.7cm}{\raggedright Computed from the ground-state projector and group-theoretical analysis. See Eq.~(\ref{eq_GSD}).} 
		\\[10pt]
		Excitations & \parbox{6.7cm}{\raggedright Represented by the inequivalent gauge-invariant Wilson operators. See Ref.~\cite{Zhang2023fusion}.} & \parbox{6.7cm}{\raggedright Represented by the string operators in Eq.~(\ref{eq_particle}) or thickened cylindrical membrane operators in Eq.~(\ref{eq_loop}).}
		\\[15pt]
		Fusion rules & \parbox{6.7cm}{\raggedright Computed from the correlation function in Eq.~(\ref{eq_fusion_TQFT}). Satisfy the fusion--shrinking consistency condition in Eq.~(\ref{eq_consistent_4D}). See Refs.~\cite{Zhang2023fusion, Huang_2025}.} & \parbox{6.7cm}{\raggedright Computed from the tensor product of irreducible representations of $DG$. See Tables~\ref{tab_D4_fusion1} and~\ref{tab_D4_fusion2}.} 
		\\[20pt]
		Shrinking rules & \parbox{6.7cm}{\raggedright Computed from the correlation function in Eq.~(\ref{eq_shrinking_TQFT}). Satisfy the fusion--shrinking consistency condition in Eq.~(\ref{eq_consistent_4D}). See Refs.~\cite{Zhang2023fusion, Huang_2025}.} & \parbox{6.7cm}{\raggedright Computed from the induced representation $\operatorname{Ind}_{Z_r}^{G}(R)$. See Table~\ref{tab_D4_shrinking}.} 
		\\[20pt]
		Particle--loop braiding & \parbox{6.7cm}{\raggedright Computed from the correlation function $\braket{\mathcal{O} _{\mathsf{P}}\mathcal{O} _{\mathsf{L}}}$. See Ref.~\cite{PhysRevLett.121.061601}.}  & \parbox{6.7cm}{\raggedright Computed from the braiding operator $O_{\left(R^{\prime}\right), \left(C, R\right)}^{\mathrm{PL}}$ shown in Eq.~(\ref{eq_operator_O}). See Table~\ref{tab_D4_braiding}.} 
		\\[10pt]
		Borromean-Rings braiding & \parbox{6.7cm}{\raggedright Computed from the correlation function $\braket{\mathcal{O} _{\mathsf{P}}\mathcal{O} _{\mathsf{L}_1}\mathcal{O} _{\mathsf{L}_2}}$. See Ref.~\cite{PhysRevLett.121.061601}.} & \parbox{6.7cm}{\raggedright Computed from the braiding operator $O^{\mathrm{BR}}_{\left(R^{\prime}\right), \left(C_1, R_1\right), \left(C_2, R_2\right)}$ shown in Eq.~(\ref{eq_BR_operator}). See Table~\ref{tab_D4_braiding}.} 
		\\[10pt]
		
	\end{tabular}
\end{adjustbox}
    \par\smallskip\hrule\smallskip
\end{table*}

In this section,  we present the 3D $\mathbb{D}_4$ quantum double model as a microscopic construction of  the $BF$ theory with an $AAB$ twist. The fact that all nontrivial fusion rules,  shrinking rules and braiding statistics in the $BF$ theory with an $AAB$ twist are captured by the $\mathbb{D}_4$ quantum double model allows us to establish a concrete correspondence between these two theories, as summarized in Table~\ref{tab_compare}. This result not only confirms the data derived from the continuum field theory but also demonstrates how these data are encoded in the microscopic lattice model.
 
\subsection{$BF$ field theory with an $AAB$ twisted term as a Borromean-Rings topological order}\label{Sec.6.3}
Here,  we briefly review the $BF$ field theory with an $AAB$ twist. We will discuss how to label topological excitations by Wilson operators and calculate fusion rules,  shrinking rules,  and braiding statistics from correlation functions.  

Given a  gauge group $G=\prod_{i=1}^3{\mathbb{Z} _{N_i}}$,  this theory describes three $3$D $\mathbb{Z}_{N_i}$ toric codes coupled in a nontrivial way,  via the topological action
\begin{align}
    S_{\text{BR}}=\int\sum_{i=1}^{3}\frac{N_{i}}{2\pi}B^{i}dA^{i}+qA^{1}A^{2}B^{3}\, , \label{eq_action_AAB}
\end{align}
where $A^{i}$ and $B^{i}$ are $1$- and $2$-form $U\left(1\right)$
gauge fields respectively. Here,  the $BF$ terms serve as the effective description of the $\mathbb{Z}_{N_i}$ toric codes,  and the $AAB$ twist encodes the coupling among them. As we will show in what follows,  with the simplest $G=\left(\mathbb{Z}_{2}\right)^{3}$,  this theory shares the same topological data as the $3$D $\mathbb{D}_{4}$ quantum double model.  

The coefficient $q=\frac{pN_{1}N_{2}N_{3}}{\left(2\pi\right)^{2}N_{123}}$ is  quantized by large gauge invariance,  where $p\in\mathbb{Z}_{N_{123}}$ and $N_{123}$ is the greatest common divisor of $N_{1}$,  $N_{2}$ and $N_{3}$. Lagrange multipliers $B^{1}$,  $B^{2}$,  and $A^{3}$ enforce the flat-connection conditions: $dA^{1}=0$,  $dA^{2}=0$,  and $dB^{3}=0$. The gauge transformations are given by
\begin{align}
    A^1\rightarrow& A^1+d\chi ^1, \nonumber \\
    A^2\rightarrow& A^2+d\chi ^2, \nonumber \\
    B^3\rightarrow& B^3+dV^3, \nonumber \\
    A^3\rightarrow& A^3+d\chi ^3 +\frac{2\pi q}{N_3}\left( \chi ^2A^1-\chi ^1A^2+\frac{1}{2}\chi ^2d\chi ^1-\frac{1}{2}\chi ^1d\chi ^2 \right) , \nonumber \\
    B^1\rightarrow& B^1+dV^1 -\frac{2\pi q}{N_1}\left( \chi ^2B^3-A^2V^3+\chi ^2dV^3 \right) , \nonumber \\
    B^2\rightarrow& B^2+dV^2 +\frac{2\pi q}{N_2}\left( \chi ^1B^3-A^1V^3+\chi ^1dV^3 \right) ,
\end{align}
where $\chi^{i}$ and $V^{i}$ are $0$- and $1$-form gauge parameters with $\int d\chi^{i}\in2\pi\mathbb{Z}$ and $\int dV^{i}\in2\pi\mathbb{Z}$.

Topological excitations in this continuum field theory are represented by gauge-invariant Wilson operators. We denote a particle excitation by $\mathsf{P}_{n_1n_2n_3}$; in the case of $G = \left(\mathbb{Z}_2\right)^{3}$,  $n_1, n_2, n_3=0, 1$,    corresponding to $n_1$,  $n_2$,  and $n_3$ units of $\mathbb{Z}_{N_1}$,  $\mathbb{Z}_{N_2}$,  and $\mathbb{Z}_{N_3}$ gauge charge. The gauge charges are minimally coupled to $1$-form fields $A^{i}$. For example,  $\mathsf{P}_{100}$ denotes a particle carrying only one unit of $\mathbb{Z}_{N_1}$ gauge charge,  and it corresponds to a Wilson operator
\begin{align}
    \mathcal{O}_{\mathsf{P}_{100}}=\exp\left(\mathrm{i}\int_{\gamma}A^{1}\right)\, , \label{eq_123}
\end{align}
where $\gamma=S^1$ is the world-line of the particle. $\mathsf{P}_{100}$ is an Abelian particle excitation. A typical non-Abelian particle in this continuum field theory is $\mathsf{P}_{001}$,  whose gauge-invariant Wilson operator is 
\begin{align}
    \mathcal{O}_{\mathsf{P}_{001}}&=2\exp\left[\mathrm{i}\int_{\gamma}A^{3}+\frac{1}{2}\frac{2\pi q}{N_{3}}\left(d^{-1}A^{1}A^{2}-d^{-1}A^{2}A^{1}\right)\right] \times\delta\left(\int_{\gamma}A^{1}\right)\delta\left(\int_{\gamma}A^{2}\right)\, , \label{eq_124}
\end{align}
where the normalization factor $2$ multiplying $\mathcal{O}_{\mathsf{P}_{001}}$ ensures that all fusion and shrinking coefficients are integers. One can verify that the full expression in the exponent is gauge-invariant,  while $e^{{\rm i}\int_{\gamma} A^3}$ alone is not. Here,  $d^{-1}A^{1}$ and $d^{-1}A^{2}$ are defined through auxiliary fields,  $d^{-1}A^{1}=\alpha^1$ and $d^{-1}A^2=\alpha^{2}$,  such that $d\alpha^1 = A^1$ and $d\alpha^2 = A^2$. Since on the closed manifold $\gamma$ one has $\int_{\gamma} d\alpha^{1, 2} = 0$,  we need to introduce two delta functions, 
\begin{align}
    \delta\left(\int_{\gamma}A^{1}\right)=\begin{cases}
1,  & \int_{\gamma}A^{1}=0\mod2\pi\\
0,  & \text{else }
\end{cases} \\
    \delta\left(\int_{\gamma}A^{2}\right)=\begin{cases}
1,  & \int_{\gamma}A^{2}=0\mod2\pi\\
0,  & \text{else }
\end{cases}
\end{align}
to ensure gauge invariance. 

A loop excitation can carry gauge fluxes that are minimally coupled to $2$-form fields $B^{i}$ and can be decorated by gauge charges that are minimally coupled to $1$-form fields $A^{i}$. We use $\mathsf{L}_{n_1n_2n_3}^{m_1m_2m_3}$ to represent a loop,  where $n_i, m_i=0, 1$. The subscript $n_1n_2n_3$ denotes $n_1$,  $n_2$,  and $n_3$ units of $\mathbb{Z}_{N_1}$,  $\mathbb{Z}_{N_2}$,  and $\mathbb{Z}_{N_3}$ gauge flux. The superscript $m_1m_2m_3$ denotes $m_1$,  $m_2$,  and $m_3$ units of $\mathbb{Z}_{N_1}$,  $\mathbb{Z}_{N_2}$,  and $\mathbb{Z}_{N_3}$ gauge charge decoration. For example,  the gauge-invariant Wilson operator for the Abelian loop $\mathsf{L}_{001}^{000}$ is
\begin{align}
    \mathcal{O}_{\mathsf{L}_{001}}=\exp\left(\mathrm{i}\int_{\sigma}B^{3}\right)\, ,
\end{align}
where $\mathsf{L}_{001}$ is an abbreviation for $\mathsf{L}_{001}^{000}$. $\sigma=S^1\times S^1$ is the world-sheet of the loop. Analogous to the non-Abelian particle $\mathsf{P}_{001}$,  the gauge-invariant Wilson operator for a non-Abelian loop must contain additional nontrivial terms and delta function constraints. For instance,  the Wilson operator for the non-Abelian decorated loop $\mathsf{L}_{100}^{001}$ is given by
\begin{align}
    \mathcal{O} _{\mathsf{L}_{100}^{001}}&=2\exp \left[ \mathrm{i}\int_{\sigma}{B^1}+\frac{1}{2}\frac{2\pi q}{N_1}\left( d^{-1}A^2B^3+d^{-1}B^3A^2 \right) \right. \nonumber \\
    &\left. +\mathrm{i}\int_{\gamma}{A^3}+\frac{1}{2}\frac{2\pi q}{N_3}\left( d^{-1}A^1A^2-d^{-1}A^2A^1 \right) \right] \times \delta \left( \int_{\gamma}{A^2} \right) \delta \left( \int_{\sigma}{B^3}-\int_{\gamma}{A^1} \right)\, , \label{eq_L100001}
\end{align}
where the normalization factor $2$ ensures that all fusion and shrinking coefficients are integers. $d^{-1}B^{3}$ is defined as $d^{-1}B^{3} = \beta^{3}$ such that $d\beta^{3} = B^{3}$. Here $\gamma$ is a closed loop on the world-sheet $\sigma$ ($\gamma\subset\sigma$), representing the world-line of the charge decoration carried by the loop. Choices of $\gamma$ that are homologous on $\sigma$ define the same operator, so the physical result is independent of this choice. The delta functions are introduced to ensure gauge invariance of the operator in Eq.~(\ref{eq_L100001}). In Ref.~\cite{Zhang2023fusion},  the delta function constraints in Eq.~(\ref{eq_L100001}) are originally written as $\delta \left( \int_{\gamma}{A^2} \right) \delta \left( \int_{\sigma}{B^3} \right) \delta \left( \int_{\gamma}{A^1} \right)$. This set of conditions,  however,  is stronger than necessary and can be relaxed to $\delta \left( \int_{\gamma}{A^2} \right) \delta \left( \int_{\sigma}{B^3}-\int_{\gamma}{A^1} \right)$ while preserving gauge invariance of Eq.~(\ref{eq_L100001}). The delta function $\delta \left( \int_{\sigma}{B^3}-\int_{\gamma}{A^1} \right)$ is defined as
\begin{align}
    \delta\left(\int_{\sigma}B^{3}-\int_{\gamma}{A^1}\right)=\begin{cases}
    1,  & \int_{\sigma}B^{3}-\int_{\gamma}{A^1}=0\mod2\pi\\
    0,  & \text{else }
    \end{cases}\, .
\end{align}
The form of the delta function constraint in Eq.~(\ref{eq_L100001}) corresponds to the case $N_i=2$. For general $N_i$ the corresponding delta function constraint is written as \[
\delta\left(\frac{N_{2}}{N_{123}}\int_{\gamma}A^{2}\right)\delta\left(\frac{N_{3}}{N_{123}}\int_{\sigma}B^{3}-\frac{N_{1}}{N_{123}}\int_{\gamma}A^{1}\right)
\], which reduces to the form above when $N_i=2$. A detailed analysis of general $N_i$ can be found in Ref.~\cite{zhang2026gym}.

The delta functions lead to possible equivalence relations between Wilson operators,  which are essential for the identification of all topological excitations of this theory. We say two Wilson operators $\mathcal{O}_{\mathsf{a}}$ and $\mathcal{O}_{\mathsf{b}}$ are equivalent if the correlation function $\braket{\mathcal{O}_{\mathsf{a}}\mathcal{Q}}=\braket{\mathcal{O}_{\mathsf{b}}\mathcal{Q}}$ holds for an arbitrary operator $\mathcal{Q}$. In this case,  the excitations $\mathsf{a}$ and $\mathsf{b}$ should be considered as the same excitation because no operator can distinguish them. When computing fusion and shrinking rules,  we only need to consider distinct Wilson operators. Take the particle $\mathsf{P}_{101}$ as an example. The gauge-invariant Wilson operator is given by
\begin{align}
    \mathcal{O}_{\mathsf{P}_{101}}&=2\exp\left[\mathrm{i}\int_{\gamma}A^{1}+\mathrm{i}\int_{\gamma}A^{3} +\frac{1}{2}\frac{2\pi q}{N_{3}}\left(d^{-1}A^{1}A^{2}-d^{-1}A^{2}A^{1}\right)\right]\nonumber \\
    &\times\delta\left(\int_{\gamma}A^{1}\right)\delta\left(\int_{\gamma}A^{2}\right)\, .
\end{align}
Since the delta function $\delta\left(\int_{\gamma}A^{1}\right)$ enforces that $\int_{\gamma}A^{1}=0\mod2\pi$,  the term $\int_{\gamma}A^{1}$ does not contribute to any correlation function $\braket{\mathcal{O}_{\mathsf{P}_{101}}\mathcal{Q}}$ and we can simply drop it. Thus,  we have $\braket{\mathcal{O}_{\mathsf{P}_{101}}\mathcal{Q}}=\braket{\mathcal{O}_{\mathsf{P}_{001}}\mathcal{Q}}$ and $\mathsf{P}_{101}$ is equivalent to $\mathsf{P}_{001}$. Similarly,  the delta function $\delta\left(\int_{\gamma}A^{2}\right)$ also leads to an equivalence relation. We conclude that $\mathsf{P}_{001}=\mathsf{P}_{101}=\mathsf{P}_{011}=\mathsf{P}_{111}$,  where $=$ indicates equivalence. These four notations form an equivalence class,  and they all represent the same excitation. In the following discussion,  we select $\mathsf{P}_{001}$ as the representative of this equivalence class.

Our analysis,  based on gauge invariance and a careful treatment of delta functions,  reveals that $S_{\text{BR}}$ with a gauge group $G=(\mathbb{Z}_2)^3$ admits $22$ distinct Wilson operators for excitations\footnote{This count differs from the $19$ excitations originally reported in Ref.~\cite{Zhang2023fusion}. The discrepancy arises because the delta function constraints in the Wilson operators for the loops $\mathsf{L}_{100}^{001}$,  $\mathsf{L}_{010}^{001}$,  and $\mathsf{L}_{110}^{001}$ were overly restrictive in Ref.~\cite{Zhang2023fusion}. By relaxing these constraints as demonstrated for $\mathsf{L}_{100}^{001}$ in Eq.~(\ref{eq_L100001}),  we obtain the present count.}.

Having identified the complete set,  we then compute their correlation functions as in Eqs.~(\ref{eq_fusion_TQFT}) and~(\ref{eq_shrinking_TQFT}) to derive all fusion and shrinking rules. Here we present some typical examples. 

First,  we consider the fusion rules. Particles $\mathsf{P}_{100}$ and $\mathsf{P}_{010}$ are both Abelian excitations; the Wilson operator for $\mathsf{P}_{010}$ is obtained by simply replacing $A^1$ with $A^2$ in Eq.~(\ref{eq_123}). The fusion of $\mathsf{P}_{100}$ and $\mathsf{P}_{010}$ produces another Abelian particle $\mathsf{P}_{110}$,  whose gauge charges are a combination of the gauge charges carried by $\mathsf{P}_{100}$ and $\mathsf{P}_{010}$. This result is also indicated by the Wilson operator: $\mathcal{O}_{\mathsf{P}_{110}}=\exp\left(\mathrm{i}\int_{\gamma}A^{1}+A^{2}\right)$.

For a non-Abelian excitation,  the corresponding Wilson operator carries delta functions,  which enforce some integrals to be trivial. A typical example is  the particle $\mathsf{P}_{001}$,  whose Wilson operator (see Eq.~(\ref{eq_124})) carries two delta functions $\delta\left(\int_{\gamma}A^{1}\right)$ and $\delta\left(\int_{\gamma}A^{2}\right)$. Thus,  the fusion processes $\mathsf{P}_{001}\otimes\mathsf{P}_{100}$ and $\mathsf{P}_{001}\otimes\mathsf{P}_{010}$ are given by 
\begin{align}
    \mathsf{P}_{001}\otimes\mathsf{P}_{100}=\mathsf{P}_{001}\otimes\mathsf{P}_{010}=\mathsf{P}_{001}\label{eq_131}
\end{align}
because $\int_{\gamma}A^{1}$ and $\int_{\gamma}A^{2}$ do not contribute to the correlation functions. When we bring the particles $\mathsf{P}_{001}$ and $\mathsf{P}_{100}$ ($\mathsf{P}_{010}$) together,  the gauge charge carried by $\mathsf{P}_{100}$ ($\mathsf{P}_{010}$) becomes trivial.

The fusion rules mentioned above are Abelian fusion rules because they have a single fusion channel. We can obtain a non-Abelian fusion rule by fusing a non-Abelian pure loop $\mathsf{L}_{100}$ and a non-Abelian particle $\mathsf{P}_{001}$,  where the Wilson operator for $\mathsf{L}_{100}$ is given by 
\begin{align}
    \mathcal{O} _{\mathsf{L}_{100}}&=2\exp \left[ \mathrm{i}\int_{\sigma}{B^1}+\frac{1}{2}\frac{2\pi q}{N_1}\left( d^{-1}A^2B^3+d^{-1}B^3A^2 \right) \right] \times \delta \left( \int_{\gamma}{A^2} \right) \delta \left( \int_{\sigma}{B^3} \right)\, .\label{eq_132}
\end{align}
Using the Wilson operator for $\mathsf{L}_{100}^{101}$:
\begin{align}
    \mathcal{O} _{\mathsf{L}_{100}^{101}}&=2\exp \left[ \mathrm{i}\int_{\sigma}{B^1}+\frac{1}{2}\frac{2\pi q}{N_1}\left( d^{-1}A^2B^3+d^{-1}B^3A^2 \right) \right. \nonumber \\
    &\left. +\mathrm{i}\int_{\gamma}{A^1+A^3}+\frac{1}{2}\frac{2\pi q}{N_3}\left( d^{-1}A^1A^2-d^{-1}A^2A^1 \right) \right] \times \delta \left( \int_{\gamma}{A^2} \right) \delta \left( \int_{\sigma}{B^3}-\int_{\gamma}{A^1} \right)\, ,
\end{align}
we compute the correlation function 
\begin{align}
    \braket{\mathcal{O} _{\mathsf{L}_{100}}\times\mathcal{O} _{\mathsf{P}_{001}}}=\braket{\mathcal{O} _{\mathsf{L}_{100}^{001}}+\mathcal{O} _{\mathsf{L}_{100}^{101}}}\, , \label{eq_134}
\end{align}
where the Wilson operators for $\mathsf{P}_{001}$ and $\mathsf{L}_{100}^{001}$ are shown in Eqs.~(\ref{eq_124}) and~(\ref{eq_L100001}). Thus,  we conclude that the non-Abelian fusion rule is given by $\mathsf{L}_{100}\otimes\mathsf{P}_{001}=\mathsf{L}_{100}^{001}\oplus \mathsf{L}_{100}^{101}$. The two fusion channels mean that if we measure the fusion output,  there are two possible results corresponding to $\mathsf{L}_{100}^{001} $ and $ \mathsf{L}_{100}^{101}$.

After deriving all fusion rules,  as illustrated in Sec.~\ref{Sec.4.3},  we obtain all quantum dimensions of excitations by computing the largest eigenvalues of the corresponding fusion coefficient matrices. In the $BF$ theory with an $AAB$ twist and the gauge group $\left(\mathbb{Z}_2\right)^3$,  any Abelian excitation $\mathsf{a}$ has quantum dimension $d_{\mathsf{a}}=1$ and any non-Abelian excitation $\mathsf{b}$ has $d_{\mathsf{b}}=2$. 

Now,  we present some typical shrinking rules. Consider an Abelian pure loop $\mathsf{L}_{001}$. Its shrinking rule is simply $\mathcal{S}\left(\mathsf{L}_{001}\right)=\mathsf{1}$ because when the manifold $\sigma$ is shrunk to $\gamma$,  the integral $\int_\sigma{B^3}$ does not contribute to the correlation function. If we decorate a pure loop with a nontrivial charge,  then the shrinking process annihilates the flux while leaving the decoration intact. For example,  $\mathcal{S}\left(\mathsf{L}_{001}^{100}\right)=\mathsf{P}_{100}$.

However,  for a non-Abelian pure loop such as $\mathsf{L}_{100}$,  the shrinking process can lead to a nontrivial gauge charge because the Wilson operator in Eq.~(\ref{eq_132}) carries a delta function $\delta\left(\int_{\gamma}A^{2}\right)$. When the manifold $\sigma$ is shrunk to $\gamma$,  all integrals over $\sigma$ in Eq.~(\ref{eq_132}) vanish. Thus,  $\braket{\mathcal{S}\left(\mathcal{O} _{\mathsf{L}_{100}}\right)}=\braket{2\delta\left(\int_{\gamma}A^{2}\right)}$. Note that when the gauge group is $\left(\mathbb{Z}_2\right)^3$,  we can expand the delta function as $\delta\left(\int_{\gamma}A^{2}\right)=\frac{1}{2}\left[1+\exp\left(\int_{\gamma}A^{2}\right)\right]$,  which leads to the shrinking rule 
\begin{align}
    \mathcal{S}\left(\mathsf{L}_{100}\right)=\mathsf{1}\oplus\mathsf{P}_{010}\, .\label{eq_135}
\end{align}
If we measure the result of shrinking $\mathsf{L}_{100}$,  we can obtain either the trivial particle $\mathsf{1}$ or the nontrivial particle $\mathsf{P}_{010}$. The general shrinking rule Eq.~(\ref{eq_135}) obtained from the continuum field theory only shows all possible shrinking channels. However,  we cannot determine which channel will be selected in a specific shrinking process. In Sec.~\ref{Sec.6.4},  we will show that the 3D $\mathbb{D}_4$ quantum double model is a microscopic construction of the $BF$ theory with an $AAB$ twist and the gauge group $\left(\mathbb{Z}_2\right)^3$. In this microscopic model,  by tuning internal degrees of freedom in operators for loops,  we can control which shrinking channel occurs in a specific shrinking process,  as shown in Sec.~\ref{Sec.5.4}.

We can verify that the fusion and shrinking rules satisfy the consistency condition shown in Eq.~(\ref{eq_consistent_4D}). Take $\mathsf{L}_{100}$ and $\mathsf{P}_{001}$ as examples. Eq.~(\ref{eq_134}) shows that fusing $\mathsf{L}_{100}$ and $\mathsf{P}_{001}$ first produces $\mathsf{L}_{100}^{001}\oplus\mathsf{L}_{100}^{101}$. By computing the correlation function in Eq.~(\ref{eq_shrinking_TQFT}),  we conclude that $\mathcal{S}\left(\mathsf{L}_{100}^{001}\right)=\mathcal{S}\left(\mathsf{L}_{100}^{101}\right)=\mathsf{P}_{001}$. Thus,  fusing  $\mathsf{L}_{100}$ and $\mathsf{P}_{001}$ followed by shrinking is given by
\begin{align}
    \mathcal{S}\left(\mathsf{L}_{100}\otimes\mathsf{P}_{001}\right)&=\mathcal{S}\left(\mathsf{L}_{100}^{001}\oplus\mathsf{L}_{100}^{101}\right)=\mathsf{P}_{001}\oplus\mathsf{P}_{001} =2\cdot\mathsf{P}_{001}\, ,
\end{align}
where the result $2\cdot\mathsf{P}_{001}$ means that in the whole process,  there are two channels pointing to the excitation $\mathsf{P}_{001}$. On the other hand,  shrinking $\mathsf{L}_{100}$ and $\mathsf{P}_{001}$ followed by fusion is given by
\begin{align}
    \mathcal{S}\left(\mathsf{L}_{100}\right)\otimes\mathcal{S}\left(\mathsf{P}_{001}\right)&=\left(\mathsf{1}\oplus\mathsf{P}_{010}\right)\otimes\mathsf{P}_{001}=\mathsf{P}_{001}\oplus\mathsf{P}_{001}=2\cdot\mathsf{P}_{001}\, ,
\end{align}
where we have used Eqs.~(\ref{eq_131}) and~(\ref{eq_135}). Thus,  fusion followed by shrinking yields the same results as shrinking followed by fusion.

Finally,  we consider the particle--loop braiding and the Borromean-Rings braiding. In the continuum field theory,  braiding statistics are evaluated through the correlation functions of the gauge-invariant Wilson operators in the path integral formalism. We first consider the braiding between an Abelian particle $\mathsf{P}_{100}$ and a non-Abelian loop $\mathsf{L}_{100}$. In $(3+1)$D spacetime,  this process is represented by the correlation function 
\begin{align}
    \braket{\mathcal{O} _{\mathsf{P}_{100}}\mathcal{O} _{\mathsf{L}_{100}}}\, .
\end{align}
Following Refs.~\cite{PhysRevLett.121.061601, Zhang:2021ycl},  we conclude that this braiding is determined by the $BF$ term $\frac{N_1}{2\pi} B^1 dA^1$ in the action and we have
\begin{align}
    \braket{\mathcal{O} _{\mathsf{P}_{100}}\mathcal{O} _{\mathsf{L}_{100}}} = \exp\left(\mathrm{i} \frac{2\pi}{N_1} \text{Link}(\gamma,  \sigma) \right)\, ,
\end{align}
where $\gamma$ and $\sigma$ are the world-line and the world-sheet of the particle and the loop. When we braid the particle around the loop once,  the linking number is $\text{Link}(\gamma,  \sigma) = 1$. For the gauge group $(\mathbb{Z}_2)^3$,  where $N_1 = 2$,  the correlation function yields:
\begin{align}
    \exp \left(\mathrm{i}\theta_{\mathsf{P}_{100}, \mathsf{L}_{100}}^{\mathrm{PL}}\right) = \exp\left(\mathrm{i}\pi \right)=-1\, .
\end{align}
This result means that $\mathsf{P}_{100}$ and $\mathsf{L}_{100}$ exhibit the mutual statistics present in the $\mathbb{Z}_2$ toric code. Similarly,  we obtain
\begin{align}
    \braket{\mathcal{O} _{\mathsf{P}_{010}}\mathcal{O} _{\mathsf{L}_{010}}}=\braket{\mathcal{O} _{\mathsf{P}_{001}}\mathcal{O} _{\mathsf{L}_{001}}}=-1\, .
\end{align}
Thus,  we obtain the particle--loop braiding phases $\exp \left(\mathrm{i}\theta_{\mathsf{P}_{010}, \mathsf{L}_{010}}^{\mathrm{PL}}\right)=\exp \left(\mathrm{i}\theta_{\mathsf{P}_{001}, \mathsf{L}_{001}}^{\mathrm{PL}}\right)=-1$,  which means that $\mathsf{P}_{010}$ and $\mathsf{L}_{010}$ as well as $\mathsf{P}_{001}$ and $\mathsf{L}_{001}$ also exhibit nontrivial mutual statistics.

A more exotic feature of the $BF$ theory with an $AAB$ twist and the gauge group $\left(\mathbb{Z}_2\right)^3$ is that it supports Borromean-Rings braiding. This process involves a particle $\mathsf{P}_{001}$ and two loops $\mathsf{L}_{100}$ and $\mathsf{L}_{010}$. We consider a configuration where any two of the three excitations are pairwise unlinked,  yet they are collectively linked. Although the particle trajectory does not wind around either loop individually (i.e.,  the Hopf linking numbers $\text{Link}(\gamma,  \sigma_1) =\text{Link}(\gamma,  \sigma_2)=0$ and also the two loops are not linked,  $\text{Link}(\sigma_1,  \sigma_2)=0$),  the triple linking number,  known as Milnor's triple linking number $\bar{\mu}(\gamma,  \sigma_1,  \sigma_2)$,  can be nontrivial. Here we consider $\bar{\mu}=1$. The braiding phase is captured by the $AAB$ twisted term. To compute it,  we consider the gauge-invariant correlation function of the three Wilson operators:
\begin{align}
    \braket{\mathcal{O} _{\mathsf{P}_{001}}\mathcal{O} _{\mathsf{L}_{100}}\mathcal{O} _{\mathsf{L}_{010}}}\, ,
\end{align}
where $\mathcal{O} _{\mathsf{P}_{001}}$ is given in Eq.~(\ref{eq_124}),  $\mathcal{O} _{\mathsf{L}_{100}}$ in Eq.~(\ref{eq_132}),  and $\mathcal{O} _{\mathsf{L}_{010}}$ is analogous with indices $1\leftrightarrow2$. A detailed path integral calculation (see Ref.~\cite{PhysRevLett.121.061601}) shows that after integrating out the Lagrange multipliers,  the effective action acquires an additional term:
\begin{align}
    S_{\text{BR}} = \frac{2\pi }{N_{123}} \bar{\mu}(\gamma,  \sigma_1,  \sigma_2)\, ,
\end{align}
where $N_{123}=\text{gcd}(N_1, N_2, N_3)$. Since we consider $N_i=2$ here,  $N_{123}=2$. Thus,  the braiding phase is
\begin{align}
    \exp \left(\mathrm{i}\theta^{\mathrm{BR}}_{\mathsf{P}_{001}, \mathsf{L}_{100}, \mathsf{L}_{010}}\right)=\exp \left(\mathrm{i}\pi\right)=-1\, .
\end{align}
We conclude that unlike the particle--loop braiding phase,  the Borromean-Rings braiding does not arise from any pairwise linking but from the higher-order linking encoded in the $AAB$ twisted term.

\subsection{Topological data of the microscopic lattice model}\label{Sec.6.2}
After reviewing the $BF$ field theory with an $AAB$ twist and the gauge group $\left( \mathbb{Z} _2 \right) ^3$,  in this section we present excitations,  fusion rules,  shrinking rules,  and braiding statistics in the $\mathbb{D}_4$ quantum double model. Denoting the two generators of $G=\mathbb{D}_4$ by $r$ and $t$, we have $r^4=t^2=e$ and $rt=t\bar{r}=tr^3$. We label the five conjugacy classes as
$
    C_e=\left\{ e \right\}\, , \,C_{r^2}=\left\{ r^2 \right\}\,, \,C_r=\left\{ r, r^3 \right\} \,, \,
C_t=\left\{ t, tr^2 \right\}\,, \,C_{tr}=\left\{ tr, tr^3 \right\}\,$,
where $e$,  $r$,  $t$,  $tr$,  and $r^2$ serve as the representatives of the classes $C_e$,  $C_r$,  $C_t$,  $C_{tr}$ and $C_{r^2}$ respectively. Their corresponding centralizers are 
 $Z_e=Z_{r^2}=\mathbb{D} _4\,,  \,
    Z_r=\left\{ e, r, r^2, r^3 \right\} \simeq \mathbb{Z} _4\,, \,
    Z_t=\left\{ e, t, r^2, tr^2 \right\} \simeq \mathbb{Z} _2\times \mathbb{Z} _2\,, \,
    Z_{tr}=\left\{ e, r^2, tr, tr^3 \right\} \simeq \mathbb{Z} _2\times \mathbb{Z} _2\, $,
where $\simeq$ denotes isomorphism. For $Q_g=G/Z_g$, we have
 $Q_e=Q_{r^2}=\left\{ e \right\} \,,\, Q_r=\left\{ e, t \right\}\,,\, Q_t=Q_{tr}=\left\{ e, r \right\} \,.$
We choose $q_e=q_r=q_t=q_{tr}=q_{r^2}=e$,  $q_{r^3}=t$,  and $q_{tr^2}=q_{tr^3}=r$. We label the five irreps of $\mathbb{D}_4$ as $Id$,  $\mathcal{R}$,  $\mathcal{T}$,  $\mathcal{RT}$,  and $\mathcal{B}$,  where $\mathcal{B}$ is a $2d$ irrep and the others are $1d$ irreps. The representation matrices of the generators are given by
\begin{align}
    &\Gamma ^{Id}\left( g \right) =1, \quad &&\Gamma ^{\mathcal{R}}\left( r \right) =-1, \nonumber \\
    &\Gamma ^{\mathcal{R}}\left( t \right) =1, \quad &&\Gamma ^{\mathcal{T}}\left( r \right) =1, \nonumber \\
    &\Gamma ^{\mathcal{T}}\left( t \right) =-1, \quad &&\Gamma ^{\mathcal{RT}}\left( r \right) =\Gamma ^{\mathcal{RT}}\left( t \right) =-1, \nonumber \\
    &\Gamma ^{\mathcal{B}}\left( r \right) =\left( \begin{matrix}
	    0&		-1\\
	    1&		0\\
    \end{matrix} \right) \, , \quad &&\Gamma ^{\mathcal{B}}\left( t \right) =\left( \begin{matrix}
	    1&		0\\
	    0&		-1\\
    \end{matrix} \right) \, .&&\label{eq_143}
\end{align}
The four $1d$ irreps of $\mathbb{Z} _4=\left\{ e, r, r^2, r^3 \right\}$ are labeled by $Id$,  $\omega$,  $\omega^2$,  and $\omega^3$. The representation matrices of the generator are given by
\begin{align}
    \Gamma ^{Id}\left( g \right) =1, \quad\Gamma ^{\omega^n}\left( r \right) =\omega^n\, ,
\end{align}
where $n=1, 2, 3$ and $\omega=\exp\left(\pi \mathrm{i}/2\right)=\mathrm{i}$. The four $1d$ irreps of $\mathbb{Z} _2\times \mathbb{Z} _2=\left\{ e, a, b, ab \right\}$ are labeled by $Id$,  $a$,  $b$,  and $ab$.  The representation matrices of the generators are given by
\begin{align}
    &\Gamma ^{Id}\left( g \right)=\Gamma ^{a}\left( b \right)=\Gamma ^{b}\left( a \right) =1, \nonumber \\
    &\Gamma ^{a}\left( a \right) =\Gamma ^{b}\left( b \right)=\Gamma ^{ab}\left( a \right)=\Gamma ^{ab}\left( b \right)=-1\, .
\end{align}
Here, the generators $a$ and $b$ are identified with the group elements as follows: for $Z_t$ one has $a=r^2$ and $b=t$, while for $Z_{tr}$ one has $a=r^2$ and $b=tr$.

Recall that each excitation is labeled by a pair $\left(C, R\right)$. Thus,  we conclude that there are $5+5+4+4+4=22$ excitations in the $\mathbb{D}_4$ quantum double model. When $C=C_e$,  the five irreps of the group $\mathbb{D}_4$ correspond to the five particles labeled by
\begin{align}
    \left[ Id \right]\, ,  \quad \left[ \mathcal{R} \right]\, ,  \quad \left[ \mathcal{T} \right]\, ,  \quad \left[ \mathcal{RT} \right]\, ,  \quad \left[ \mathcal{B} \right] \, ,
\end{align}
where we omit $C_e$ in the notation $\left[C, R\right]$. For the non-trivial conjugacy classes $C_r$,  $C_t$,  $C_{tr}$,  and $C_{r^2}$,  the numbers of associated inequivalent irreps are four,  four,  four,  and five respectively. The resulting $4+4+4+5 = 17$ distinct pairs $(C, R)$ correspond to loop excitations,  labeled as follows:
\begin{align}
    &\left[ C_r, Id \right]\, ,   && \left[ C_r, \omega \right]\, ,   && \left[ C_r, \omega ^2 \right]\, ,    &&\left[ C_r, \omega ^3 \right]\, ,  &&\nonumber \\
    &\left[ C_t, Id \right]\, ,   && \left[ C_t, a \right]\, ,    &&\left[ C_t, b \right] \, ,   &&\left[ C_t, ab \right]\, ,  &&\nonumber \\
    &\left[ C_{tr}, Id \right]\, ,    &&\left[ C_{tr}, a \right]\, ,    &&\left[ C_{tr}, b \right]\, ,    &&\left[ C_{tr}, ab \right]\, ,  &&\nonumber \\
    &\left[ C_{r^2}, Id \right]\, ,    &&\left[ C_{r^2}, \mathcal{R} \right] \, , && \left[ C_{r^2}, \mathcal{T} \right]\, ,    &&\left[ C_{r^2}, \mathcal{RT} \right] \, , \nonumber \\
    &\left[ C_{r^2}, \mathcal{B} \right] \, .
\end{align}
The irrep $R$ encodes charge decorations carried by loops. A loop with the trivial irrep $Id$ is a pure loop,  whereas a loop with any nontrivial irrep is a decorated loop carrying additional gauge charge.

Using the same strategy illustrated in Sec.~\ref{sec4} and Sec.~\ref{sec5},  we derive some typical fusion and shrinking rules as examples. Consider three particles $\left[ \mathcal{R} \right]$,  $\left[ \mathcal{T} \right]$,  and $\left[ \mathcal{RT} \right]$. They are associated with three $1d$ irreps $\mathcal{R}$,  $\mathcal{T}$,  and $\mathcal{RT}$ of the group $G=\mathbb{D}_4$. Recall that the general form of the creation operator for a particle is given by Eq.~(\ref{eq_particle}). Thus,  the corresponding operators for $\left[ \mathcal{R} \right]$,  $\left[ \mathcal{T} \right]$ and $\left[ \mathcal{RT} \right]$ have no internal degrees of freedom,  which indicates that they are Abelian particles. According to the discussion in Sec.~\ref{Sec.4.1},  a particle labeled by $R$ is associated with a local Hilbert space $V_{\left(R\right)}$ spanned by $\left\{\ket{j }_{\left(R\right)} \mid j=1, \cdots, n_R\right\}$,  where $n_R$ denotes the dimension of the irrep $R$. Thus,  the one-dimensional Hilbert spaces for $\left[ \mathcal{R} \right]$,  $\left[ \mathcal{T} \right]$,  and $\left[ \mathcal{RT} \right]$ are spanned by $ \left\{\ket{1 }_{\left(\mathcal{R}\right)} \right\}$,  $ \left\{\ket{1 }_{\left(\mathcal{T}\right)} \right\}$,  and $ \left\{\ket{1 }_{\left(\mathcal{RT}\right)} \right\}$. When we bring $\left[ \mathcal{R} \right]$ and $\left[ \mathcal{T} \right]$ to a site $s$ and fuse them,  the local Hilbert space on the site $s$ is spanned by $\left\{\ket{1 }_{\left(\mathcal{R}\right)}\otimes\ket{1 }_{\left(\mathcal{T}\right)} \right\}$. The local operation $D_{\left(h, g\right)}\equiv B^hA^g$ (we omit the labels $p$,  $s$,  and $v$ for the plaquette,  site,  and vertex here for simplicity) acts on the local state $\ket{1 }_{\left(\mathcal{R}\right)}\otimes\ket{1 }_{\left(\mathcal{T}\right)}$ as
\begin{align}
    \sum_{k\in G}{D_{\left( h\bar{k}, g \right)}\ket{1 }_{\left(\mathcal{R}\right)}\otimes D_{\left( k, g \right)}\ket{1 }_{\left(\mathcal{T}\right)}} &=\delta_{h, e}\Gamma^{\mathcal{R}}\left(g\right)\ket{1 }_{\left(\mathcal{R}\right)}\otimes \Gamma^{\mathcal{T}}\left(g\right)\ket{1 }_{\left(\mathcal{T}\right)}\, ,
\end{align}
where we have used Eq.~(\ref{eq_action_DG}). From Eq.~(\ref{eq_143}) we conclude that $\Gamma^{\mathcal{R}}\left(g\right)\Gamma^{\mathcal{T}}\left(g\right)=\Gamma^{\mathcal{RT}}\left(g\right)$. Comparing this with the action of $D_{\left(h, g\right)}$ on the local state $\ket{1 }_{\left(\mathcal{RT}\right)}$:
\begin{align}
    D_{\left(h, g\right)}\ket{1 }_{\left(\mathcal{RT}\right)}=\delta_{h, e}\Gamma^{\mathcal{RT}}\left(g\right)\ket{1 }_{\left(\mathcal{RT}\right)}\, ,
\end{align}
we conclude that the spaces spanned by $\left\{\ket{1 }_{\left(\mathcal{R}\right)}\otimes\ket{1 }_{\left(\mathcal{T}\right)} \right\}$ and $\left\{\ket{1 }_{\left(\mathcal{RT}\right)} \right\}$ correspond to equivalent irreps of the quantum double algebra,  which leads to the fusion rule  
\begin{align}
    \left[ \mathcal{R} \right]\otimes\left[ \mathcal{T} \right]=\left[ \mathcal{RT} \right]\, .
\end{align}
The charge carried by $\left[ \mathcal{RT} \right]$ is simply a combination of the charges carried by $\left[ \mathcal{R} \right]$ and $\left[ \mathcal{T} \right]$. This is similar to the case in Sec.~\ref{Sec.6.3},  where $\mathsf{P}_{100}\otimes\mathsf{P}_{010}=\mathsf{P}_{110}$.

The particle $\left[ \mathcal{B} \right]$ in the $\mathbb{D}_4$ quantum double model is a non-Abelian particle because it is associated with a $2d$ irrep $\mathcal{B}$ of the group $G=\mathbb{D}_4$. From Eq.~(\ref{eq_particle}) we can see that the creation operator for $\left[ \mathcal{B} \right]$ has internal degrees of freedom labeled by the row and column indices of the representation matrix. The corresponding local Hilbert space $V_{\left(\mathcal{B}\right)}$ is spanned by $\left\{\ket{j }_{\left(\mathcal{B}\right)} \mid j=1, 2\right\}$,  which indicates that the quantum dimension of $\left[ \mathcal{B} \right]$ is $d_{\left[ \mathcal{B} \right]}=2$. If we consider bringing $\left[ \mathcal{B} \right]$ and $\left[ \mathcal{R} \right]$ to the same site $s$,  then the local Hilbert space on $s$ is spanned by $\left\{\ket{1 }_{\left(\mathcal{R}\right)}\otimes\ket{j }_{\left(\mathcal{B}\right)} \mid j=1, 2\right\}$. The local operation $D_{\left(h, g\right)}\equiv B^hA^g$ acts on $\ket{1 }_{\left(\mathcal{R}\right)}\otimes\ket{j }_{\left(\mathcal{B}\right)}$ as
\begin{align}
    \sum_{k\in G}{D_{\left( h\bar{k}, g \right)}\ket{1 }_{\left(\mathcal{R}\right)}\otimes D_{\left( k, g \right)}\ket{j }_{\left(\mathcal{B}\right)}} &=\delta_{h, e}\Gamma^{\mathcal{R}}\left(g\right)\ket{1 }_{\left(\mathcal{R}\right)}\otimes \sum_{i=1}^2\left[\Gamma^{\mathcal{B}}\left(g\right)\right]_{ij}\ket{i }_{\left(\mathcal{B}\right)}\nonumber \\
    &=\delta_{h, e}\sum_{i=1}^2\left[\tilde\Gamma^{\mathcal{B}}\left(g\right)\right]_{ij}\ket{1 }_{\left(\mathcal{R}\right)}\otimes \ket{i }_{\left(\mathcal{B}\right)}\, , \label{eq_155}
\end{align}
where we have used Eqs.~(\ref{eq_action_DG}) and~(\ref{eq_143}),  $\tilde\Gamma^{\mathcal{B}}\left(g\right)=U^\dagger\Gamma^{\mathcal{B}}\left(g\right)U$,  $U$ is a unitary matrix given by $U=\left( \begin{matrix}
	1&		0\\
	0&		-1\\
\end{matrix} \right) $. Comparing Eq.~(\ref{eq_155}) with the action of $D_{\left(h, g\right)}$ on the local state $\ket{j }_{\left(\mathcal{B}\right)}$:
\begin{align}
    D_{\left(h, g\right)}\ket{j }_{\left(\mathcal{B}\right)}=\delta_{h, e}\sum_{i=1}^2\left[\Gamma^{\mathcal{B}}\left(g\right)\right]_{ij} \ket{i }_{\left(\mathcal{B}\right)}\, ,
\end{align}
we conclude that the spaces spanned by $\left\{\ket{1 }_{\left(\mathcal{R}\right)}\otimes\ket{j }_{\left(\mathcal{B}\right)} \mid j=1, 2\right\}$ and $\left\{\ket{j }_{\left(\mathcal{B}\right)} \mid j=1, 2\right\}$ correspond to equivalent irreps of the quantum double algebra,  which leads to the fusion rule
\begin{align}
    \left[ \mathcal{R} \right]\otimes\left[ \mathcal{B} \right]=\left[ \mathcal{B} \right]\, .
\end{align}
Similarly,  we have
\begin{align}
    \left[ \mathcal{T} \right]\otimes\left[ \mathcal{B} \right]=\left[ \mathcal{B} \right]\, .
\end{align}
These results show that when we bring $\left[ \mathcal{B} \right]$ together with either $\left[ \mathcal{R} \right]$ or $\left[ \mathcal{T} \right]$,  the charge carried by the latter becomes trivial. Comparing with the fusion rules in Eq.~(\ref{eq_131}) in Sec.~\ref{Sec.6.3},  we can see that $\left[ \mathcal{B} \right]$ has the same fusion behavior as $\mathsf{P}_{001}$ in the $BF$ field theory. In Sec.~\ref{Sec.6.4},  we will confirm that $\mathsf{P}_{100}$,  $\mathsf{P}_{010}$,  $\mathsf{P}_{110}$,  and $\mathsf{P}_{001}$  can be mapped to $\left[ \mathcal{R} \right]$,  $\left[ \mathcal{T} \right]$,  $\left[ \mathcal{RT} \right]$,  and $\left[ \mathcal{B} \right]$ without altering the fusion rules, shrinking rules, or braiding statistics.

The fusion rules mentioned above are all Abelian. Now,  we consider a typical example of a non-Abelian fusion rule. In Sec.~\ref{Sec.6.3},  we have obtained a typical non-Abelian fusion rule as $\mathsf{L}_{100}\otimes\mathsf{P}_{001}=\mathsf{L}_{100}^{001}\oplus \mathsf{L}_{100}^{101}$. Its counterpart in the 3D $\mathbb{D}_4$ quantum double model is given by
\begin{align}
    \left[ C_r, Id \right]\otimes\left[ \mathcal{B} \right]=\left[ C_r, \omega \right]\oplus\left[ C_r, \omega ^3 \right]\, , \label{eq_160}
\end{align}
where $\left[ C_r, Id \right]$ is a non-Abelian pure loop,  $\left[ C_r, \omega \right] $ and $ \left[ C_r, \omega ^3 \right]$ are non-Abelian decorated loops. These three non-Abelian loops are associated with the conjugacy class $C_r$,  where $\left|C_r\right|=2$. Since the irreps $Id$,  $\omega$,  and $\omega^3$ of the centralizer $Z_r$ are one-dimensional,  according to Eq.~(\ref{eq_loop}),  the internal degrees of freedom in the creation operators for these loops are labeled by the elements in the conjugacy class $C_r$. Thus,  the local Hilbert spaces corresponding to these three loops are two-dimensional,  which indicates that their quantum dimensions are two. For $\left[ C_r, Id \right]$,  the corresponding local Hilbert space is spanned by $\left\{\ket{r }_{\left(C_r, Id\right)}, \ket{r^3 }_{\left(C_r, Id\right)}\right\}$. Fusing $\left[ C_r, Id \right]$ and $\left[ \mathcal{B} \right]$ leads to a local Hilbert space $V_{\left(C_r, Id\right)}\otimes V_{\left(\mathcal{B}\right)}$ spanned by $\left\{\ket{g }_{\left(C_r, Id\right)}\otimes\ket{j }_{\left(\mathcal{B}\right)} \mid g=r, r^3;\, j=1, 2\right\}$. Using Eq.~(\ref{eq_fusion_coe}),  we conclude that $V_{\left(C_r, Id\right)}\otimes V_{\left(\mathcal{B}\right)}$ can be decomposed as $V_{\left(C_r, Id\right)}\otimes V_{\left(\mathcal{B}\right)}=V_{\left(C_r, \omega\right)}\oplus V_{\left(C_r, \omega^3\right)}$,  which leads to the fusion rule Eq.~(\ref{eq_160}). Bringing $\left[ \mathcal{B} \right]$ and $\left[ C_r, Id \right]$ together does not produce a definite fusion result but rather two possible channels. 

Now,  we present some typical shrinking rules. As in the $BF$ field theory with an $AAB$ twist and the gauge group $\left(\mathbb{Z}_2\right)^3$,  the $\mathbb{D}_4$ quantum double model only has one nontrivial Abelian pure loop $\left[ C_{r^2}, Id \right]$,  whose operator is given by $W_M\left( C_{r^2}, Id \right) =\sum_{g\in \mathbb{D} _4}{L_{M}^{r^2}T_{P}^{g}}$. Following the discussion in Sec.~\ref{sec5},  after the shrinking process,  the operator becomes $\sum_{g\in \mathbb{D} _4}{T_{P}^{g}}=W_P\left( Id \right)$. Thus,  we conclude that
\begin{align}
    \mathcal{S}\left(\left[ C_{r^2}, Id \right]\right)=\left[ Id \right]\, .
\end{align}
If we decorate $\left[ C_{r^2}, Id \right]$ with a charge $\left[ \mathcal{R} \right]$,  we obtain another Abelian decorated loop $\left[ C_{r^2}, \mathcal{R} \right]$,  which is the counterpart of $\mathsf{L}_{001}^{100}$ in Sec.~\ref{Sec.6.3}. Using Eq.~(\ref{eq_loop}),  we obtain the operator for $\left[ C_{r^2}, \mathcal{R} \right]$ as $\sum_{g\in \mathbb{D} _4}{\Gamma^{\mathcal{R}} \left(g\right)L_{M}^{r^2}T_{P}^{g}}$. After the shrinking process,  the operator becomes $\sum_{g\in \mathbb{D} _4}{\Gamma^{\mathcal{R}} \left(g\right)T_{P}^{g}}=W_P\left( \mathcal{R} \right)$. Thus,  we conclude that
\begin{align}
    \mathcal{S}\left(\left[ C_{r^2}, \mathcal{R} \right]\right)=\left[ \mathcal{R} \right]\, .
\end{align}
The charge decoration on the Abelian loop is unaffected by the shrinking process. 

Only non-Abelian pure loops can have non-Abelian shrinking rules. Take the counterpart of $\mathsf{L}_{100}$,  i.e.,  $\left[ C_r, Id \right]$,  as an example. The corresponding operators are given by
\begin{align}
    W_M\left( C_r, Id;r, r \right) &=\sum_{g\in Z_r}{L_{M}^{r}T_{P}^{g}}\, , \label{eq_170} \\
    W_M\left( C_r, Id;r^3, r \right) &=\sum_{g\in Z_r}{L_{M}^{r^3}T_{P}^{tg}}\, .\label{eq_171}
\end{align}
After the shrinking process,  the operators become
\begin{align}
    \mathcal{S}\left(W_M\left( C_r, Id;r, r \right)\right) &=\sum_{g\in Z_r}{T_{P}^{g}} =\frac{1}{2}\left[W_P\left( Id \right)+W_P\left( \mathcal{T} \right)\right]\, , \\
    \mathcal{S}\left(W_M\left( C_r, Id;r^3, r \right)\right) &=\sum_{g\in Z_r}{T_{P}^{tg}} =\frac{1}{2}\left[W_P\left( Id \right)-W_P\left( \mathcal{T} \right)\right]\, .
\end{align}
Thus,  the local Hilbert space $\mathcal{S}\left(V_{\left( C_r, Id\right)}\right)$ spanned by 
\begin{align}
    \left\{\mathcal{S}\left(\ket{r}_{\left(C_r, Id\right)}\right), \mathcal{S}\left(\ket{r^3}_{\left(C_r, Id\right)}\right)\right\} &=\left\{\frac{1}{2}\left[\ket{1}_{\left(Id\right)}+\ket{1}_{\left(\mathcal{T}\right)}\right], \frac{1}{2}\left[\ket{1}_{\left(Id\right)}-\ket{1}_{\left(\mathcal{T}\right)}\right]\right\}
\end{align}
can be decomposed as $\mathcal{S}\left(V_{\left( C_r, Id\right)}\right)=V_{\left( Id\right)}\oplus V_{\left( \mathcal{T}\right)}$,  which leads to the shrinking rule 
\begin{align}
    \mathcal{S}\left(\left[ C_r, Id \right]\right)=\left[ Id \right]\oplus\left[ \mathcal{T} \right]\, .\label{eq_176}
\end{align}
This shrinking rule coincides with Eq.~(\ref{eq_135}). It is possible to observe a nontrivial charge after shrinking the pure loop $\left[ C_r, Id \right]$.

In the continuum field theory,  the shrinking rules like Eq.~(\ref{eq_135}) only show all possible shrinking channels without providing information about how these channels are selected. One advantage of our microscopic lattice model is that we can control the non-Abelian shrinking channels by tuning the internal degrees of freedom in Eq.~(\ref{eq_loop}). Take the pure loop $\left[ C_r, Id \right]$ as an example. Following the discussion in Sec.~\ref{Sec.5.4},  we can linearly combine the operators shown in Eqs.~(\ref{eq_170}) and~(\ref{eq_171}) to obtain another legitimate creation operator for $\left[ C_r, Id \right]$ because this new operator still carries the conserved labels $C_r$ and $Id$. When we linearly combine the operators in Eqs.~(\ref{eq_170}) and~(\ref{eq_171}) as
\begin{align}
    W_M\left( C_r, Id;r, r \right) +W_M\left( C_r, Id;r^3, r \right) &=\sum_{g\in Z_r}{L_{M}^{r}T_{P}^{g}}+\sum_{g\in Z_r}{L_{M}^{r^3}T_{P}^{tg}}\, ,
\end{align}
the shrinking process leads to
\begin{align}
    \mathcal{S}\left(\sum_{g\in Z_r}{L_{M}^{r}T_{P}^{g}}+\sum_{g\in Z_r}{L_{M}^{r^3}T_{P}^{tg}}\right)&=\sum_{g\in \mathbb{D} _4}{T_{P}^{g}}=W_P\left( Id \right)\, .
\end{align}
In this case,  shrinking the loop $\left[ C_r, Id \right]$ produces a definite result $\left[Id\right]$. On the other hand,  when we linearly combine the operators in Eqs.~(\ref{eq_170}) and~(\ref{eq_171}) as
\begin{align}
    W_M\left( C_r, Id;r, r \right) -W_M\left( C_r, Id;r^3, r \right) &=\sum_{g\in Z_r}{L_{M}^{r}T_{P}^{g}}-\sum_{g\in Z_r}{L_{M}^{r^3}T_{P}^{tg}}\nonumber \\
    &=\sum_{g\in Z_r}{\Gamma^{\mathcal{T}}\left(g\right)L_{M}^{r}T_{P}^{g}}+\sum_{g\in Z_r}{\Gamma^{\mathcal{T}}\left(tg\right) L_{M}^{r^3}T_{P}^{tg}}\, ,
\end{align}
the shrinking process leads to
\begin{align}
    \mathcal{S}\left(\sum_{g\in Z_r}{\Gamma^{\mathcal{T}}\left(g\right)L_{M}^{r}T_{P}^{g}}+\sum_{g\in Z_r}{\Gamma^{\mathcal{T}}\left(tg\right) L_{M}^{r^3}T_{P}^{tg}}\right) &=\sum_{g\in \mathbb{D} _4}{\Gamma^{\mathcal{T}}\left(g\right)T_{P}^{g}}=W_P\left( \mathcal{T} \right)\, .
\end{align}
In this case,  shrinking the loop $\left[ C_r, Id \right]$ produces a definite result $\left[\mathcal{T}\right]$.

The fusion and shrinking rules obtained from our $\mathbb{D}_4$ quantum double model satisfy the consistency condition in Eq.~(\ref{eq_consistent_4D}). For example,  we can fuse $\left[ C_r, Id \right]$ and $\left[ \mathcal{B} \right]$ as shown in Eq.~(\ref{eq_160}) first. The subsequent shrinking process leads to
\begin{align}
    \mathcal{S}\left(\left[ C_r, Id \right]\otimes\left[ \mathcal{B} \right]\right)&=\mathcal{S}\left(\left[ C_r, \omega \right]\oplus\left[ C_r, \omega ^3 \right]\right) =\left[ \mathcal{B} \right]\oplus\left[ \mathcal{B} \right]=2\cdot\left[ \mathcal{B} \right]\, ,
\end{align}
where we have used $\mathcal{S}\left(\left[ C_r, \omega \right]\right)=\mathcal{S}\left(\left[ C_r, \omega^3 \right]\right)=\left[ \mathcal{B} \right]$. If we shrink $\left[ C_r, Id \right]$ and $\left[ \mathcal{B} \right]$ first and then perform fusion,  we have
\begin{align}
    \mathcal{S}\left(\left[ C_r, Id \right]\right)\otimes\mathcal{S}\left(\left[ \mathcal{B} \right]\right)&=\left(\left[ Id \right]\oplus\left[ \mathcal{T} \right]\right)\otimes\left[ \mathcal{B} \right] =\left[ \mathcal{B} \right]\oplus\left[ \mathcal{B} \right]=2\cdot\left[ \mathcal{B} \right]\, ,
\end{align}
which has the same result as fusion followed by shrinking.

Since the particle $\left[ \mathcal{R} \right]$ is an Abelian particle,  according to Sec.~\ref{Sec.8.1},  the particle--loop braiding of $\left[ \mathcal{R} \right]$ and $\left[ C_r, Id \right]$ simply gives a phase
\begin{align}
    \exp \left(\mathrm{i}\theta_{\left(\mathcal{R}\right), \left(C_r, Id\right)}^{\mathrm{PL}}\right)=\Gamma^{\mathcal{R}^{\ast}}\left( r \right)=-1\, .
\end{align}
Similarly,  we can derive the particle--loop braiding phases
\begin{align}
    \exp \left(\mathrm{i}\theta_{\left(\mathcal{T}\right), \left(C_t, Id\right)}^{\mathrm{PL}}\right)&=\Gamma^{\mathcal{T}^{\ast}}\left( t \right)=-1\, , \\
    \exp \left(\mathrm{i}\theta_{\left(\mathcal{B}\right), \left(C_{r^2}, Id\right)}^{\mathrm{PL}}\right)&=\Gamma_{11}^{\mathcal{B}^{\ast}}\left( r^2 \right)=-1\, ,
\end{align}
which are the counterparts of $\exp \left(\mathrm{i}\theta_{\mathsf{P}_{010}, \mathsf{L}_{010}}^{\mathrm{PL}}\right)=\exp \left(\mathrm{i}\theta_{\mathsf{P}_{001}, \mathsf{L}_{001}}^{\mathrm{PL}}\right)=-1$ shown in Sec.~\ref{Sec.6.3}. 

The Borromean-Rings braiding in the $\mathbb{D}_4$ quantum double is very special because when a particle carries a non-Abelian charge and two loops carry two non-Abelian fluxes (see \textbf{Definition}~\ref{def:abelian_charge} and \textbf{Definition}~\ref{def:abelian_flux}), the Borromean-Rings braiding process of these three non-Abelian excitations still produces a phase rather than an operator. For the particle $\left[ \mathcal{B} \right]$ and the loops $\left[ C_r, Id \right]$ and $\left[ C_t, Id \right]$, we can verify that
\begin{align}
    \left( g_1:\bar{c}_1 \right) \left( g_2:\bar{c}_2 \right) \left( g_1:c_1 \right) \left( g_2:c_2 \right)=r^2\, ,
\end{align}
where $c_1\in C_r$ and $c_2\in C_t$. The equation above is independent of the choices of $g_1$ and $g_2$. Thus,  according to the discussion in Sec.~\ref{Sec.8.2},  this Borromean-Rings braiding leads to a phase
\begin{align}
    \exp \left(\mathrm{i}\theta^{\mathrm{BR}}_{\left(\mathcal{B}\right), \left(C_r, Id\right), \left(C_t, Id\right)}\right)=\Gamma_{11}^{\mathcal{B}^{\ast}}\left( r^2 \right)=-1\, .
\end{align}
This nontrivial phase is independent of the internal degrees of freedom associated with the non-Abelian excitations. According to \textbf{Theorem}~\ref{theorem_bring},  nontrivial Borromean-Rings braiding is possible only when the charge carried by the particle and two fluxes carried by the two loops are all non-Abelian. This nontrivial phase indicates that the particle--loop braiding between $\left[ \mathcal{B} \right]$ and $\left[ C_r, Id \right]$ and the particle--loop braiding between $\left[ \mathcal{B} \right]$ and $\left[ C_t, Id \right]$ are non-Abelian.

Finally,  we present all the fusion rules in Tables~\ref{tab_D4_fusion1} and~\ref{tab_D4_fusion2} as well as all the shrinking rules in Table~\ref{tab_D4_shrinking}. From Eqs.~(\ref{eq_operator_O}) and~(\ref{eq_BR_operator}) we can see that charge decorations on loops do not affect the particle--loop and the Borromean-Rings braiding. Thus,  we only need to consider pure loops when computing all the braiding processes. We derive all braiding phases in Table~\ref{tab_D4_braiding}. In Sec.~\ref{Sec.6.4},  by comparing the fusion, shrinking, and braiding tables of the 3D $\mathbb{D}_4$ quantum double model with those of the $BF$ field theory with an $AAB$ twist and gauge group $(\mathbb{Z}_2)^3$,  we establish an isomorphism between the excitations in these two theories. 
\begin{table*}
	\scriptsize
	\caption{\label{tab_D4_fusion1}The first part of the fusion table for the $\mathbb{D}_4$ quantum double model. The complete fusion table is a $22\times22$ table. This table shows columns 1 to 11; columns 12 to 22 are shown in Table~\ref{tab_D4_fusion2}.}
	\centering
    \par\smallskip\hrule\smallskip
	\begin{adjustbox}{max width=\linewidth,max totalheight=.82\textheight}
\begin{tabular}{cccccccccccc}
		\noalign{\vspace{3pt}}
		$\otimes$ & $\left[ Id \right]$ & $\left[ \mathcal{R} \right]$ & $\left[ \mathcal{T} \right]$ & $\left[ \mathcal{RT} \right]$ & $\left[ \mathcal{B} \right]$ & $\left[ C_r, \!Id \right]$ & $\left[ C_r, \!\omega \right]$ & $\left[ C_r, \!\omega ^2 \right]$ & $\left[ C_r, \!\omega ^3 \right]$ & $\left[ C_t, \!Id \right]$ & $\left[ C_t, \!a \right]$
		\\[3pt]
        \hline
        \noalign{\vspace{3pt}}
		$\left[ Id \right]$ & $\left[ Id \right]$ & $\left[ \mathcal{R} \right]$ & $\left[ \mathcal{T} \right]$ & $\left[ \mathcal{RT} \right]$ & $\left[ \mathcal{B} \right]$ & $\left[ C_r, \!Id \right]$ & $\left[ C_r, \!\omega \right]$ & $\left[ C_r, \!\omega ^2 \right]$ & $\left[ C_r, \!\omega ^3 \right]$ & $\left[ C_t, \!Id \right]$ & $\left[ C_t, \!a \right]$ \\[5pt]
		 
		$\left[ \mathcal{R} \right]$ & $\left[ \mathcal{R} \right]$ & $\left[ Id \right]$ & $\left[ \mathcal{RT} \right]$ & $\left[ \mathcal{T} \right]$ & $\left[ \mathcal{B} \right]$ & $\left[ C_r, \!\omega ^2 \right]$ & $\left[ C_r, \!\omega ^3 \right]$ & $\left[ C_r, \!Id \right]$ & $\left[ C_r, \!\omega \right]$ & $\left[ C_t, \!Id \right]$ & $\left[ C_t, \!a \right]$ \\[5pt]
		 
		$\left[ \mathcal{T} \right]$ & $\left[ \mathcal{T} \right]$ & $\left[ \mathcal{RT} \right]$ & $\left[ Id \right]$ & $\left[ \mathcal{R} \right]$ & $\left[ \mathcal{B} \right]$ & $\left[ C_r, \!Id \right]$ & $\left[ C_r, \!\omega \right]$ & $\left[ C_r, \!\omega ^2 \right]$ & $\left[ C_r, \!\omega ^3 \right]$ & $\left[ C_t, \!b \right]$ & $\left[ C_t, \!ab \right]$ \\[5pt]
		 
		$\left[ \mathcal{RT} \right]$ & $\left[ \mathcal{RT} \right]$ & $\left[ \mathcal{T} \right]$ & $\left[ \mathcal{R} \right]$ & $\left[ Id \right]$ & $\left[ \mathcal{B} \right]$ & $\left[ C_r, \!\omega ^2 \right]$ & $\left[ C_r, \!\omega ^3 \right]$ & $\left[ C_r, \!Id \right]$ & $\left[ C_r, \!\omega \right]$ & $\left[ C_t, \!b \right]$ & $\left[ C_t, \!ab \right]$ \\[5pt]
		 
		$\left[ \mathcal{B} \right]$ & $\left[ \mathcal{B} \right]$ & $\left[ \mathcal{B} \right]$ & $\left[ \mathcal{B} \right]$ & $\left[ \mathcal{B} \right]$ & \makecell{$\left[ Id \right]$\\$\oplus \left[ \mathcal{R} \right]$\\$\oplus \left[ \mathcal{T} \right]$\\$\oplus \left[ \mathcal{RT} \right]$} & \makecell{$\left[ C_r, \!\omega ^3 \right]$\\$\oplus \left[ C_r, \!\omega \right]$} & \makecell{$\left[ C_r, \!\omega ^2 \right]$\\$\oplus \left[ C_r, \!Id \right]$} & \makecell{$\left[ C_r, \!\omega ^3 \right]$\\$\oplus \left[ C_r, \!\omega \right]$} & \makecell{$\left[ C_r, \!\omega ^2 \right]$\\$\oplus \left[ C_r, \!Id \right]$} & \makecell{$\left[ C_t, \!ab \right]$\\$\oplus \left[ C_t, \!a \right]$} & \makecell{$\left[ C_t, \!Id \right]$\\$\oplus \left[ C_t, \!b \right]$} \\[15pt]
		 
		$\left[ C_r, \!Id \right]$ & $\left[ C_r, \!Id \right]$ & $\left[ C_r, \!\omega ^2 \right]$ & $\left[ C_r, \!Id \right]$ & $\left[ C_r, \!\omega ^2 \right]$ & \makecell{$\left[ C_r, \!\omega ^3 \right]$\\$\oplus \left[ C_r, \!\omega \right]$} & \makecell{$\left[ C_{r^2}, \!Id \right]$\\$\oplus \left[ C_{r^2}, \!\mathcal{T} \right]$\\$\oplus \left[ \mathcal{T} \right]$\\$\oplus \left[ Id \right]$} & \makecell{$\left[ C_{r^2}, \!\mathcal{B} \right]$\\$\oplus \left[ \mathcal{B} \right]$} & \makecell{$\left[ C_{r^2}, \!\mathcal{RT} \right]$\\$\oplus \left[ C_{r^2}, \!\mathcal{R} \right]$\\$\oplus \left[ \mathcal{RT} \right]$\\$\oplus \left[ \mathcal{R} \right]$} & \makecell{$\left[ C_{r^2}, \!\mathcal{B} \right]$\\$\oplus \left[ \mathcal{B} \right]$} & \makecell{$\left[ C_{tr}, \!Id \right]$\\$\oplus \left[ C_{tr}, \!b \right]$} & \makecell{$\left[ C_{tr}, \!ab \right]$\\$\oplus \left[ C_{tr}, \!a \right]$} \\[15pt]
		 
		$\left[ C_r, \!\omega \right]$ & $\left[ C_r, \!\omega \right]$ & $\left[ C_r, \!\omega ^3 \right]$ & $\left[ C_r, \!\omega \right]$ & $\left[ C_r, \!\omega ^3 \right]$ & \makecell{$\left[ C_r, \!\omega ^2 \right]$\\$\oplus \left[ C_r, \!Id \right]$} & \makecell{$\left[ C_{r^2}, \!\mathcal{B} \right]$\\$\oplus \left[ \mathcal{B} \right]$} & \makecell{$\left[ C_{r^2}, \!\mathcal{RT} \right]$\\$\oplus \left[ C_{r^2}, \!\mathcal{R} \right]$\\$\oplus \left[ \mathcal{T} \right]$\\$\oplus \left[ Id \right]$} & \makecell{$\left[ C_{r^2}, \!\mathcal{B} \right]$\\$\oplus \left[ \mathcal{B} \right]$} & \makecell{$\left[ C_{r^2}, \!Id \right]$\\$\oplus \left[ C_{r^2}, \!\mathcal{T} \right]$\\$\oplus \left[ \mathcal{RT} \right]$\\$\oplus \left[ \mathcal{R} \right]$} & \makecell{$\left[ C_{tr}, \!ab \right]$\\$\oplus \left[ C_{tr}, \!a \right]$} & \makecell{$\left[ C_{tr}, \!Id \right]$\\$\oplus \left[ C_{tr}, \!b \right]$} \\[15pt]
		 
		$\left[ C_r, \!\omega ^2 \right]$ & $\left[ C_r, \!\omega ^2 \right]$ & $\left[ C_r, \!Id \right]$ & $\left[ C_r, \!\omega ^2 \right]$ & $\left[ C_r, \!Id \right]$ & \makecell{$\left[ C_r, \!\omega ^3 \right]$\\$\oplus \left[ C_r, \!\omega \right]$} & \makecell{$\left[ C_{r^2}, \!\mathcal{RT} \right]$\\$\oplus \left[ C_{r^2}, \!\mathcal{R} \right]$\\$\oplus \left[ \mathcal{RT} \right]$\\$\oplus \left[ \mathcal{R} \right]$} & \makecell{$\left[ C_{r^2}, \!\mathcal{B} \right]$\\$\oplus \left[ \mathcal{B} \right]$} & \makecell{$\left[ C_{r^2}, \!Id \right]$\\$\oplus \left[ C_{r^2}, \!\mathcal{T} \right]$\\$\oplus \left[ \mathcal{T} \right]$\\$\oplus \left[ Id \right]$} & \makecell{$\left[ C_{r^2}, \!\mathcal{B} \right]$\\$\oplus \left[ \mathcal{B} \right]$} & \makecell{$\left[ C_{tr}, \!Id \right]$\\$\oplus \left[ C_{tr}, \!b \right]$} & \makecell{$\left[ C_{tr}, \!ab \right]$\\$\oplus \left[ C_{tr}, \!a \right]$} \\[15pt]
		 
		$\left[ C_r, \!\omega ^3 \right]$ & $\left[ C_r, \!\omega ^3 \right]$ & $\left[ C_r, \!\omega \right]$ & $\left[ C_r, \!\omega ^3 \right]$ & $\left[ C_r, \!\omega \right]$ & \makecell{$\left[ C_r, \!\omega ^2 \right]$\\$\oplus \left[ C_r, \!Id \right]$} & \makecell{$\left[ C_{r^2}, \!\mathcal{B} \right]$\\$\oplus \left[ \mathcal{B} \right]$} & \makecell{$\left[ C_{r^2}, \!Id \right]$\\$\oplus \left[ C_{r^2}, \!\mathcal{T} \right]$\\$\oplus \left[ \mathcal{RT} \right]$\\$\oplus \left[ \mathcal{R} \right]$} & \makecell{$\left[ C_{r^2}, \!\mathcal{B} \right]$\\$\oplus \left[ \mathcal{B} \right]$} & \makecell{$\left[ C_{r^2}, \!\mathcal{RT} \right]$\\$\oplus \left[ C_{r^2}, \!\mathcal{R} \right]$\\$\oplus \left[ \mathcal{T} \right]$\\$\oplus \left[ Id \right]$} & \makecell{$\left[ C_{tr}, \!ab \right]$\\$\oplus \left[ C_{tr}, \!a \right]$} & \makecell{$\left[ C_{tr}, \!Id \right]$\\$\oplus \left[ C_{tr}, \!b \right]$} \\[15pt]
		 
		$\left[ C_t, \!Id \right]$ & $\left[ C_t, \!Id \right]$ & $\left[ C_t, \!Id \right]$ & $\left[ C_t, \!b \right]$ & $\left[ C_t, \!b \right]$ & \makecell{$\left[ C_t, \!ab \right]$\\$\oplus \left[ C_t, \!a \right]$} & \makecell{$\left[ C_{tr}, \!Id \right]$\\$\oplus \left[ C_{tr}, \!b \right]$} & \makecell{$\left[ C_{tr}, \!ab \right]$\\$\oplus \left[ C_{tr}, \!a \right]$} & \makecell{$\left[ C_{tr}, \!Id \right]$\\$\oplus \left[ C_{tr}, \!b \right]$} & \makecell{$\left[ C_{tr}, \!ab \right]$\\$\oplus \left[ C_{tr}, \!a \right]$} & \makecell{$\left[ C_{r^2}, \!Id \right]$\\$\oplus \left[ C_{r^2}, \!\mathcal{R} \right]$\\$\oplus \left[ \mathcal{R} \right]$\\$\oplus \left[ Id \right]$} & \makecell{$\left[ C_{r^2}, \!\mathcal{B} \right]$\\$\oplus \left[ \mathcal{B} \right]$} \\[15pt]
		 
		$\left[ C_t, \!a \right]$ & $\left[ C_t, \!a \right]$ & $\left[ C_t, \!a \right]$ & $\left[ C_t, \!ab \right]$ & $\left[ C_t, \!ab \right]$ & \makecell{$\left[ C_t, \!Id \right]$\\$\oplus \left[ C_t, \!b \right]$} & \makecell{$\left[ C_{tr}, \!ab \right]$\\$\oplus \left[ C_{tr}, \!a \right]$} & \makecell{$\left[ C_{tr}, \!Id \right]$\\$\oplus \left[ C_{tr}, \!b \right]$} & \makecell{$\left[ C_{tr}, \!ab \right]$\\$\oplus \left[ C_{tr}, \!a \right]$} & \makecell{$\left[ C_{tr}, \!Id \right]$\\$\oplus \left[ C_{tr}, \!b \right]$} & \makecell{$\left[ C_{r^2}, \!\mathcal{B} \right]$\\$\oplus \left[ \mathcal{B} \right]$} & \makecell{$\left[ C_{r^2}, \!\mathcal{RT} \right]$\\$\oplus \left[ C_{r^2}, \!\mathcal{T} \right]$\\$\oplus \left[ \mathcal{R} \right]$\\$\oplus \left[ Id \right]$} \\[15pt]
		 
		$\left[ C_t, \!b \right]$ & $\left[ C_t, \!b \right]$ & $\left[ C_t, \!b \right]$ & $\left[ C_t, \!Id \right]$ & $\left[ C_t, \!Id \right]$ & \makecell{$\left[ C_t, \!ab \right]$\\$\oplus \left[ C_t, \!a \right]$} & \makecell{$\left[ C_{tr}, \!Id \right]$\\$\oplus \left[ C_{tr}, \!b \right]$} & \makecell{$\left[ C_{tr}, \!ab \right]$\\$\oplus \left[ C_{tr}, \!a \right]$} & \makecell{$\left[ C_{tr}, \!Id \right]$\\$\oplus \left[ C_{tr}, \!b \right]$} & \makecell{$\left[ C_{tr}, \!ab \right]$\\$\oplus \left[ C_{tr}, \!a \right]$} & \makecell{$\left[ C_{r^2}, \!\mathcal{RT} \right]$\\$\oplus \left[ C_{r^2}, \!\mathcal{T} \right]$\\$\oplus \left[ \mathcal{RT} \right]$\\$\oplus \left[ \mathcal{T} \right]$} & \makecell{$\left[ C_{r^2}, \!\mathcal{B} \right]$\\$\oplus \left[ \mathcal{B} \right]$} \\[15pt]
		 
		$\left[ C_t, \!ab \right]$ & $\left[ C_t, \!ab \right]$ & $\left[ C_t, \!ab \right]$ & $\left[ C_t, \!a \right]$ & $\left[ C_t, \!a \right]$ & \makecell{$\left[ C_t, \!Id \right]$\\$\oplus \left[ C_t, \!b \right]$} & \makecell{$\left[ C_{tr}, \!ab \right]$\\$\oplus \left[ C_{tr}, \!a \right]$} & \makecell{$\left[ C_{tr}, \!Id \right]$\\$\oplus \left[ C_{tr}, \!b \right]$} & \makecell{$\left[ C_{tr}, \!ab \right]$\\$\oplus \left[ C_{tr}, \!a \right]$} & \makecell{$\left[ C_{tr}, \!Id \right]$\\$\oplus \left[ C_{tr}, \!b \right]$} & \makecell{$\left[ C_{r^2}, \!\mathcal{B} \right]$\\$\oplus \left[ \mathcal{B} \right]$} & \makecell{$\left[ C_{r^2}, \!Id \right]$\\$\oplus \left[ C_{r^2}, \!\mathcal{R} \right]$\\$\oplus \left[ \mathcal{RT} \right]$\\$\oplus \left[ \mathcal{T} \right]$} \\[15pt]
		 
		$\left[ C_{tr}, \!Id \right]$ & $\left[ C_{tr}, \!Id \right]$ & $\left[ C_{tr}, \!b \right]$ & $\left[ C_{tr}, \!b \right]$ & $\left[ C_{tr}, \!Id \right]$ & \makecell{$\left[ C_{tr}, \!ab \right]$\\$\oplus \left[ C_{tr}, \!a \right]$} & \makecell{$\left[ C_t, \!Id \right]$\\$\oplus \left[ C_t, \!b \right]$} & \makecell{$\left[ C_t, \!ab \right]$\\$\oplus \left[ C_t, \!a \right]$} & \makecell{$\left[ C_t, \!Id \right]$\\$\oplus \left[ C_t, \!b \right]$} & \makecell{$\left[ C_t, \!ab \right]$\\$\oplus \left[ C_t, \!a \right]$} & \makecell{$\left[ C_r, \!\omega ^2 \right]$\\$\oplus \left[ C_r, \!Id \right]$} & \makecell{$\left[ C_r, \!\omega ^3 \right]$\\$\oplus \left[ C_r, \!\omega \right]$} \\[15pt]
		 
		$\left[ C_{tr}, \!a \right]$ & $\left[ C_{tr}, \!a \right]$ & $\left[ C_{tr}, \!ab \right]$ & $\left[ C_{tr}, \!ab \right]$ & $\left[ C_{tr}, \!a \right]$ & \makecell{$\left[ C_{tr}, \!Id \right]$\\$\oplus \left[ C_{tr}, \!b \right]$} & \makecell{$\left[ C_t, \!ab \right]$\\$\oplus \left[ C_t, \!a \right]$} & \makecell{$\left[ C_t, \!Id \right]$\\$\oplus \left[ C_t, \!b \right]$} & \makecell{$\left[ C_t, \!ab \right]$\\$\oplus \left[ C_t, \!a \right]$} & \makecell{$\left[ C_t, \!Id \right]$\\$\oplus \left[ C_t, \!b \right]$} & \makecell{$\left[ C_r, \!\omega ^3 \right]$\\$\oplus \left[ C_r, \!\omega \right]$} & \makecell{$\left[ C_r, \!\omega ^2 \right]$\\$\oplus \left[ C_r, \!Id \right]$} \\[15pt]
		 
		$\left[ C_{tr}, \!b \right]$ & $\left[ C_{tr}, \!b \right]$ & $\left[ C_{tr}, \!Id \right]$ & $\left[ C_{tr}, \!Id \right]$ & $\left[ C_{tr}, \!b \right]$ & \makecell{$\left[ C_{tr}, \!ab \right]$\\$\oplus \left[ C_{tr}, \!a \right]$} & \makecell{$\left[ C_t, \!Id \right]$\\$\oplus \left[ C_t, \!b \right]$} & \makecell{$\left[ C_t, \!ab \right]$\\$\oplus \left[ C_t, \!a \right]$} & \makecell{$\left[ C_t, \!Id \right]$\\$\oplus \left[ C_t, \!b \right]$} & \makecell{$\left[ C_t, \!ab \right]$\\$\oplus \left[ C_t, \!a \right]$} & \makecell{$\left[ C_r, \!\omega ^2 \right]$\\$\oplus \left[ C_r, \!Id \right]$} & \makecell{$\left[ C_r, \!\omega ^3 \right]$\\$\oplus \left[ C_r, \!\omega \right]$} \\[15pt]
		 
		$\left[ C_{tr}, \!ab \right]$ & $\left[ C_{tr}, \!ab \right]$ & $\left[ C_{tr}, \!a \right]$ & $\left[ C_{tr}, \!a \right]$ & $\left[ C_{tr}, \!ab \right]$ & \makecell{$\left[ C_{tr}, \!Id \right]$\\$\oplus \left[ C_{tr}, \!b \right]$} & \makecell{$\left[ C_t, \!ab \right]$\\$\oplus \left[ C_t, \!a \right]$} & \makecell{$\left[ C_t, \!Id \right]$\\$\oplus \left[ C_t, \!b \right]$} & \makecell{$\left[ C_t, \!ab \right]$\\$\oplus \left[ C_t, \!a \right]$} & \makecell{$\left[ C_t, \!Id \right]$\\$\oplus \left[ C_t, \!b \right]$} & \makecell{$\left[ C_r, \!\omega ^3 \right]$\\$\oplus \left[ C_r, \!\omega \right]$} & \makecell{$\left[ C_r, \!\omega ^2 \right]$\\$\oplus \left[ C_r, \!Id \right]$} \\[15pt]
		
		$\left[ C_{r^2}, \!Id \right]$ & $\left[ C_{r^2}, \!Id \right]$ & $\left[ C_{r^2}, \!\mathcal{R} \right]$ & $\left[ C_{r^2}, \!\mathcal{T} \right]$ & $\left[ C_{r^2}, \!\mathcal{RT} \right]$ & $\left[ C_{r^2}, \!\mathcal{B} \right]$ & $\left[ C_r, \!Id \right]$ & $\left[ C_r, \!\omega ^3 \right]$ & $\left[ C_r, \!\omega ^2 \right]$ & $\left[ C_r, \!\omega \right]$ & $\left[ C_t, \!Id \right]$ & $\left[ C_t, \!ab \right]$ \\[5pt]
		
		$\left[ C_{r^2}, \!\mathcal{R} \right]$ & $\left[ C_{r^2}, \!\mathcal{R} \right]$ & $\left[ C_{r^2}, \!Id \right]$ & $\left[ C_{r^2}, \!\mathcal{RT} \right]$ & $\left[ C_{r^2}, \!\mathcal{T} \right]$ & $\left[ C_{r^2}, \!\mathcal{B} \right]$ & $\left[ C_r, \!\omega ^2 \right]$ & $\left[ C_r, \!\omega \right]$ & $\left[ C_r, \!Id \right]$ & $\left[ C_r, \!\omega ^3 \right]$ & $\left[ C_t, \!Id \right]$ & $\left[ C_t, \!ab \right]$ \\[5pt]
		
		$\left[ C_{r^2}, \!\mathcal{T} \right]$ & $\left[ C_{r^2}, \!\mathcal{T} \right]$ & $\left[ C_{r^2}, \!\mathcal{RT} \right]$ & $\left[ C_{r^2}, \!Id \right]$ & $\left[ C_{r^2}, \!\mathcal{R} \right]$ & $\left[ C_{r^2}, \!\mathcal{B} \right]$ & $\left[ C_r, \!Id \right]$ & $\left[ C_r, \!\omega ^3 \right]$ & $\left[ C_r, \!\omega ^2 \right]$ & $\left[ C_r, \!\omega \right]$ & $\left[ C_t, \!b \right]$ & $\left[ C_t, \!a \right]$ \\[5pt]
		
		$\left[ C_{r^2}, \!\mathcal{RT} \right]$ & $\left[ C_{r^2}, \!\mathcal{RT} \right]$ & $\left[ C_{r^2}, \!\mathcal{T} \right]$ & $\left[ C_{r^2}, \!\mathcal{R} \right]$ & $\left[ C_{r^2}, \!Id \right]$ & $\left[ C_{r^2}, \!\mathcal{B} \right]$ & $\left[ C_r, \!\omega ^2 \right]$ & $\left[ C_r, \!\omega \right]$ & $\left[ C_r, \!Id \right]$ & $\left[ C_r, \!\omega ^3 \right]$ & $\left[ C_t, \!b \right]$ & $\left[ C_t, \!a \right]$ \\[5pt]
		
		$\left[ C_{r^2}, \!\mathcal{B} \right]$ & $\left[ C_{r^2}, \!\mathcal{B} \right]$ & $\left[ C_{r^2}, \!\mathcal{B} \right]$ & $\left[ C_{r^2}, \!\mathcal{B} \right]$ & $\left[ C_{r^2}, \!\mathcal{B} \right]$ & \makecell{$\left[ C_{r^2}, \!\mathcal{RT} \right]$\\$\oplus \left[ C_{r^2}, \!\mathcal{T} \right]$\\$\oplus \left[ C_{r^2}, \!\mathcal{R} \right]$\\$\oplus \left[ C_{r^2}, \!Id \right]$} & \makecell{$\left[ C_r, \!\omega ^3 \right]$\\$\oplus \left[ C_r, \!\omega \right]$} & \makecell{$\left[ C_r, \!\omega ^2 \right]$\\$\oplus \left[ C_r, \!Id \right]$} & \makecell{$\left[ C_r, \!\omega ^3 \right]$\\$\oplus \left[ C_r, \!\omega \right]$} & \makecell{$\left[ C_r, \!\omega ^2 \right]$\\$\oplus \left[ C_r, \!Id \right]$} & \makecell{$\left[ C_t, \!ab \right]$\\$\oplus \left[ C_t, \!a \right]$} & \makecell{$\left[ C_t, \!Id \right]$\\$\oplus \left[ C_t, \!b \right]$} \\[15pt]
		
	\end{tabular}
\end{adjustbox}
    \par\smallskip\hrule\smallskip
\end{table*}
\begin{table*}
	\scriptsize
	\caption{\label{tab_D4_fusion2}The second part of the fusion table for the $\mathbb{D}_4$ quantum double model. The complete fusion table is a $22\times22$ table. This table shows columns 12 to 22.}
	\centering
    \par\smallskip\hrule\smallskip
	\begin{adjustbox}{max width=\linewidth,max totalheight=.82\textheight}
\begin{tabular}{cccccccccccc}
		\noalign{\vspace{3pt}}
		$\otimes$ & $\left[ C_t, \!b \right]$ & $\left[ C_t, \!ab \right]$ & $\left[ C_{tr}, \!Id \right]$ & $\left[ C_{tr}, \!a \right]$ & $\left[ C_{tr}, \!b \right]$ & $\left[ C_{tr}, \!ab \right]$ & $\left[ C_{r^2}, \!Id \right]$ & $\left[ C_{r^2}, \!\mathcal{R} \right]$ & $\left[ C_{r^2}, \!\mathcal{T} \right]$ & $\left[ C_{r^2}, \!\mathcal{RT} \right]$ & $\left[ C_{r^2}, \!\mathcal{B} \right]$
		\\[3pt]
        \hline\noalign{\vspace{3pt}}
		$\left[ Id \right]$ & $\left[ C_t, \!b \right]$ & $\left[ C_t, \!ab \right]$ & $\left[ C_{tr}, \!Id \right]$ & $\left[ C_{tr}, \!a \right]$ & $\left[ C_{tr}, \!b \right]$ & $\left[ C_{tr}, \!ab \right]$ & $\left[ C_{r^2}, \!Id \right]$ & $\left[ C_{r^2}, \!\mathcal{R} \right]$ & $\left[ C_{r^2}, \!\mathcal{T} \right]$ & $\left[ C_{r^2}, \!\mathcal{RT} \right]$ & $\left[ C_{r^2}, \!\mathcal{B} \right]$ \\[5pt]
		
		$\left[ \mathcal{R} \right]$ & $\left[ C_t, \!b \right]$ & $\left[ C_t, \!ab \right]$ & $\left[ C_{tr}, \!b \right]$ & $\left[ C_{tr}, \!ab \right]$ & $\left[ C_{tr}, \!Id \right]$ & $\left[ C_{tr}, \!a \right]$ & $\left[ C_{r^2}, \!\mathcal{R} \right]$ & $\left[ C_{r^2}, \!Id \right]$ & $\left[ C_{r^2}, \!\mathcal{RT} \right]$ & $\left[ C_{r^2}, \!\mathcal{T} \right]$ & $\left[ C_{r^2}, \!\mathcal{B} \right]$ \\[5pt]
		
		$\left[ \mathcal{T} \right]$ & $\left[ C_t, \!Id \right]$ & $\left[ C_t, \!a \right]$ & $\left[ C_{tr}, \!b \right]$ & $\left[ C_{tr}, \!ab \right]$ & $\left[ C_{tr}, \!Id \right]$ & $\left[ C_{tr}, \!a \right]$ & $\left[ C_{r^2}, \!\mathcal{T} \right]$ & $\left[ C_{r^2}, \!\mathcal{RT} \right]$ & $\left[ C_{r^2}, \!Id \right]$ & $\left[ C_{r^2}, \!\mathcal{R} \right]$ & $\left[ C_{r^2}, \!\mathcal{B} \right]$ \\[5pt]
		
		$\left[ \mathcal{RT} \right]$ & $\left[ C_t, \!Id \right]$ & $\left[ C_t, \!a \right]$ & $\left[ C_{tr}, \!Id \right]$ & $\left[ C_{tr}, \!a \right]$ & $\left[ C_{tr}, \!b \right]$ & $\left[ C_{tr}, \!ab \right]$ & $\left[ C_{r^2}, \!\mathcal{RT} \right]$ & $\left[ C_{r^2}, \!\mathcal{T} \right]$ & $\left[ C_{r^2}, \!\mathcal{R} \right]$ & $\left[ C_{r^2}, \!Id \right]$ & $\left[ C_{r^2}, \!\mathcal{B} \right]$ \\[5pt]
		
		$\left[ \mathcal{B} \right]$ & \makecell{$\left[ C_t, \!ab \right]$\\$\oplus \left[ C_t, \!a \right]$} & \makecell{$\left[ C_t, \!Id \right]$\\$\oplus \left[ C_t, \!b \right]$} & \makecell{$\left[ C_{tr}, \!ab \right]$\\$\oplus \left[ C_{tr}, \!a \right]$} & \makecell{$\left[ C_{tr}, \!Id \right]$\\$\oplus \left[ C_{tr}, \!b \right]$} & \makecell{$\left[ C_{tr}, \!ab \right]$\\$\oplus \left[ C_{tr}, \!a \right]$} & \makecell{$\left[ C_{tr}, \!Id \right]$\\$\oplus \left[ C_{tr}, \!b \right]$} & $\left[ C_{r^2}, \!\mathcal{B} \right]$ & $\left[ C_{r^2}, \!\mathcal{B} \right]$ & $\left[ C_{r^2}, \!\mathcal{B} \right]$ & $\left[ C_{r^2}, \!\mathcal{B} \right]$ & \makecell{$\left[ C_{r^2}, \!\mathcal{RT} \right]$\\$\oplus \left[ C_{r^2}, \!\mathcal{T} \right]$\\$\oplus \left[ C_{r^2}, \!\mathcal{R} \right]$\\$\oplus \left[ C_{r^2}, \!Id \right]$} \\[15pt]
		
		$\left[ C_r, \!Id \right]$ & \makecell{$\left[ C_{tr}, \!Id \right]$\\$\oplus \left[ C_{tr}, \!b \right]$} & \makecell{$\left[ C_{tr}, \!ab \right]$\\$\oplus \left[ C_{tr}, \!a \right]$} & \makecell{$\left[ C_t, \!Id \right]$\\$\oplus \left[ C_t, \!b \right]$} & \makecell{$\left[ C_t, \!ab \right]$\\$\oplus \left[ C_t, \!a \right]$} & \makecell{$\left[ C_t, \!Id \right]$\\$\oplus \left[ C_t, \!b \right]$} & \makecell{$\left[ C_t, \!ab \right]$\\$\oplus \left[ C_t, \!a \right]$} & $\left[ C_r, \!Id \right]$ & $\left[ C_r, \!\omega ^2 \right]$ & $\left[ C_r, \!Id \right]$ & $\left[ C_r, \!\omega ^2 \right]$ & \makecell{$\left[ C_r, \!\omega ^3 \right]$\\$\oplus \left[ C_r, \!\omega \right]$} \\[10pt]
		
		$\left[ C_r, \!\omega \right]$ & \makecell{$\left[ C_{tr}, \!ab \right]$\\$\oplus \left[ C_{tr}, \!a \right]$} & \makecell{$\left[ C_{tr}, \!Id \right]$\\$\oplus \left[ C_{tr}, \!b \right]$} & \makecell{$\left[ C_t, \!ab \right]$\\$\oplus \left[ C_t, \!a \right]$} & \makecell{$\left[ C_t, \!Id \right]$\\$\oplus \left[ C_t, \!b \right]$} & \makecell{$\left[ C_t, \!ab \right]$\\$\oplus \left[ C_t, \!a \right]$} & \makecell{$\left[ C_t, \!Id \right]$\\$\oplus \left[ C_t, \!b \right]$} & $\left[ C_r, \!\omega ^3 \right]$ & $\left[ C_r, \!\omega \right]$ & $\left[ C_r, \!\omega ^3 \right]$ & $\left[ C_r, \!\omega \right]$ & \makecell{$\left[ C_r, \!\omega ^2 \right]$\\$\oplus \left[ C_r, \!Id \right]$} \\[10pt]
		
		$\left[ C_r, \!\omega ^2 \right]$ & \makecell{$\left[ C_{tr}, \!Id \right]$\\$\oplus \left[ C_{tr}, \!b \right]$} & \makecell{$\left[ C_{tr}, \!ab \right]$\\$\oplus \left[ C_{tr}, \!a \right]$} & \makecell{$\left[ C_t, \!Id \right]$\\$\oplus \left[ C_t, \!b \right]$} & \makecell{$\left[ C_t, \!ab \right]$\\$\oplus \left[ C_t, \!a \right]$} & \makecell{$\left[ C_t, \!Id \right]$\\$\oplus \left[ C_t, \!b \right]$} & \makecell{$\left[ C_t, \!ab \right]$\\$\oplus \left[ C_t, \!a \right]$} & $\left[ C_r, \!\omega ^2 \right]$ & $\left[ C_r, \!Id \right]$ & $\left[ C_r, \!\omega ^2 \right]$ & $\left[ C_r, \!Id \right]$ & \makecell{$\left[ C_r, \!\omega ^3 \right]$\\$\oplus \left[ C_r, \!\omega \right]$} \\[10pt]
		
		$\left[ C_r, \!\omega ^3 \right]$ & \makecell{$\left[ C_{tr}, \!ab \right]$\\$\oplus \left[ C_{tr}, \!a \right]$} & \makecell{$\left[ C_{tr}, \!Id \right]$\\$\oplus \left[ C_{tr}, \!b \right]$} & \makecell{$\left[ C_t, \!ab \right]$\\$\oplus \left[ C_t, \!a \right]$} & \makecell{$\left[ C_t, \!Id \right]$\\$\oplus \left[ C_t, \!b \right]$} & \makecell{$\left[ C_t, \!ab \right]$\\$\oplus \left[ C_t, \!a \right]$} & \makecell{$\left[ C_t, \!Id \right]$\\$\oplus \left[ C_t, \!b \right]$} & $\left[ C_r, \!\omega \right]$ & $\left[ C_r, \!\omega ^3 \right]$ & $\left[ C_r, \!\omega \right]$ & $\left[ C_r, \!\omega ^3 \right]$ & \makecell{$\left[ C_r, \!\omega ^2 \right]$\\$\oplus \left[ C_r, \!Id \right]$} \\[10pt]
		
		$\left[ C_t, \!Id \right]$ & \makecell{$\left[ C_{r^2}, \!\mathcal{RT} \right]$\\$\oplus \left[ C_{r^2}, \!\mathcal{T} \right]$\\$\oplus \left[ \mathcal{RT} \right]$\\$\oplus \left[ \mathcal{T} \right]$} & \makecell{$\left[ C_{r^2}, \!\mathcal{B} \right]$\\$\oplus \left[ \mathcal{B} \right]$} & \makecell{$\left[ C_r, \!\omega ^2 \right]$\\$\oplus \left[ C_r, \!Id \right]$} & \makecell{$\left[ C_r, \!\omega ^3 \right]$\\$\oplus \left[ C_r, \!\omega \right]$} & \makecell{$\left[ C_r, \!\omega ^2 \right]$\\$\oplus \left[ C_r, \!Id \right]$} & \makecell{$\left[ C_r, \!\omega ^3 \right]$\\$\oplus \left[ C_r, \!\omega \right]$} & $\left[ C_t, \!Id \right]$ & $\left[ C_t, \!Id \right]$ & $\left[ C_t, \!b \right]$ & $\left[ C_t, \!b \right]$ & \makecell{$\left[ C_t, \!ab \right]$\\$\oplus \left[ C_t, \!a \right]$} \\[15pt]
		
		$\left[ C_t, \!a \right]$ & \makecell{$\left[ C_{r^2}, \!\mathcal{B} \right]$\\$\oplus \left[ \mathcal{B} \right]$} & \makecell{$\left[ C_{r^2}, \!Id \right]$\\$\oplus \left[ C_{r^2}, \!\mathcal{R} \right]$\\$\oplus \left[ \mathcal{RT} \right]$\\$\oplus \left[ \mathcal{T} \right]$} & \makecell{$\left[ C_r, \!\omega ^3 \right]$\\$\oplus \left[ C_r, \!\omega \right]$} & \makecell{$\left[ C_r, \!\omega ^2 \right]$\\$\oplus \left[ C_r, \!Id \right]$} & \makecell{$\left[ C_r, \!\omega ^3 \right]$\\$\oplus \left[ C_r, \!\omega \right]$} & \makecell{$\left[ C_r, \!\omega ^2 \right]$\\$\oplus \left[ C_r, \!Id \right]$} & $\left[ C_t, \!ab \right]$ & $\left[ C_t, \!ab \right]$ & $\left[ C_t, \!a \right]$ & $\left[ C_t, \!a \right]$ & \makecell{$\left[ C_t, \!Id \right]$\\$\oplus \left[ C_t, \!b \right]$} \\[15pt]
		
		$\left[ C_t, \!b \right]$ & \makecell{$\left[ C_{r^2}, \!Id \right]$\\$\oplus \left[ C_{r^2}, \!\mathcal{R} \right]$\\$\oplus \left[ \mathcal{R} \right]$\\$\oplus \left[ Id \right]$} & \makecell{$\left[ C_{r^2}, \!\mathcal{B} \right]$\\$\oplus \left[ \mathcal{B} \right]$} & \makecell{$\left[ C_r, \!\omega ^2 \right]$\\$\oplus \left[ C_r, \!Id \right]$} & \makecell{$\left[ C_r, \!\omega ^3 \right]$\\$\oplus \left[ C_r, \!\omega \right]$} & \makecell{$\left[ C_r, \!\omega ^2 \right]$\\$\oplus \left[ C_r, \!Id \right]$} & \makecell{$\left[ C_r, \!\omega ^3 \right]$\\$\oplus \left[ C_r, \!\omega \right]$} & $\left[ C_t, \!b \right]$ & $\left[ C_t, \!b \right]$ & $\left[ C_t, \!Id \right]$ & $\left[ C_t, \!Id \right]$ & \makecell{$\left[ C_t, \!ab \right]$\\$\oplus \left[ C_t, \!a \right]$} \\[15pt]
		
		$\left[ C_t, \!ab \right]$ & \makecell{$\left[ C_{r^2}, \!\mathcal{B} \right]$\\$\oplus \left[ \mathcal{B} \right]$} & \makecell{$\left[ C_{r^2}, \!\mathcal{RT} \right]$\\$\oplus \left[ C_{r^2}, \!\mathcal{T} \right]$\\$\oplus \left[ \mathcal{R} \right]$\\$\oplus \left[ Id \right]$} & \makecell{$\left[ C_r, \!\omega ^3 \right]$\\$\oplus \left[ C_r, \!\omega \right]$} & \makecell{$\left[ C_r, \!\omega ^2 \right]$\\$\oplus \left[ C_r, \!Id \right]$} & \makecell{$\left[ C_r, \!\omega ^3 \right]$\\$\oplus \left[ C_r, \!\omega \right]$} & \makecell{$\left[ C_r, \!\omega ^2 \right]$\\$\oplus \left[ C_r, \!Id \right]$} & $\left[ C_t, \!a \right]$ & $\left[ C_t, \!a \right]$ & $\left[ C_t, \!ab \right]$ & $\left[ C_t, \!ab \right]$ & \makecell{$\left[ C_t, \!Id \right]$\\$\oplus \left[ C_t, \!b \right]$} \\[15pt]
		
		$\left[ C_{tr}, \!Id \right]$ & \makecell{$\left[ C_r, \!\omega ^2 \right]$\\$\oplus \left[ C_r, \!Id \right]$} & \makecell{$\left[ C_r, \!\omega ^3 \right]$\\$\oplus \left[ C_r, \!\omega \right]$} & \makecell{$\left[ C_{r^2}, \!\mathcal{RT} \right]$\\$\oplus \left[ C_{r^2}, \!Id \right]$\\$\oplus \left[ \mathcal{RT} \right]$\\$\oplus \left[ Id \right]$} & \makecell{$\left[ C_{r^2}, \!\mathcal{B} \right]$\\$\oplus \left[ \mathcal{B} \right]$} & \makecell{$\left[ C_{r^2}, \!\mathcal{T} \right]$\\$\oplus \left[ C_{r^2}, \!\mathcal{R} \right]$\\$\oplus \left[ \mathcal{T} \right]$\\$\oplus \left[ \mathcal{R} \right]$} & \makecell{$\left[ C_{r^2}, \!\mathcal{B} \right]$\\$\oplus \left[ \mathcal{B} \right]$} & $\left[ C_{tr}, \!Id \right]$ & $\left[ C_{tr}, \!b \right]$ & $\left[ C_{tr}, \!b \right]$ & $\left[ C_{tr}, \!Id \right]$ & \makecell{$\left[ C_{tr}, \!ab \right]$\\$\oplus \left[ C_{tr}, \!a \right]$} \\[15pt]
		
		$\left[ C_{tr}, \!a \right]$ & \makecell{$\left[ C_r, \!\omega ^3 \right]$\\$\oplus \left[ C_r, \!\omega \right]$} & \makecell{$\left[ C_r, \!\omega ^2 \right]$\\$\oplus \left[ C_r, \!Id \right]$} & \makecell{$\left[ C_{r^2}, \!\mathcal{B} \right]$\\$\oplus \left[ \mathcal{B} \right]$} & \makecell{$\left[ C_{r^2}, \!\mathcal{T} \right]$\\$\oplus \left[ C_{r^2}, \!\mathcal{R} \right]$\\$\oplus \left[ \mathcal{RT} \right]$\\$\oplus \left[ Id \right]$} & \makecell{$\left[ C_{r^2}, \!\mathcal{B} \right]$\\$\oplus \left[ \mathcal{B} \right]$} & \makecell{$\left[ C_{r^2}, \!\mathcal{RT} \right]$\\$\oplus \left[ C_{r^2}, \!Id \right]$\\$\oplus \left[ \mathcal{T} \right]$\\$\oplus \left[ \mathcal{R} \right]$} & $\left[ C_{tr}, \!ab \right]$ & $\left[ C_{tr}, \!a \right]$ & $\left[ C_{tr}, \!a \right]$ & $\left[ C_{tr}, \!ab \right]$ & \makecell{$\left[ C_{tr}, \!Id \right]$\\$\oplus \left[ C_{tr}, \!b \right]$} \\[15pt]
		
		$\left[ C_{tr}, \!b \right]$ & \makecell{$\left[ C_r, \!\omega ^2 \right]$\\$\oplus \left[ C_r, \!Id \right]$} & \makecell{$\left[ C_r, \!\omega ^3 \right]$\\$\oplus \left[ C_r, \!\omega \right]$} & \makecell{$\left[ C_{r^2}, \!\mathcal{T} \right]$\\$\oplus \left[ C_{r^2}, \!\mathcal{R} \right]$\\$\oplus \left[ \mathcal{T} \right]$\\$\oplus \left[ \mathcal{R} \right]$} & \makecell{$\left[ C_{r^2}, \!\mathcal{B} \right]$\\$\oplus \left[ \mathcal{B} \right]$} & \makecell{$\left[ C_{r^2}, \!\mathcal{RT} \right]$\\$\oplus \left[ C_{r^2}, \!Id \right]$\\$\oplus \left[ \mathcal{RT} \right]$\\$\oplus \left[ Id \right]$} & \makecell{$\left[ C_{r^2}, \!\mathcal{B} \right]$\\$\oplus \left[ \mathcal{B} \right]$} & $\left[ C_{tr}, \!b \right]$ & $\left[ C_{tr}, \!Id \right]$ & $\left[ C_{tr}, \!Id \right]$ & $\left[ C_{tr}, \!b \right]$ & \makecell{$\left[ C_{tr}, \!ab \right]$\\$\oplus \left[ C_{tr}, \!a \right]$} \\[15pt]
		
		$\left[ C_{tr}, \!ab \right]$ & \makecell{$\left[ C_r, \!\omega ^3 \right]$\\$\oplus \left[ C_r, \!\omega \right]$} & \makecell{$\left[ C_r, \!\omega ^2 \right]$\\$\oplus \left[ C_r, \!Id \right]$} & \makecell{$\left[ C_{r^2}, \!\mathcal{B} \right]$\\$\oplus \left[ \mathcal{B} \right]$} & \makecell{$\left[ C_{r^2}, \!\mathcal{RT} \right]$\\$\oplus \left[ C_{r^2}, \!Id \right]$\\$\oplus \left[ \mathcal{T} \right]$\\$\oplus \left[ \mathcal{R} \right]$} & \makecell{$\left[ C_{r^2}, \!\mathcal{B} \right]$\\$\oplus \left[ \mathcal{B} \right]$} & \makecell{$\left[ C_{r^2}, \!\mathcal{T} \right]$\\$\oplus \left[ C_{r^2}, \!\mathcal{R} \right]$\\$\oplus \left[ \mathcal{RT} \right]$\\$\oplus \left[ Id \right]$} & $\left[ C_{tr}, \!a \right]$ & $\left[ C_{tr}, \!ab \right]$ & $\left[ C_{tr}, \!ab \right]$ & $\left[ C_{tr}, \!a \right]$ & \makecell{$\left[ C_{tr}, \!Id \right]$\\$\oplus \left[ C_{tr}, \!b \right]$} \\[15pt]
		
		$\left[ C_{r^2}, \!Id \right]$ & $\left[ C_t, \!b \right]$ & $\left[ C_t, \!a \right]$ & $\left[ C_{tr}, \!Id \right]$ & $\left[ C_{tr}, \!ab \right]$ & $\left[ C_{tr}, \!b \right]$ & $\left[ C_{tr}, \!a \right]$ & $\left[ Id \right]$ & $\left[ \mathcal{R} \right]$ & $\left[ \mathcal{T} \right]$ & $\left[ \mathcal{RT} \right]$ & $\left[ \mathcal{B} \right]$ \\[5pt]
		
		$\left[ C_{r^2}, \!\mathcal{R} \right]$ & $\left[ C_t, \!b \right]$ & $\left[ C_t, \!a \right]$ & $\left[ C_{tr}, \!b \right]$ & $\left[ C_{tr}, \!a \right]$ & $\left[ C_{tr}, \!Id \right]$ & $\left[ C_{tr}, \!ab \right]$ & $\left[ \mathcal{R} \right]$ & $\left[ Id \right]$ & $\left[ \mathcal{RT} \right]$ & $\left[ \mathcal{T} \right]$ & $\left[ \mathcal{B} \right]$ \\[5pt]
		
		$\left[ C_{r^2}, \!\mathcal{T} \right]$ & $\left[ C_t, \!Id \right]$ & $\left[ C_t, \!ab \right]$ & $\left[ C_{tr}, \!b \right]$ & $\left[ C_{tr}, \!a \right]$ & $\left[ C_{tr}, \!Id \right]$ & $\left[ C_{tr}, \!ab \right]$ & $\left[ \mathcal{T} \right]$ & $\left[ \mathcal{RT} \right]$ & $\left[ Id \right]$ & $\left[ \mathcal{R} \right]$ & $\left[ \mathcal{B} \right]$ \\[5pt]
		
		$\left[ C_{r^2}, \!\mathcal{RT} \right]$ & $\left[ C_t, \!Id \right]$ & $\left[ C_t, \!ab \right]$ & $\left[ C_{tr}, \!Id \right]$ & $\left[ C_{tr}, \!ab \right]$ & $\left[ C_{tr}, \!b \right]$ & $\left[ C_{tr}, \!a \right]$ & $\left[ \mathcal{RT} \right]$ & $\left[ \mathcal{T} \right]$ & $\left[ \mathcal{R} \right]$ & $\left[ Id \right]$ & $\left[ \mathcal{B} \right]$ \\[5pt]
		
		$\left[ C_{r^2}, \!\mathcal{B} \right]$ & \makecell{$\left[ C_t, \!ab \right]$\\$\oplus \left[ C_t, \!a \right]$} & \makecell{$\left[ C_t, \!Id \right]$\\$\oplus \left[ C_t, \!b \right]$} & \makecell{$\left[ C_{tr}, \!ab \right]$\\$\oplus \left[ C_{tr}, \!a \right]$} & \makecell{$\left[ C_{tr}, \!Id \right]$\\$\oplus \left[ C_{tr}, \!b \right]$} & \makecell{$\left[ C_{tr}, \!ab \right]$\\$\oplus \left[ C_{tr}, \!a \right]$} & \makecell{$\left[ C_{tr}, \!Id \right]$\\$\oplus \left[ C_{tr}, \!b \right]$} & $\left[ \mathcal{B} \right]$ & $\left[ \mathcal{B} \right]$ & $\left[ \mathcal{B} \right]$ & $\left[ \mathcal{B} \right]$ & \makecell{$\left[ Id \right]$\\$\oplus \left[ \mathcal{R} \right]$\\$\oplus \left[ \mathcal{T} \right]$\\$\oplus \left[ \mathcal{RT} \right]$} \\[15pt]
		
	\end{tabular}
\end{adjustbox}
    \par\smallskip\hrule\smallskip
\end{table*}

\begin{table*}
	\caption{\label{tab_D4_shrinking}Shrinking table for the 3D $\mathbb{D}_4$ quantum double model.}
	\centering
    \renewcommand{\arraystretch}{1.3}  
    \par\smallskip\hrule\smallskip
	\begin{adjustbox}{max width=\linewidth,max totalheight=.82\textheight}
\begin{tabular}{cccccccccc}
		\noalign{\vspace{3pt}}
		Loop & $\left[ C_r, Id \right]$ & $\left[ C_r, \omega \right]$ & $\left[ C_r, \omega ^2 \right]$ & $\left[ C_r, \omega ^3 \right]$ & $\left[ C_t, Id \right]$ & $\left[ C_t, a \right]$ & $\left[ C_t, b \right]$ & $\left[ C_t, ab \right]$ & $\left[ C_{tr}, Id \right]$
		\\[3pt]
        \hline\noalign{\vspace{3pt}}
		$\mathcal{S}\left(\text{Loop}\right)$ & $\left[ Id \right]\oplus\left[ \mathcal{T} \right]$ & $\left[ \mathcal{B} \right]$ & $\left[ \mathcal{R} \right]\oplus\left[ \mathcal{RT} \right]$ & $\left[ \mathcal{B} \right]$ & $\left[ Id \right]\oplus\left[ \mathcal{R} \right]$ & $\left[ \mathcal{B} \right]$ & $\left[ \mathcal{T} \right]\oplus\left[ \mathcal{RT} \right]$ & $\left[ \mathcal{B} \right]$ & $\left[ Id \right]\oplus\left[ \mathcal{RT} \right]$  \\[15pt]
		Loop & $\left[ C_{tr}, a \right]$ & $\left[ C_{tr}, b \right]$ & $\left[ C_{tr}, ab \right]$ & $\left[ C_{r^2}, Id \right]$ & $\left[ C_{r^2}, \mathcal{R} \right] $ & $\left[ C_{r^2}, \mathcal{T} \right] $ & $\left[ C_{r^2}, \mathcal{RT} \right] $ & $\left[ C_{r^2}, \mathcal{B} \right] $ &
		\\[5pt]
		$\mathcal{S}\left(\text{Loop}\right)$ & $\left[ \mathcal{B} \right]$ & $\left[ \mathcal{R} \right]\oplus\left[ \mathcal{T} \right]$ & $\left[ \mathcal{B} \right]$ & $\left[ Id \right]$ & $\left[ \mathcal{R} \right]$ & $\left[ \mathcal{T} \right]$ & $\left[ \mathcal{RT} \right]$ & $\left[ \mathcal{B} \right]$ &  \\[5pt]
		
\end{tabular}
\end{adjustbox}
\par\smallskip\hrule\smallskip
\end{table*}

\begin{table*}
	\caption{\label{tab_D4_braiding}Braiding statistics table for the 3D $\mathbb{D}_4$ quantum double model. There are three non-Abelian particle--loop braiding processes,  i.e.,  braiding of $\left[ \mathcal{B} \right]$ and $\left[ C_r, Id \right]$,  braiding of $\left[ \mathcal{B} \right]$ and $\left[ C_t, Id \right]$,  braiding of $\left[ \mathcal{B} \right]$ and $\left[ C_{tr}, Id \right]$,  which lead to operators rather than phases. }

	\centering
	\renewcommand{\arraystretch}{1.3}  
    \par\smallskip\hrule\smallskip
	\begin{adjustbox}{max width=\linewidth,max totalheight=.82\textheight}
\begin{tabular}{cccccccccc}
		\noalign{\vspace{3pt}}
		Particle $\left[R^{\prime}\right]$ & $\left[ \mathcal{R} \right]$ & $\left[ \mathcal{R} \right]$ & $\left[ \mathcal{R} \right]$ & $\left[ \mathcal{R} \right]$ & $\left[ \mathcal{T} \right]$ & $\left[ \mathcal{T} \right]$ & $\left[ \mathcal{T} \right]$ & $\left[ \mathcal{T} \right]$ &
		\\[3pt]
        \hline\noalign{\vspace{3pt}}
		Loop $\left[C, R\right]$ & $\left[ C_r, Id \right]$ & $\left[ C_t, Id \right]$ & $\left[ C_{tr}, Id \right]$ & $\left[ C_{r^2}, Id \right]$ &  $\left[ C_r, Id \right]$ & $\left[ C_t, Id \right]$ & $\left[ C_{tr}, Id \right]$ & $\left[ C_{r^2}, Id \right]$ &
		\\[5pt]
		$\exp \left(\mathrm{i}\theta^{\mathrm{PL}}_{\left(R^{\prime}\right), \left(C, R\right)}\right)$ & $-1$ & $1$ & $-1$ & $1$ & $1$ & $-1$ & $-1$ & $1$ &
		\\[15pt]
		Particle $\left[R^{\prime}\right]$ & $\left[ \mathcal{RT} \right]$ & $\left[ \mathcal{RT} \right]$ & $\left[ \mathcal{RT} \right]$ & $\left[ \mathcal{RT} \right]$ & $\left[ \mathcal{B} \right]$ & $\left[ \mathcal{B} \right]$ & $\left[ \mathcal{B} \right]$ & $\left[ \mathcal{B} \right]$ &
		\\[5pt]
		Loop $\left[C, R\right]$  & $\left[ C_r, Id \right]$ & $\left[ C_t, Id \right]$ & $\left[ C_{tr}, Id \right]$ & $\left[ C_{r^2}, Id \right]$ &  $\left[ C_r, Id \right]$ & $\left[ C_t, Id \right]$ & $\left[ C_{tr}, Id \right]$ & $\left[ C_{r^2}, Id \right]$ &
		\\[5pt]
		$\exp \left(\mathrm{i}\theta^{\mathrm{PL}}_{\left(R^{\prime}\right), \left(C, R\right)}\right)$ & $-1$ & $-1$ & $1$ & $1$ & \text{non-Abelian} & \text{non-Abelian} & \text{non-Abelian} & $-1$ &
		\\[15pt]
		Particle $\left[R^{\prime}\right]$ & $\left[ \mathcal{B} \right]$ & $\left[ \mathcal{B} \right]$ & $\left[ \mathcal{B} \right]$ & $\left[ \mathcal{B} \right]$ & $\left[ \mathcal{B} \right]$ & $\left[ \mathcal{B} \right]$ & $\left[ \mathcal{B} \right]$ & $\left[ \mathcal{B} \right]$ & $\left[ \mathcal{B} \right]$
		\\[5pt]
		Loop 1 $\left[C_1, R_1\right]$ & $\left[ C_r, Id \right]$ & $\left[ C_r, Id \right]$ & $\left[ C_r, Id \right]$ & $\left[ C_t, Id \right]$ & $\left[ C_t, Id \right]$ & $\left[ C_t, Id \right]$ & $\left[ C_{tr}, Id \right]$ & $\left[ C_{tr}, Id \right]$ & $\left[ C_{tr}, Id \right]$
		\\[5pt]
		Loop 2 $\left[C_2, R_2\right]$ & $\left[ C_r, Id \right]$ & $\left[ C_t, Id \right]$ & $\left[ C_{tr}, Id \right]$ & $\left[ C_r, Id \right]$ & $\left[ C_t, Id \right]$ & $\left[ C_{tr}, Id \right]$ & $\left[ C_r, Id \right]$ & $\left[ C_t, Id \right]$ & $\left[ C_{tr}, Id \right]$
		\\[5pt]
		$\exp \left(\mathrm{i}\theta^{\mathrm{BR}}_{\left(R^{\prime}\right), \left(C_1, R_1\right), \left(C_2, R_2\right)}\right)$ & $1$ & $-1$ & $-1$ & $-1$ & $1$ & $-1$ & $-1$ & $-1$ & $1$
		\\[5pt]

	\end{tabular}
\end{adjustbox}
    \par\smallskip\hrule\smallskip
\end{table*}

\subsection{Isomorphism between excitations of microscopic construction and continuum field theories}\label{Sec.6.4}
\begin{table*}
	\caption{\label{tab_f}Mapping table of $f$. We write $\mathsf{L}_{n_1n_2n_3}^{000}$ as $\mathsf{L}_{n_1n_2n_3}$ for simplicity.}
	\centering
    \renewcommand{\arraystretch}{1.3}  
    \par\smallskip\hrule\smallskip
	\begin{adjustbox}{max width=\linewidth,max totalheight=.82\textheight}
\begin{tabular}{cccccccccccc}
		\noalign{\vspace{3pt}}
		$\mathsf{a}$ & $\mathsf{1}$ & $\mathsf{P}_{100}$ & $\mathsf{P}_{010}$ & $\mathsf{P}_{110}$ & $\mathsf{P}_{001}$ & $\mathsf{L}_{100}$ & $\mathsf{L}_{100}^{001}$ & $\mathsf{L}_{100}^{100}$ & $\mathsf{L}_{100}^{101}$ & $\mathsf{L}_{010}$ & $\mathsf{L}_{010}^{001}$ 
		\\[3pt]
        \hline\noalign{\vspace{3pt}}
		$f\left(\mathsf{a}\right)$ & $\left[ Id \right]$ & $\left[ \mathcal{R} \right]$ & $\left[ \mathcal{T} \right]$ & $\left[ \mathcal{RT} \right]$ & $\left[ \mathcal{B} \right]$ & $\left[ C_r, Id \right]$ & $\left[ C_r, \omega \right]$ & $\left[ C_r, \omega ^2 \right]$ & $\left[ C_r, \omega ^3 \right]$ & $\left[ C_t, Id \right]$ & $\left[ C_t, a \right]$
		\\[15pt]
		$\mathsf{a}$ & $\mathsf{L}_{010}^{010}$ & $\mathsf{L}_{010}^{011}$ & $\mathsf{L}_{110}$ & $\mathsf{L}_{110}^{001}$ & $\mathsf{L}_{110}^{100}$ & $\mathsf{L}_{110}^{101}$ & $\mathsf{L}_{001}$ & $\mathsf{L}_{001}^{100}$ & $\mathsf{L}_{001}^{010}$ & $\mathsf{L}_{001}^{110}$ & $\mathsf{L}_{001}^{001}$
        \\[5pt]
        $f\left(\mathsf{a}\right)$ & $\left[ C_t, b \right]$ & $\left[ C_t, ab \right]$ & $\left[ C_{tr}, Id \right]$ & $\left[ C_{tr}, a \right]$ & $\left[ C_{tr}, b \right]$ & $\left[ C_{tr}, ab \right]$ & $\left[ C_{r^2}, Id \right]$ & $\left[ C_{r^2}, \mathcal{R} \right] $ & $\left[ C_{r^2}, \mathcal{T} \right] $ & $\left[ C_{r^2}, \mathcal{RT} \right] $ & $\left[ C_{r^2}, \mathcal{B} \right] $
        \\[5pt]
		
	\end{tabular}
\end{adjustbox}
    \par\smallskip\hrule\smallskip
\end{table*}

\begin{table*}
	\caption{\label{tab_tilde_f}Mapping table of $\tilde{f}\equiv f\circ F$,  where the mapping $F$ is defined as $F\left(\mathsf{P}_{n_1n_2n_3}\right)=\mathsf{P}_{n_2n_1n_3}$ and $F\left(\mathsf{L}_{n_1n_2n_3}^{m_1m_2m_3}\right)=\mathsf{L}_{n_2n_1n_3}^{m_2m_1m_3}$.}
	\centering
    \renewcommand{\arraystretch}{1.3}  
    \par\smallskip\hrule\smallskip
	\begin{adjustbox}{max width=\linewidth,max totalheight=.82\textheight}
\begin{tabular}{cccccccccccc}
		\noalign{\vspace{3pt}}
		$\mathsf{a}$ & $\mathsf{1}$ & $\mathsf{P}_{100}$ & $\mathsf{P}_{010}$ & $\mathsf{P}_{110}$ & $\mathsf{P}_{001}$ & $\mathsf{L}_{100}$ & $\mathsf{L}_{100}^{001}$ & $\mathsf{L}_{100}^{100}$ & $\mathsf{L}_{100}^{101}$ & $\mathsf{L}_{010}$ & $\mathsf{L}_{010}^{001}$ 
		\\[3pt]
        \hline\noalign{\vspace{3pt}}
		$\tilde{f}\left(\mathsf{a}\right)$ & $\left[ Id \right]$ & $\left[ \mathcal{T} \right]$ & $\left[ \mathcal{R} \right]$ & $\left[ \mathcal{RT} \right]$ & $\left[ \mathcal{B} \right]$ & $\left[ C_t, Id \right]$ & $\left[ C_t, a \right]$ & $\left[ C_t, b \right]$ & $\left[ C_t, ab \right]$ & $\left[ C_r, Id \right]$ & $\left[ C_r, \omega \right]$
		\\[15pt]
		$\mathsf{a}$ & $\mathsf{L}_{010}^{010}$ & $\mathsf{L}_{010}^{011}$ & $\mathsf{L}_{110}$ & $\mathsf{L}_{110}^{001}$ & $\mathsf{L}_{110}^{100}$ & $\mathsf{L}_{110}^{101}$ & $\mathsf{L}_{001}$ & $\mathsf{L}_{001}^{100}$ & $\mathsf{L}_{001}^{010}$ & $\mathsf{L}_{001}^{110}$ & $\mathsf{L}_{001}^{001}$
        \\[5pt]
        $\tilde{f}\left(\mathsf{a}\right)$ & $\left[ C_r, \omega^2 \right]$ & $\left[ C_r, \omega^3 \right]$ & $\left[ C_{tr}, Id \right]$ & $\left[ C_{tr}, a \right]$ & $\left[ C_{tr}, b \right]$ & $\left[ C_{tr}, ab \right]$ & $\left[ C_{r^2}, Id \right]$ & $\left[ C_{r^2}, \mathcal{T} \right] $ & $\left[ C_{r^2}, \mathcal{R} \right] $ & $\left[ C_{r^2}, \mathcal{RT} \right] $ & $\left[ C_{r^2}, \mathcal{B} \right] $
        \\[5pt]

	\end{tabular}
\end{adjustbox}
    \par\smallskip\hrule\smallskip
\end{table*}

We establish that the $\mathbb{D}_4$ quantum double model matches the  $BF$ field theory with an $AAB$ twist and gauge group $G=(\mathbb{Z}_2)^3$,  by constructing an explicit isomorphism between their excitations. A direct comparison of the fusion and shrinking tables of these two theories confirms that an isomorphism $f:\mathsf{a}\mapsto\left[C, R\right]$ preserves the fusion and shrinking structures,  satisfying
\begin{align}
    f\left(\mathsf{a}\right)\otimes f\left(\mathsf{b}\right)&=f\left(\mathsf{a}\otimes\mathsf{b}\right)\, , \label{eq_d4aab_fusion} \\
    f\left(\mathcal{S}\left(\mathsf{a}\right)\right)&=\mathcal{S}\left(f\left(\mathsf{a}\right)\right)\, , \label{eq_d4aab_shrinking}
\end{align}
where $\mathsf{a}$ and $\mathsf{b}$ denote excitations in the  $BF$ field theory with an $AAB$ twist and gauge group $G=(\mathbb{Z}_2)^3$,  and $\left[C, R\right]$ denotes an excitation in the $\mathbb{D}_4$ quantum double model. Besides,  when the particle--loop braiding or the Borromean-Rings braiding only produces phases,  the isomorphism $f$ preserves the braiding phases:
\begin{align}
    \exp \left(\mathrm{i}\theta^{\mathrm{PL}}_{f\left(\mathsf{P}\right), f\left(\mathsf{L}\right)}\right)&=\exp \left(\mathrm{i}\theta^{\mathrm{PL}}_{\mathsf{P}, \mathsf{L}}\right)\, , \label{eq_d4aab_PL} \\
    \exp \left(\mathrm{i}\theta^{\mathrm{BR}}_{f\left(\mathsf{P}\right), f\left(\mathsf{L}_1\right), f\left(\mathsf{L}_2\right)}\right)&=\exp \left(\mathrm{i}\theta^{\mathrm{BR}}_{\mathsf{P}, \mathsf{L}_1, \mathsf{L}_2}\right)\, , \label{eq_d4aab_BR}
\end{align}
where $\mathsf{P}$ and $\mathsf{L}$ denote a particle and a loop in the $BF$ field theory with an $AAB$ twist and gauge group $G=(\mathbb{Z}_2)^3$ respectively.

For example,  consider the following mapping of excitations:
\begin{align}
    f\left(\mathsf{1}\right)&=\left[ Id \right]\, , \nonumber \\
    f\left(\mathsf{P}_{010}\right)&=\left[ \mathcal{T} \right]\, , \nonumber \\
    f\left(\mathsf{L}_{100}\right)&=\left[ C_r, Id \right]\, , \nonumber \\
    f\left(\mathsf{P}_{001}\right)&=\left[ \mathcal{B} \right]\, , \nonumber \\
    f\left(\mathsf{L}_{100}^{001}\right)&=\left[ C_r, \omega \right]\, , \nonumber \\
    f\left(\mathsf{L}_{100}^{101}\right)&=\left[ C_r, \omega ^3 \right]\, , \nonumber \\
    f\left(\mathsf{L}_{010}\right)&=\left[ C_t, Id \right]\, .\label{eq_f_example}
\end{align}
By using Eqs.~(\ref{eq_134}) and~(\ref{eq_160}),  we can verify that
\begin{align}
    f\left(\mathsf{L}_{100}\right)\otimes f\left(\mathsf{P}_{001}\right)&=\left[ C_r, Id \right]\otimes \left[ \mathcal{B} \right] =\left[ C_r, \omega \right]\oplus\left[ C_r, \omega ^3 \right]
\end{align}
and
\begin{align}
    f\left(\mathsf{L}_{100}\otimes \mathsf{P}_{001}\right)&=f\left(\mathsf{L}_{100}^{001}\oplus\mathsf{L}_{100}^{101}\right) =\left[ C_r, \omega \right]\oplus\left[ C_r, \omega ^3 \right] = f\left(\mathsf{L}_{100}\right)\otimes f\left(\mathsf{P}_{001}\right)\, .
\end{align}
The isomorphism $f$ preserves the fusion rules. As for the shrinking rules,  we have
\begin{align}
    f\left(\mathcal{S}\left(\mathsf{L}_{100}\right)\right)&=f\left(\mathsf{1}\oplus\mathsf{P}_{010}\right)=\left[ Id \right]\oplus\left[ \mathcal{T} \right]
\end{align}
and
\begin{align}
    \mathcal{S}\left(f\left(\mathsf{L}_{100}\right)\right)&=\mathcal{S}\left(\left[ C_r, Id \right]\right)=\left[ Id \right]\oplus\left[ \mathcal{T} \right]=f\left(\mathcal{S}\left(\mathsf{L}_{100}\right)\right)\, ,
\end{align}
where we have used Eqs.~(\ref{eq_135}) and~(\ref{eq_176}). The isomorphism $f$ preserves the shrinking rules. Finally,  we verify that
\begin{align}
    \exp \left(\mathrm{i}\theta^{\mathrm{PL}}_{\mathsf{P}_{010}, \mathsf{L}_{010}}\right)&=\exp \left(\mathrm{i}\theta^{\mathrm{PL}}_{\left(\mathcal{T}\right), \left( C_t, Id \right)}\right)=-1\, , \\
    \exp \left(\mathrm{i}\theta^{\mathrm{BR}}_{\mathsf{P}_{001}, \mathsf{L}_{100},  \mathsf{L}_{010}}\right)&=\exp \left(\mathrm{i}\theta^{\mathrm{BR}}_{\left(\mathcal{B}\right), \left( C_r, Id \right), \left( C_t, Id \right)}\right)=-1\, .
\end{align}
The isomorphism $f$ preserves the braiding phases.

We present the explicit form of this isomorphism $f$ in Table~\ref{tab_f}. Under the isomorphism $f$,  the gauge fluxes that are minimally coupled to 2-form fields $B^1$,  $B^2$ and $B^3$ are related to the conjugacy classes $C_r$,  $C_t$ and $C_{r^2}$ respectively. The gauge charges that are minimally coupled to 1-form fields $A^1$,  $A^2$ and $A^3$ are related to the irreps $\mathcal{R}$,  $\mathcal{T}$,  and $\mathcal{B}$ respectively.

Note that the choice of isomorphism from the  $BF$ field theory with an $AAB$ twist and gauge group $G=(\mathbb{Z}_2)^3$ to the $\mathbb{D}_4$ quantum double model is not unique. An alternative isomorphism $\tilde{f}:\mathsf{a}\mapsto\left[C, R\right]$ can be constructed by $\tilde{f}=f\circ F$,  where $F$ permutes excitations in the  $BF$ field theory with an $AAB$ twist and gauge group $G=(\mathbb{Z}_2)^3$ as 
\begin{align}
    F\left(\mathsf{P}_{n_1n_2n_3}\right)&=\mathsf{P}_{n_2n_1n_3}\, , \nonumber \\
    F\left(\mathsf{L}_{n_1n_2n_3}^{m_1m_2m_3}\right)&=\mathsf{L}_{n_2n_1n_3}^{m_2m_1m_3}\, .
\end{align}
Under the isomorphism $\tilde{f}$,  the examples shown in Eq.~(\ref{eq_f_example}) now become
\begin{align}
    \tilde{f}\left(\mathsf{1}\right)&=f\left(\mathsf{1}\right)=\left[ Id \right]\, , \nonumber \\
    \tilde{f}\left(\mathsf{P}_{010}\right)&=f\left(\mathsf{P}_{100}\right)=\left[ \mathcal{R} \right]\, , \nonumber \\
    \tilde{f}\left(\mathsf{L}_{100}\right)&=f\left(\mathsf{L}_{010}\right)=\left[ C_t, Id \right]\, , \nonumber \\
    \tilde{f}\left(\mathsf{P}_{001}\right)&=f\left(\mathsf{P}_{001}\right)=\left[ \mathcal{B} \right]\, , \nonumber \\
    \tilde{f}\left(\mathsf{L}_{100}^{001}\right)&=f\left(\mathsf{L}_{010}^{001}\right)=\left[ C_t, a \right]\, , \nonumber \\
    \tilde{f}\left(\mathsf{L}_{100}^{101}\right)&=f\left(\mathsf{L}_{010}^{011}\right)=\left[ C_t, ab \right]\, , \nonumber \\
    \tilde{f}\left(\mathsf{L}_{010}\right)&=f\left(\mathsf{L}_{100}\right)=\left[ C_r, Id \right]\, ,
\end{align}
where we have used Table~\ref{tab_f} for the isomorphism $f$. From Tables~\ref{tab_D4_fusion1} and~\ref{tab_D4_shrinking},  we obtain the following fusion and shrinking rules
\begin{align}
    \left[ C_t, Id \right]\otimes\left[ \mathcal{B} \right]&=\left[ C_t, a \right]\oplus\left[ C_t, ab \right]\, , \\
    \mathcal{S}\left(\left[ C_t, Id \right]\right)&=\left[ Id \right]\oplus \left[ \mathcal{R} \right]\, .
\end{align}
Thus,  we can verify that the isomorphism $\tilde{f}$ still preserves the fusion rules
\begin{align}
    \tilde{f}\left(\mathsf{L}_{100}\right)\otimes \tilde{f}\left(\mathsf{P}_{001}\right)&=\left[ C_t, Id \right]\otimes \left[ \mathcal{B} \right] =\left[ C_t, a \right]\oplus\left[ C_t, ab \right]\, , \\
    \tilde{f}\left(\mathsf{L}_{100}\otimes \mathsf{P}_{001}\right)&=\tilde{f}\left(\mathsf{L}_{100}^{001}\oplus\mathsf{L}_{100}^{101}\right) =\left[ C_t, a \right]\oplus\left[ C_t, ab \right] = \tilde{f}\left(\mathsf{L}_{100}\right)\otimes \tilde{f}\left(\mathsf{P}_{001}\right)\, ,
\end{align}
and the shrinking rules
\begin{align}
    \tilde{f}\left(\mathcal{S}\left(\mathsf{L}_{100}\right)\right)&=\tilde{f}\left(\mathsf{1}\oplus\mathsf{P}_{010}\right)=\left[ Id \right]\oplus\left[ \mathcal{R} \right]\, , \\
    \mathcal{S}\left(\tilde{f}\left(\mathsf{L}_{100}\right)\right)&=\mathcal{S}\left(\left[ C_t, Id \right]\right)=\left[ Id \right]\oplus\left[ \mathcal{R} \right]=\tilde{f}\left(\mathcal{S}\left(\mathsf{L}_{100}\right)\right)\, ,
\end{align}
and the braiding phases
\begin{align}
    \exp \left(\mathrm{i}\theta^{\mathrm{PL}}_{\mathsf{P}_{010}, \mathsf{L}_{010}}\right)&=\exp \left(\mathrm{i}\theta^{\mathrm{PL}}_{\left(\mathcal{R}\right), \left( C_r, Id \right)}\right)=-1\, , \\
    \exp \left(\mathrm{i}\theta^{\mathrm{BR}}_{\mathsf{P}_{001}, \mathsf{L}_{100},  \mathsf{L}_{010}}\right)&=\exp \left(\mathrm{i}\theta^{\mathrm{BR}}_{\left(\mathcal{B}\right), \left( C_t, Id \right), \left( C_r, Id \right)}\right)=-1\, .
\end{align}
The explicit form of the isomorphism $\tilde{f}$ is shown in Table~\ref{tab_tilde_f}. The isomorphism $\tilde{f}$ relates the gauge fluxes that are minimally coupled to 2-form fields $B^1$,  $B^2$ and $B^3$ to conjugacy classes $C_t$,  $C_r$ and $C_{r^2}$ respectively. Gauge charges that are minimally coupled to 1-form fields $A^1$,  $A^2$ and $A^3$ are related to the irreps $\mathcal{T}$,  $\mathcal{R}$,  and $\mathcal{B}$ respectively.

Eqs.~(\ref{eq_d4aab_fusion}), ~(\ref{eq_d4aab_shrinking}), ~(\ref{eq_d4aab_PL}) and~(\ref{eq_d4aab_BR}) demonstrate that the $\mathbb{D}_4$ quantum double model and the  $BF$ field theory with an $AAB$ twist and gauge group $G=(\mathbb{Z}_2)^3$ share identical fusion and shrinking rules and braiding phases. The isomorphism $f$ establishes the $\mathbb{D}_4$ quantum double model as the microscopic construction of the $BF$ field theory with an $AAB$ twist and gauge group $G = (\mathbb{Z}_2)^3$. This correspondence bridges two complementary approaches to topological order: the long-wavelength continuum field theory and the microscopic lattice model. The agreement of their topological data not only provides strong support for the field-theoretic derivation of the fusion rules,  shrinking rules,  and braiding statistics,  but also reveals how these rules are encoded in the microscopic degrees of freedom of a local lattice Hamiltonian. Finally,  we summarize the comparison between field theory and the microscopic lattice approach in Table~\ref{tab_compare}.

\section{Summary and outlook}\label{sec7} 

\subsection{Summary}

Understanding higher-dimensional non-Abelian topological orders requires a framework that connects continuum topological field theories with explicit microscopic constructions. 
In this work, we establish such a connection in three dimensions by constructing microscopic lattice operators for particle and loop excitations and by using them to realize fusion, shrinking, particle--loop braiding, and Borromean-Rings braiding. 
Together with the microscopic verification of fusion--shrinking consistency, these results place several field-theoretical principles of three-dimensional topological order on a concrete lattice footing.

Concretely, we construct lattice creation operators for all topological excitations in the three-dimensional quantum double model and establish the corresponding connecting rules, which allow excitations to be transported and deformed on the lattice. 
These operators provide a common microscopic platform for implementing the basic topological operations studied in this work. 
The topological type of an excitation is specified by the pair $(C,R)$, where $C$ labels the flux sector and $R$ labels the charge sector, while the remaining internal labels can be changed by local operations.

Using this operator formalism, we derive the microscopic fusion rules. 
Fusion is realized by bringing excitations together and decomposing the tensor product of their local Hilbert spaces, with the resulting direct-sum decomposition defining the fusion coefficients. 
This gives a lattice realization of the fusion algebra expected from the continuum description.

We then construct microscopic shrinking processes for loop excitations. 
Shrinking is realized by contracting thickened membrane operators into lower-dimensional operators, and the shrinking rules are obtained from the corresponding decomposition of the local Hilbert space after shrinking.

Our explicit lattice construction also reveals a microscopic mechanism for controlling non-Abelian shrinking: by selecting the internal degrees of freedom of non-Abelian loop operators, one can selectively realize specific shrinking channels and obtain definite particle outcomes after shrinking. 
This shows that non-Abelian shrinking is not merely a formal decomposition rule, but an operator-level process whose individual channels can be accessed and controlled microscopically.

Moreover, by analyzing the fusion and shrinking tables of the quantum double model, we verify that shrinking rules are fully consistent with fusion rules: performing fusion followed by shrinking gives the same result as shrinking first and then fusing. 
This fusion--shrinking consistency condition, previously derived in continuum field-theoretical analyses, is therefore realized directly at the lattice-operator level.

We further implement particle--loop braiding and Borromean-Rings braiding microscopically by intertwining particle string operators with loop membrane operators. 
In general, both particle--loop braiding and Borromean-Rings braiding give rise to nontrivial operators acting on the internal degrees of freedom of the participating excitations. 
When either the charge carried by the particle or the flux carried by the loop is Abelian, the corresponding braiding operator reduces to a phase. 
For Borromean-Rings braiding, a nontrivial operator appears only when the charge carried by the particle and the two fluxes carried by the two loops are all non-Abelian; in special cases, including the $\mathbb{D}_4$ and $\mathbb{Q}_8$ quantum double models, this operator further reduces to a nontrivial phase.

We apply this operator formalism to the three-dimensional $\mathbb{D}_3$ and $\mathbb{D}_4$ quantum double models, obtaining their fusion rules, shrinking rules, particle--loop braiding, and Borromean-Rings braiding data. 
The $\mathbb{D}_4$ model plays a central role in this work. 
Although the continuum theory is formulated as a $(3+1)$D $BF$ field theory with an $AAB$ twist and Abelian gauge group $G=(\mathbb{Z}_2)^3$, its excitation spectrum and its fusion, shrinking, and braiding structures are non-Abelian and are isomorphic to those of the three-dimensional $\mathbb{D}_4$ quantum double model. 
In particular, the isomorphism preserves not only the excitation labels, but also the full set of topological operations analyzed in this work: fusion, shrinking, particle--loop braiding, and Borromean-Rings braiding. 
This exact correspondence establishes the $\mathbb{D}_4$ quantum double model as a concrete microscopic realization of the continuum topological field theory originally introduced in connection with Borromean-Rings braiding~\cite{PhysRevLett.121.061601}. 
It also resolves the question of whether the exotic $BF{+}AAB$ field theory is physically relevant and microscopically realizable by a local lattice model with a tensor-product local Hilbert space and short-range interactions.

Finally, our microscopic results are consistent with the diagrammatic framework developed in Ref.~\cite{Huang_2025}, where fusion and shrinking processes are represented by diagrams and constrained by algebraic relations such as fusion pentagon equations and (hierarchical) shrinking--fusion hexagon equations. 
In particular, the lattice realization of fusion--shrinking consistency shows that the consistency relation underlying this framework can be implemented and verified in an explicit microscopic model. 
This provides a concrete setting for using such algebraic relations as consistency conditions for anomaly-free higher-dimensional topological data, and may offer a computationally efficient route toward systematic categorical classifications~\cite{PhysRevB.71.045110, cat1, delcamp2018gauge, kong1, delcamp20192, hank23, debray2025buildanomalous31dtopological, kong2014braidedfusioncategoriesgravitational}.

\subsection{Outlook}

Several promising research directions emerge from this work. 
First, the interplay among braiding, fusion, and shrinking suggests that higher-dimensional topological orders require a richer set of topological data than their two-dimensional counterparts. 
While the present work establishes and microscopically verifies the consistency between fusion and shrinking, it remains an open question whether analogous consistency relations arise when braiding statistics are incorporated. 
For example, one may ask how particle--loop braiding, multi-loop braiding, and Borromean-Rings braiding are constrained by fusion and shrinking in three-dimensional quantum double models and in the Hamiltonian formulation of $(3+1)$D Dijkgraaf--Witten theory~\cite{PhysRevB.92.045101, PhysRevB.110.035117}. 
Constructing and analyzing combined braiding--fusion--shrinking processes may reveal deeper topological consistency conditions beyond those studied here.

Motivated by the exact correspondence between the $\mathbb{D}_4$ quantum double model and the $BF$ field theory with an $AAB$ twist and gauge group $G=(\mathbb{Z}_2)^3$, it is natural to ask whether this relation reflects a more general algebraic mechanism. 
In the continuum theory, the $AAB$ twist involves two $1$-form gauge fields $A$ and one $2$-form gauge field $B$. 
This structure suggests a decomposition into a $(\mathbb{Z}_2)^2$ sector associated with the two $A$ fields and a $\mathbb{Z}_2$ sector associated with the $B$ field. 
Developing this connection systematically could provide a route to identifying microscopic lattice models for more general $BF$ field theories with $AAB$ twists and gauge groups 
$\mathbb{Z}_{N_1}\times\mathbb{Z}_{N_2}\times\mathbb{Z}_{N_3}$. Interestingly, the $\mathbb{Q}_8$ quantum double model also exhibits nontrivial Borromean-Rings braiding, making it worthwhile to investigate its corresponding twisted $BF$ field theory. Such a study may further deepen our understanding of the mechanisms underlying nontrivial Borromean-Rings braiding in continuum field theory.

As all topological orders considered in this work are bosonic, an important future direction is to extend the microscopic construction to three-dimensional topological orders with fermionic excitations. 
Continuum descriptions of such phases include $BF$ theories with a $BB$ twist~\cite{PhysRevB.99.235137, bti2, Kapustin:2014gua, PhysRevResearch.5.043111}. 
More generally, fermionic topological orders can be constructed and classified by gauging fermionic SPT phases~\cite{PhysRevX.8.011055, 2025arXiv251225069N, PhysRevX.10.031055, PhysRevX.15.031029, PhysRevB.93.075135, Kapustin:2015aa, PhysRevX.8.011054, PhysRevB.111.205102, PhysRevB.90.115141, Zhou:2021aa, PhysRevResearch.3.013056}. 
Developing explicit lattice operators for fermionic particles, loops, and their fusion, shrinking, and braiding processes would extend the correspondence between microscopic constructions and continuum topological field theories established here beyond bosonic topological orders. 
Fermionic excitations may also change the finite-temperature stability of topological order. 
Recent work has shown that thermal states of the three-dimensional fermionic toric code remain long-range entangled below a finite temperature~\cite{zhou2025finite}. 
Such finite-temperature topological order in fewer than four spatial dimensions has been linked to anomalous two-form symmetries associated with emergent fermionic particle excitations. 
It would be interesting to understand these phenomena from both microscopic lattice models and continuum field theories, and to explore whether the diagrammatic framework of fusion, shrinking, and braiding can be extended to include fermionic structures.

Like conventional topological orders, fracton ordered phases exhibit long-range entanglement, but they are non-liquid phases whose universal properties are not captured by ordinary topological quantum field theories. 
Instead, they require unconventional continuum descriptions involving, for example, higher-rank gauge fields, subsystem structures, or other nonstandard field-theoretical ingredients~\cite{PhysRevB.92.235136, PhysRevB.105.195124, PhysRevLett.126.101603, li2026infinitecomponentbffieldtheory, PhysRevB.110.205108, PhysRevB.101.245134, PhysRevB.104.235127, PhysRevB.107.115169}. 
Given the bridge established here between microscopic constructions and continuum field theories of three-dimensional topological orders, it would be natural to ask whether analogous operator-level dictionaries can be constructed for fracton phases. 
In particular, one may seek explicit correspondences between lattice fracton models, their continuum field-theoretical descriptions, and the associated fusion, mobility, and dimensional-reduction structures of fractonic excitations.

Finally, these theoretical advances point toward possible applications to topological quantum computation in three dimensions. 
In two dimensions, topological quantum computation is based on the creation, fusion, and braiding of anyons, and recent experiments on programmable quantum processors based on trapped ions and superconducting qubits have demonstrated the preparation of non-Abelian topologically ordered states together with controlled anyon fusion and braiding~\cite{iqbal2024non, xu2024non, google2023non, PRXQuantum.3.040315, Minev:2025aa}. 
Recent progress using sequential quantum circuits~\cite{PhysRevLett.95.110503, PhysRevA.75.032311, PhysRevA.77.052306, PRXQuantum.3.040315, PhysRevLett.128.010607, PRXQuantum.2.010342, PhysRevB.109.075116, d8gs-fnwt} has further enabled concrete protocols for implementing ribbon operators of non-Abelian anyons~\cite{lyons2025protocolscreatinganyonsdefects}, providing a pathway toward universal topological quantum computation in two-dimensional quantum double models~\cite{lo2025universalquantumcomputations3}. 
The operator-level framework developed here suggests a natural three-dimensional extension of this program. 
One may seek quantum circuits, measurement protocols, or higher-order cellular automata~\cite{zhang2025programmableanyonmobilityhigher, PRXQuantum.5.030342} that create, transport, fuse, shrink, and braid particles, loops, and possibly membrane excitations in higher-dimensional topological orders. 
In particular, non-Abelian shrinking introduces a topological operation without a direct two-dimensional anyon analogue, and its controllable channels may provide new resources for manipulating topological information. 
Understanding the computational power of such fusion--shrinking--braiding operations remains an intriguing direction for future work.

\acknowledgments 
This work was supported in part by the National Natural Science Foundation of China (NSFC)  under Grant Nos. 12474149 and 12274250, the Research Center for Magnetoelectric Physics of Guangdong Province under Grant No. 2024B0303390001, and the Guangdong Provincial Key Laboratory of Magnetoelectric Physics and Devices under Grant No. 2022B1212010008.

\appendix
\section{Diagrammatics, pentagon equations, and hexagon equations}\label{ap3}
In this appendix,  we introduce unitary $F$- and $\Delta$-symbols to transform diagrams. They satisfy a series of stringent algebraic equations known as the pentagon equation and the shrinking--fusion hexagon equation~\cite{Huang_2025}. Any violation of these equations indicates a quantum anomaly in 3D topological orders.

\begin{figure}
	\centering
	\includegraphics[scale=0.5, keepaspectratio,max width=\linewidth]{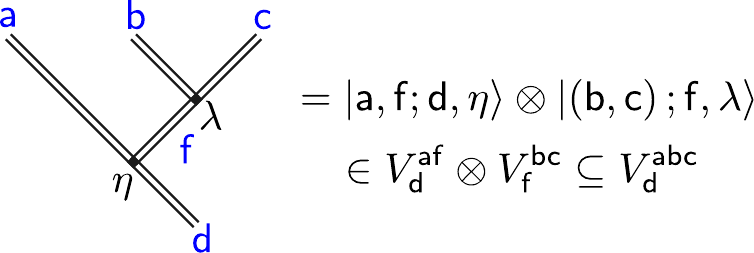}
	\caption{\textbf{An alternative diagram for fusing three excitations in 3D.} The associativity of fusion rules guarantees that this diagram describes the same physics as the diagram shown in Fig.~\ref{fig_3fusion4d_1}. The only difference between them is just a change of basis.}
	\label{fig_3fusion4d_2}
\end{figure}

\begin{figure}
	\centering
	\includegraphics[scale=0.5, keepaspectratio,max width=\linewidth]{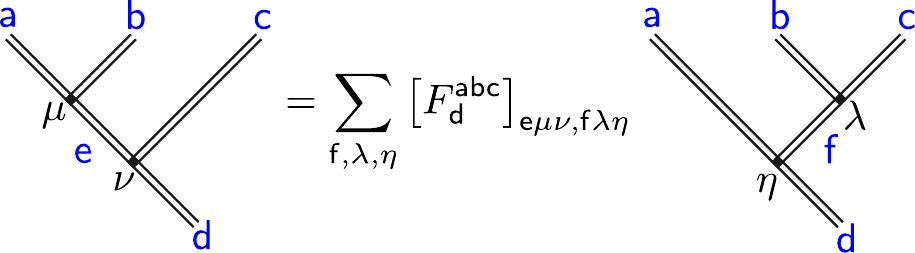}
	\caption{\textbf{Definition of the $F$-symbol in 3D.} The left and right diagrams represent different bases for the same space $V_{\mathsf{d}}^{\mathsf{abc}}$,  and we use the unitary $F$-symbol to change the basis. Transforming these diagrams into each other does not change any physics.}
	\label{fig_Fmatrix4d_2}
\end{figure}

In Sec.~\ref{sec2.2},  we illustrate how to construct more complicated diagrams by stacking basic diagrams. Besides Fig.~\ref{fig_3fusion4d_1},  we can also construct the diagram that describes the fusion of three excitations as shown in Fig.~\ref{fig_3fusion4d_2}. The difference between the diagrams shown in Figs.~\ref{fig_3fusion4d_1} and~\ref{fig_3fusion4d_2} is the order of the fusion process. Physically,  fusion rules satisfy associativity,  i.e.,  fusing $\mathsf{b}$ and $\mathsf{c}$ first should give the same final output as fusing $\mathsf{a}$ and $\mathsf{b}$ first. Therefore,  the diagram shown in Fig.~\ref{fig_3fusion4d_2} also represents a set of basis vectors that span the same space $V_{\mathsf{d}}^{\mathsf{abc}}$. The two different bases represented by Fig.~\ref{fig_3fusion4d_1} and Fig.~\ref{fig_3fusion4d_2} can be transformed into each other by a unitary matrix known as the $F$-symbol,  which is defined in Fig.~\ref{fig_Fmatrix4d_2}. We can express the equation in Fig.~\ref{fig_Fmatrix4d_2} as follows:
\begin{align}
    \ket{\left( \mathsf{a}, \mathsf{b} \right) ;\mathsf{e}, \mu} \otimes \ket{\mathsf{e}, \mathsf{c};\mathsf{d}, \nu} &=\sum_{\mathsf{f}, \lambda , \eta}{\left[ F_{\mathsf{d}}^{\mathsf{abc}} \right] _{\mathsf{e}\mu \nu , \mathsf{f}\lambda \eta}}\ket{\mathsf{a}, \mathsf{f};\mathsf{d}, \eta} \otimes \ket{\left( \mathsf{b}, \mathsf{c} \right) ;\mathsf{f}, \lambda}\, .
\end{align}
The summation over $\mathsf{f}$ exhausts all excitations in the set $\Phi _{0}^{3+1}$,  $\lambda =\left\{ 1, 2, \cdots , N_{\mathsf{f}}^{\mathsf{bc}} \right\} $ and $\eta=\left\{ 1, 2, \cdots , N_{\mathsf{d}}^{\mathsf{af}} \right\}$. Since the $F$-symbol is unitary,  we have
\begin{gather}
	\sum_{\mathsf{f}, \lambda,  \eta}{\left[ F_{\mathsf{d}}^{\mathsf{abc}} \right] _{\mathsf{e}\mu \nu , \mathsf{f}\lambda \eta}}\left( \left[ F_{\mathsf{d}}^{\mathsf{abc}} \right] _{\mathsf{e}^{\prime}\mu^{\prime}\nu ^{\prime}, \mathsf{f}\lambda \eta} \right) ^{\ast}=\delta _{\mathsf{ee}^{\prime}}\delta _{\mu \mu ^{\prime}}\delta _{\nu \nu ^{\prime}}\, 
\end{gather}
which allows us to implement the inverse transformation. Since the total space $V_{\mathsf{d}}^{\mathsf{abc}}$ is isomorphic to $\oplus _{\mathsf{f}}V_{\mathsf{d}}^{\mathsf{af}}\otimes V_{\mathsf{f}}^{\mathsf{bc}}$ in Fig.~\ref{fig_3fusion4d_2},  the dimension of the total space is given by
\begin{align}
    \text{dim}(V_{\mathsf{d}}^{\mathsf{a}\mathsf{b}\mathsf{c}})=\sum_{\mathsf{f}}{N_{\mathsf{d}}^{\mathsf{af}}}N_{\mathsf{f}}^{\mathsf{bc}}\, .\label{eq_dim2}
\end{align}
By comparing Eq.~(\ref{eq_dim1}) and Eq.~(\ref{eq_dim2}),  we obtain a constraint on fusion coefficients:
\begin{align}
    \sum_{\mathsf{e}}{N_{\mathsf{e}}^{\mathsf{ab}}}N_{\mathsf{d}}^{\mathsf{ec}}=\sum_{\mathsf{f}}{N_{\mathsf{d}}^{\mathsf{af}}}N_{\mathsf{f}}^{\mathsf{bc}}\, .\label{eq_91}
\end{align}

\begin{figure*}
	\centering
	\includegraphics[scale=0.55, keepaspectratio,max width=\linewidth]{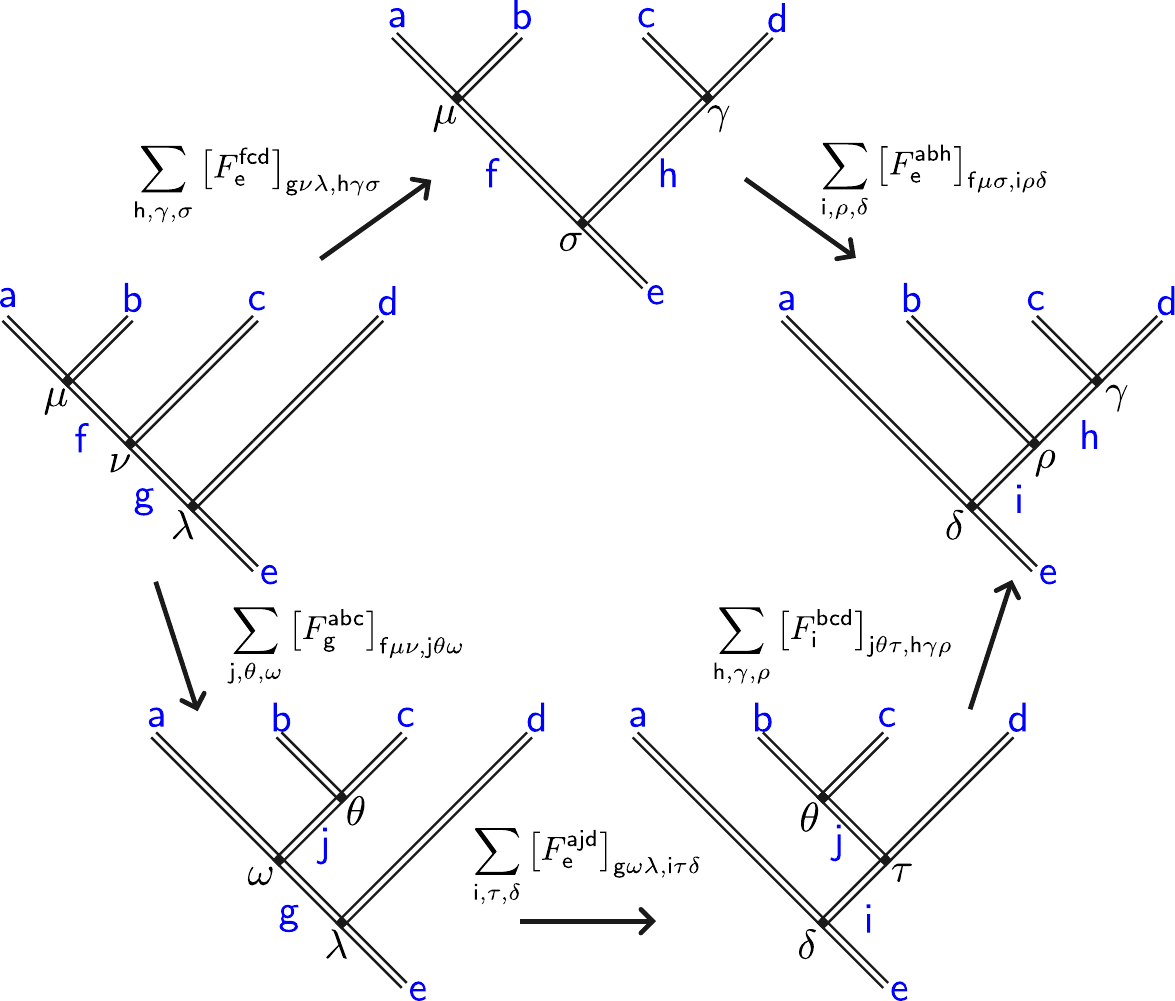}
	\caption{\textbf{Diagrammatic representation of the pentagon equation (\ref{3eq_pentagon}).} Starting from the leftmost diagram,  we can go to the rightmost diagram through either the upper path or the lower path. Comparing these two paths,  we derive the pentagon equation.}
	\label{fig_pentagon4d}
\end{figure*}

We employ a similar tensor product approach to construct fusion diagrams that involve more excitations. Applying the $F$-symbols inside these diagrams leads to a basis transformation. For example, when fusing four excitations, we have two different ways to transform the leftmost diagram to the rightmost diagram as shown in Fig.~\ref{fig_pentagon4d},  which gives a very strong algebraic constraint on the $F$-symbols,  known as the pentagon equation:
\begin{align}
    &\sum_{\sigma =1}^{N_{\mathsf{e}}^{\mathsf{fh}}}{\left[ F_{\mathsf{e}}^{\mathsf{fcd}} \right] _{\mathsf{g}\nu \lambda , \mathsf{h} \gamma \sigma}}\left[ F_{\mathsf{e}}^{\mathsf{abh}} \right] _{\mathsf{f}\mu \sigma , \mathsf{i} \rho \delta}\nonumber \\
    &=\sum_{\mathsf{j}}{\sum_{\omega =1}^{N_{\mathsf{g}}^{\mathsf{aj}}}{\sum_{\theta =1}^{N_{\mathsf{j}}^{\mathsf{bc}}}{\sum_{\tau =1}^{N_{\mathsf{i}}^{\mathsf{jd}}}{\left[ F_{\mathsf{g}}^{\mathsf{abc}} \right] _{\mathsf{f}\mu \nu , \mathsf{j} \theta \omega}\left[ F_{\mathsf{e}}^{\mathsf{ajd}} \right] _{\mathsf{g}\omega \lambda , \mathsf{i} \tau \delta}\left[ F_{\mathsf{i}}^{\mathsf{bcd}} \right] _{\mathsf{j}\theta \tau , \mathsf{h} \gamma \rho}}}}}.
	\label{3eq_pentagon}
\end{align}
A similar pentagon equation can also be derived in the diagrammatic representations of 2D anyons. Actually,  if we draw all previous fusion diagrams in a single-line form,  our fusion diagrams automatically reduce to 2D anyonic fusion diagrams. Considering fusion diagrams for more excitations does not give any independent constraint beyond the pentagon equation. 

\begin{figure}
	\centering
	\includegraphics[scale=0.5, keepaspectratio,max width=\linewidth]{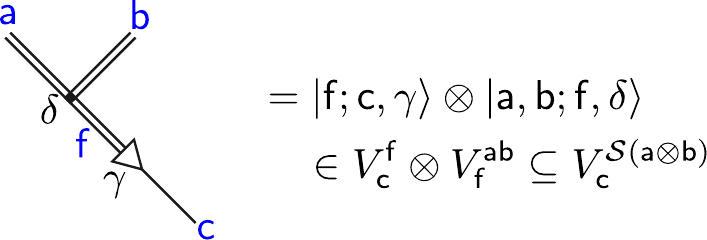}
	\caption{\textbf{The diagram describing $\mathcal{S} \left( \mathsf{a}\otimes \mathsf{b} \right) $ in 3D.} This diagram is constructed by stacking a basic fusion diagram and a basic shrinking diagram. The corresponding vector is obtained by tensor product construction. 
	}
	\label{fig_shrinking_fusion}
\end{figure}

Now we construct diagrams that incorporate both fusion and shrinking processes. For $\mathcal{S} \left( \mathsf{a} \right) \otimes \mathcal{S} \left( \mathsf{b} \right) $ and $\mathcal{S} \left( \mathsf{a}\otimes \mathsf{b} \right) $, the corresponding diagrams are shown in Figs.~\ref{fig_sa_times_sb} and~\ref{fig_shrinking_fusion} respectively. Since we have the consistency condition in Eq.~(\ref{eq_consistent_4D}),  these two processes have the same final output and we consider them as different bases in the same space. The diagram in Fig.~\ref{fig_sa_times_sb} is defined as $\ket{\mathsf{d}, \mathsf{e};\mathsf{c}, \lambda}\otimes\ket{\mathsf{b};\mathsf{e}, \nu}\otimes\ket{\mathsf{a};\mathsf{d}, \mu}$,  where different $\mathsf{d}$,  $\mathsf{e}$,  $\mu$,  $\nu$,  and $\lambda$ label different orthogonal vectors. This set of orthogonal vectors spans the space denoted by $V_{\mathsf{c}}^{\mathcal{S} \left( \mathsf{a} \right) \otimes \mathcal{S} \left( \mathsf{b} \right)}$,  which is isomorphic to $\oplus _{\mathsf{d}, \mathsf{e}}V_{\mathsf{c}}^{\mathsf{de}}\otimes V_{\mathsf{e}}^{\mathsf{b}}\otimes V_{\mathsf{d}}^{\mathsf{a}}$ due to our tensor product construction. Similarly,  the diagram shown in Fig.~\ref{fig_shrinking_fusion} is defined as $\ket{\mathsf{f};\mathsf{c}, \gamma}\otimes\ket{\mathsf{a}, \mathsf{b};\mathsf{f}, \delta}$,  where $\mathsf{f}$,  $\delta$,  and $\gamma$ label different orthogonal vectors. The corresponding space $V_{\mathsf{c}}^{\mathcal{S} \left( \mathsf{a}\otimes \mathsf{b} \right)}=V_{\mathsf{c}}^{\mathcal{S} \left( \mathsf{a} \right) \otimes \mathcal{S} \left( \mathsf{b} \right)}$ is isomorphic to $\oplus _{\mathsf{f}}V_{\mathsf{c}}^{\mathsf{f}}\otimes V_{\mathsf{f}}^{\mathsf{ab}}$,  which recovers the constraint in Eq.~(\ref{eq_84}):
\begin{align}
    \mathrm{dim}\left( V_{\mathsf{c}}^{\mathcal{S} \left( \mathsf{a} \right) \otimes \mathcal{S} \left( \mathsf{b} \right)} \right) =\mathrm{dim}\left( V_{\mathsf{c}}^{\mathcal{S} \left( \mathsf{a}\otimes \mathsf{b} \right)} \right) &=\sum_{\mathsf{d}, \mathsf{e}}{N_{\mathsf{c}}^{\mathsf{de}}S_{\mathsf{e}}^{\mathsf{b}}S_{\mathsf{d}}^{\mathsf{a}}}=\sum_{\mathsf{f}}{S_{\mathsf{c}}^{\mathsf{f}}N_{\mathsf{f}}^{\mathsf{ab}}}\, .\label{eq_re}
\end{align}

\begin{figure}
	\centering
	\includegraphics[scale=0.5, keepaspectratio,max width=\linewidth]{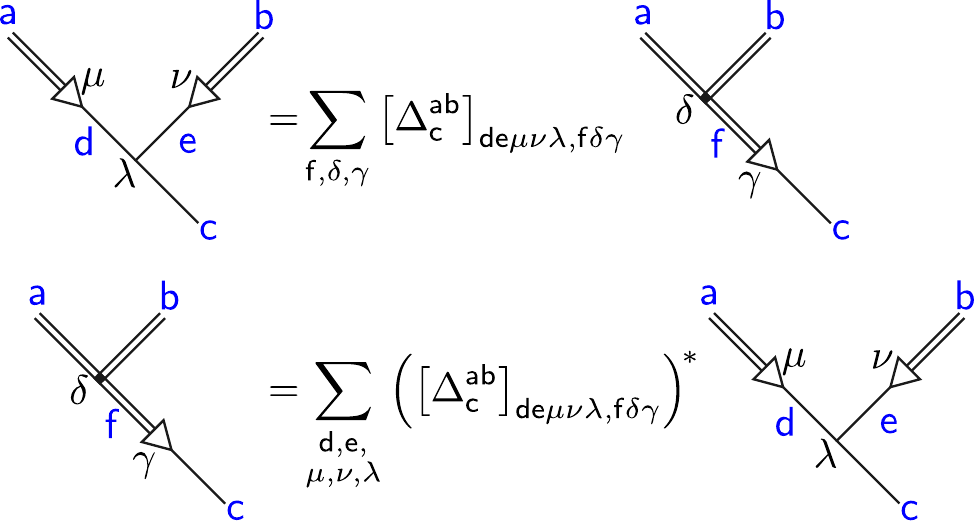}
	\caption{\textbf{Definition of the $\Delta$-symbol in 3D.} The left diagram and the right diagram describe the same physics in different bases. Like the $F$-symbol,  the $\Delta$-symbol is also unitary and we use it to change bases. }
	\label{fig_delta_matrix}
\end{figure}
The two sets of vectors shown in Fig.~\ref{fig_shrinking_fusion} correspond to two different bases of the total space $V_{\mathsf{c}}^{\mathcal{S} \left( \mathsf{a} \right) \otimes \mathcal{S} \left( \mathsf{b} \right)}=V_{\mathsf{c}}^{\mathcal{S} \left( \mathsf{a}\otimes \mathsf{b} \right)}$. We expect that these two bases can be transformed into each other by a unitary matrix,  called the $\Delta$-symbol. The definition of the $\Delta$-symbol is shown in Fig.~\ref{fig_delta_matrix}. We  explicitly write down the transformations in Fig.~\ref{fig_delta_matrix} as follows:
\begin{align}
    \ket{\mathsf{d}, \mathsf{e};\mathsf{c}, \lambda}\otimes\ket{\mathsf{b};\mathsf{e}, \nu}\otimes\ket{\mathsf{a};\mathsf{d}, \mu} &=\sum_{\mathsf{f}, \delta , \gamma}{\left[ \Delta _{\mathsf{c}}^{\mathsf{ab}} \right] _{\mathsf{de}\mu \nu \lambda , \mathsf{f}\delta \gamma}\ket{\mathsf{f};\mathsf{c}, \gamma}\otimes\ket{\mathsf{a}, \mathsf{b};\mathsf{f}, \delta}}\, , \\
    \ket{\mathsf{f};\mathsf{c}, \gamma}\otimes\ket{\mathsf{a}, \mathsf{b};\mathsf{f}, \delta} &=\sum_{\substack{\mathsf{d}, \mathsf{e}, \\ \mu , \nu , \lambda}}{\left( \left[ \Delta _{\mathsf{c}}^{\mathsf{ab}} \right] _{\mathsf{de}\mu \nu \lambda , \mathsf{f}\delta \gamma } \right) ^{\ast}\ket{\mathsf{d}, \mathsf{e};\mathsf{c}, \lambda}\otimes\ket{\mathsf{b};\mathsf{e}, \nu}\otimes\ket{\mathsf{a};\mathsf{d}, \mu}}\, .
\end{align}
Unitarity of the $\Delta$-symbol demands:
\begin{align}
    \sum_{\mathsf{f}, \delta , \gamma}{\left[ \Delta _{\mathsf{c}}^{\mathsf{ab}} \right] _{\mathsf{de}\mu \nu \lambda , \mathsf{f}\delta \gamma}\left( \left[ \Delta _{\mathsf{c}}^{\mathsf{ab}} \right] _{\mathsf{d}^{\prime}\mathsf{e}^{\prime}\mu^{\prime}\nu^{\prime}\lambda^{\prime}, \mathsf{f}\delta \gamma} \right)}^{\ast} &=\delta _{\mathsf{dd}^{\prime}}\delta _{\mathsf{ee}^{\prime}}\delta _{\mu \mu^{\prime}}\delta _{\nu \nu^{\prime}}\delta _{\lambda \lambda^{\prime}}\, , \\
    \sum_{\substack{\mathsf{d}, \mathsf{e}, \\ \mu , \nu , \lambda}}{\left[ \Delta _{\mathsf{c}}^{\mathsf{ab}} \right] _{\mathsf{de}\mu \nu \lambda , \mathsf{f}\delta \gamma}\left( \left[ \Delta _{\mathsf{c}}^{\mathsf{ab}} \right] _{\mathsf{de}\mu \nu \lambda , \mathsf{f}^{\prime}\delta ^{\prime}\gamma ^{\prime}} \right) ^{\ast}} &=\delta _{\mathsf{ff}^{\prime}}\delta _{\delta \delta ^{\prime}}\delta _{\gamma \gamma ^{\prime}}\, .
\end{align}

\begin{figure*}
	\centering
	\includegraphics[scale=0.55, keepaspectratio,max width=\linewidth]{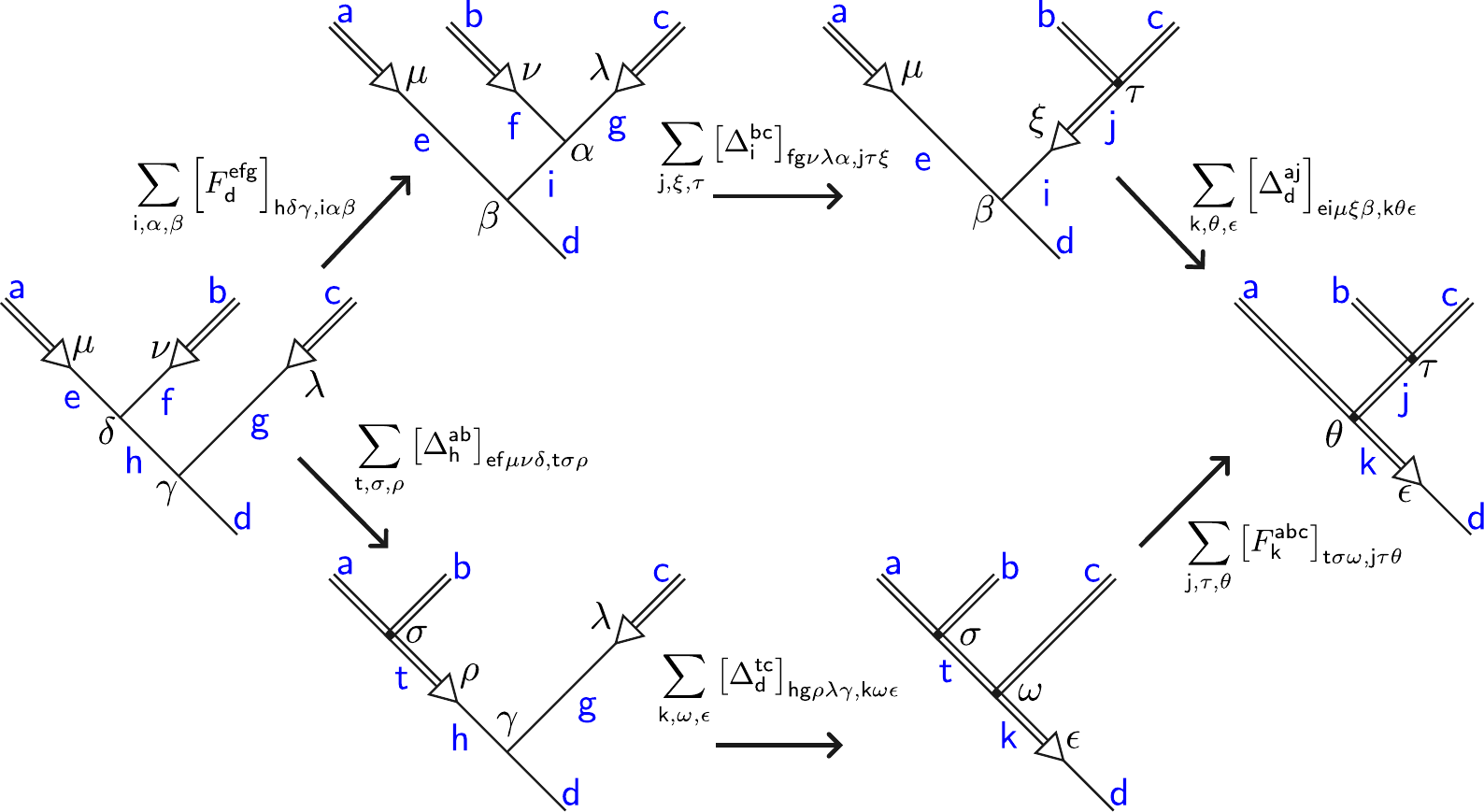}
	\caption{\textbf{Diagrammatic representations of the shrinking--fusion hexagon equation (\ref{eq_shrinking_fusion_hexagon}).} Two different paths can be constructed to transform the leftmost diagram to the rightmost diagram. Comparing these two paths,  we obtain the shrinking--fusion hexagon equation.}
	\label{fig_shrinking_fusion_hexagon}
\end{figure*}
The consistency condition between the fusion and shrinking rules further leads to a strong constraint on both $F$-symbols and $\Delta$-symbols called the shrinking--fusion hexagon equation. Consider three excitations going through fusion and shrinking processes as shown in Fig.~\ref{fig_shrinking_fusion_hexagon}. By applying the $F$-symbols and $\Delta$-symbols within the diagrams,  we find two paths to transform the leftmost diagram to the rightmost diagram. By comparing the coefficients calculated from the upper and lower paths in Fig.~\ref{fig_shrinking_fusion_hexagon},  we obtain the shrinking--fusion hexagon equation:
    \begin{align}
    &\sum_{\mathsf{i}}{\sum_{\alpha =1}^{N_{\mathsf{i}}^{\mathsf{fg}}}{\sum_{\beta =1}^{N_{\mathsf{d}}^{\mathsf{ei}}}{\sum_{\xi =1}^{S_{\mathsf{i}}^{\mathsf{j}}}{}}}}\left[ F_{\mathsf{d}}^{\mathsf{efg}} \right] _{\mathsf{h}\delta \gamma , \mathsf{i}\alpha \beta}\left[ \Delta _{\mathsf{i}}^{\mathsf{bc}} \right] _{\mathsf{fg}\nu \lambda \alpha , \mathsf{j}\tau \xi}\left[ \Delta _{\mathsf{d}}^{\mathsf{aj}} \right] _{\mathsf{ei}\mu \xi \beta , \mathsf{k}\theta \epsilon}\nonumber \\
    &=\sum_{\mathsf{t}}{\sum_{\sigma =1}^{N_{\mathsf{t}}^{\mathsf{ab}}}{\sum_{\rho =1}^{S_{\mathsf{h}}^{\mathsf{t}}}{\sum_{\omega =1}^{N_{\mathsf{k}}^{\mathsf{tc}}}{}}}}\left[ \Delta _{\mathsf{h}}^{\mathsf{ab}} \right] _{\mathsf{ef}\mu \nu \delta , \mathsf{t}\sigma \rho}\left[ \Delta _{\mathsf{d}}^{\mathsf{tc}} \right] _{\mathsf{hg}\rho \lambda \gamma , \mathsf{k}\omega \epsilon}\left[ F_{\mathsf{k}}^{\mathsf{abc}} \right] _{\mathsf{t}\sigma \omega , \mathsf{j}\tau \theta}\, .
	\label{eq_shrinking_fusion_hexagon}
\end{align}
The shrinking--fusion hexagon equation is understood as the consistency relation between $F$-symbols and $\Delta$-symbols. We can further consider applying $F$-symbols and $\Delta$-symbols to transform diagrams that involve more excitations,  such as $\mathcal{S} \left( \mathsf{a} \right) \otimes \mathcal{S} \left( \mathsf{b} \right) \otimes \mathcal{S} \left( \mathsf{c} \right) \otimes \mathcal{S} \left( \mathsf{d} \right) $. However,  no additional independent equations are obtained from these more complicated diagrams.

Since using the $F$-symbols and $\Delta$-symbols to change bases implies the constraints on fusion and shrinking coefficients (i.e.,  Eq.~(\ref{eq_84}) and Eq.~(\ref{eq_91}) hold),  we conclude that the pentagon Eq.~(\ref{3eq_pentagon}) and shrinking--fusion hexagon Eq.~(\ref{eq_shrinking_fusion_hexagon}) are stronger constraints. These equations describe not only the behavior of the fusion and shrinking coefficients but also the basis transformations.  We conjecture that the pentagon equation and shrinking--fusion hexagon equation are universal for all anomaly-free 3D topological orders. Our construction of the diagrammatics for 3D topological orders can be generalized to higher dimensions, and the details can be found in Ref.~\cite{Huang_2025}.

\section{Details of the basic operators}\label{ap2}
In this appendix,  we provide some technical details of the basic operators introduced in Sec.~\ref{Sec.3}.

\subsection{Proof of Eq.~(\ref{eq_connect_order})}\label{ap2.1}
     Consider a string $L=L_1\cup L_2\cup L_3$. If we connect $L_2$ and $L_3$ first,  by using Eq.~(\ref{eq_T_connect}),  we have
\begin{align}
    T_{L_1\cup\left( L_2\cup L_3 \right)}^{g}&=\sum_h{T_{L_2\cup L_3}^{\bar{h}g}T_{L_1}^{h}}=\sum_{h, m}{T_{L_3}^{\bar{m}\bar{h}g}T_{L_2}^{m}T_{L_1}^{h}}\, ,
\end{align}
where the parentheses $\left( L_2\cup L_3 \right)$ are introduced to emphasize that $L_2$ and $L_3$ are connected first. If we connect $L_1$ and $L_2$ first,  then we have
\begin{align}
    T_{\left( L_1\cup L_2 \right)\cup L_3}^{g}&=\sum_m{T_{L_3}^{\bar{m}g}T_{L_1\cup L_2}^{m}}=\sum_{h, m}{T_{L_3}^{\bar{m}g}T_{L_2}^{\bar{h}m}T_{L_1}^{h}}\, .
\end{align}
Rewriting $T_{L_3}^{\bar{m}g}$ as $T_{L_3}^{\bar{m}h\bar{h}g}$ and replacing the sum over $m$ with the sum over $\bar{h}m$, we have
\begin{align}
    T_{\left( L_1\cup L_2 \right)\cup L_3}^{g}&=\sum_h{\sum_{\bar{h}m}{T_{L_3}^{\overline{\left(\bar{h}m\right)}\bar{h}g}T_{L_2}^{\left(\bar{h}m\right)}T_{L_1}^{h}}}=\sum_{h, m^{\prime}}{T_{L_3}^{\bar{m}^{\prime}\bar{h}g}T_{L_2}^{m^{\prime}}T_{L_1}^{h}}=T_{L_1\cup \left( L_2\cup L_3 \right)}^{g}\, ,
\end{align}
where the second line is obtained by relabeling $\bar{h}m$ as $m^{\prime}$. Thus,  we conclude that the order of connection is irrelevant.

\subsection{Proof of Eqs.~(\ref{eq_31}), ~(\ref{eq_32}) and~(\ref{eq_33})}\label{ap2.2}
    Suppose $L=\left( v_0, v_1, \cdots , v_n \right) $ starts at the vertex $v_0$,  passes through $n$ edges denoted by $v_iv_{i+1}$ and finally ends at $v_n$. We choose the arrow of $v_iv_{i+1}$ to point from $v_i$ to $v_{i+1}$. Then we write $T_{L}^{g}$ as
\begin{align}
    T_{L}^{g}=\sum_{h_1, h_2, \cdots h_{n-1}}{T_{v_{n-1}v_n}^{\bar{h}_{n-1}g}\cdots T_{v_iv_{i+1}}^{\bar{h}_ih_{i+1}}\cdots T_{v_1v_2}^{\bar{h}_1h_2}T_{v_0v_1}^{h_1}}\, .
\end{align}
Notice that the vertex operator $A_{v_i}^{k}$ commutes with all $T_{v_jv_{j+1}}^{g}$ when $j\ne i-1, i$ because they act on different edges. Since the action of $A_{v_i}^{k}$ on the edges $v_{i-1}v_i$ and $v_iv_{i+1}$ is given by $L_{v_{i-1}v_i}^k=L_k^-$ and $L_{v_iv_{i+1}}^k=L_k^+$,  from Eq.~(\ref{eq_tl}) we can see
\begin{align}
    A_{v_i}^{k}T_{v_iv_{i+1}}^{\bar{h}_ih_{i+1}}&=T_{v_iv_{i+1}}^{k\bar{h}_ih_{i+1}}A_{v_i}^{k}\, , \label{eq_atta1} \\
    A_{v_i}^{k}T_{v_{i-1}v_i}^{\bar{h}_{i-1}h_i}&=T_{v_{i-1}v_i}^{\bar{h}_{i-1}h_i\bar{k}}A_{v_i}^{k}\, .\label{eq_atta2}
\end{align}
Thus,  we have
\begin{align}
    A_{v_i}^{k}T_{L}^{g} &=\sum_{\substack{h_1, \cdots , h_i, \\ \cdots , h_{n-1}}}{T_{v_{n-1}v_n}^{\bar{h}_{n-1}g}\cdots A_{v_i}^{k}T_{v_iv_{i+1}}^{\bar{h}_ih_{i+1}}T_{v_{i-1}v_i}^{\bar{h}_{i-1}h_i}\cdots T_{v_0v_1}^{h_1}}\nonumber \\
    &=\sum_{\substack{h_1, \cdots , h_i, \\\cdots , h_{n-1}}}{T_{v_{n-1}v_n}^{\bar{h}_{n-1}g}\cdots T_{v_iv_{i+1}}^{k\bar{h}_ih_{i+1}}T_{v_{i-1}v_i}^{\bar{h}_{i-1}h_i\bar{k}}A_{v_i}^{k}\cdots T_{v_0v_1}^{h_1}}\nonumber \\
    &=\left[ \sum_{\substack{h_1, \cdots , h_i , \\ \cdots , h_{n-1}}}{T_{v_{n-1}v_n}^{\bar{h}_{n-1}g}\cdots T_{v_iv_{i+1}}^{k\bar{h}_ih_{i+1}}T_{v_{i-1}v_i}^{\bar{h}_{i-1}h_i\bar{k}}\cdots T_{v_0v_1}^{h_1}} \right] A_{v_i}^{k}\,.
\end{align}
By replacing $\sum_{h_i}$ with $\sum_{h_i\bar{k}}$, we obtain
\begin{align}
    A_{v_i}^{k}T_{L}^{g} &=\left[ \sum_{\substack{h_1, \cdots , \left( h_i\bar{k} \right) , \\ \cdots , h_{n-1}}}{T_{v_{n-1}v_n}^{\bar{h}_{n-1}g}\cdots T_{v_iv_{i+1}}^{k\bar{h}_ih_{i+1}}T_{v_{i-1}v_i}^{\bar{h}_{i-1}h_i\bar{k}}\cdots T_{v_0v_1}^{h_1}} \right] A_{v_i}^{k} =T_{L}^{g}A_{v_i}^{k}
\end{align}
where $i\ne0, n$, thus proving Eq.~(\ref{eq_31}).

When $i=0$ or $i=n$,  similar to Eqs.~(\ref{eq_atta1}) and~(\ref{eq_atta2}),  we obtain
\begin{align}
    A_{v_0}^{k}T_{v_0v_1}^{h_1}&=T_{v_0v_{1}}^{kh_1}A_{v_0}^{k}\, , \\
    A_{v_n}^{k}T_{v_{n-1}v_n}^{\bar{h}_{n-1}g}&=T_{v_{n-1}v_n}^{\bar{h}_{n-1}g\bar{k}}A_{v_n}^{k}\, .
\end{align}
Thus,  we have
\begin{align}
    A_{v_0}^{k}T_{L}^{g} &=\sum_{\substack{h_1, \cdots , h_j, \\ \cdots , h_{n-1}}}{T_{v_{n-1}v_n}^{\bar{h}_{n-1}g}\cdots T_{v_1v_2}^{\bar{h}_1h_2}A_{v_0}^{k}T_{v_0v_1}^{h_1}} =\sum_{\substack{h_1, \cdots , h_j, \\ \cdots , h_{n-1}}}{T_{v_{n-1}v_n}^{\bar{h}_{n-1}g}\cdots T_{v_1v_2}^{\bar{h}_1h_2}T_{v_iv_{i+1}}^{kh_1}A_{v_0}^{k}}\, ,
\end{align}
By replacing $\sum_{h_i}$ with $\sum_{kh_i}$, where $i=1,\cdots,n-1$, we obtain
\begin{align}
    A_{v_0}^{k}T_{L}^{g} &=\left[ \sum_{\substack{\left( kh_1 \right) , \cdots , \left( kh_j \right) , \\ \cdots , \left( kh_{n-1} \right)}}{T_{v_{n-1}v_n}^{\overline{kh_{n-1}}kg}\cdots T_{v_iv_{i+1}}^{\overline{kh_i}kh_{i+1}}\cdots T_{v_0v_1}^{kh_1}} \right] A_{v_0}^{k} =T_{L}^{kg}A_{v_0}^{k}\, ,
\end{align}
thus proving Eq.~(\ref{eq_32}). Similarly, we have
\begin{align}
    A_{v_n}^{k}T_{L}^{g}&=\sum_{h_1, \cdots , h_j, \cdots , h_{n-1}}{A_{v_n}^{k}T_{v_{n-1}v_n}^{\bar{h}_{n-1}g}\cdots T_{v_0v_1}^{h_1}} =\sum_{h_1, \cdots , h_j, \cdots , h_{n-1}}{T_{v_{n-1}v_n}^{\bar{h}_{n-1}g\bar{k}}\cdots T_{v_0v_1}^{h_1}A_{v_n}^{k}}\nonumber \\
    &=T_{L}^{g\bar{k}}A_{v_n}^{k}\, ,
\end{align}
thus proving Eq.~(\ref{eq_33}). Since the vertex term $A_{v}=\frac{1}{|G|}\sum_{k}{A_{v}^{k}}$,  we conclude that $T_{L}^{g}$ does not commute with $A_{v_0}$ and $A_{v_n}$:
\begin{align}
    A_{v_0}T_{L}^{g}&=\frac{1}{|G|}\sum_{k}{A_{v_0}^{k}T_{L}^{g}}=\frac{1}{|G|}\sum_{k}{T_{L}^{kg}A_{v_0}^{k}}\ne T_{L}^{g}A_{v_0}\, , \\
    A_{v_n}T_{L}^{g}&=\frac{1}{|G|}\sum_{k}{A_{v_n}^{k}T_{L}^{g}}=\frac{1}{|G|}\sum_{k}{T_{L}^{g\bar{k}}A_{v_n}^{k}}\ne T_{L}^{g}A_{v_n}\, .
\end{align}

\subsection{Proof of Eqs.~(\ref{eq_39}), ~(\ref{eq_40}) and~(\ref{eq_41})}\label{ap2.3}    
     We consider Eq.~(\ref{eq_39}) first. Using Eq.~(\ref{eq_particle}),  we have
    \begin{align}
    A_{v=\partial _0L}^{g}\ket{R;i, j} &=A_{v=\partial _0L}^{g}W_L\left( R;i, j \right) \ket{\mathrm{GS}} =A_{v=\partial _0L}^{g}\sum_{h\in G}{\Gamma _{ij}^{R^{\ast}}\left( h \right) T_{L}^{h}}\ket{\mathrm{GS}}\, .
\end{align}
    Using Eq.~(\ref{eq_32}),  we obtain
    \begin{align}
    A_{v=\partial _0L}^{g}\ket{R;i, j} &=\sum_{h\in G}{\Gamma _{ij}^{R^{\ast}}\left( h \right) T_{L}^{gh}}A_{v=\partial _0L}^{g}\ket{\mathrm{GS}} =\sum_{h\in G}{\Gamma _{ij}^{R^{\ast}}\left( h \right) T_{L}^{gh}}\ket{\mathrm{GS}} \, ,
\end{align}
    where we absorb the $A_{v=\partial _0L}^{g}$ into the ground state in the last step. By writing the matrix element $\Gamma _{ij}^{R^{\ast}}\left( h \right)$ as $\sum_k{\Gamma _{ik}^{R^{\ast}}\left( \bar{g} \right) \Gamma _{kj}^{R^{\ast}}\left( gh \right)}$,  we can finally prove Eq.~(\ref{eq_39}):
    \begin{align}
    A_{v=\partial _0L}^{g}\ket{R;i, j} &=\sum_k{\Gamma _{ik}^{R^{\ast}}\left( \bar{g} \right)}\sum_{h\in G}{\Gamma _{kj}^{R^{\ast}}\left( gh \right) T_{L}^{gh}\ket{\mathrm{GS}}} =\sum_k{\Gamma _{ik}^{R^{\ast}}\left( \bar{g} \right) W_L\left( R;k, j \right)}\ket{\mathrm{GS}} \nonumber \\
    &=\sum_k{\Gamma _{ki}^{R}\left( g \right)}\ket{R;k, j} \, ,
\end{align}
    where we have used $\Gamma _{ik}^{R^{\ast}}\left( \bar{g} \right) =\Gamma _{ki}^{R}\left( g \right) $. Similarly,  we prove Eq.~(\ref{eq_40}) as follows:
    \begin{align}
    A_{v=\partial _1L}^{g}\ket{R;i, j} &=A_{v=\partial _1L}^{g}\sum_{h\in G}{\Gamma _{ij}^{R^{\ast}}\left( h \right) T_{L}^{h}}\ket{\mathrm{GS}} =\sum_{h\in G}{\Gamma _{ij}^{R^{\ast}}\left( h \right) T_{L}^{h\bar{g}}}A_{v=\partial _1L}^{g}\ket{\mathrm{GS}}  \nonumber \\
    &=\sum_k{\sum_{h\in G}{\Gamma _{kj}^{R^{\ast}}\left( g \right) \Gamma _{ik}^{R^{\ast}}\left( h\bar{g} \right) T_{L}^{h\bar{g}}}}\ket{\mathrm{GS}} =\sum_k{\Gamma _{kj}^{R^{\ast}}\left( g \right) W_L\left( R;i, k \right)}\ket{\mathrm{GS}}  \nonumber \\
    &=\sum_k{\Gamma _{kj}^{R^{\ast}}\left( g \right)}\ket{R;i, k} \, .
\end{align}
    As for Eq.~(\ref{eq_41}),  notice that the plaquette operator $B_{p, s}^{h}$ is a projector,  while the $T_{L}^{h}$ term does not change any group element. Thus,  the plaquette operator $B_{p, s}^{h}$ automatically commutes with the string operator $W_L\left( R;i, j \right)$ for all $\left(p, s\right)$. We can evaluate the plaquette operator $B_{p, s}^{h}$ on the ground state to obtain
    \begin{align}
    B_{p, s}^{h}\ket{R;i, j}&=B_{p, s}^{h}W_L\left( R;i, j \right) \ket{\mathrm{GS}}=W_L\left( R;i, j \right) \delta _{h, e}\ket{\mathrm{GS}}=\delta _{h, e}\ket{R;i, j}\, .
\end{align}

\subsection{Proof of Eq.~(\ref{eq_move_particle})}\label{ap2.4}
    By expanding $W_{L_2}\left( R;k, j \right) $ and $W_{L_1}\left( R;i, k \right)$ as linear combinations of $T$-operators as shown in Eq.~(\ref{eq_particle}), we have
    \begin{align}
    \sum_k{W_{L_2}\left( R;k, j \right) W_{L_1}\left( R;i, k \right)} &=\sum_k{\sum_{g, h\in G}{\Gamma _{kj}^{R^{\ast}}\left( h \right) T_{L_2}^{h}\Gamma _{ik}^{R^{\ast}}\left( g \right) T_{L_1}^{g}}}\nonumber \\
    &=\sum_{g, h}{\Gamma _{ij}^{R^{\ast}}\left( gh \right) T_{L_2}^{h}T_{L_1}^{g}} =\sum_m{\sum_g{\Gamma _{ij}^{R^{\ast}}\left( m \right) T_{L_2}^{\bar{g}m}T_{L_1}^{g}}}\, .
\end{align}
    Using the connecting rule in Eq.~(\ref{eq_T_connect}) for $T$-operators,  we have
    \begin{align}
    \sum_m{\sum_g{\Gamma _{ij}^{R^{\ast}}\left( m \right) T_{L_2}^{\bar{g}m}T_{L_1}^{g}}}&=\sum_m{\Gamma _{ij}^{R^{\ast}}\left( m \right) T_{L_1\cup L_2}^{m}} =W_{L=L_1\cup L_2}\left( R ;i, j \right)\, .
\end{align}
    Thus,  we obtain the connecting rule of $W_{L=L_1\cup L_2}\left( R ;i, j \right)$:
    \begin{align}
    W_{L_1\cup L_2}\left( R;i, j \right) &=\sum_k{W_{L_2}\left( R;k, j \right) W_{L_1}\left( R;i, k \right)}\, .
\end{align}

\subsection{The commutation relations between $B_p$ and $L_{M}^{c}$}\label{ap2.5}
    When $p$ lies in the bulk of the membrane $M$,  we have $\left[ B_p, L_{M}^{c} \right] =0$. For simplicity,  we consider the plaquette $v_0v_1v_1^{\prime}v_0^{\prime}$ shown in Fig.~\ref{fig_connect1_L}. If the edges $v_0v_0^{\prime}$, $v_0v_1$, and $v_1v_1^{\prime}$ are assigned group elements $g_1$, $g_2$, and $g_3$, respectively, then we label this configuration as $\ket{g_1, g_2, g_3}$. Before the action of the operator $L_{M_2}^{h}$,  the ordered product on the path $v_0^{\prime}v_0v_1v_1^{\prime}$ is $\bar{g}_1g_2g_3$. After the action of the operator $L_{M_2}^{h}$,  the configuration becomes
    \begin{align}
    L_{M_2}^{h}\ket{g_1, g_2, g_3}&=\sum_{g_0}{L_{v_1v_{1}^{\prime}}^{\bar{g}_0hg_0}T_{v_0v_1}^{g_0}L_{v_0v_{0}^{\prime}}^{h}}\ket{g_1, g_2, g_3}=\sum_{g_0}\delta_{g_0, g_2}\ket{hg_1, g_0, \bar{g}_0hg_0g_3}\nonumber \\
    &=\ket{hg_1, g_2, \bar{g}_2hg_2g_3}\, .
\end{align}
    Now,  the ordered product becomes
    \begin{align}
    \overline{hg_1}g_2\bar{g}_2hg_2g_3=\bar{g}_1\bar{h}g_2\bar{g}_2hg_2g_3=\bar{g}_1g_2g_3\, .
\end{align}
Since the operator $L_{M_2}^{h}$ does not change the ordered product on the path $v_0^{\prime}v_0v_1v_1^{\prime}$,  it automatically commutes with the plaquette term $B_{p=v_0v_1v_1^{\prime}v_0^{\prime}}$. For an arbitrary plaquette $p$ lying in the bulk of a membrane $M$,  we can similarly prove that $\left[ B_p, L_{M}^{c} \right] =0$.

\subsection{The commutation relations between $A_{v}$ and $L_{M}^{c}$}\label{ap2.6}
    When $v\ne v_0$ and $c\notin Z(G)$,  where $L_{M}^{c}$ is constructed by applying $L_c^{\pm}$ to $v_0v_0^{\prime}$ first,  we have $\left[ A_{v}, L_{M}^{c} \right] =0$. From Eq.~(\ref{eq_L_connect1}),  we can see that any $L_{M}^{c}$ has the form
    \begin{align}
    L_{M}^{c}=\sum{\cdots \left( \sum_g{L_{vv^{\prime}}^{\bar{g}cg}T_{v_0v}^{g}} \right) \cdots L_{v_0v_{0}^{\prime}}^{c}}\, , \label{eq_B22}
\end{align}
    where $\cdots$ represents some terms that commute with $A_{v}^h$ because they do not act on edges that contain vertex $v$. $T_{v_0v}^{g}$ acts on an arbitrary string with endpoints denoted by $v_0$ and $v$. Thus,  we only need to consider the term $\sum_g{L_{vv^{\prime}}^{\bar{g}cg}T_{v_0v}^{g}}$. Denote an arbitrary state on the edge $vv^{\prime}$ by $\ket{k}$ and suppose that the arrow of $vv^{\prime}$ points from $v$ to $v^{\prime}$. Then we have
    \begin{align}
    L_{vv^{\prime}}^{c}A_{v}^{h}\ket{k}&=\ket{chk}=\ket{h\bar{h}chk}=A_{v}^{h}L_{vv^{\prime}}^{\bar{h}ch}\ket{k}\, , \\
    L_{vv^{\prime}}^{c}A_{v^{\prime}}^{h}\ket{k}&=\ket{ck\bar{h}}=A_{v^{\prime}}^{h}L_{vv^{\prime}}^{c}\ket{k}\, .
\end{align}
    Thus,  we conclude that $L_{vv^{\prime}}^{c}A_{v}^{h}=A_{v}^{h}L_{vv^{\prime}}^{\bar{h}ch}$ and $\left[L_{vv^{\prime}}^{c}, A_{v^{\prime}}^{h}\right]=0$,  where $v$ lives on the direct part of the membrane $M$. By using Eq.~(\ref{eq_33}),  we verify that $A_{v}^{h}$ commutes with  $\sum_g{L_{vv^{\prime}}^{\bar{g}cg}T_{v_0v}^{g}}$ as follows:
    \begin{align}
    \sum_{g\in G}{L_{vv^{\prime}}^{\bar{g}cg}T_{v_0v}^{g}}A_{v}^{h}&=\sum_{g\in G}{L_{vv^{\prime}}^{\bar{g}cg}A_{v}^{h}T_{v_0v}^{gh}} =A_{v}^{h}\sum_{g\in G}{L_{vv^{\prime}}^{\bar{h}\bar{g}cgh}T_{v_0v}^{gh}}\, .
\end{align}
    By replacing $\sum_{g}$ with $\sum_{gh}$, we obtain
    \begin{align}
    \sum_{g\in G}{L_{vv^{\prime}}^{\bar{g}cg}T_{v_0v}^{g}}A_{v}^{h}&=A_{v}^{h}\sum_{gh\in G}{L_{vv^{\prime}}^{\bar{h}\bar{g}cgh}T_{v_0v}^{gh}} =A_{v}^{h}\sum_{m\in G}{L_{vv^{\prime}}^{\bar{m}cm}T_{v_0v}^{m}}\, .
\end{align}
Thus,  we have $\left[ A_{v}, L_{M}^{c} \right] =0$,  where $v\ne v_0$.

\subsection{Proof of Eqs.~(\ref{eq_av}), ~(\ref{eq_bp}), ~(\ref{eq_av1}) and~(\ref{eq_bp1})}\label{ap2.7}
    First, consider the case in which $v_0v$ in Eq.~(\ref{eq_B22}) is an edge. Then $L_{M}^{c}$ can be written as
    \begin{align}
    L_{M}^{c}=\sum{\cdots \left( \sum_{g_0}{L_{vv^{\prime}}^{\bar{g}_0cg_0} T_{v_0v}^{g_0} } \right) L_{v_0v_{0}^{\prime}}^{c} }\, .
\end{align}
    By using Eq.~(\ref{eq_32}) and $A_{v_0}^{g}L_{v_0v_{0}^{\prime}}^{c}=L_{v_0v_{0}^{\prime}}^{gc\bar{g}}A_{v_0}^{g}$,  we have
    \begin{align}
    A_{v_0}^{g}L_{M}^{c}&=\sum{\cdots \left( \sum_{g_0}{L_{vv^{\prime}}^{\bar{g}_0cg_0} A_{v_0}^{g}T_{v_0v}^{g_0} } \right) L_{v_0v_{0}^{\prime}}^{c} } =\sum{\cdots \left( \sum_{g_0}{L_{vv^{\prime}}^{\bar{g}_0cg_0} T_{v_0v}^{gg_0} } \right) L_{v_0v_{0}^{\prime}}^{gc\bar{g}} }A_{v_0}^{g}\nonumber \\
    &=\left[\sum{\cdots \left( \sum_{gg_0}{L_{vv^{\prime}}^{\bar{g}_0\bar{g}gc\bar{g}gg_0} T_{v_0v}^{gg_0} } \right) L_{v_0v_{0}^{\prime}}^{gc\bar{g}} }\right]A_{v_0}^{g} =L_{M}^{gc\bar{g}}A_{v_0}^{g}\, .\label{eq_B25}
\end{align}

    \begin{figure}
	\centering
	\includegraphics[scale=2.6, keepaspectratio,max width=\linewidth]{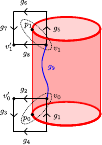}
	\caption{\textbf{Two sites $s_0=\left(v_0, p_0\right)$ and $s_1=\left(v_1, p_1\right)$ on the boundaries of $M$.}  $v_0v_1$ denotes a string that starts and ends at $v_0$ and $v_1$. The ordered product of elements along this string is labeled by $g_9$. 
	}
	\label{fig_BL}
    \end{figure}
    As shown in Fig.~\ref{fig_BL},  we consider a cylindrical membrane $M$ and construct the operator $L_{M}^{c}$ as 
    \begin{align}
    L_{M}^{c}=\sum{ \left( \sum_{g_0}{L_{v_1v_1^{\prime}}^{\bar{g}_0cg_0} T_{v_0v_1}^{g_0} } \right) \cdots L_{v_0v_{0}^{\prime}}^{c} }\, ,
\end{align}
    where the plaquette $p_0$ ($p_1$) and the vertex $v_0$ ($v_1$) form a site $s_0$ ($s_1$). Before we apply the operator $L_{M}^{c}$,  an arbitrary configuration on plaquette $p_0$ is labeled by $\ket{g_1, g_2, g_3, g_4}$, while an arbitrary configuration on plaquette $p_1$ and the ordered product on the string with endpoints $v_0$ and $v_1$ are labeled by $\ket{g_5, g_6, g_7, g_8, g_9}$. Since $L_{M}^{c}$ acts on $\ket{g_1, g_2, g_3, g_4}$ as $L_{v_0v_{0}^{\prime}}^{c}$,  we consider applying $B_{p_0, s_0}^{h}$ and $L_{M}^{c}$ to $\ket{g_1, g_2, g_3, g_4}$ as
    \begin{align}
    B_{p_0, s_0}^{h}L_{M}^{c}\ket{g_1, g_2, g_3, g_4 } &=B_{p_0, s_0}^{h}L_{v_0v_{0}^{\prime}}^{c}\ket{g_1, g_2, g_3, g_4 } =B_{p_0, s_0}^{h}\ket{g_1, cg_2, g_3, g_4 } \nonumber \\
    &=\delta _{h, cg_2g_3\bar{g}_4g_1}\ket{g_1, cg_2, g_3, g_4 } \, .
\end{align}
    Notice that
    \begin{align}
    L_{M}^{c}B_{p_0, s_0}^{\bar{c}h}\ket{g_1, g_2, g_3, g_4 } &=L_{M}^{c}\delta _{\bar{c}h, g_2g_3\bar{g}_4g_1}\ket{g_1, g_2, g_3, g_4 } \nonumber \\
    &=\delta _{\bar{c}h, g_2g_3\bar{g}_4g_1}L_{v_0v_{0}^{\prime}}^{c}\ket{g_1, g_2, g_3, g_4 } =\delta _{\bar{c}h, g_2g_3\bar{g}_4g_1}\ket{g_1, cg_2, g_3, g_4 } \nonumber \\
    &=\delta _{h, cg_2g_3\bar{g}_4g_1}\ket{g_1, cg_2, g_3, g_4 } \, ,
\end{align}
    we conclude that
    \begin{align}
    B_{p_0, s_0}^{h}L_{M}^{c}=L_{M}^{c}B_{p_0, s_0}^{\bar{c}h}\, .\label{eq_B29}
\end{align}
    
    Now we consider applying $B_{p_1, s_1}^{h}$ and $L_{M}^{c}T_{v_0v_1}^{g}$ to $\ket{g_5, g_6, g_7, g_8, g_9}$. Since $L_{M}^{c}$ acts on $\ket{g_5, g_6, g_7, g_8, g_9}$ as $\sum_{g_0}{L_{v_1v_{1}^{\prime}}^{\bar{g}_0cg_0}T_{v_0v_1}^{g_0}}$, we have
    \begin{align}
    B_{p_1, s_1}^{h}L_{M}^{c}T_{v_0v_1}^{g}\ket{g_5, g_6, g_7, g_8, g_9} &=B_{p_1, s_1}^{h}\sum_{g_0}{L_{v_1v_{1}^{\prime}}^{\bar{g}_0cg_0}T_{v_0v_1}^{g_0}}T_{v_0v_1}^{g}\ket{g_5, g_6, g_7, g_8, g_9}\, .
\end{align}
    Note that $T$-operators are projectors, which leads to $T_{v_0v_1}^{g_0}T_{v_0v_1}^{g}=\delta_{g_0,g}T_{v_0v_1}^{g}$. Thus, the above equation can be written as
    \begin{align}
        &B_{p_1, s_1}^{h}L_{M}^{c}T_{v_0v_1}^{g}\ket{g_5, g_6, g_7, g_8, g_9}\nonumber
        \\
        &=B_{p_1, s_1}^{h}\sum_{g_0}{\delta _{g_0, g}\delta _{g_9, g}}\ket{g_5, g_6, g_7, \bar{g}_0cg_0g_8, g_0} \nonumber
        \\
        &=\delta _{g_9, g}B_{p_1, s_1}^{h}\ket{g_5, g_6, g_7, \bar{g}_9cg_9g_8, g_9}\nonumber
        \\
        &=\delta _{g_9, g}\delta _{h, g_5g_6g_7\overline{\bar{g}_9cg_9g_8}}\ket{g_5, g_6, g_7, \bar{g}_9cg_9g_8, g_9}\, .
    \end{align}

    Similarly, when $B_{p_1, s_1}^{h\bar{g}cg}$ acts on the state $\ket{g_5, g_6, g_7, g_8, g_9}$ first, we have
    \begin{align}
    L_{M}^{c}T_{v_0v_1}^{g}B_{p_1, s_1}^{h\bar{g}cg}\ket{g_5, g_6, g_7, g_8, g_9} &=L_{M}^{c}\delta _{g_9, g}\delta _{h\bar{g}cg, g_5g_6g_7\bar{g}_8}\ket{g_5, g_6, g_7, g_8, g_9} \nonumber \\
    &=\delta _{g_9, g}\delta _{h\bar{g}cg, g_5g_6g_7\bar{g}_8}\sum_{g_0}{L_{v_1v_{1}^{\prime}}^{\bar{g}_0cg_0}T_{v_0v_1}^{g_0}}\ket{g_5, g_6, g_7, g_8, g_9}\nonumber \\
    &=\delta _{g_9, g}\delta _{h\bar{g}cg, g_5g_6g_7\bar{g}_8}\ket{g_5, g_6, g_7, \bar{g}_9cg_9g_8, g_9}\nonumber \\
    &=\delta _{g_9, g}\delta _{h, g_5g_6g_7\overline{\bar{g}_9cg_9g_8}}\ket{g_5, g_6, g_7, \bar{g}_9cg_9g_8, g_9}\, ,
\end{align}
    we conclude that
    \begin{align}
    B_{p_1, s_1}^{h}L_{M}^{c}T_{v_0v_1}^{g}=L_{M}^{c}T_{v_0v_1}^{g}B_{p_1, s_1}^{h\bar{g}cg}\, .\label{eq_B32}
\end{align}
    
    By using Eqs.~(\ref{eq_32}) and~(\ref{eq_B25}),  we prove Eq.~(\ref{eq_av}) as follows:
    \begin{align}
        &A_{v_0}^{g}W_M\left( C, R;c, j;c^{\prime}, j^{\prime} \right) \ket{\mathrm{GS}} \nonumber
        \\
        &=\sum_{h\in Z_r}{\Gamma _{jj^{\prime}}^{R^{\ast}}\left( h \right) A_{v_0}^{g}L_{M}^{c}T_{P}^{q_ch\bar{q}_{c^{\prime}}}}\ket{\mathrm{GS}} \nonumber
        \\
        &=\sum_{h\in Z_r}{\Gamma _{jj^{\prime}}^{R^{\ast}}\left( h \right) L_{M}^{gc\bar{g}}T_{P}^{gq_ch\bar{q}_{c^{\prime}}}}A_{v_0}^{g}\ket{\mathrm{GS}}\nonumber 
        \\
        &=\sum_{h\in Z_r}{\Gamma _{jj^{\prime}}^{R^{\ast}}\left( h \right) L_{M}^{gc\bar{g}}T_{P}^{gq_ch\bar{q}_{c^{\prime}}}}\ket{\mathrm{GS}} \, .
    \end{align}
    For any $g\in G$,  we have
    \begin{align}
    \left( \bar{q}_{gc\bar{g}}gq_c \right) r\left( \overline{\bar{q}_{gc\bar{g}}gq_c} \right) &=\left(\bar{q}_{gc\bar{g}}gq_c \right) r\left( \bar{q}_{c}\bar{g}q_{gc\bar{g}} \right)=\bar{q}_{gc\bar{g}}gc\bar{g}q_{gc\bar{g}}=r\, ,
\end{align}
    where we have used 
    \begin{align}
    q_cr\bar{q}_{c}=c\,  \Longrightarrow \,  r=\bar{q}_{c}cq_c, \, \, \,  \forall q_c\in G, \, \forall c\in C_r \, .
\end{align}
    Thus, we conclude that $\bar{q}_{gc\bar{g}}gq_c\in Z_r$. Since $h\in Z_r$,  we have $\bar{q}_{gc\bar{g}}gq_ch\in Z_r$. By writing $T_{P}^{gq_ch\bar{q}_{c^{\prime}}}$ as $T_{P}^{q_{gc\bar{g}}\bar{q}_{gc\bar{g}}gq_ch\bar{q}_{c^{\prime}}}$ and expanding $\Gamma _{jj^{\prime}}^{R^{\ast}}\left( h \right) $ as
    \begin{align}
    \Gamma _{jj^{\prime}}^{R^{\ast}}\left( h \right) =\sum_i{\Gamma _{ji}^{R^{\ast}}\left( \bar{q}_c\bar{g}q_{gc\bar{g}} \right) \Gamma _{ij^{\prime}}^{R^{\ast}}\left( \bar{q}_{gc\bar{g}}gq_ch \right)}\, ,
\end{align}
    we have
    \begin{align}
    \sum_{h\in Z_r}{\Gamma _{jj^{\prime}}^{R^{\ast}}\left( h \right) L_{M}^{gc\bar{g}}T_{P}^{gq_ch\bar{q}_{c^{\prime}}}}\ket{\mathrm{GS}} &=\sum_{h\in Z_r}{\sum_i{\Gamma _{ji}^{R^{\ast}}\left( \bar{q}_c\bar{g}q_{gc\bar{g}} \right) \Gamma _{ij^{\prime}}^{R^{\ast}}\left( \bar{q}_{gc\bar{g}}gq_ch \right)}}\nonumber \\
    \times {L_{M}^{gc\bar{g}}T_{P}^{q_{gc\bar{g}}\bar{q}_{gc\bar{g}}gq_ch\bar{q}_{c^{\prime}}}}\ket{\mathrm{GS}} &=\sum_i{\Gamma _{ji}^{R^{\ast}}\left( \bar{q}_c\bar{g}q_{gc\bar{g}} \right)}\sum_{h^{\prime}\in Z_r}{\Gamma _{ij^{\prime}}^{R^{\ast}}\left( h^{\prime} \right) L_{M}^{gc\bar{g}}T_{P}^{q_{gc\bar{g}}h^{\prime}\bar{q}_{c^{\prime}}}}\ket{\mathrm{GS}} \nonumber \\
    &=\sum_i{\Gamma _{ji}^{R^{\ast}}\left( \bar{q}_c\bar{g}q_{gc\bar{g}} \right)}W_M\left( C, R;gc\bar{g}, i;c^{\prime}, j^{\prime} \right) \ket{\mathrm{GS}} \, ,
\end{align}
    where $h^{\prime}$ is defined as $\bar{q}_{gc\bar{g}}gq_ch$. Using $\Gamma _{ji}^{R^{\ast}}\left( \bar{q}_c\bar{g}q_{gc\bar{g}} \right) =\Gamma _{ij}^{R}\left( \bar{q}_{gc\bar{g}}gq_c \right) $, we obtain
    \begin{align}
    A_{v_0}^{g}W_M\left( C, R;c, j;c^{\prime}, j^{\prime} \right) \ket{\mathrm{GS}} &=\sum_i{\Gamma _{ij}^{R}\left( \bar{q}_{gc\bar{g}}gq_c \right)}W_M\left( C, R;gc\bar{g}, i;c^{\prime}, j^{\prime} \right) \ket{\mathrm{GS}} \, .
\end{align}
    
    Now,  by using Eq.~(\ref{eq_B29}),  we prove Eq.~(\ref{eq_bp}) as follows:
    \begin{align}
    B_{p_0, s_0}^{h}W_M\left( C, R;c, j;c^{\prime}, j^{\prime} \right) \ket{\mathrm{GS}} &=\sum_{g\in Z_r}{\Gamma _{jj^{\prime}}^{R^{\ast}}\left( g \right) B_{p_0, s_0}^{h}L_{M}^{c}T_{P}^{q_cg\bar{q}_{c^{\prime}}}}\ket{\mathrm{GS}}\nonumber \\
    &=\sum_{g\in Z_r}{\Gamma _{jj^{\prime}}^{R^{\ast}}\left( g \right) L_{M}^{c}T_{P}^{q_cg\bar{q}_{c^{\prime}}}}B_{p_0, s_0}^{\bar{c}h}\ket{\mathrm{GS}}\, .
\end{align}
    All plaquettes on a ground state satisfy the zero-flux condition, i.e., we have $B_{p_0, s_0}^{\bar{c}h}\ket{\mathrm{GS}}=\delta_{\bar{c}h,e}\ket{\mathrm{GS}}=\delta_{c,h}\ket{\mathrm{GS}}$. Thus, we conclude that
    \begin{align}
    B_{p_0, s_0}^{h}W_M\left( C, R;c, j;c^{\prime}, j^{\prime} \right) \ket{\mathrm{GS}} &=\delta _{c, h}W_M\left( C, R;c, j;c^{\prime}, j^{\prime} \right) \ket{\mathrm{GS}}\, .
\end{align}

    Similarly,  we use Eq.~(\ref{eq_33}) to prove Eq.~(\ref{eq_av1}):
    \begin{align}
        &A_{v_1}^{g}W_M\left( C, R;c, j;c^{\prime}, j^{\prime} \right) \ket{\mathrm{GS}}\nonumber
        \\
        &=\sum_{h\in Z_r}{\Gamma _{jj^{\prime}}^{R^{\ast}}\left( h \right) A_{v_1}^{g}L_{M}^{c}T_{P}^{q_ch\bar{q}_{c^{\prime}}}}\ket{\mathrm{GS}}\nonumber
        \\
        &=\sum_{h\in Z_r}{\Gamma _{jj^{\prime}}^{R^{\ast}}\left( h \right) L_{M}^{c}T_{P}^{q_ch\bar{q}_{c^{\prime}}\bar{g}}}A_{v_1}^{g}\ket{\mathrm{GS}}\nonumber
        \\
        &=\sum_{h\in Z_r}{\Gamma _{jj^{\prime}}^{R^{\ast}}\left( h \right) L_{M}^{c}T_{P}^{q_ch\bar{q}_{c^{\prime}}\bar{g}}}\ket{\mathrm{GS}}\, . 
    \end{align}
    By writing $\Gamma _{jj^{\prime}}^{R^{\ast}}\left( h \right)$ as \[
\sum_i{\Gamma _{ji}^{R^{\ast}}\left( h\bar{q}_{c^{\prime}}\bar{g}q_{gc^{\prime}\bar{g}} \right) \Gamma _{ij^{\prime}}^{R^{\ast}}\left( \bar{q}_{gc^{\prime}\bar{g}}gq_{c^{\prime}} \right)}
\] and replacing $\sum_{h\in Z_r}$ with $\sum_{h\bar{q}_{c^{\prime}}\bar{g}q_{gc^{\prime}\bar{g}}\in Z_r}$, we have
    \begin{align}
    A_{v_1}^{g}W_M\left( C, R;c, j;c^{\prime}, j^{\prime} \right) \ket{\mathrm{GS}} &=\sum_{h\bar{q}_{c^{\prime}}\bar{g}q_{gc^{\prime}\bar{g}}\in Z_r}{\sum_i{\Gamma _{ji}^{R^{\ast}}\left( h\bar{q}_{c^{\prime}}\bar{g}q_{gc^{\prime}\bar{g}} \right) \Gamma _{ij^{\prime}}^{R^{\ast}}\left( \bar{q}_{gc^{\prime}\bar{g}}gq_{c^{\prime}} \right)}}\nonumber \\
    \times L_{M}^{c}T_{P}^{q_ch\bar{q}_{c^{\prime}}\bar{g}q_{gc^{\prime}\bar{g}}\bar{q}_{gc^{\prime}\bar{g}}}\ket{\mathrm{GS}} &=\sum_i{\Gamma _{ij^{\prime}}^{R^{\ast}}\left( \bar{q}_{gc^{\prime}\bar{g}}gq_{c^{\prime}} \right)}W_M\left( C, R;c, j;gc^{\prime}\bar{g}, i \right) \ket{\mathrm{GS}}\, .
\end{align}

    As for Eq.~(\ref{eq_bp1}),  we use Eq.~(\ref{eq_B32}) to obtain that
    \begin{align}
    B_{p_1, s_1}^{h}W_M\left( C, R;c, j;c^{\prime}, j^{\prime} \right) \ket{\mathrm{GS}} &=\sum_{g\in Z_r}{\Gamma _{jj^{\prime}}^{R^{\ast}}\left( g \right) B_{p_1, s_1}^{h}L_{M}^{c}T_{P}^{q_cg\bar{q}_{c^{\prime}}}}\ket{\mathrm{GS}}\nonumber \\
    &=\sum_{g\in Z_r}{\Gamma _{jj^{\prime}}^{R^{\ast}}\left( g \right) L_{M}^{c}T_{P}^{q_cg\bar{q}_{c^{\prime}}}}B_{p_1, s_1}^{h\overline{q_cg\bar{q}_{c^{\prime}}}cq_cg\bar{q}_{c^{\prime}}}\ket{\mathrm{GS}}\nonumber \\
    &=\sum_{g\in Z_r}{\Gamma _{jj^{\prime}}^{R^{\ast}}\left( g \right) L_{M}^{c}T_{P}^{q_cg\bar{q}_{c^{\prime}}}}\delta _{h\overline{q_cg\bar{q}_{c^{\prime}}}cq_cg\bar{q}_{c^{\prime}}, e}\ket{\mathrm{GS}}\, .
\end{align}
    Using $c=q_cr\bar{q}_c$ and $g\in Z_r$, we have
    \begin{align}
    \delta _{hq_{c^{\prime}}\bar{g}\bar{q}_ccq_cg\bar{q}_{c^{\prime}}, e}&=\delta _{hq_{c^{\prime}}\bar{g}rg\bar{q}_{c^{\prime}}, e} =\delta _{hq_{c^{\prime}}r\bar{q}_{c^{\prime}}, e}=\delta _{hc^{\prime}, e}\, .
\end{align}
    Thus,  we conclude that
    \begin{align}
    B_{p_1, s_1}^{h}W_M\left( C, R;c, j;c^{\prime}, j^{\prime} \right) \ket{\mathrm{GS}} &=\delta _{h, \bar{c}^{\prime}}W_M\left( C, R;c, j;c^{\prime}, j^{\prime} \right) \ket{\mathrm{GS}}\, .
\end{align}

\section{Details of shrinking rules in the $\mathbb{D}_3$ quantum double model}\label{ap4}
In this appendix,  we provide technical details about computing shrinking rules in the $\mathbb{D}_3$ quantum double model.
In the following discussion, we omit $C_e$ for particles and omit the $c$ and $j$ labels when they have only one possible value. For $\left[ C_r, \omega \right]$,  the operators are given by
\begin{align}
    W_M\left( C_r, \omega ;r, r \right) &=\sum_{g\in Z_r}{\Gamma ^{\omega^{\ast}}\left( g \right) L_{M}^{r}T_{P}^{g}} =L_{M}^{r}T_{P}^{e}+\omega^2 L_{M}^{r}T_{P}^{r}+\omega L_{M}^{r}T_{P}^{r^2}\, , \\
    W_M\left( C_r, \omega ;r^2, r \right) &=\sum_{g\in Z_r}{\Gamma ^{\omega^{\ast}}\left( g \right) L_{M}^{r^2}T_{P}^{tg}} =L_{M}^{r^2}T_{P}^{t}+\omega^2 L_{M}^{r^2}T_{P}^{tr}+\omega L_{M}^{r^2}T_{P}^{tr^2}\, ,
\end{align}
where $\omega =\exp \left( \frac{2\pi \mathrm{i}}{3} \right)$ and $\omega^{\ast}=\omega^2$. Note that only the first degree of freedom,  i.e.,  $c$ in Eq.~(\ref{eq_loop}),  is related to the local space $V_{\left( C, R \right)}$ on the boundary $\partial_0M$. Thus,  we need to vary the first degree of freedom over all allowed values. Here we fix the second degree of freedom to be $c^{\prime}=r$. One can also consider $c^{\prime}=r^2$,  which does not affect the final result. After the shrinking process,  we have
\begin{align}
    \mathcal{S} \left( W_M\left( C_r, \omega ;r, r \right) \right) &=T_{P}^{e}+\omega^2 T_{P}^{r}+\omega T_{P}^{r^2} =W_P\left( B;1, 1 \right) \, , \\
    \mathcal{S} \left( W_M\left( C_r, \omega ;r^2, r \right) \right) &=T_{P}^{t}+\omega^2 T_{P}^{tr}+\omega T_{P}^{tr^2} =W_P\left( B;2, 1 \right) \, ,
\end{align}
where the operators for the particle $\left[B\right]$ are given by 
\begin{align}
    W_P\left( B;1,1 \right) &=\sum_{g\in G}{\Gamma _{11}^{B^{\ast}}\left( g \right) L_{M}^{e}T_{P}^{g}}=T_{P}^{e}+\omega^2 T_{P}^{r}+\omega T_{P}^{r^2}\, , \\
    W_P\left( B;2,2 \right) &=\sum_{g\in G}{\Gamma _{22}^{B^{\ast}}\left( g \right) L_{M}^{e}T_{P}^{g}}=T_{P}^{e}+\omega T_{P}^{r}+\omega^2 T_{P}^{r^2}\, , \\
    W_P\left( B;1,2 \right) &=\sum_{g\in G}{\Gamma _{12}^{B^{\ast}}\left( g \right) L_{M}^{e}T_{P}^{g}}=T_{P}^{t}+\omega T_{P}^{tr}+\omega^2 T_{P}^{tr^2}\, , \\
    W_P\left( B;2,1 \right) &=\sum_{g\in G}{\Gamma _{21}^{B^{\ast}}\left( g \right) L_{M}^{e}T_{P}^{g}}=T_{P}^{t}+\omega^2 T_{P}^{tr}+\omega T_{P}^{tr^2}\, .
\end{align}
The two operators $W_P\left( B;1, 1 \right)$ and $W_P\left( B;2, 1 \right)$ are related to the vectors $\ket{1}_{\left(B\right)}$ and $\ket{2}_{\left(B\right)}$,  which span the space $V_{\left( B \right)}$. Thus,  we have
\begin{align}
    \mathcal{S}\left(V_{\left( C_r, \omega \right)}\right)=V_{\left( B \right)}\, ,
\end{align}
which gives the Abelian shrinking rule 
\begin{align}
    \mathcal{S} \left( \left[ C_r, \omega \right] \right) =\left[ B \right]\, .
\end{align}

For $\left[ C_r, \omega^2 \right]$,  we consider
\begin{align}
    W_M\left( C_r, \omega ^2;r, r \right) &=\sum_{g\in Z_r}{\Gamma ^{\omega }\left( g \right) L_{M}^{r}T_{P}^{g}} =L_{M}^{r}T_{P}^{e}+\omega L_{M}^{r}T_{P}^{r}+\omega^2 L_{M}^{r}T_{P}^{r^2}\, , \\
    W_M\left( C_r, \omega ^2;r^2, r \right) &=\sum_{g\in Z_r}{\Gamma ^{\omega }\left( g \right) L_{M}^{r^2}T_{P}^{tg}} =L_{M}^{r^2}T_{P}^{t}+\omega L_{M}^{r^2}T_{P}^{tr}+\omega^2 L_{M}^{r^2}T_{P}^{tr^2}\, .
\end{align}
After the shrinking process,  we have 
\begin{align}
    \mathcal{S} \left( W_M\left( C_r, \omega ^2;r, r \right) \right) &=T_{P}^{e}+\omega T_{P}^{r}+\omega^2 T_{P}^{r^2} =W_P\left( B;2, 2 \right) \, , \\
    \mathcal{S} \left( W_M\left( C_r, \omega ^2;r^2, r \right) \right) &=T_{P}^{t}+\omega T_{P}^{tr}+\omega^2 T_{P}^{tr^2} =W_P\left( B;1, 2 \right)  \, .
\end{align}
The two operators $W_P\left( B;2, 2 \right)$ and $W_P\left( B;1, 2 \right)$ are also related to the vectors $\ket{2}_{\left(B\right)}$ and $\ket{1}_{\left(B\right)}$. Thus,  we have
\begin{align}
    \mathcal{S}\left(V_{\left( C_r, \omega ^2 \right)}\right)=V_{\left( B \right)}\, ,
\end{align}
which leads to the shrinking rule
\begin{align}
    \mathcal{S} \left( \left[ C_r, \omega ^2 \right] \right) =\left[ B \right]\, .
\end{align} 

For $\left[ C_t, Id \right]$,  consider
\begin{align}
    W_M\left( C_t, Id;t, t \right) &=\sum_{g\in Z_t}{\Gamma ^{Id}\left( g \right) L_{M}^{t}T_{P}^{g}} =L_{M}^{t}T_{P}^{e}+L_{M}^{t}T_{P}^{t}\, , \\
    W_M\left( C_t, Id;tr, t \right) &=\sum_{g\in Z_t}{\Gamma ^{Id}\left( g \right) L_{M}^{tr}T_{P}^{rg}} =L_{M}^{tr}T_{P}^{r}+L_{M}^{tr}T_{P}^{tr^2}\, , \\
    W_M\left( C_t, Id;tr^2, t \right) &=\sum_{g\in Z_t}{\Gamma ^{Id}\left( g \right) L_{M}^{tr^2}T_{P}^{r^2g}} =L_{M}^{tr^2}T_{P}^{r^2}+L_{M}^{tr^2}T_{P}^{tr}\, .
\end{align}
After the shrinking process,  we have 
\begin{align}
    \mathcal{S} \left( W_M\left( C_t, Id;t, t \right) \right) =T_{P}^{e}+T_{P}^{t} &=\frac{1}{3}\left(W_P\left( B;1, 1 \right) +W_P\left( B;1, 2 \right) +W_P\left( B;2, 1 \right) \right.\nonumber \\
    &\left.+W_P\left( B;2, 2 \right) +W_P\left( Id \right)\right) \, , \\
    &\mathcal{S} \left( W_M\left( C_t, Id;tr, t \right) \right) =T_{P}^{r}+T_{P}^{tr^2}\nonumber \\
    &=\frac{1}{3}\left(\omega W_P\left( B;1, 1 \right) +\omega W_P\left( B;1, 2 \right) +\omega^2 W_P\left( B;2, 1 \right)\right.\nonumber \\
    &\left.+\omega^2 W_P\left( B;2, 2 \right) +W_P\left( Id \right)\right) \, , \\
    &\mathcal{S} \left( W_M\left( C_t, Id;tr^2, t \right) \right) =T_{P}^{r^2}+T_{P}^{tr}\nonumber \\
    &=\frac{1}{3}\left(\omega^2 W_P\left( B;1, 1 \right) +\omega^2 W_P\left( B;1, 2 \right) +\omega W_P\left( B;2, 1 \right)\right.\nonumber \\
    &\left.+\omega W_P\left( B;2, 2 \right) +W_P\left( Id \right)\right)  \, .
\end{align}
Thus,  $\mathcal{S}\left(V_{\left( C_t, Id \right)}\right)$ has two invariant subspaces $V_{\left( Id \right)}$ and $ V_{\left( B \right)}$,  which leads to the decomposition
\begin{align}
    \mathcal{S}\left(V_{\left( C_t, Id \right)}\right)=V_{\left( Id \right)}\oplus V_{\left( B \right)}\, ,
\end{align}
and the shrinking rule
\begin{align}
    \mathcal{S} \left( \left[ C_t, Id \right] \right) =\left[ Id \right] \oplus \left[ B \right]\, .
\end{align} 

For $\left[ C_t, - \right]$,  consider
\begin{align}
    W_M\left( C_t, -;t, t \right) &=\sum_{g\in Z_t}{\Gamma ^{-^{\ast}}\left( g \right) L_{M}^{t}T_{P}^{g}} =L_{M}^{t}T_{P}^{e}-L_{M}^{t}T_{P}^{t}\, , \\
    W_M\left( C_t, -;tr, t \right) &=\sum_{g\in Z_t}{\Gamma ^{-^{\ast}}\left( g \right) L_{M}^{tr}T_{P}^{rg}} =L_{M}^{tr}T_{P}^{r}-L_{M}^{tr}T_{P}^{tr^2}\, , \\
    W_M\left( C_t, -;tr^2, t \right) &=\sum_{g\in Z_t}{\Gamma ^{-^{\ast}}\left( g \right) L_{M}^{tr^2}T_{P}^{r^2g}} =L_{M}^{tr^2}T_{P}^{r^2}-L_{M}^{tr^2}T_{P}^{tr}\, .
\end{align}
After the shrinking process,  we have
\begin{align}
	&\mathcal{S} \left( W_M\left( C_t, -;t, t \right) \right) =T_{P}^{e}-T_{P}^{t}\nonumber
	\\
	&=\frac{1}{3}\left(W_P\left( B;1, 1 \right) -W_P\left( B;1, 2 \right) -W_P\left( B;2, 1 \right)\right. \nonumber
    \\
    &\left.+W_P\left( B;2, 2 \right) +W_P\left( A \right)\right) \, , 
	\\[10pt]
	&\mathcal{S} \left( W_M\left( C_t, -;tr, t \right) \right) =T_{P}^{r}-T_{P}^{tr^2}\nonumber
	\\
	&=\frac{1}{3}\left(\omega W_P\left( B;1, 1 \right) -\omega W_P\left( B;1, 2 \right) -\omega^2 W_P\left( B;2, 1 \right)\right. \nonumber
    \\
    &+\left.\omega^2 W_P\left( B;2, 2 \right) +W_P\left( A \right)\right) \, , 
	\\[10pt]
	&\mathcal{S} \left( W_M\left( C_t, -;tr^2, t \right) \right) =T_{P}^{r^2}-T_{P}^{tr}\nonumber
	\\
	&=\frac{1}{3}\left(\omega^2 W_P\left( B;1, 1 \right) -\omega^2 W_P\left( B;1, 2 \right) -\omega W_P\left( B;2, 1 \right)\right. \nonumber
    \\
    &+\left.\omega W_P\left( B;2, 2 \right) +W_P\left( A \right)\right) \, .
\end{align}
The space $\mathcal{S}\left(V_{\left(  C_t, - \right)}\right)$ decomposes as
\begin{align}
    \mathcal{S}\left(V_{\left(  C_t, - \right)}\right)=V_{\left( A \right)}\oplus V_{\left( B \right)}\, ,
\end{align}
which leads to the shrinking rule
\begin{align}
    \mathcal{S} \left( \left[  C_t, - \right] \right) =\left[ A \right] \oplus \left[ B \right]\, .
\end{align} 
All the shrinking rules in the 3D $\mathbb{D}_3$ quantum double model are shown in Table~\ref{tab_shrinking_d3}.

\section{Details of braiding in the 3D quantum double model}\label{ap5}
\subsection{Proof of Eq.~(\ref{eq_operator_O})}\label{ap5.1}
\begin{figure}
	\centering
	\includegraphics[scale=1.9, keepaspectratio,max width=\linewidth]{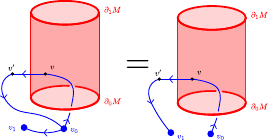}
	\caption{\textbf{Deformation of the string $P$ in Fig.~\ref{fig_particle_loop_braiding}.} We apply the membrane operator to the ground state to create the loops first. Then,  we create the particles by subsequently applying the string operator. The string $P$ in Fig.~\ref{fig_particle_loop_braiding} is composed of a closed part $v_{0}v\cup vv^{\prime}\cup v^{\prime}v_{0}$ and an open part $v_{0}v_{1}$. Since $v^{\prime}v_{0}\cup v_{0}v_{1}$ does not intersect with the membrane $M$,  we can freely deform it to the string $v^{\prime}v_{1}$ as shown in the right-hand diagram. Thus,  the original string $P$ now becomes an open string $\tilde{P}=v_{0}v\cup vv^{\prime}\cup v^{\prime}v_{1}$.}
	\label{fig_particle_loop_braiding_closed}
\end{figure}
In this appendix,  we prove that the particle--loop braiding leads to the operator shown in Eq.~(\ref{eq_operator_O}). Before we present the proof,  we first deform the string $P=v_{0}v\cup vv^{\prime}\cup v^{\prime}v_{0}\cup v_{0}v_{1}$ in Fig.~\ref{fig_particle_loop_braiding} to the string $\tilde{P}=v_{0}v\cup vv^{\prime}\cup v^{\prime}v_{1}$ as shown in Fig.~\ref{fig_particle_loop_braiding_closed}. Since deforming $v^{\prime}v_{0}\cup v_{0}v_{1}$ to $v^{\prime}v_{1}$ does not cut any excitation,  this deformation does not change the state and it simplifies the structure of the string.

Now we introduce a useful operator defined as 
\begin{align}
    \tilde{A}^x_{v_0v_1}=\sum_{y\in G}{T_{v_{0}v_{1}}^{y}A_{v_{0}}^{yx\bar{y}}}\, ,
\end{align}
where $x$ labels a group element. $v_{0}v_{1}$ is a string with all of the arrows on it aligned with the direction that points from $v_0$ to $v_1$. Notice that this operator acts on the ground state as the identity:
\begin{align}
    \sum_{y\in G}{T_{v_{0}v_{1}}^{y}A_{v_{0}}^{yx\bar{y}}}\ket{\mathrm{GS}}&=\sum_{y\in G}{T_{v_{0}v_{1}}^{y}}\ket{\mathrm{GS}}=\ket{\mathrm{GS}}\, ,
\end{align}
where we have used $\sum_{y\in G}{T_{v_{0}v_{1}}^{y}}=1$. To prove Eq.~(\ref{eq_operator_O}),  we need to first prove the following:
\begin{align}
    \left[ \sum_{g, h\in G}{n\left( g \right) T_{v_1v_2}^{x\bar{h}g}T_{v_0v_1}^{h}} \right] \tilde{A}^x_{v_0v_1}=\tilde{A}^x_{v_0v_1}\left[\sum_{g\in G}{n\left( g \right) T_{v_0v_2}^{g}}\right]\, , \label{eq_Ax_commutation}
\end{align}
where $n\left( g \right)\in \mathbb{C}$ is a function of $g$. To prove Eq.~(\ref{eq_Ax_commutation}),  we consider an arbitrary state $\ket{k, l}$,  where $k$ and $l$ label the ordered products of group elements on the path $v_{0}v_{1}$ and $v_{1}v_{2}$ respectively. Then we have
\begin{align}
	&\sum_{g, h\in G}{n\left( g \right) T_{v_1v_2}^{x\bar{h}g}T_{v_0v_1}^{h}}\sum_{y\in G}{T_{v_0v_1}^{y}A_{v_0}^{yx\bar{y}}}\ket{k, l}\nonumber
	\\
    &=\sum_{g, h\in G}{n\left( g \right) T_{v_1v_2}^{x\bar{h}g}T_{v_0v_1}^{h}}\sum_{y\in G}{T_{v_0v_1}^{yx}A_{v_0}^{yx\bar{y}}}\ket{k, l}\nonumber
    \\
	&=\sum_{g, h\in G}{n\left( g \right) T_{v_1v_2}^{x\bar{h}g}T_{v_0v_1}^{h}}\sum_{y\in G}{T_{v_0v_1}^{yx}}\ket{yx\bar{y}k, l}\nonumber
	\\
	&=\sum_{g, h\in G}{n\left( g \right) T_{v_1v_2}^{x\bar{h}g}T_{v_0v_1}^{h}}\sum_{y\in G}{\delta _{yx\bar{y}k, yx}}\ket{yx\bar{y}k, l}\nonumber
	\\
	&=\sum_{g, h\in G}{n\left( g \right) T_{v_1v_2}^{x\bar{h}g}T_{v_0v_1}^{h}}\ket{kx, l}\nonumber
	\\
	&=\sum_{g, h\in G}{n\left( g \right) \delta _{kx, h}\delta _{l, x\bar{h}g}}\ket{kx, l}=n\left( kl \right) \ket{kx, l}\, , 
\end{align}
and
\begin{align}
    \sum_{y\in G}{T_{v_0v_1}^{y}A_{v_0}^{yx\bar{y}}}\sum_{g\in G}{n\left( g \right) T_{v_0v_2}^{g}}\ket{k, l} &=\sum_{y\in G}{T_{v_0v_1}^{yx}A_{v_0}^{yx\bar{y}}}\sum_{g\in G}{n\left( g \right) \delta _{kl, g}}\ket{k, l}\nonumber \\
    &=\sum_{y\in G}{T_{v_0v_1}^{yx}A_{v_0}^{yx\bar{y}}}n\left( kl \right) \ket{k, l} =\sum_{y\in G}{T_{v_0v_1}^{yx}}n\left( kl \right) \ket{yx\bar{y}k, l}\nonumber \\
    &=\sum_{y\in G}{\delta _{yx\bar{y}k, yx}}n\left( kl \right) \ket{yx\bar{y}k, l}=n\left( kl \right) \ket{kx, l}\, ,
\end{align}
which proves Eq.~(\ref{eq_Ax_commutation}). 

Now,  consider the left diagram shown in Fig.~\ref{fig_particle_loop_braiding},  which represents the process:
\begin{align}
    W_P\left( R^{\prime};i, i^{\prime} \right)W_M\left( C, R;c, j;c^{\prime}, j^{\prime} \right)\ket{\mathrm{GS}}\, .\label{eq_E6}
\end{align}
By using the connecting rule in Eq.~(\ref{eq_T_connect}) and Fig.~\ref{fig_particle_loop_braiding_closed},  we can expand the operator for the particle as follows:
\begin{align}
    W_P\left( R^{\prime};i, i^{\prime} \right) &=W_{\tilde{P}}\left( R^{\prime};i, i^{\prime} \right)= \sum_{g\in G}{\Gamma _{ii^{\prime}}^{R^{\prime \ast}}\left( g \right) T_{v_{0}v_{1}}^{g}}  \nonumber \\
    &= \sum_{k_1, k_2, g\in G}{\Gamma _{ii^{\prime}}^{R^{\prime \ast}}\left( g \right) T_{v^{\prime}v_{1}}^{\bar{k}_2g}T_{vv^{\prime}}^{\bar{k}_1k_2}T_{v_{0}v}^{k_1}} \, .
\end{align}
Similarly,  by using the connecting rule in Eq.~(\ref{eq_L_connect2}),  we can expand the operator for the loop as follows:
\begin{align}
    &W_M\left( C, R;c, j;c^{\prime}, j^{\prime} \right)= \sum_{h\in Z_r}{\Gamma _{jj^{\prime}}^{R^{\ast}}\left( h \right) L^{c}_{M}T_{v_{2}v_{3}}^{q_ch\bar{q}_{c^{\prime}}}} \nonumber \\
    &=\sum_{\substack{h\in Z_r\\k_3\in G}}{\Gamma _{jj^{\prime}}^{R^{\ast}}\left( h \right) T_{v_{2}v_{3}}^{q_ch\bar{q}_{c^{\prime}}}     \cdots L^{\bar{k}_3ck_3}_{vv^{\prime}}T^{k_3}_{v_{2}v}\cdots L^{c}_{v_{2}v^{\prime}_{2}}}\, ,
\end{align}
where we omit some $T$- and $L$-operators in the above equation because they commute with the operator $W_P\left( R^{\prime};i, i^{\prime} \right)$. Thus,  Eq.~(\ref{eq_E6}) becomes:
\begin{align}
    W_P\left( R^{\prime};i, i^{\prime} \right)W_M\left( C, R;c, j;c^{\prime}, j^{\prime} \right)\ket{\mathrm{GS}} &=\sum_{\substack{k_1, k_2, k_3, g\in G\\h\in Z_r}}{\Gamma _{ii^{\prime}}^{R^{\prime \ast}}\left( g \right) \Gamma _{jj^{\prime}}^{R^{\ast}}\left( h \right)T_{v^{\prime}v_{1}}^{\bar{k}_2g}T_{vv^{\prime}}^{\bar{k}_1k_2}T_{v_{0}v}^{k_1}}\nonumber \\
    &\times { T_{v_{2}v_{3}}^{q_ch\bar{q}_{c^{\prime}}}     \cdots L^{\bar{k}_3ck_3}_{vv^{\prime}}T^{k_3}_{v_{2}v}\cdots L^{c}_{v_{2}v^{\prime}_{2}}}\ket{\mathrm{GS}}\nonumber \\
    &=\sum_{\substack{k_1, k_2, k_3, g\in G\\h\in Z_r}}{\Gamma _{ii^{\prime}}^{R^{\prime \ast}}\left( g \right) \Gamma _{jj^{\prime}}^{R^{\ast}}\left( h \right)}{ T_{v_{2}v_{3}}^{q_ch\bar{q}_{c^{\prime}}}     \cdots L^{\bar{k}_3ck_3}_{vv^{\prime}}T^{k_3}_{v_{2}v}}\nonumber \\
    \times \cdots L^{c}_{v_{2}v^{\prime}_{2}}T_{v^{\prime}v_{1}}^{\bar{k}_2g}T_{vv^{\prime}}^{\bar{k}_3\bar{c}k_3\bar{k}_1k_2}T_{v_{0}v}^{k_1}\ket{\mathrm{GS}} &=\sum_{\substack{k_1, k_3, g\in G\\h\in Z_r}}{\Gamma _{ii^{\prime}}^{R^{\prime \ast}}\left( g \right) \Gamma _{jj^{\prime}}^{R^{\ast}}\left( h \right)}{ T_{v_{2}v_{3}}^{q_ch\bar{q}_{c^{\prime}}}     \cdots L^{\bar{k}_3ck_3}_{vv^{\prime}}T^{k_3}_{v_{2}v}}\nonumber \\
    &\times \cdots L^{c}_{v_{2}v^{\prime}_{2}}T_{vv_{1}}^{\bar{k}_3\bar{c}k_3\bar{k}_1g}T_{v_{0}v}^{k_1}\ket{\mathrm{GS}}\, , \label{eq_E9}
\end{align}
where we have used Eqs.~(\ref{eq_tl}) and~(\ref{eq_T_connect}). Since the operator $\tilde{A}^x_{v_0v}$ acts on the ground state as the identity,  we can replace $\ket{\mathrm{GS}}$ with 
\begin{align}
    \tilde{A}^{\bar{k}_3\bar{c}k_3}_{v_0v}\ket{\mathrm{GS}}&=\sum_{y\in G}{T_{v_{0}v}^{y}A_{v_{0}}^{y\bar{k}_3\bar{c}k_3\bar{y}}}\ket{\mathrm{GS}} =\sum_{y\in G}{T_{v_{0}v}^{y\bar{k}_3\bar{c}k_3}A_{v_{0}}^{y\bar{k}_3\bar{c}k_3\bar{y}}}\ket{\mathrm{GS}}
\end{align}
in Eq.~(\ref{eq_E9}). By using Eq.~(\ref{eq_Ax_commutation}),  we have
\begin{align}
    &W_P\left( R^{\prime};i, i^{\prime} \right)W_M\left( C, R;c, j;c^{\prime}, j^{\prime} \right)\ket{\mathrm{GS}}\nonumber \\
    &=\sum_{\substack{k_1, k_3, g, y\in G\\h\in Z_r}}{\Gamma _{ii^{\prime}}^{R^{\prime \ast}}\left( g \right) \Gamma _{jj^{\prime}}^{R^{\ast}}\left( h \right)}{ T_{v_{2}v_{3}}^{q_ch\bar{q}_{c^{\prime}}}     \cdots L^{\bar{k}_3ck_3}_{vv^{\prime}}T^{k_3}_{v_{2}v}}\nonumber \\
    &\times \cdots L^{c}_{v_{2}v^{\prime}_{2}}T_{vv_{1}}^{\bar{k}_3\bar{c}k_3\bar{k}_1g}T_{v_{0}v}^{k_1}{T_{v_{0}v}^{y\bar{k}_3\bar{c}k_3}A_{v_{0}}^{y\bar{k}_3\bar{c}k_3\bar{y}}}\ket{\mathrm{GS}}\nonumber \\
    &=\sum_{\substack{k_3, y\in G\\h\in Z_r}}{ \Gamma _{jj^{\prime}}^{R^{\ast}}\left( h \right)}{ T_{v_{2}v_{3}}^{q_ch\bar{q}_{c^{\prime}}}     \cdots L^{\bar{k}_3ck_3}_{vv^{\prime}}T^{k_3}_{v_{2}v}}\cdots L^{c}_{v_{2}v^{\prime}_{2}}\nonumber \\
    &\times T_{v_{0}v}^{y\bar{k}_3\bar{c}k_3}A_{v_{0}}^{y\bar{k}_3\bar{c}k_3\bar{y}}\sum_{g\in G}\Gamma _{ii^{\prime}}^{R^{\prime \ast}}\left( g \right)T_{v_{0}v_{1}}^{g}\ket{\mathrm{GS}}\nonumber \\
    &=\sum_{\substack{k_3, y\in G\\h\in Z_r}}{ \Gamma _{jj^{\prime}}^{R^{\ast}}\left( h \right)}{ T_{v_{2}v_{3}}^{q_ch\bar{q}_{c^{\prime}}}     \cdots L^{\bar{k}_3ck_3}_{vv^{\prime}}T^{k_3}_{v_{2}v}}\cdots L^{c}_{v_{2}v^{\prime}_{2}} \times T_{v_{0}v}^{y\bar{k}_3\bar{c}k_3}A_{v_{0}}^{y\bar{k}_3\bar{c}k_3\bar{y}}W_P\left( R^{\prime};i, i^{\prime} \right)\ket{\mathrm{GS}}\, .\label{eq_E10}
\end{align}
Notice that in Fig.~\ref{fig_particle_loop_braiding},  the paths $v_{0}v$ (blue line),  $v_{2}v$ (black dashed line),  and $v_{0}v_{2}$ (gray dashed line) together form a contractible closed path that does not link with any loop excitation. Thus,  the ordered product of the group elements along this closed path must be the identity $e$. To be more specific,  we have
\begin{align}
    T^{k_3}_{v_{2}v}T_{v_{0}v}^{y\bar{k}_3\bar{c}k_3}=T^{k_3}_{v_{2}v}T_{v_{0}v_{2}}^{y\bar{k}_3\bar{c}}\, .
\end{align}
Since $T^{k_3}_{v_{2}v}$ commutes with $A_{v_{0}}^{y\bar{k}_3\bar{c}k_3\bar{y}}$ in Eq.~(\ref{eq_E10}),  we have
\begin{align}
    &W_P\left( R^{\prime};i, i^{\prime} \right)W_M\left( C, R;c, j;c^{\prime}, j^{\prime} \right)\ket{\mathrm{GS}}\nonumber \\
    &=\sum_{\substack{k_3, y\in G\\h\in Z_r}}{ \Gamma _{jj^{\prime}}^{R^{\ast}}\left( h \right)}{ T_{v_{2}v_{3}}^{q_ch\bar{q}_{c^{\prime}}}     \cdots L^{\bar{k}_3ck_3}_{vv^{\prime}}T^{k_3}_{v_{2}v}}\cdots L^{c}_{v_{2}v^{\prime}_{2}} \times T_{v_{0}v_{2}}^{y\bar{k}_3\bar{c}}A_{v_{0}}^{y\bar{k}_3\bar{c}k_3\bar{y}}W_P\left( R^{\prime};i, i^{\prime} \right)\ket{\mathrm{GS}}\nonumber \\
    &=\sum_{\substack{k_3, y\in G\\h\in Z_r}}{ \Gamma _{jj^{\prime}}^{R^{\ast}}\left( h \right)}{ T_{v_{0}v_{2}}^{q_ch\bar{q}_{c^{\prime}}}     \cdots L^{\bar{k}_3ck_3}_{vv^{\prime}}T^{k_3}_{v_{2}v}}\cdots L^{c}_{v_{2}v^{\prime}_{2}} \times T_{v_{0}v_{2}}^{y\bar{c}}A_{v_{0}}^{y\bar{c}\bar{y}}W_P\left( R^{\prime};i, i^{\prime} \right)\ket{\mathrm{GS}}\nonumber \\
    &=W_M\left( C, R;c, j;c^{\prime}, j^{\prime} \right) \times\sum_{y\in G}T_{v_{0}v_{2}}^{y\bar{c}}A_{v_{0}}^{y\bar{c}\bar{y}}W_P\left( R^{\prime};i, i^{\prime} \right)\ket{\mathrm{GS}}\, .
\end{align}
Since $\sum_{y\in G}T_{v_{0}v_{2}}^{y\bar{c}}A_{v_{0}}^{y\bar{c}\bar{y}}=\sum_{y\in G}T_{v_{0}v_{2}}^{y}A_{v_{0}}^{y\bar{c}\bar{y}}$ and $W_M\left( C, R;c, j;c^{\prime}, j^{\prime} \right)$ do not overlap,  they commute with each other. By using Eq.~(\ref{eq_bp}),  we finally obtain that
\begin{align}
    W_P\left( R^{\prime};i, i^{\prime} \right)W_M\left( C, R;c, j;c^{\prime}, j^{\prime} \right)\ket{\mathrm{GS}} &=\sum_{x, y\in G}\delta_{x, c}T_{v_{0}v_{2}}^{y}A_{v_{0}}^{y\bar{x}\bar{y}}\nonumber \\
    \times W_M\left( C, R;c, j;c^{\prime}, j^{\prime} \right)W_P\left( R^{\prime};i, i^{\prime} \right)\ket{\mathrm{GS}} &=\sum_{x, y\in G}B^{x}_{p_{2}, s_{2}}T_{v_{0}v_{2}}^{y}A_{v_{0}}^{y\bar{x}\bar{y}}\nonumber \\
    \times W_M\left( C, R;c, j;c^{\prime}, j^{\prime} \right)W_P\left( R^{\prime};i, i^{\prime} \right)\ket{\mathrm{GS}} &=\sum_{x, y\in G}T_{v_{0}v_{2}}^{y}B^{x}_{p_{2}, s_{2}}A_{v_{0}}^{y\bar{x}\bar{y}}\nonumber \\
    &\times W_M\left( C, R;c, j;c^{\prime}, j^{\prime} \right)W_{P^{\prime}}\left( R^{\prime};i, i^{\prime} \right)\ket{\mathrm{GS}}\, .
\end{align}

\subsection{Proof of Eq.~(\ref{eq_BR_operator})}\label{ap5.2}
From Appendix~\ref{ap5.1} we can see that each intersection between the string and the membrane leads to an operator $O_{\left(R^{\prime}\right), \left(C, R\right)}^{\mathrm{PL}}$. In Fig.~\ref{fig_BR_braiding},  there are four intersections between the string $P$ and the membranes $M_1$ and $M_2$. Thus,  the operator $O^{\mathrm{BR}}_{\left(R^{\prime}\right), \left(C_1, R_1\right), \left(C_2, R_2\right)}$ in Eq.~(\ref{eq_BR_operator}) is given by
\begin{align}
    O^{\mathrm{BR}}_{\left(R^{\prime}\right), \left(C_1, R_1\right), \left(C_2, R_2\right)} &= O_{\left(R^{\prime}\right), (C_2, R_2)}^{\mathrm{PL}}\left[O_{\left(R^{\prime}\right), (C_1, R_1)}^{\mathrm{PL}}\right]^{-1} \left[O_{\left(R^{\prime}\right), (C_2, R_2)}^{\mathrm{PL}}\right]^{-1}\nonumber \\
    \times O_{\left(R^{\prime}\right), (C_1, R_1)}^{\mathrm{PL}} &=\sum_{\substack{c_1, c_2, g_1, \\g_2, g_3, g_4\in G}}{B_{p_{2}, s_{2}}^{c_2}B_{p_{1}, s_{1}}^{c_1}T_{v_0v_{2}}^{g_4\bar{c}_2}A_{v_0}^{g_4:\bar{c}_2}}\nonumber \\
    &\times T_{v_0v_{1}}^{g_3c_1}A_{v_0}^{g_3:c_1}T_{v_0v_{2}}^{g_2c_2}A_{v_0}^{g_2:c_2}T_{v_0v_{1}}^{g_1\bar{c}_1}A_{v_0}^{g_1:\bar{c}_1}\, ,
\end{align}
where $g:h=gh\bar{g}$ and we have used
\begin{align}
    B_{p_{i}, s_{i}}^{c_i^{\prime}}B_{p_{i}, s_{i}}^{c_i}=B_{p_{i}, s_{i}}^{c_i}\delta_{c_i, c_i^{\prime}}\, ,
\end{align}
where $i=1, 2$. By using Eqs.~(\ref{eq_Av_mutliplication}) and~(\ref{eq_32}),  we have
\begin{align}
	&O^{\mathrm{BR}}_{\left(R^{\prime}\right), \left(C_1, R_1\right), \left(C_2, R_2\right)}\nonumber
    \\
    &=\sum_{\substack{c_1, c_2, g_1, \\g_2, g_3, g_4\in G}}B_{p_{2}, s_{2}}^{c_2}B_{p_{1}, s_{1}}^{c_1}T_{v_0v_{2}}^{g_4\bar{c}_2}T_{v_0v_{1}}^{\left( g_4:\bar{c}_2 \right) g_3c_1}\nonumber
	\\
	&\times T_{v_0v_{2}}^{\left( g_4:\bar{c}_2 \right) \left( g_3:c_1 \right) g_2c_2}T_{v_0v_{1}}^{\left( g_4:\bar{c}_2 \right) \left( g_3:c_1 \right) \left( g_2:c_2 \right) g_1\bar{c}_1}\nonumber
	\\
	&\times A_{v_0}^{g_4:\bar{c}_2}A_{v_0}^{g_3:c_1}A_{v_0}^{g_2:c_2}A_{v_0}^{g_1:\bar{c}_1}\nonumber
	\\
	&=\sum_{\substack{c_1, c_2, g_1, \\g_2, g_3, g_4\in G}}B_{p_{2}, s_{2}}^{c_2}B_{p_{1}, s_{1}}^{c_1}\delta _{g_4\bar{c}_2, \left( g_4:\bar{c}_2 \right) \left( g_3:c_1 \right) g_2c_2}\nonumber
	\\
	&\times\delta _{\left( g_4:\bar{c}_2 \right) g_3c_1, \left( g_4:\bar{c}_2 \right) \left( g_3:c_1 \right) \left( g_2:c_2 \right) g_1\bar{c}_1}\nonumber
	\\
	&\times T_{v_0v_{2}}^{g_4\bar{c}_2}T_{v_0v_{1}}^{\left( g_4:\bar{c}_2 \right) g_3c_1}A_{v_0}^{\left( g_4:\bar{c}_2 \right) \left( g_3:c_1 \right) \left( g_2:c_2 \right) \left( g_1:\bar{c}_1 \right)}\nonumber
	\\
	&=\sum_{\substack{c_1, c_2, \\g_3, g_4\in G}}B_{p_{2}, s_{2}}^{c_2}B_{p_{1}, s_{1}}^{c_1}\nonumber
	\\
	&\times T_{v_0v_{2}}^{g_4\bar{c}_2}T_{v_0v_{1}}^{\left( g_4:\bar{c}_2 \right) g_3c_1}A_{v_0}^{\left( g_4:\bar{c}_2 \right) \left( g_3:\bar{c}_1 \right) \left( g_4:c_2 \right) \left( g_3:c_1 \right)}\, .
\end{align}

\clearpage
\section{Summary of notation}\label{ap_notation_summary}
Table~\ref{tab_notation_summary} collects frequently used symbols, their mathematical or physical meanings, and the equations or sections where they are first introduced.

\begin{longtable}{l@{\hspace{13pt}}l@{\hspace{13pt}}l}
	\caption{Summary of frequently used notation. This table lists the symbols, their mathematical or physical meanings, and the equation or section where they are first introduced.}
	\label{tab_notation_summary}
	\\
	\hline\hline\noalign{\vspace{3pt}}
	\textbf{Notation} & \textbf{Mathematical/Physical Meaning} & \textbf{First introduced} \\ \hline\noalign{\vspace{3pt}}
	\endfirsthead
	
	\multicolumn{3}{c}%
	{{\bfseries Table \thetable{} -- continued from previous page}} \\
	\hline\hline\noalign{\vspace{3pt}}
	\textbf{Notation} & \textbf{Mathematical/Physical Meaning} & \textbf{First introduced} \\ \hline\noalign{\vspace{3pt}}
	\endhead
	
	\hline\hline
	\multicolumn{3}{r}{{Continued on next page...}} \\
	\endfoot
	
	\hline\hline
	\endlastfoot
	\parbox{2.7cm}{\raggedright $\mathsf{a},\mathsf{b},\cdots$} & \parbox{7.1cm}{\raggedright Topological excitations in continuum field theory.} & \parbox{2.7cm}{\raggedright Sec.~\ref{Sec.2.2}, Eq.~(\ref{eq_fusion})}  
	\\ [8pt]
	$N_{\mathsf{c}}^{\mathsf{ab}}$ & \parbox{7.1cm}{\raggedright Fusion coefficient.} & Sec.~\ref{Sec.2.2}, Eq.~(\ref{eq_fusion}) 
	\\ [8pt]
	$S_{\mathsf{b}}^{\mathsf{a}}$ & \parbox{7.1cm}{\raggedright Shrinking coefficient.} & Sec.~\ref{Sec.2.2}, Eq.~(\ref{eq_shrinking}) 
	\\ [8pt]
	$\mathcal{S}$ & \parbox{7.1cm}{\raggedright Shrinking operation.} & Sec.~\ref{Sec.2.2}, Eq.~(\ref{eq_shrinking})
	\\ [8pt]
	$A^i$ & \parbox{7.1cm}{\raggedright $1$-form $U(1)$ gauge field.} & Sec.~\ref{Sec.2.2}, Eq.~(\ref{eq_BF_BdA}) 
	\\ [8pt]
	$B^i$ & \parbox{7.1cm}{\raggedright $2$-form $U(1)$ gauge field.} & Sec.~\ref{Sec.2.2}, Eq.~(\ref{eq_BF_BdA}) 
	\\ [8pt]
	$\mathcal{O}_{\mathsf{a}}$ & \parbox{7.1cm}{\raggedright Gauge-invariant Wilson operator representing a topological excitation $\mathsf{a}$ in continuum field theory.} & Sec.~\ref{Sec.2.2} 
	\\ [8pt]
	$\gamma$ & \parbox{7.1cm}{\raggedright World-line of particles in continuum field theory.} & Sec.~\ref{Sec.2.2} 
	\\ [8pt]
	$\sigma$ & \parbox{7.1cm}{\raggedright World-sheet of loops in continuum field theory.} & Sec.~\ref{Sec.2.2} 
	\\ [8pt]
	$G$ & \parbox{7.1cm}{\raggedright Finite group specifying the quantum double model.} & Sec.~\ref{Sec.3.1} 
	\\ [8pt]
	$g, h, k, \cdots$ & \parbox{7.1cm}{\raggedright Elements of the finite group $G$.} & Sec.~\ref{Sec.3.1} 
	\\ [8pt]
	$\bar{g}$ & \parbox{7.1cm}{\raggedright Inverse of a group element $g$.} & Sec.~\ref{Sec.3.1}  
	\\ [8pt]
	$v$ & \parbox{7.1cm}{\raggedright A vertex on the cubic lattice.} & Sec.~\ref{Sec.3.1}
	\\ [8pt]
	$p$ & \parbox{7.1cm}{\raggedright A plaquette on the cubic lattice.} & Sec.~\ref{Sec.3.1}
	\\ [8pt]
	$s = (v, p)$ & \parbox{7.1cm}{\raggedright A site, consisting of a vertex $v$ and an adjacent plaquette $p$.} & Sec.~\ref{Sec.3.1}
	\\ [8pt]
	$T_g^{\pm}$ & \parbox{7.1cm}{\raggedright Basic projection operators on an edge.} & Sec.~\ref{Sec.3.1}, Eq.~(\ref{eq_t}) 
	\\ [8pt]
	$L_g^{\pm}$ & \parbox{7.1cm}{\raggedright Basic multiplication operators on an edge.} & Sec.~\ref{Sec.3.1}, Eq.~(\ref{eq_l}) 
	\\ [8pt]
	$B_{p,s}^h$ & \parbox{7.1cm}{\raggedright Plaquette operator that projects the ordered product around $p$ to $h$. The ordered product generally depends on the choice of the site $s$.} & Sec.~\ref{Sec.3.1}, Fig.~\ref{fig_Bp}
	\\ [8pt]
	$A_v^g$ & \parbox{7.1cm}{\raggedright A $g$-labeled vertex operator that acts on edges attached to the vertex $v$.} & Sec.~\ref{Sec.3.1}, Fig.~\ref{fig_Av} 
	\\ [8pt]
	$A_v$ & \parbox{7.1cm}{\raggedright Vertex term in the quantum double Hamiltonian; $A_v = \frac{1}{|G|}\sum_{g\in G} A_v^g$.} & Sec.~\ref{Sec.3.1}, Eq.~(\ref{eq_ham}) 
	\\ [8pt]
	$B_p$ & \parbox{7.1cm}{\raggedright Plaquette term in the quantum double Hamiltonian; $B_p = B_{p,s}^e$.} & Sec.~\ref{Sec.3.1}, Eq.~(\ref{eq_ham}) 
	\\ [8pt]
	$L$ or $v_1v_2$ & \parbox{7.1cm}{\raggedright An oriented string on the lattice, consisting of a sequence of edges. We sometimes use $v_1v_2$ to emphasize that the endpoints of the string are vertices $v_1$ and $v_2$.} & Sec.~\ref{Sec.3.2} 
	\\ [11pt]
	$T_L^g$ & \parbox{7.1cm}{\raggedright String operator projecting the ordered product along the string $L$ to $g$.} & Sec.~\ref{Sec.3.2}, Eq.~(\ref{eq_T_connect}) 
	\\ [8pt]
	$\partial_0 L$, $\partial_1 L$ & \parbox{7.1cm}{\raggedright The start and end vertices of an oriented string $L$.} & Sec.~\ref{Sec.3.2} 
	\\ [8pt]
	$\Gamma^{R}_{ij}(g)$ & \parbox{7.1cm}{\raggedright Representation matrix of an irreducible representation $R$ of $G$ (or of a subgroup $Z_r$) evaluated at the group element $g$. $i$ and $j$ are row and column indices.} & Sec.~\ref{Sec.3.2}, Eq.~(\ref{eq_particle}) 
	\\ [13pt]
	$\ast$ & \parbox{7.1cm}{\raggedright Complex conjugation.} & Sec.~\ref{Sec.3.2}, Eq.~(\ref{eq_particle}) 
	\\ [8pt]
	$W_L(R;i,j)$ & \parbox{7.1cm}{\raggedright Creation operator for a particle labeled by an irrep $R$ of $G$.} & Sec.~\ref{Sec.3.2}, Eq.~(\ref{eq_particle}) 
	\\ [8pt]
	$\ket{R;i,j}$ & \parbox{7.1cm}{\raggedright Particle state created by $W_L(R;i,j)$.} & Sec.~\ref{Sec.3.2}, Eq.~(\ref{eq_particle_state}) 
	\\ [8pt]
	$DG$ & \parbox{7.1cm}{\raggedright Quantum double algebra of the finite group $G$.} & Sec.~\ref{Sec.3.2}
	\\ [8pt]
	$M$ & \parbox{7.1cm}{\raggedright A thickened membrane, consisting of dual and direct parts.} & Sec.~\ref{Sec.3.3} 
	\\ [8pt]
	$L_M^h$ & \parbox{7.1cm}{\raggedright Membrane operator that creates a flux $h$.} & Sec.~\ref{Sec.3.3}, Eq.~(\ref{eq_L_connect1}) 
	\\ [8pt]
	$P$ & \parbox{7.1cm}{\raggedright An open string that lives on the direct part of a thickened membrane $M$, ending at the boundaries of $M$.} & Sec.~\ref{Sec.3.3}, Eq.~(\ref{eq_loop}) 
	\\ [8pt]
	$\partial_0 M$, $\partial_1 M$ & \parbox{7.1cm}{\raggedright The two boundaries of a cylindrical membrane $M$.} & Sec.~\ref{Sec.3.3} 
	\\ [8pt]
	$Z_r$ & \parbox{7.1cm}{\raggedright Centralizer subgroup of a representative element $r \in C$, where $C$ is a conjugacy class of $G$.} & Sec.~\ref{Sec.3.3} 
	\\ [8pt]
	$q_c$ & \parbox{7.1cm}{\raggedright A group element satisfying $c = q_c r \bar{q}_c$ for a given representative $r \in C$.} & Sec.~\ref{Sec.3.3}, Eq.~(\ref{eq_loop}) 		
	\\ [8pt]
	$\left(C,R\right)$ & \parbox{7.1cm}{\raggedright A pair of data in the quantum double model. $C$ is a conjugacy class of $G$ and $R$ is an irreducible representation (irrep) of the centralizer $Z_r$.} & Sec.~\ref{Sec.3.3} 
	\\ [8pt]
	$W_M(C,R;c,j;c^{\prime},j^{\prime})$ & \parbox{7.1cm}{\raggedright Creation operator for a loop excitation labeled by a conjugacy class $C$ and an irrep $R$ of the centralizer $Z_r$.} & Sec.~\ref{Sec.3.3}, Eq.~(\ref{eq_loop})
	\\ [8pt]
	$[C,R]$ & \parbox{7.1cm}{\raggedright An excitation in the quantum double model. A particle $[C_e,R^{\prime}]$ is usually abbreviated as $[R^{\prime}]$, where $R^{\prime}$ is an irrep of $G$.} & Sec.~\ref{Sec.4.1}
	\\ [8pt]
	$V_{(C,R)}$ & \parbox{7.1cm}{\raggedright Local Hilbert space of an excitation $[C,R]$.} & Sec.~\ref{Sec.4.1} 
	\\ [8pt]
	$\ket{c,j}_{(C,R)}$ & \parbox{7.1cm}{\raggedright A basis state in the local Hilbert space $V_{(C,R)}$, labeling internal degrees of freedom of a loop.} & Sec.~\ref{Sec.4.1} 
	\\ [8pt]
	$\Pi^{(C,R)}$ & \parbox{7.1cm}{\raggedright Irreducible representation of the quantum double algebra $DG$ labeled by $(C,R)$.} & Sec.~\ref{Sec.4.1}, Eq.~(\ref{eq_action_DG})
	\\ [8pt]
	$\operatorname{Ind}_{Z_r}^G$ & \parbox{7.1cm}{\raggedright Induction functor from a subgroup $Z_r$ to the full group $G$, used to compute shrinking rules.} & Sec.~\ref{Sec.5.1} 
	\\ [8pt]
	$O^{\mathrm{PL}}_{(R^{\prime}),(C,R)}$ & \parbox{7.1cm}{\raggedright Operator resulting from the braiding of a particle $[R^{\prime}]$ around a loop $[C,R]$.} & Sec.~\ref{Sec.8.1}, Eq.~(\ref{eq_operator_O}) 
	\\ [8pt]
	$O^{\mathrm{BR}}_{(R^{\prime}),(C_1,R_1),(C_2,R_2)}$ & \parbox{7.1cm}{\raggedright Operator resulting from the Borromean-Rings braiding of a particle and two loops.} & Sec.~\ref{Sec.8.2}, Eq.~(\ref{eq_BR_operator}) 
	\\ [8pt]
	$\mathsf{P}_{n_1n_2n_3}$ & \parbox{7.1cm}{\raggedright Particle excitation in the $BF+AAB$ theory, carrying $n_i$ units of gauge charge minimally coupled to $A^i$.} & Sec.~\ref{Sec.6.3}
	\\ [8pt]
	$\mathsf{L}_{n_1n_2n_3}^{m_1m_2m_3}$ & \parbox{7.1cm}{\raggedright Loop excitation in the $BF+AAB$ theory, with $n_i$ units of gauge flux minimally coupled to $B^i$ and $m_i$ units of gauge charge decoration minimally coupled to $A^i$.} & Sec.~\ref{Sec.6.3}
	\\ [12pt]
	$\mathsf{1}$ & \parbox{7.1cm}{\raggedright Trivial excitation (vacuum).} & Sec.~\ref{Sec.6.3}
	\\ [8pt]
	$f\left(\mathsf{a}\right)$ & \parbox{7.1cm}{\raggedright An isomorphism mapping an excitation $\mathsf{a}$ in the $BF+AAB$ field theory to an excitation $\left[C,R\right]$ in the $\mathbb{D}_4$ quantum double model.} & Sec.~\ref{Sec.6.4}, Eq.~(\ref{eq_d4aab_fusion}) 
	\\ [8pt]
\end{longtable}

\bibliographystyle{JHEP}

\providecommand{\href}[2]{#2}\begingroup\raggedright\endgroup

\end{document}